\documentclass[instruments,review,accept,pdftex,moreauthors]{Definitions/mdpi} 
\usepackage{mathrsfs}
\usepackage{mathtools}
\usepackage{longtable}
\usepackage{array}
\firstpage{1} 
\pubvolume{1}
\issuenum{1}
\articlenumber{0}
\pubyear{2024}
\copyrightyear{2024}
\externaleditor{Academic Editor: }
\datereceived{31 July 2024} 
\daterevised{30 September 2024} % Comment out if no revised date
\dateaccepted{17 October 2024} 
\datepublished{ } 
\hreflink{https://doi.org/} % If needed use \linebreak
\Title{Transition Edge Sensors: Physics and Applications}

\TitleCitation{Transition Edge Sensors: Physics and Applications}

\Author{Mario De Lucia %MDPI: Please carefully check the accuracy of names and affiliations.
 $^{1,2,*}$\orcidA{}, Paolo Dal Bo $^{2, 3}$\orcidB{}, Eugenia Di Giorgi $^{2,3}$\orcidC{}, Tommaso Lari $^{1,2}$\orcidD{}, Claudio Puglia $^{2}$\orcidE{}  
\mbox{and  Federico Paolucci} $^{1,2}$\orcidH{} }

\AuthorNames{Mario De Lucia, Paolo Dal Bo, Eugenia Di Giorgi, Tommaso Lari, Federico Paolucci, Claudio Puglia, Andrea Tartari}

\AuthorCitation{De Lucia, %MDPI: Please carefully check the accuracy of names.
 M.; Dal Bo, P.; Di Giorgi, E.; Lari, T.; Paolucci, F.; Puglia, C.; Tartari, A.}
\address{%
$^{1}$ \quad Dipartimento di Fisica, Università di Pisa, Largo Bruno Pontecorvo 3, %MDPI: Newly added comma, please confirm. Same below.
 56127 Pisa, PI,
 Italy; 
\\
$^{2}$ \quad INFN Sezione di Pisa, Largo Bruno Pontecorvo 3, 56127 Pisa, PI, Italy; 
$^{3}$ \quad Dipartimento di Fisica, Università di Trento, Via Sommarive 14, 38123 Trento, TN, Italy}
\corres{Correspondence: mario.delucia@unipi.it}

\abstract{Transition Edge Sensors (TESs) are amongst the most sensitive cryogenic detectors and can be easily optimized for the detection of massive particles or photons ranging from X-rays all the way down to millimetre radiation. Furthermore, TESs exhibit unmatched energy resolution while being easily frequency domain multiplexed in arrays of several hundred pixels. Such great performance, along with rather simple and sturdy readout and amplification chains make TESs extremely compelling for applications in {many} fields of scientific endeavour. While the first part of this article is an in-depth discussion on the working principles of Transition Edge Sensors, the remainder of this review article focuses on the applications of Transition Edge Sensors in advanced scientific instrumentation serving as an  accessible and thorough list of possible starting points for more comprehensive literature research. }


\begin{document}

%%%%%%%%%%%%%%%%%%%%%%%%%%%%%%%%%%%%%%%%%%
%\setcounter{section}{-1} %% Remove this when starting to work on the template.
%\section{How to Use this Template}

%The template details the sections that can be used in a manuscript. Note that the order and names of article sections may differ from the requirements of the journal (e.g., the positioning of the Materials and Methods section). Please check the instructions on the authors' page of the journal to verify the correct order and names. For any questions, please contact the editorial office of the journal or support@mdpi.com. For LaTeX-related questions please contact latex@mdpi.com.%\endnote{This is an endnote.} % To use endnotes, please un-comment \printendnotes below (before References). Only journal Laws uses \footnote.

% The order of the section titles is different for some journals. Please refer to the "Instructions for Authors” on the journal homepage.

\section{Introduction}

Advanced scientific instrumentation and up-and-coming technologies require detectors with exceptional performance. Making use of detectors with excellent resolving powers, extremely low noise and limited spurious and dark counts will allow for ambitious results in a large number of scientific applications including, but~not limited to, astronomy, cosmology, particle physics, biophysics, chemistry and quantum computing. Arguably, a~very promising way to achieve such demanding performances is to rely on low-temperature superconducting detectors. At~low temperatures, spurious thermal dark counts become negligible and, if~paired with single photon sensitivity, such detectors can be optimized to be excellent photon counters at all wavelengths of the energy spectrum. Naturally, they can also be deployed as highly sensitive~calorimeters.

As we shall discuss in this review, among~the plethora of superconducting detectors Transition Edge Sensors (TESs) are the most widely used because of their excellent energy resolution, their relatively easy fabrication process and their capability of being frequency domain multiplexed. TESs are extremely sensitive superconducting thermometers and their working principle is surprisingly simple.

For most  materials, as~their temperature decreases, so does their electrical resistivity, but~those that are know{n} to exhibit a superconducting phase-transition feature a sudden drop to zero in resistivity when they are cooled below a certain temperature, known as superconducting critical temperature ($T_C$), specific to each superconductor. 
In the proximity of $T_C$,  the~resistance vs temperature curve can be very steep, leading to a variation of up to several Ohms ($\Omega$) in a range of a few milli-Kelvins (mK). Further details and a reading list on the phenomenology of superconductivity can be found in Section~\ref{sec:Superconductivity}.

In a nutshell, a~TES can be imagined as a superconductor  kept in contact with a bath (reservoir) at temperature  $T_C$ to which a voltage bias is applied. Upon~the striking of a particle which deposits its energy in the TES, it is then temporarily heated up to a temperature $T$ > $T_C$ before cooling down again to the reservoir temperature. In~this process, the~resistance of the TES increases with its temperature and therefore with the energy deposited in the detector itself. In~such a picture, the~steeper the superconducting transition the more sensitive the TES is, in~a trade-off that links the sensitivity to the dynamic range of the detector.

Further details on the working principle of a Transition Edge Sensor and its electro-thermal feedback  can be found in Section~\ref{sec:TES} along with an in-depth discussion on its multiplexing, the~amplification chains, its noise and cross-talk sources.

{In Section~\ref{sec:Applications}, we present the most prominent applications for Transition Edge Sensors in the field of astronomy and astrophysics, while Section~\ref{sec:particle} will discuss the applications of TESs for nuclear, particle  and~astroparticle  physics including dark matter detection.}

The development of TESs for applications as bio-imagers is discussed in  Section~\ref{sec:bio}, while Section~\ref{sec:quantum} describes the applications of Transition Edge Sensors in the field of quantum optics and quantum communications. TES arrays see a wide application as X-ray detectors; their applications as spectroscopes and X-ray imagers are discussed in Section~\ref{sec:X-ray}. Finally, Section~\ref{sec:comparison} contains a brief introduction to the working principles of other superconducting detectors and a discussion on how they compare with~TESs. 

{While the physics and application of TESs have been covered in other review articles and books have been published over the last two decades, they often appear to be either too broad or extremely niche-specific. We would still like to acknowledge their contribution and present them to the interested reader. The~two publications by \citet{enss2005cryogenic} (Ed.) and \citet{enss2008physical} mostly focus on the physics of low temperature micro-calorimeters, \citet{gottardi2021review} present the applications of TESs in the field of astronomy and astroparticle physics, \citet{nucciotti2016use}, \citet{pirro2017advances}, \citet{poda2017low}, and~\citet{poda2021scintillation} describe the use of TESs for applications in nuclear and particle physics, while \citet{koehler2021low} and    \citet{ullom2015review} discuss the use of TESs as X-ray detectors.}
Differently, the~purpose of this review paper is to provide a useful handbook on TESs starting from their working principles and their main figures of merit in order to convey their critical importance in many scientific applications. {In order to facilitate the use of this article, a~table of contents is available at the end of the text.}

\section{Superconductivity}\label{sec:Superconductivity}
Superconductivity is the prototypical example of macroscopic quantum effect, since {some} of the electrons are condensed into a superfluid state extending {through} the entire system. The~condensation occurs for electronic temperatures ($T$) lower than a critical value, known as critical temperature ($T_c$). As \textit{T}
 decreases, the~number of condensed electrons increases until full condensation is reached at $T=0$. Indeed, a superconductor can be represented by means of a two fluids model including normal electrons, known as \emph{quasiparticles} and coupled condensed electrons, known as Cooper Pairs.

\subsection{Phenomenology}
Despite it being usually identified with only zero resistivity, superconductivity is experimentally described by the all following phenomena~\cite{Tinkham2004}.

\textbf{Zero resistivity.}
The DC electrical resistivity of a superconductor shows a steep transition to zero at $T\lesssim T_c$ \cite{Tinkham2004}. Zero resistivity is preserved for injected currents lower than a threshold value, known as \emph{{critical current}}  
($I_c$). At~$T=0$, $R=0$ is true for AC signals up to a critical frequency $\omega_c\approx 3.5 k_B T_c/\hbar$ (with $k_B$ the Boltzmann constant and $\hbar$ the reduced Planck constant), thus indicating a lower limit to excite electrons from the condensed to the normal~state. 

\textbf{Response to magnetic fields.}
A bulk superconductor exhibits the \emph{Meissner effect} \cite{Meissner1933}, {which is the expulsion of an external magnetic field}. To~exclude the magnetic field from its bulk, a~superconductor increases its free energy till reaching the normal state value at the \emph{critical field} ($H_c$) \cite{London1948}, which is maximum at $T=0$ and is fully suppressed at $T=T_c$. Finally, the~magnetic flux passing through a cylindrical superconductor shows a $hc/2e$ (with $h$ being the Planck constant, $c$ the speed of light and $e$ the electron charge) \cite{Doll,Deaver}, where the factor 2 arises from the double charge of a Cooper~pair.

\textbf{Specific heat.}
The specific heat of a superconductor shows a discontinuity with respect to the normal state value at $T_c$. The~{$1.43$} times larger value of the superconductor specific heat is strongly connected to formation of a condensate and its associated condensation energy~\cite{schrieffer1999theory}. In~addition, the~specific heat decreases exponentially with temperature \mbox{($\propto e^{-T_c/T}$)} differently from its linearity in the normal state ($\propto T$); thus, a superconductor shows a smaller specific heat for $T\ll T_c$ \cite{KEESOM1934175}. This behaviour is due to the fact that Cooper pairs (condensed electrons) are not able to store thermal~energy. 

\textbf{Energy gap.}
As already introduced for the AC electrical response of a superconductor, there is a forbidden energy range, known as the \emph{superconducting energy gap} ($\Delta$), between~the superconducting condensate and the normal electrons~\cite{Tinkham2004,schrieffer1999theory}. In~particular, $\Delta$  increases monotonically as  T decreases; thus, the superconducting state becomes more favourable at low temperatures. The~energy gap can be measured through electron tunnel spectroscopy, since single electron transport is allowed only for voltages $V \geq \Delta/e$ \cite{Giaever}. Differently, a~superconductor can absorb radiation of frequency $\nu \geq 2\Delta/h$ \cite{biondi}, since both electrons forming a Cooper pair need energy $\Delta$ to switch to the normal state~\cite{glover}.

\subsection{Brief Introduction to BCS~Theory}
The Bardeen--Cooper--Schrieffer (\emph{BCS}) theory provides the microscopic explanation of the superconductivity~\cite{BCS}. In~particular, BCS describes the superconducting state as a boson condensate living at energies lower than the metallic normal state.
This is possible by assuming that at $T=0$ the fundamental state of a metal (Fermi sea) is unstable for the addition of a weak attractive interaction between a couple of electrons, i.e.,~the Cooper pair~\cite{Cooper1956}. 

To provide a simple explanation of the BCS theory, we can perform a \emph{Gedankenexperiment} {(thought experiment)}: we add two electrons [$\overrightarrow{k_1}$, $E(\overrightarrow{k_1})$] and [$\overrightarrow{k_2}$, $E(\overrightarrow{k_2})$] to the fundamental state of a gas of non-interacting electrons. These electrons occupy previously empty states above the Fermi level ($E_F$). For~the Pauli exclusion principle, the two electrons cannot occupy states with $|\overrightarrow{k}|<k_F$ (with $k_F$ the Fermi wavevector); thus, a weak attractive potential needs to build up. On~the one hand, the~two electrons continuously change their wavevector; on~the other hand, the~total momentum is conserved
\begin{equation}
   \overrightarrow{k_1}+\overrightarrow{k_2}=\overrightarrow{k_1'}+\overrightarrow{k_2'}=\overrightarrow{K},
     \label{Eq:KCoppie}
\end{equation}
\noindent
where $\overrightarrow{K}$ is the momentum of the Cooper pair. Indeed, the~interaction is allowed in the intersection between the spherical cortex of thickness $\hbar \omega_D$ (with $\omega_D$ the Debye frequency) around $\overrightarrow{k_1}$ and $\overrightarrow{k_2}$. The~maximum attractive force is provided by $\overrightarrow{K}=0$; thus, the two electrons forming the Cooper pair have wavevectors of the same absolute value and opposite direction ($\overrightarrow{k_1}=-\overrightarrow{k_2}=\overrightarrow{k}$). The~Schr\"{o}dinger equation for the Cooper pair reads
\begin{equation}
  -\frac{\hbar^2}{2m}\big(\nabla_1^2+\nabla_2^2 \big)\Psi\big(\overrightarrow{r_1},\overrightarrow{r_2} \big) + V\big(\overrightarrow{r_1},\overrightarrow{r_2} \big)\Psi\big(\overrightarrow{r_1},\overrightarrow{r_2} \big)= \big(  \varepsilon+2 E_{F,0} \big)\Psi\big(\overrightarrow{r_1},\overrightarrow{r_2} \big),
     \label{Eq:Schr}
\end{equation}
\noindent
where  $V$ is the interaction potential, $\overrightarrow{r_i}$ is the position of the $i$-th electron (with $i=1,2$), $\varepsilon$ is the energy of the electrons pair with respect to the non-interacting state ($V=0$) and~$E_{F,0}$ is the single-electron Fermi energy (the factor 2 accounts for the coupled electrons). In~the presence of {an} attractive potential, the~general solution of the Schr\"{o}dinger equation  is:
\begin{equation}
 \Psi\big(\overrightarrow{r} \big)= \frac{1}{L^3} \sum_{\substack{
   k}} g\big(\overrightarrow{k}\big)e^{i\overrightarrow{k}\overrightarrow{r}},
     \label{Eq:wave}
\end{equation}
\noindent
where $L$ is the space coordinate, $\overrightarrow{r}=\overrightarrow{r_1}-\overrightarrow{r_2}$ are the real space relative coordinates of the two electrons and~$|g\big(\overrightarrow{k}\big)|^2$ is the probability of finding an electron in $\overrightarrow{k}$ and the other in $-\overrightarrow{k}$. 
By substituting Equation~(\ref{Eq:wave}) in  Equation~(\ref{Eq:Schr}) and assuming an interaction independent {of} $\overrightarrow{k}$ {with a value of} $V_0$ (occurring only in the spherical cortex $\hbar \omega_D$), we can evaluate the energy difference of the electrons  {with} respect to the non-interacting state for weak coupling as
\begin{equation}
 \varepsilon \simeq -2\hbar \omega_D e^{-2/V_0\mathcal{N}(E_{F,0})},
     \label{Eq:En}
\end{equation}
\noindent
where $\mathcal{N}(E_{F,0})$ is the density of states at the Fermi level for 1 spin type. Equation~(\ref{Eq:En}) shows that a bounded 2-electron state of energy lower than the Fermi sea can occur at $T=0$. Thus, the~fundamental state of a non-interacting electron gas is unstable in the presence of a weakly attractive potential. It is possible to demonstrate that this instability is brought to the formation of a large density of Cooper pairs; thus, the fundamental state of lowest energy is the \emph{superconducting state}. Since the superconducting wave function is symmetric, the~Pauli exclusion principle implies that the two electrons forming the Cooper pair have opposite spin. We also note that the~Cooper pairs are bosons, thus occupying the same energy~level. 

The total energy of the fundamental superconducting state ($W_{BCS,0}$) considers all the possible configurations of coupled electrons, the~kinetic energy of the Cooper pair and energy reduction due to interaction.
\begin{equation}
 W_{BCS,0}= \sum_{\substack{k}} \xi_k\bigg( 1-\frac{\xi_k}{E_k} \bigg)-L^3\frac{\Delta^2}{V_0},
     \label{Eq:BCSEn}
\end{equation}
\noindent
where $\xi_k=\hbar^2k^2/2m-E_{F,0}$ is the kinetic energy of {a} Cooper pair,  and~$E_k=\sqrt{\xi_k^2+\Delta^2}$ is the total energy of a Cooper pair. Thus, the~condensation energy of the superconducting state, obtained by subtracting the normal-state energy ($W_{N,0}$) to $ W_{BCS,0}$, reads
\begin{equation}
 \frac{W_{BCS,0}-W_{N,0}}{L^3}=- \frac{1}{2}\mathcal{N}(E_{F,0})\Delta^2.
     \label{Eq:CondEn}
\end{equation}

Thus, the~energy gap provides a measure of the energy reduction of the superconducting state. The~first excited BCS state implies the breakage of a Cooper pair; that is, an energy $2\Delta $ is necessary ($\Delta$ for each electron). An~uncoupled electron, also known as \emph{quasiparticle}, can only occupy states of energy $\Delta$ above the fundamental state. Thus, the~quasiparticle density of states of a superconductor at $T=0$ takes the form
\begin{equation}
 \mathcal{N}_S(E)=\mathcal{N}_N(E_{F,0})\Re\Bigg(\frac{E}{\sqrt{E^2-\Delta^2}}\Bigg),
     \label{Eq:DOSsup}
\end{equation}
\noindent
where $\mathcal{N}_N(E_{F,0})$ is the normal-state density of states and $\Re$ represents the real-part component. The~density of states of a superconductor strongly depends on temperature, since the energy gap obeys  
\begin{equation}
 \frac{1}{V_0\mathcal{N}_N(E_{F,0})}=\int_0^{\hbar\omega_D}\frac{\text{d}\xi}{\sqrt{\xi^2+\Delta^2}}\bigg[1-2f\bigg(\sqrt{\xi^2+\Delta^2}+E_{F,0},T\bigg)\bigg],
     \label{Eq:GAPvsT}
\end{equation}
\noindent
where $f\big(\sqrt{\xi^2+\Delta^2}+E_{F,0},T\big)$ is the temperature-dependent Fermi function. Finally, the~critical temperature of a superconductor is related to the zero-temperature energy gap ($\Delta_0$) through
\begin{equation}
 \Delta_0= 1.764k_BT_c.
     \label{Eq:TC}
\end{equation}

\subsection{Engineering Artificial Superconductors by Proximity~Effect}\label{sec:proxeffect}
\begin{figure}[H]
    \centering
    \includegraphics[width=\textwidth]{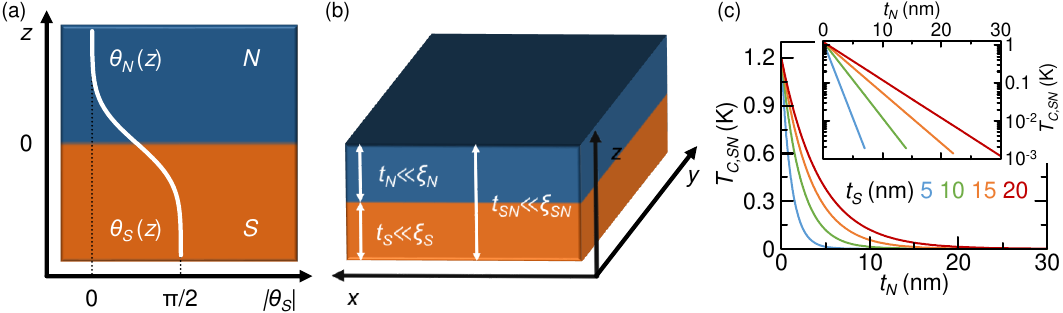}
    \caption{\textbf{Proximity effect and critical temperature engineering.}
    (\textbf{a}) Schematic representation of the $z$-dependence of the parameter $\theta_i$ describing superconductivity in an $SN$ bilayer (with $i=S,N$).
    (\textbf{b})~Schematics of the Cooper limit for an  $SN$ bilayer, where $t_i$ is the thickness of the $i$ element and $\xi_i$ is its coherence length (with $i=S,N,SN$).
    (\textbf{c}) Critical temperature of an $SN$ bilayer versus $t_N$ for different values of $t_S$ with $S$ aluminium and $N$ copper. The~parameters used  for Equation~(\ref{Eq:Usadel}) are as follows: $T_{C,S}=1.2$ K, $\omega_{D,N}=7.98\times 10^{13}$ rad/s the Debye frequency of copper, $\mathcal{N}_S=2.15\times 10^{47}$ J$^{-1}$m$^{-3}$ the density of state at the Fermi level of aluminium, $\mathcal{N}_N=1.56\times 10^{47}$ J$^{-1}$m$^{-3}$ the density of state at the Fermi level of copper and $\mathcal{T}_{SN}=1$ the transmission probability of the $SN$ interface.}
    \label{fig:ProxEff}
\end{figure}

The optimization of TES detectors for different applications requires the development of artificial superconductors with engineered $T_c$. A~rather standard approach exploits the so-called \emph{proximity effect} \cite{DeGennes}: a weakened superconducting wave function penetrates in a normal metal placed in good ohmic contact with a superconductor. As~a consequence, the~$T_c$ of a TES can be adjusted by changing the relative thickness of the superconducting ($S$) and normal metal ($N$) thin films forming a bilayer (see Figure~\ref{fig:ProxEff}a). 

The superconducting properties of an $SN$ bilayer can be predicted by means of the Usadel theory~\cite{Usadel1970}. The~Usadel theory can be applied for dirty  systems~\cite{anderson1959theory,fominov2001superconductive}, that is the diffusive charge transport both in $S$ and $N$. The~superconducting properties of the bilayer are translationally invariant in the $(x,y)$-plane, while they vary along the $z$-axis. By~exploiting a function $\theta(z)$ describing the superconducting state, the~pairing potential obeys   the coupled Usadel equations
\vspace{-6pt}
\begin{adjustwidth}{-\extralength}{0cm}
%\centering %% If there is a figure in wide page, please release command \centering
\begin{equation}
    \begin{cases}
    \frac{\hbar D_j}{2} \partial_z^2 (\theta_j)+iE\sin{\theta_j}-\bigg[ \frac{\hbar}{\tau_{sf,j}}+\frac{\hbar D_j}{2} \bigg( \partial_z(\varphi)+\frac{2e}{\hbar}A_z\bigg)^2 \bigg]\cos{\theta_j}\sin{\theta_j}+\Delta_j(T,z)\cos{\theta_j}=0\\
    \Delta_j(T,z)=\mathcal{N}_S V_{eff}\int_0^{\hbar \omega_D}{\tanh{\big(\frac{E}{2k_BT}\big)}\Im{\big(\sin{\theta_j}\big)}\text{d}E}
     \end{cases}
     \label{Eq:Usadel}
\end{equation}
\end{adjustwidth}
\noindent
where  $D_j$ is the normal-state electron diffusion constant, $E$ is the energy, $\tau_{sf,j}$ is the spin-flip time, $\varphi$ is {t}he superconducting phase, $\Delta (z,T)$ is the temperature ($T$) and $z$ dependent effective superconducting energy gap, $\mathcal{N}_S$ is the density of states at the Fermi level, $V_{eff}$ is the pairing potential  and~$j=S,N$ indicates the superconductor and normal metal side. This version of the Usadel theory does not consider the impact on the proximity effect of the electrical resistance between $N$ and $S$, because~the large contact area provides a negligibly small~contribution. 

The absolute value of the complex function $\theta_j$ obtained by solving the coupled Usadel Equations (Equation (\ref{Eq:Usadel})) ranges from 0 (normal-state) to $\pi/2$ (fully superconducting state). Figure~\ref{fig:ProxEff}a schematically shows the evolution of $|\theta_j|$ with $z$ within the $SN$ bilayer. 
The negligible $SN$ resistance implies the continuity of the $\theta_j(z)$ function, that is \linebreak$\theta_S(z=0)=\theta_N(z=0)$ and $\partial_z\theta_S(z=0)=\partial_z\theta_N(z=0)$, with~$z=0$ being  the interface between $S$ and $N$. Differently, $\theta_j(z)$ would show a discontinuity at the interface for large resistances between the two layers~\cite{Bram}. In~either case, the~superconducting energy gap and, therefore, the~critical temperature vary strongly along the $z$-axis. Consequently, the~$R$ versus $T$ characteristic of the $SN$ bilayer would show an extremely broad transition to zero~resistance. 
\newpage
Sensitive TESs require a sharp transition to the superconducting state; that is,  the $SN$ bilayer is expected to posses a superconducting energy gap constant along the $z$-axis. This is achieved for $SN$ bilayers lying in the \emph{Cooper limit} \cite{Cooper1961} (see Figure~\ref{fig:ProxEff}b), thus fulfilling the following~requirements

\begin{enumerate}
\item The thickness of the superconductor thin film ($t_S$) is lower than its superconducting coherence length (typical distance between the two paired electrons~\cite{Tinkham2004}): $t_S\ll\xi_S$;
\item The thickness of the normal metal layer ($t_N$) is smaller than its coherence length: $t_N\ll\xi_N=\sqrt{\hbar D_N/(2\pi k_B T)}$;
\item The thickness of the $SN$ bilayer ($t_{SN}$) is less than its effective coherence length ($\xi_{SN}$) $t_{SN}\ll\xi_{SN}=\sqrt{\hbar(t_SD_S+t_ND_N)/(t_{SN}\Delta_{SN})}$ (with $\Delta_{SN}$ being the effective bilayer superconducting energy gap).
\end{enumerate}

These conditions ensure that the coherence is fully preserved along the $z$-axis both in the superconducting and normal state{s}. As~a consequence, superconductivity is not affected by the non-ideality (finite charge diffusion, amorphous thin film and defects) of the normal metal thin film and the $SN$ bilayer behaves as a monolithic~superconductor.

The effective critical temperature of an $SN$ bilayer within the Cooper limit can be approximated to~\cite{MARTINIS200023}
\begin{equation}
    T_{c,SN}=T_{c,S}\Bigg[ \bigg( \frac{k_BT_{c,S}}{1.13\hbar \omega_{D,N}}\bigg)^2  + \bigg( \frac{k_BT_{c,S}\lambda_{F,N}^2(\frac{1}{t_N\mathcal{N}_N}+\frac{1}{t_S\mathcal{N}_S})}{2.26 \mathcal{T}_{SN}}\bigg)^2\Bigg]^{t_N\mathcal{N}_N/2t_S\mathcal{N}_S},
\end{equation}
\noindent
where $T_{c,S}$ is the critical temperature of $S$, $\omega_{D,N}$ is the Debye frequency of the normal metal, $\lambda_{F,N}$ is the Fermi wavelength of $N$, $\mathcal{N}_N$ is the density of states at the Fermi level of the normal metal and $\mathcal{T}_{SN}$ is the interface transmission probability. 
Figure~\ref{fig:ProxEff}c shows the dependence of the critical temperature of an Al/Cu bilayer ($T_{C,SN}$) as a function of the thickness of the Cu film ($t_N$) for selected values of the thickness of Al ($t_S$). Interestingly, $T_{C,SN}$ decreases exponentially with $t_N$, as~shown by the inset. 
Furthermore, by~decreasing $t_S$, the critical temperature of the bilayer decreases faster for increasing $t_N$, because~the superconducting wavefunction is \emph{weaker} and the proximitization of $N$ is less~efficient. 

\subsection{Thermal~Properties}\label{sec:thermalproperties}
The sensitivity of a TES strongly depends on the thermal exchange mechanisms between the different elements when power or energy are absorbed.
The various components of a TES lie within the quasi-classical diffusive limit~\cite{Tinkham2004}; that is, their physical dimensions are larger than their Fermi wavelength ($\lambda_F$) and elastic mean free path ($\lambda_{free}$) of electrons. Thus, the~electron energy distribution can be well described by the Fermi function and the thermal transport can be simply described by energy diffusion. Here, we will describe the main thermal exchange mechanisms occurring in a~TES.

The thermistor of a TES is kept at the superconducting-to-normal-state transition, thus showing the thermal properties of a normal metal. Indeed, its charge carrier thermalization is dominated by \emph{electron--electron scattering}, either due to direct Coulomb interaction or impurities~\cite{Giazotto2006}. Within~these conditions, the~electron--electron thermalization length scale reaches tens of $\upmu$m~\cite{Pekola2004}; thus, the electronic temperature ($T_e$) can be considered homogeneous all over the entire thermistor. Differently, the~thermal conductance of a superconductor ($G_{th,S}$) at $T_e\ll T_c$ is exponentially damped by the presence of the energy gap~\cite{Bardeen1959}. Indeed, the~Cooper pairs do not transport heat and the number of unpaired electrons decreases exponentially with temperature~\cite{Tinkham2004}. In~the linear response approximation, $G_{th,S}$ can be expressed as~\cite{Pekola2021}
\begin{equation}
G_{th,S}\simeq G_{th,N}\frac{6}{\pi^2}\bigg( \frac{\Delta_0}{k_B T_e}\bigg)^2 e^{-\Delta_0/k_B T_e} \text{,}
\label{Eq:GthS}
\end{equation}
\noindent
where $G_{th,N}$ is the normal-state thermal~conductance. 

One of the main thermalization channels of the charge carriers i{n} the thermistor is given by the scattering with the film phonons. In~particular, the~\emph{electron--phonon scattering} in a clean normal metal can be calculated by~\cite{Giazotto2006}
\begin{equation}
P_{e\text{-}ph,N}=\Sigma \mathcal{V} \left(T_e^5-T_{ph}^5 \right)\text{,}
\label{e-phpowerN}
\end{equation}
\noindent
where $\Sigma$ is the electron--phonon coupling constant, $\mathcal{V}$ is the volume and $T_{ph}$ is the phonon temperature. We stress that the exponent can be $n = 4.6$ for dirty metals depending on the type of impurity immersed in the metallic matrix. 
Similarly to energy transport, the~electron--phonon thermalization is exponentially damped in the superconducting state ($P_{e\text{-}ph,S}$) with respect to the normal metal value ($P_{e\text{-}ph,N}$). Indeed, within~the linear approximation limit, the~electron--phonon thermalization takes the form~\cite{Bergeret2018}
\begin{equation}
P_{e\text{-}ph,S} \propto P_{e\text{-}ph,N} e^{-\Delta_0/k_B T_e}\text{.}
\label{e-phpowerS}
\end{equation}

In full agreement with thermal conductivity, this behaviour is related to the suppression of the single-electron population in a superconductor with decreasing temperature~\cite{Tinkham2004}. Indeed, the~Cooper pairs do not exchange energy with the crystal, since their scattering with the phonons is~forbidden.

\subsubsection*{Andreev~Mirrors}

Usually, the~electrodes contacting the thermistor are made of a superconductor of critical temperature much larger than the operating temperature of the TES, that is the critical temperature of the thermistor. Indeed, these superconducting leads act as energy filters, the~so-called \emph{Andreev thermal mirrors} \cite{Andreev1964}, thus confining the electron overheating in the thermistor. In~the linear response regime, the~thermal conductance of an $SN$ contact is suppressed with respect to the conventional value given by the Wiedemann--Franz law by a factor
\begin{equation}
\delta G_{th,NS}=\bigg(\frac{p_0}{2\pi}\bigg)^2  \frac{f_0}{\sqrt{\varphi(\eta)k_BT_e/\Delta_0^3}}e^{-\Delta_0/k_B T_e}   \text{,}
\label{EQ:NSGth}
\end{equation}
\noindent
where $p_0$ is the Fermi momentum, $\eta=2H/H_c-1$ is a dimensionless factor representing the impact of an external magnetic field, $\varphi(\eta)=e^{ip_0 \hat{n} \cdot \vec{r}}\eta$ (with $\hat{n}$ being the unit vector and $\vec{r}$ the real space vector) is tabulated, and~$f_0=\int\limits_0^1\cos{\alpha} f(\cos{\alpha})\text{d}\cos{\alpha}\sim1$ (with $\alpha$ being the incidence angle of the electrons at the interface and $f$ is a slowly varying function).
Intuitively, the~strong reduction of the thermal conductance can be understood by considering that the hot electrons in the normal metal find a limited number of accessible quasiparticle states within the superconducting gap to diffuse. Thus, the~perfect Andreev mirror effect occurs only for very low values of the electronic temperature ($T_e\ll T_c$). 
By rising $T_e$, the~Fermi distribution of the hot electrons in $N$ broadens and reaches the quasiparticle peaks in the density of states of the superconductor. Therefore, hot electrons of the normal metal find available states in $S$ and $G_{th,NS}$ approaches the conventional Wiedemann--Franz value. Indeed, the~clean contact between and Al/Cu bilayer and Al lead showed efficient thermal insulation for temperatures up to about 280 mK, that is $\sim 0.22 T_c$ of the aluminium film~\cite{Paolucci2020}. 

\section{Physics of Transition Edge~Sensors}\label{sec:TES}
\begin{figure}[H]
    \centering
    \includegraphics[width=0.9\textwidth]{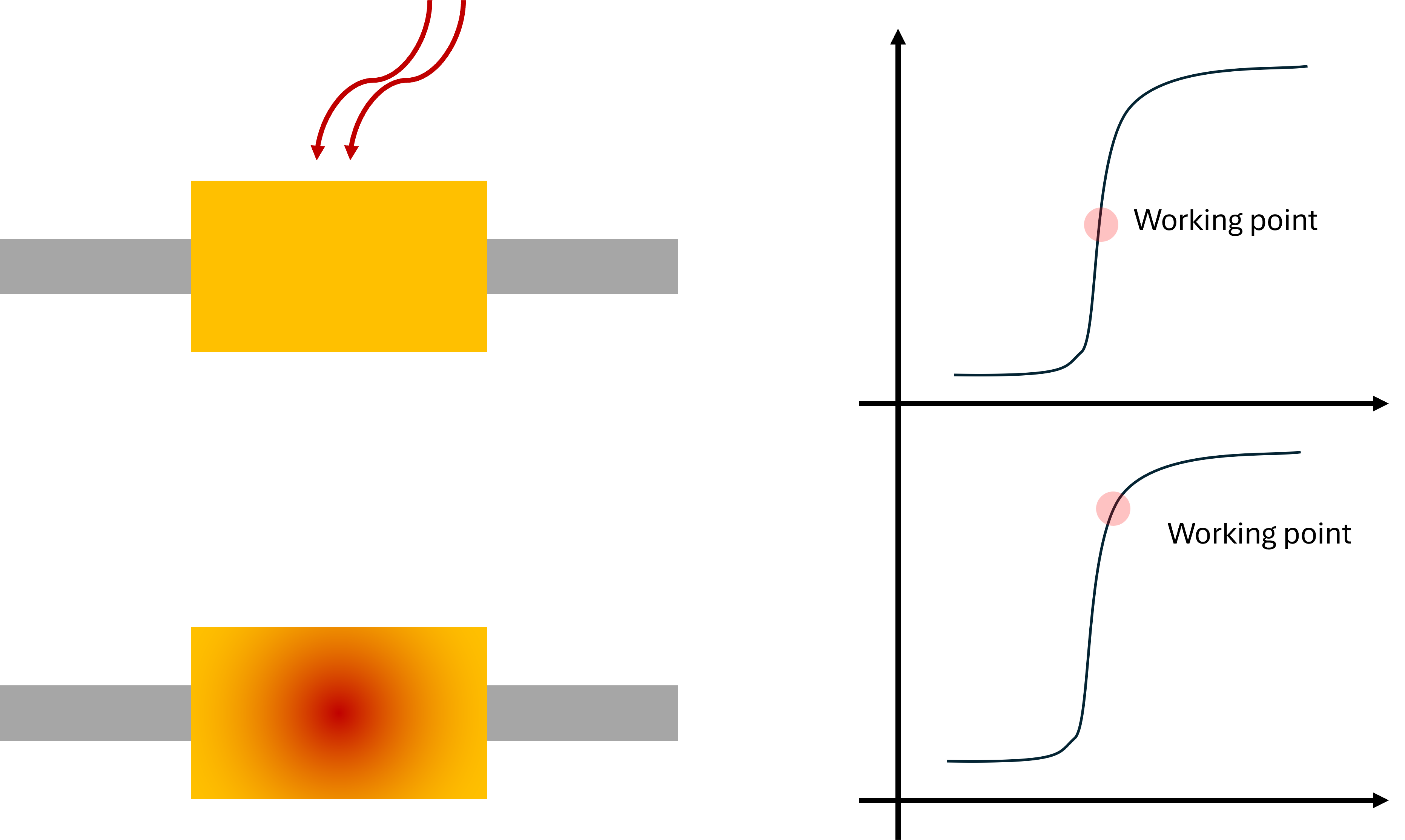}
    \caption{\textbf{{Working principle of a Transition Edge Sensor.}} %MDPI: 1. Please confirm if the explanation of the colors needs to be added in the figure caption.
    An event (e.g., photon/phonon absorption) heats the superconductor, moving its temperature from the ideal working point (center of the normal to superconducting transition).}
    \label{fig:tesfunc}
\end{figure}

The idea of exploiting a superconducting thin film at temperatures close to its critical temperature as a detector has been around since the 1940s, when proposed by \citet{Andrews_TES_1942} as infra-red radiation detectors. In~particular, the~proposed device exploits the steep change in resistance {to detect the energy deposited by a striking particle} in Figure~\ref{fig:tesfunc}.

\begin{figure}[H]
    \centering
    \includegraphics[width=1\linewidth]{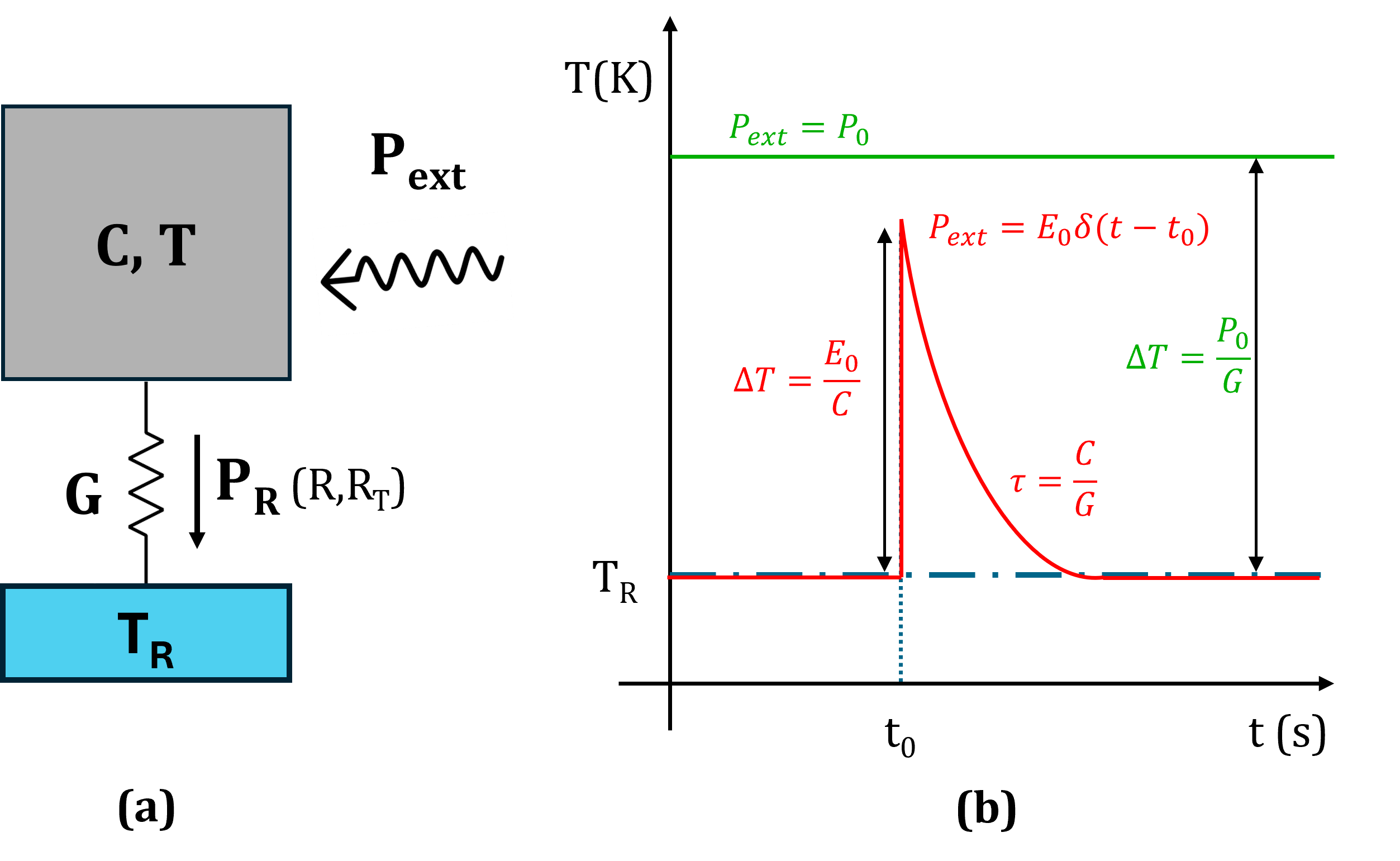}
    \caption{\textbf{{Working principle of a TES not in electro-thermal feedback.}} %MDPI: 1. Please confirm if the explanation of the colors needs to be added in the figure caption.
     (\textbf{a}) The thermal scheme of a TES with  temperature and heat capacity $C$, the~weak thermal link to the reservoir through a thermal conductance $G$ and the external power {P$_{ext}$} and power exchanged to the reservoir {$P_{R}(R,R_{T})$. (\textbf{b})} Solution to Equation~(\ref{eq:bolo}) of the TES as a function of time in the two aforementioned~cases.}
    \label{fig:schemabolometro}
\end{figure}

A Transition Edge Sensor (TES) can be modelled as a thin film with a finite thermal capacity $C$ connected to a reservoir at temperature $T_R$ as shown in Figure~\ref{fig:schemabolometro}. Any energy deposited onto the film by radiation or particles striking on the TES is converted into heat and then dissipated towards the reservoir with a characteristic time which is {inversely} proportional to the thermal conductance $G$. Considerations based on the conservation of energy result in the following equation that describes the heating of the film when irradiated or struck by a particle
\begin{equation}
    P_{ext} = P_{R}(T,T_R) + C \frac{dT}{dt},
    \label{eq:thermal}
\end{equation}
\noindent
where $P_{ext}$ is the power deposited by the particle or the radiation, $P_{R}$ represents the power dissipated towards the thermal reservoir and can be assumed to depend linearly on the temperature difference between the electronic temperature $T_e$ (see Section~\ref{sec:thermalproperties})  referred to simply as T from now on,  and~$T_R$ through the thermal conductance G, assumed constant in the first approximation. Equation~(\ref{eq:thermal}) can be, therefore, rewritten as
\begin{equation}
    P_{ext} = G(T - T_R) + C\frac{dT}{dt},
    \label{eq:bolo}
\end{equation}
which yields two particular solutions in two specific~cases
\begin{enumerate}
     \item $T = T_R + \frac{P_0}{G}$ in case $P_{ext} = P_0$ is a constant
     \item $\delta T = \frac{\delta P_{ext}}{\tau_0}\frac{1}{1-i\omega \tau}$ with $\tau=\frac{C}{G}$ with the obvious meaning of the symbols.
     
     In particular, an~instantaneous deposit of energy $E_0$ results in an instantaneous increase in temperature by $\Delta T =\frac{E_0}{C}$ with an exponential decay with time constant $\tau = \frac{C}{G}$.
\end{enumerate}

\unskip

\subsection{Negative Electro-Thermal~Feedback}%MDPI: there is no ``subsection'' before it, so we changed ``subsubsection'' to ``subsection'', please confirm. Please pay attention to the change of the section order number below.
\begin{figure}[H]
    \centering
    \includegraphics[width=\textwidth]{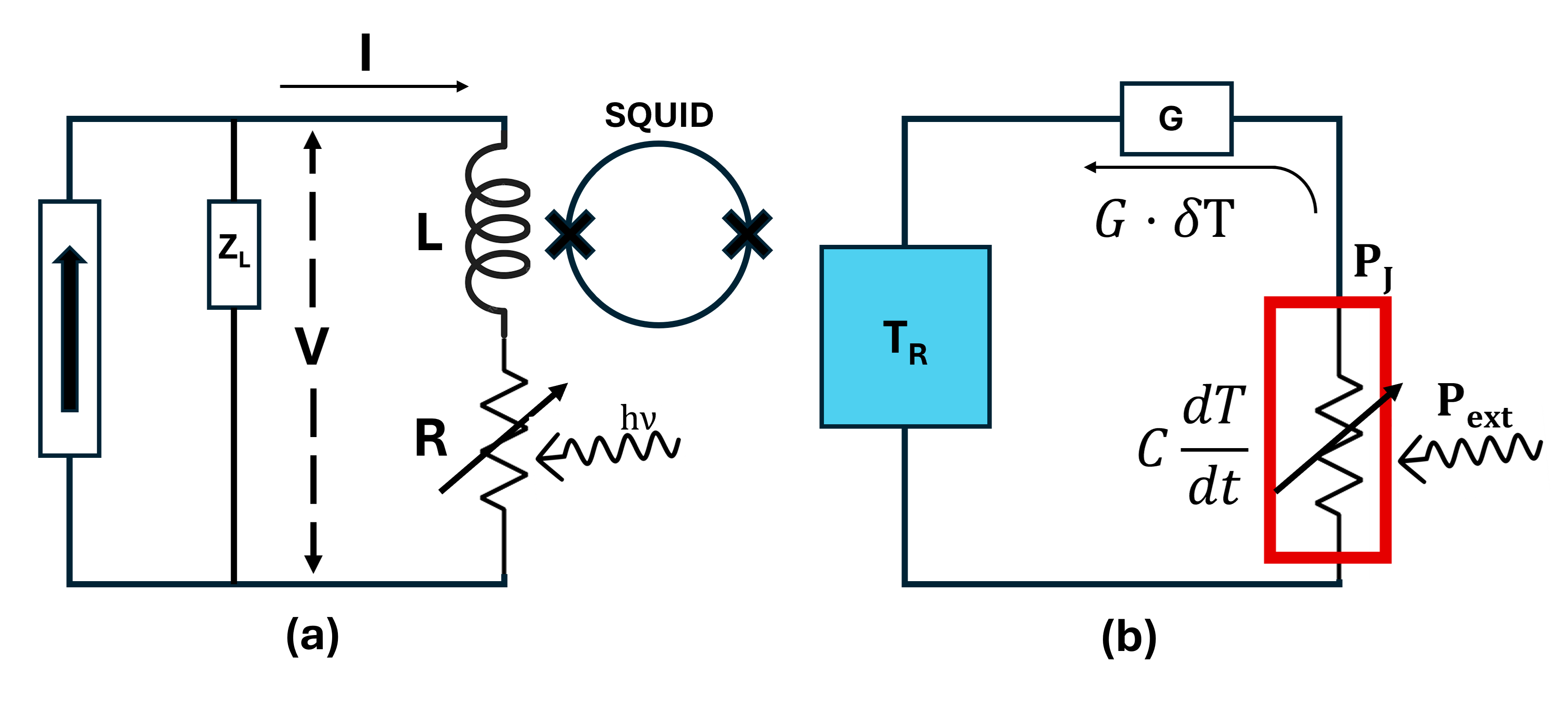}
    \caption{\textbf{{The schematic of a TES biased in negative electro-thermal feedback}}. %MDPI: 1. Please confirm if the explanation of the colors needs to be added in the figure caption.
     (\textbf{a}) Electric scheme of a TES and (\textbf{b}) thermal scheme of a TES. Both schemes contribute to the coupled differential equations in Equation~(\ref{Eq:electrofeedback}). }
    \label{fig:TESscheme}
\end{figure}

Transition Edge Sensors initially faced challenges in adoption in the scientific community mostly due to their complexity and the difficulty in keeping the TES stable in temperature. This issue was  only   solved in 1995 by Kent D. Irwin~\cite{Kent_D_Irwin_1995},  who proposed and successfully established a voltage-biased negative electro-thermal feedback (NETF) of the TES in order to keep its stability in temperature. Such a feedback is typically implemented by adding in parallel to the TES a shunt impedance $Z_{L}$ with $Z_{L} << R \left(T,I\right)$ so as to contribute to the previous thermal equation with an extra Joule component. The~electric scheme of this circuit is shown in Figure~\ref{fig:TESscheme}a, whereas Figure~\ref{fig:TESscheme}b describes its thermal behaviour. Through the effect of the Joule component, the~thermal and electrical equations are paired and thus the scheme takes the name of electro-thermal~feedback.

Two coupled differential equations can be written, one to take into account the thermal behaviour of the superconducting film and the second  to represent the electrical behaviour {of a} TES in its NETF scheme.
\begin{equation}
    \begin{cases}
    C\frac{dT}{dt}=-P_R(T,T_R)+P_J+P_{ext},\\
    L\frac{dI}{dt}=V-I Z_L-I R\left(T,I\right).
     \end{cases}
     \label{Eq:electrofeedback}
\end{equation}
where $C$ is the heat capacity of the TES and absorber subsystem, $T$ is the temperature of the TES, $P_{ext}$ is the signal power, $P_j$ is the power dissipated through the resistance of the TES, {$P_{R}$} is the power exchanged between the TES and the reservoir kept at temperature $T_R$  (thermal feedback) that contributes to cooling down the TES after detection, $L$ is the inductance, $V$ is the Thevenin-equivalent bias voltage that is applied to the TES, $Z_L$ is the shunt impedance and $R\left(T,I\right)$ is the resistance of the TES, which depends on the temperature of the TES as well as the current flowing through~it. 

\begin{figure}[H]
    
    \includegraphics[width=\linewidth]{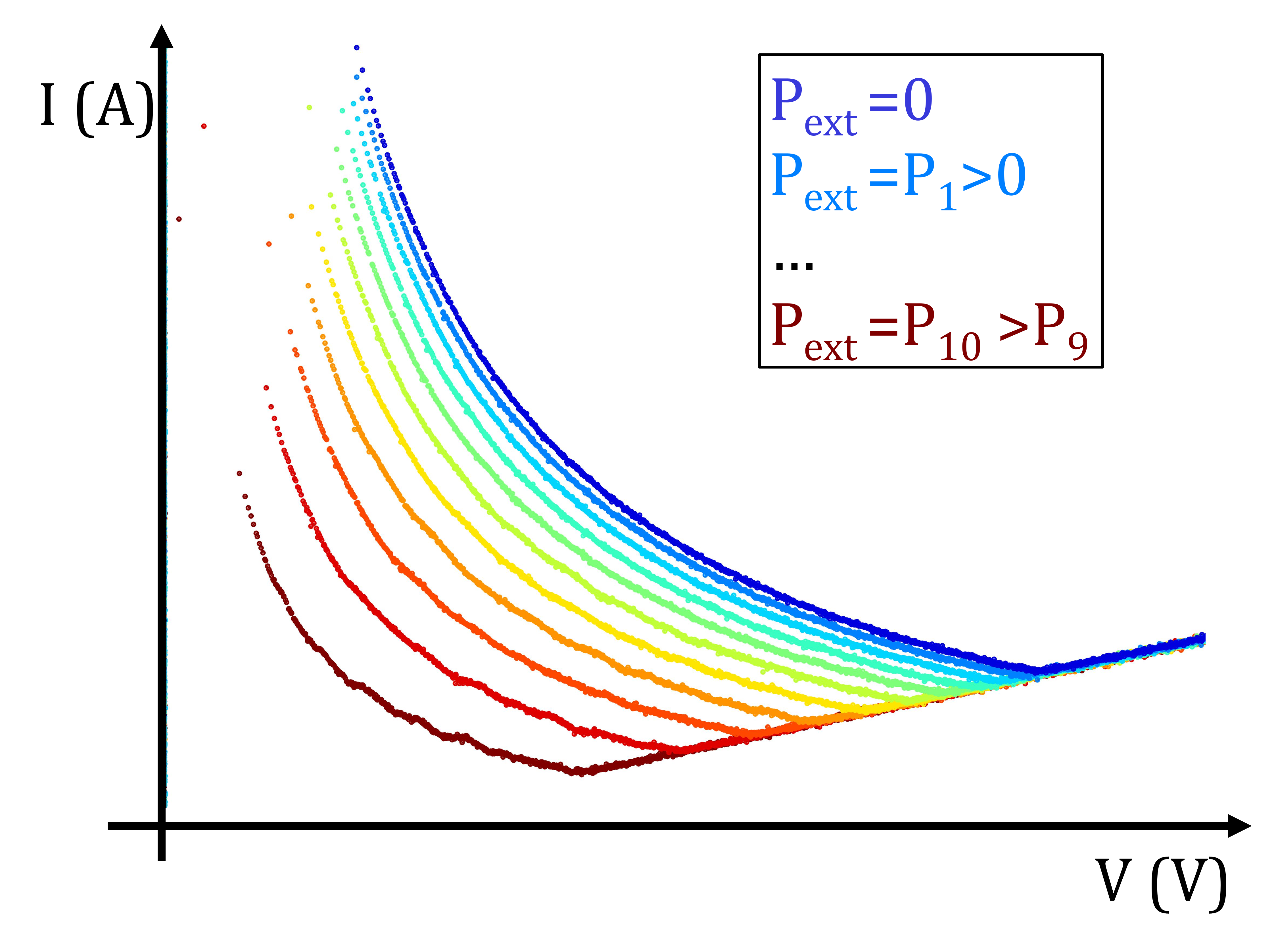}
    \caption{\textbf{Measured I--V curves of a TES}
    The~I--V characteristic of a TES can be derived as the steady-state solution of Equation~(\ref{Eq:electrofeedback}). The~top most blue curve represents the case in which no optical load heats up the TES, whereas all the other curves represent increasing values of $P_{ext}$ as   described in the~legend.}
    \label{fig:I-VTES}
\end{figure}

The I--V  characteristic of the system can be determined by solving the coupled differential equations in their steady state ( $\frac{dT}{dt}=0$, and~$\frac{dI}{dt}=0$ ). {For the sake of simplicity, we will assume that the resistance of the TES does not depend on the current flowing through it but only on its temperature ($\beta$ = $\frac{I}{R}\frac{dR}{dI}$=0). An~in-depth analysis of the $\beta \neq 0$ case is discussed by \citet{irwin2005transition}.}
In the steady state, which describes a condition of dynamic equilibrium between the TES, the~reservoir and the external radiation, we can write
\begin{equation}
    P_{ext}+ P_J = const.
    \label{eq:hyperbole}
\end{equation}
From Equation~(\ref{eq:hyperbole}), we can infer that as long as $P_J$ is constant, the~product $I\cdot V$ is kept constant; hence,  the I--V curves of a TES describe the branch of a hyperbola. This holds true until the TES is driven outside of its superconducting state by the external load or because of the Joule power dissipated through the TES itself. When the TES is driven outside of its transition, it is described by its ohmic behaviour and therefore all the hyperbola branches connect to the same straight line, the slope of which represents the normal-state resistivity of the TES. Assuming that the bias load $Z_L$ is negligible, the~typical I--V curves of a TES are shown in Figure~\ref{fig:I-VTES}a where curves of different colors represent different values of $P_{ext}$.

Once we have discussed the I--V characteristic  curves of the TES in its NETF configuration, it is worth briefly discussing the detection principle and the operation of a TES as a bolometer and as a~calorimeter.

\begin{figure}[H]
  
    \includegraphics[width=\linewidth]{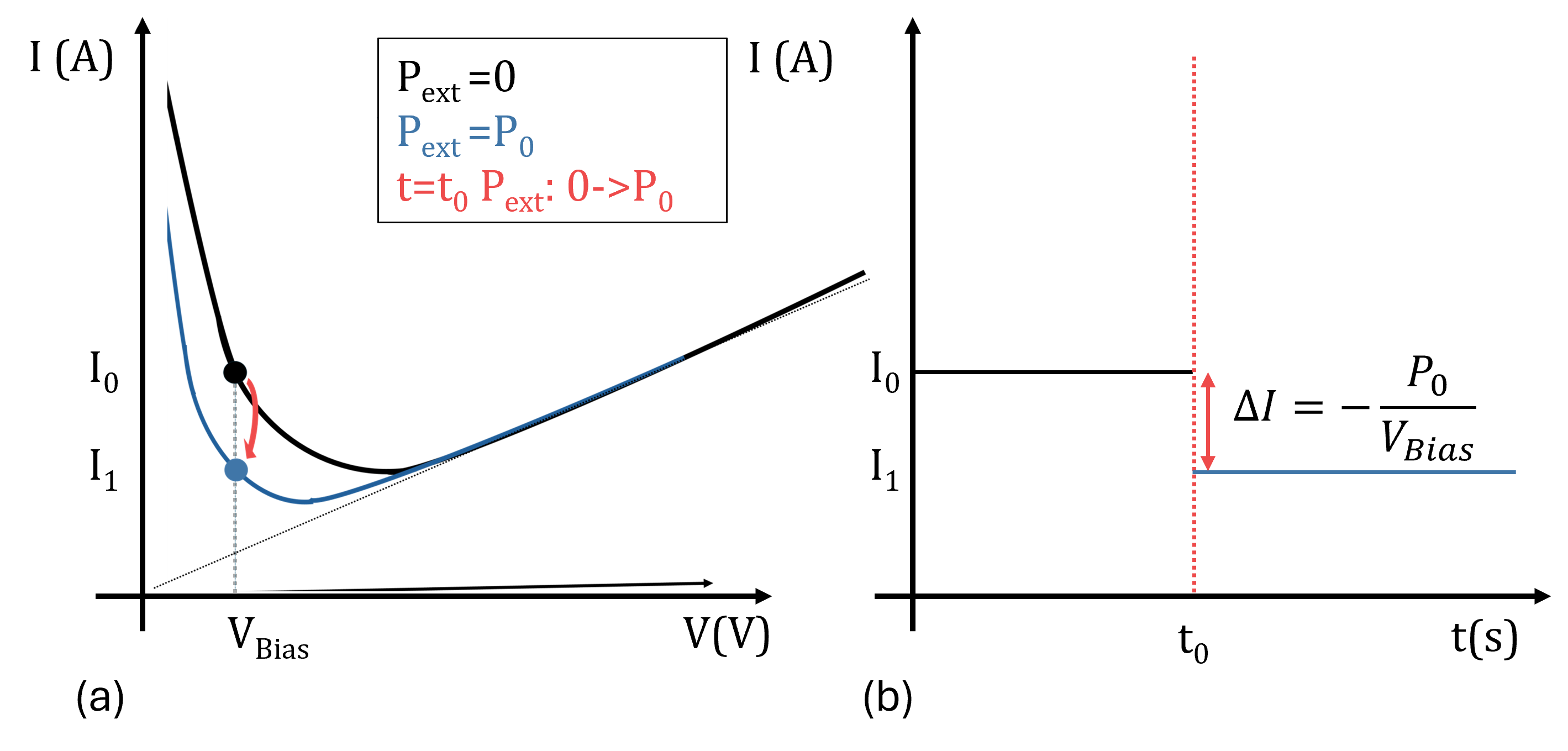}
    \caption{\textbf{Operation of a TES in {a} bolometric regime.}    (\textbf{a})  Two I--V curves relative to $P_{ext}=0$ (black) and $P_{ext}=P_0$ (blue). At~time $t=t_0$,  the external power is turned on and the TES, biased with {a} voltage $V_{Bias}$, rapidly transitions from one I--V curve to the other,  resulting in a different current flowing through the superconductor. (\textbf{b}) Response of the TES to the scenario described in (\textbf{a}) but represented as a function of~time.}
    \label{fig:tesbolo}
\end{figure}

\begin{figure}[H]
   
    \includegraphics[width=\linewidth]{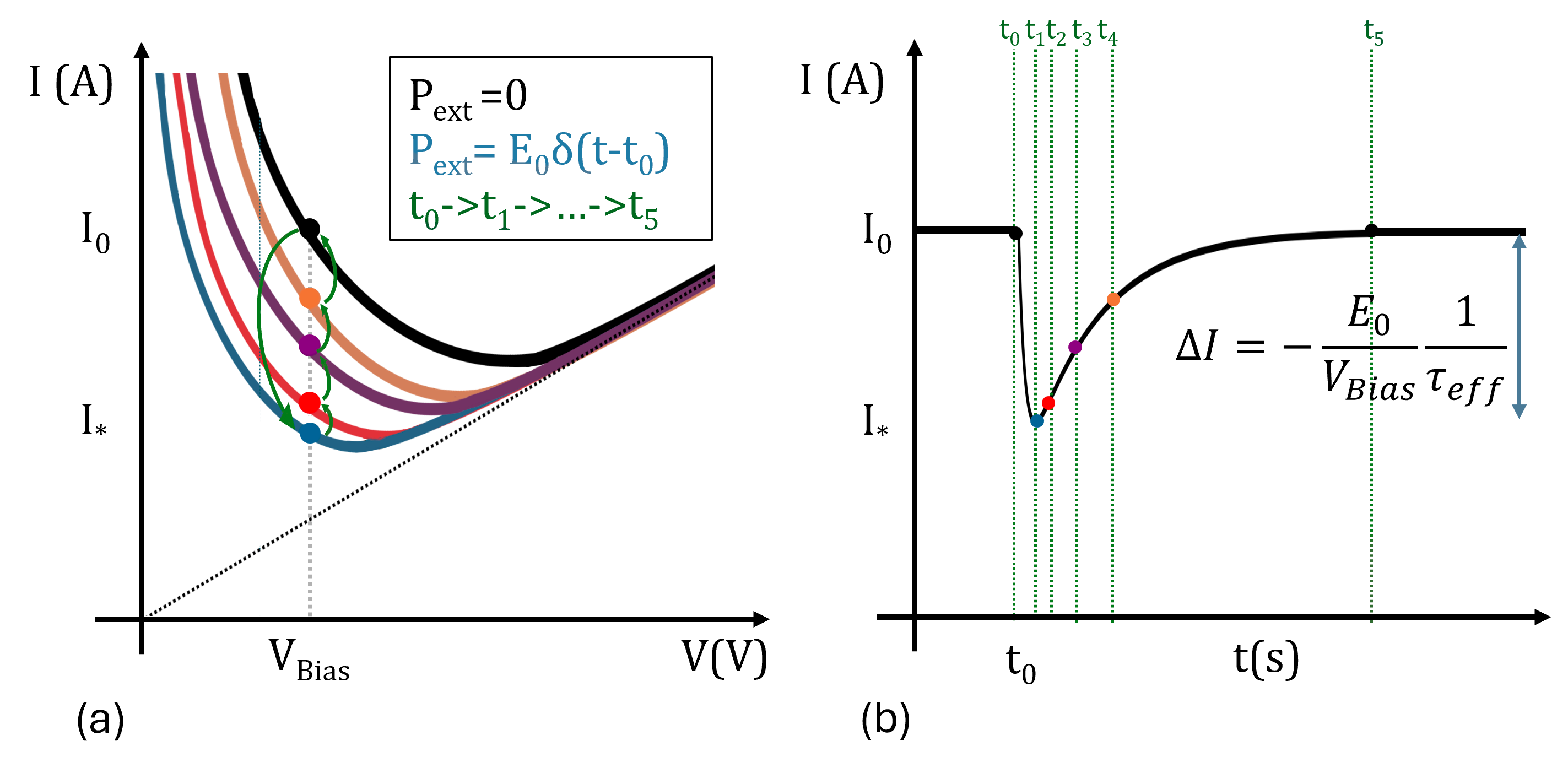}
    \caption{\textbf{Operation of a TES in {a} calorimetric regime.} %MDPI: 1. We moved the Figure after where it is first mentioned in the text. Please confirm. 2. please confirm if the explanation of the dots in colors needs to be added in the figure caption.
     (\textbf{a})  At time $t=t_0$, the external power is turned on and the TES for an infinitely small amount of time; the~TES biased with voltage $V_{Bias}$ rapidly transitions from one I--V curve to another and back to the original curve as   described by the green arrows. This results in a time-varying current flowing through the superconductor. (\textbf{b}) Response of the TES to the scenario described in (\textbf{a}) but represented as a function of~time.}
    \label{fig:tescalo}
\end{figure}

To reiterate, when a deposition of energy occurs onto the TES, its temperature increases and so does its resistance. If~biased correctly in an NETF scheme, this results in an increased current flowing through the TES (according to the I--V curves shown in Figure~\ref{fig:tesbolo}), which can be further amplified and read out. The~main difference between the two operational regimes is the nature of the impinging power: if it is transferred to the TES (or its absorber) instantaneously or with a time-scale which is much faster than the characteristic time-scale of the TES response in NETF, it is said to be operated as a calorimeter (single photon/particle detector). Differently, if~the power is delivered to the TES in a time-scale which is comparable or even larger (imagine an optical flux or the  variations thereof) than the characteristic time-scale of the TES in NETF, the~TES is said to be operated as a bolometer. From~a purely physical point of view, there is not a large difference other than the functional dependency of $P_{ext}(t)$ in Equation~(\ref{Eq:electrofeedback}).  {Figure~\ref{fig:tesbolo}a,b} show the operation of a TES as a bolometer (See Section~\ref{sec:tes-bolo}), whereas Figure~\ref{fig:tescalo}a,b show the operation of a TES as a calorimeter (See Section~\ref{sec:tes-calo}).
\unskip
\unskip
\subsection{Bolometric~Operation}\label{sec:tes-bolo}
By addressing the steady-state case in the previous section, we have effectively also discussed the bolometric operation of a TES. A~`slow' variation of $P_{ext}$ in Equation~(\ref{Eq:electrofeedback}) results in the bias of the TES on a different I--V curve. It is fair to say that as the energy flux impinging on the TES varies, slowly with respect to its characteristic time scales, the~TES moves from one I--V curve to another and the physical signal is its bias current which, at~all times, is a solution to the steady-state equations presented in Equation~(\ref{Eq:electrofeedback}).
A small change in the temperature of the TES $T \rightarrow{T+\delta T}$ results in a change in the Joule power $\delta P_j = - \frac{V^2}{R^2}\frac{dR}{dT}\delta T = - \mathcal{L} G \delta T$. Here, the~loop-gain $\mathcal{L}$ is defined as $\mathcal{L}=\frac{\alpha P_j}{G T}$ and $\alpha =\frac{dR}{dT}\frac{T}{R}$ is the a-dimensional derivative of the transition~curve.

In this configuration, when the TES is in dynamic thermal equilibrium, the~sum of Joule power and external power is conserved; hence, we can write
\begin{equation}
    P_{ext}+ P_J = const
\end{equation}
and any small fluctuation is given by:
\begin{equation}
    \delta P_{ext} + \delta I \cdot V = 0.
    \label{eq:deltaPext}
\end{equation}

Whereby, because~of the fixed voltage negative electro-thermal feedback the component, $I\cdot \delta V = 0$. From~Equation~(\ref{eq:deltaPext}), we can infer the responsivity of the bolometer as
\begin{equation}
    \delta I = -\frac{\delta P_{ext}}{V}.
\end{equation}

\subsection{Calorimetric~Operation}\label{sec:tes-calo}
The calorimetric operation occurs when the impinging power is described, ideally, as~$P(t) = E_0\, \delta(t-t_0)$. In~the first approximation, we can imagine a small fluctuation of the impinging power $\delta P_{ext} e^{-i\omega t} $ as a generic Fourier component of an impinging power $P_{ext}(t)$.  If~we solve this problem in Fourier transform space, we obtain
\begin{equation}
    \delta P_{ext} - \frac{V^2}{R^2}\frac{dR}{dT}\delta T = G\delta T - i \omega \delta T,
\end{equation}
\begin{equation}
    \delta P_{ext}= (-i\omega C + G +\mathcal{L} G\delta T) \delta T,
\end{equation}
\begin{equation}
    \delta P_{ext}= \delta T G(\mathcal{L}+1)(1-\omega\tau_{eff}) ,
\end{equation}
\noindent
where $\tau_{eff}=\frac{\tau}{(1+\mathcal{L})}$ is the new time constant of the system, which  not only depends on $C$ and $G$, but also~now  depends on the loop parameters which can significantly speed up the response of the TES. Furthermore, by~applying   Ohm's law, we can infer
\begin{equation}
    \delta I = - \frac{V}{R^2}\frac{dR}{dT}\delta T = - \frac{\mathcal{L}G}{V}\delta T.
\end{equation}

Finally, the~responsivity of the detector $S$,  defined as the current produced by a unitary change in $P_{ext}$,  can be evaluated as
\begin{equation}
    S = \frac{\delta I}{\delta P_{ext}} = - \frac{1}{V} \frac{\mathcal{L}}{(\mathcal{L}+1)}\frac{1}{1-i\omega \tau_{eff}}.
    \label{eq:TES-eq}
\end{equation}

Equation~(\ref{eq:TES-eq}) describes the response of a TES in NETF to any periodic variation of $P_{ext}$. Through it, the response of any optical input can be determined by decomposing it into its Fourier components and computing the transfer function independently before re-combining the solutions; we want to focus on the two most interesting~cases:
\begin{enumerate}
    \item For slow-varying signals ( $\omega<< 1/\tau_{eff}$) and in a strong electro-thermal feedback ($\mathcal{L}>>1$) the responsivity of the detector only depends on the bias voltage
\begin{equation}
    S \propto - \frac{1}{V},
    \end{equation}
which is consistent with what we discussed in Section~\ref{sec:tes-bolo} where we addressed the bolometeric operation of a TES. 
    \item The current response of the system to an instantaneous delta-like deposit of energy can be calculated through the 1-pole transfer function as
\begin{equation}
        I(t) =-\frac{1}{V}\frac{E_0}{\tau_{eff}}e^{-\frac{t}{\tau_{eff}}}\left(\frac{\mathcal{L}}{\mathcal{L}+1}\right),
    \end{equation}which in a strong electro-thermal feedback ($\mathcal{L}>>1$) can be further simplified as:
\begin{equation}
        I(t) =-\frac{1}{V}\frac{E_0}{\tau_{eff}}e^{-\frac{t}{\tau_{eff}}}.
    \end{equation}
\end{enumerate}

%\begin{figure}[!ht]
%    \centering
%    \includegraphics[width=\linewidth]{tessignal.png}
%    \caption{\textbf{Response of a TES in NETF operated as a bolometer (in green) and a calorimeter (in red)}. For a slow-varying power influx the response of the TES is proportional to the external power and the voltage with which the TES is biased. For a fast-varying power deposited in the TES, it responds with a fast rising pulse which decays exponentially with a time constant that is characteristic of the device and its NETF circuit.}
%    \label{fig:tessignal}
%\end{figure}

Since the currents   produced by such systems are extremely small, in~order to effectively detect the response of a TES detector, the~signal thus produced needs to be amplified before being digitized. The~most widespread approach involves a superconducting quantum interference device (SQUID) operated as an amplifier (further details in Section~\ref{sec:amp}). The~main advantage of a SQUID-based amplifier involves its capability of acting as an `ideal' trans-conductance amplifier, with~an almost zero input impedance and a large output impedance. Thus, the SQUID amplifier couples all the current produced by the TES to its input and while amplifying the signal, it converts it into a voltage turning an impedance-sensitive measurement into a more standard voltage measurement that can be performed with a {Digital-to-Analog Converter (DAC)}.

\subsection{Inductive~Bias}
As a small note, in~some applications where noise is critical, it might be worth using a purely inductive load to bias the TES $Z_L = L_L$. This results in the effective reduction of the Johnson noise component,  which is due to the finite resistance of the shunt resistance $R_L$. Further details on noise can be found in Section~\ref{sec:noise}.

\subsection{TES~Examples}%MDPI: We changed ``subsubsection'' to ``subsection'', please confirm.
\begin{figure}[H]
    \centering
    \includegraphics[width=\linewidth]{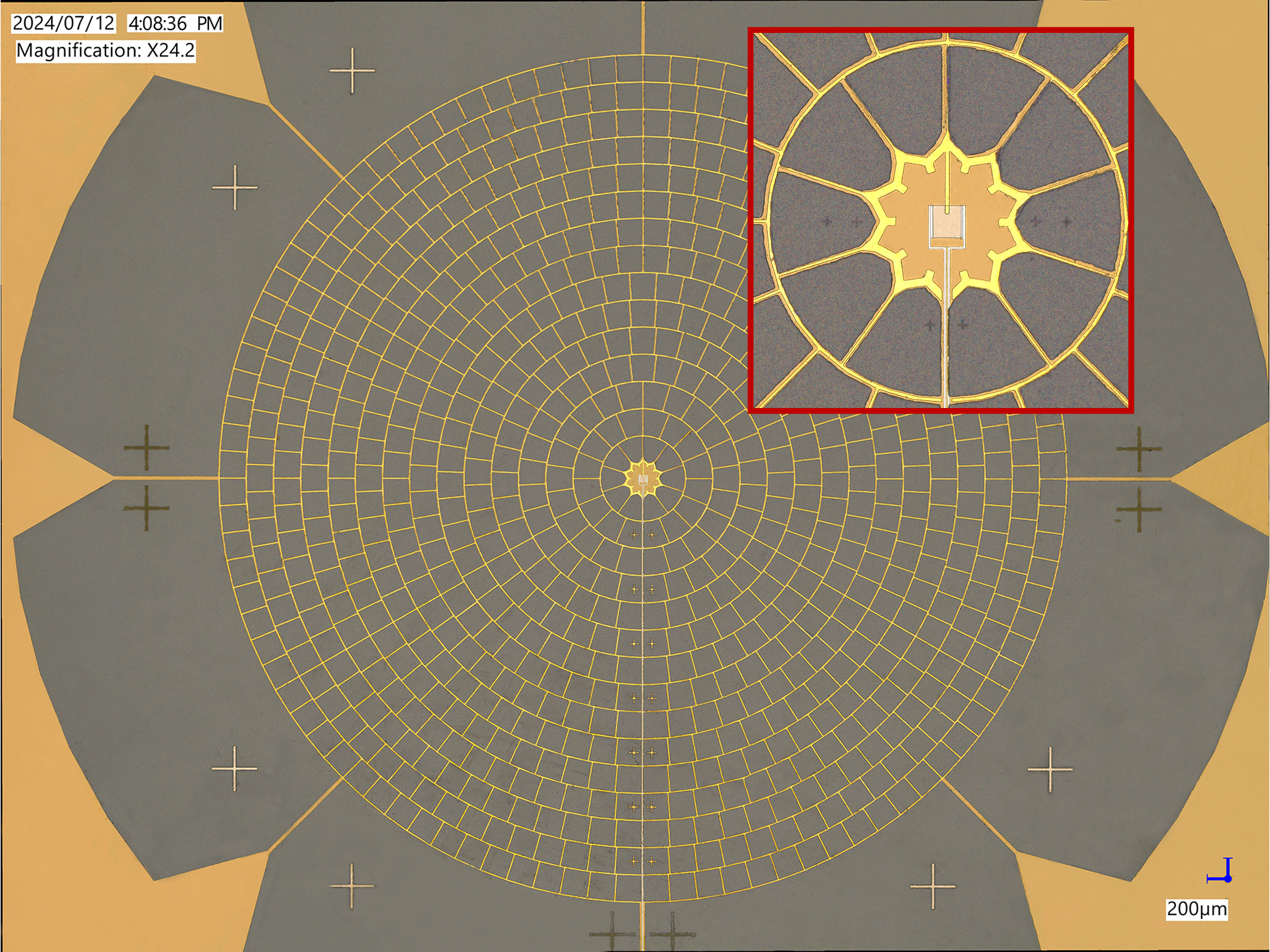}
    \caption{\textbf{A spiderweb TES} %MDPI: 1. please confirm if the explanation of the colors and symbols ``+'' needs to be added in the figure caption. 2. Please confirm if dot should be added after this sentence.
    developed during the R\&D phase of the LSPE/SWIPE~experiment.T The red inset shows a close-up on the TES thermistor }
    \label{fig:teslspe}
\end{figure}

For the sake of completeness, before~further discussions on the technological details and the applications of Transition Edge Sensors, we believe it is appropriate to show two different TESs intended for the detection of {cosmic microwave background} radiation as part of two different experiments  LSPE/SWIPE and LiteBIRD,  {described} in Sections~\ref{sec:lspe} and \ref{sec:litebird}, respectively. 
Figure~\ref{fig:teslspe} shows a Transition Edge Sensor developed during the R\&D phase of the LSPE/SWIPE experiment; it consists of a Au/Ti/Au trilayer which allows for the fine control of the critical temperature of the film (280 mK) on a SiN membrane, which decouples the TES from the Si substrate and helps {to} achieve a high sensitivity of the TES by reducing the thermal capacity. The~thermal link to the bath is achieved through a gold spiderweb which also acts as an absorber. The~size of the spiderweb is $\approx$8 mm in diameter, while the diameter of the inner core of the web, which hosts the TES (as shown in the inset of Figure~\ref{fig:teslspe}), measures $\approx 370$ 
{$\upmu$m} in~diameter. 

\begin{figure}[H]
    \centering
    \includegraphics[width=\linewidth]{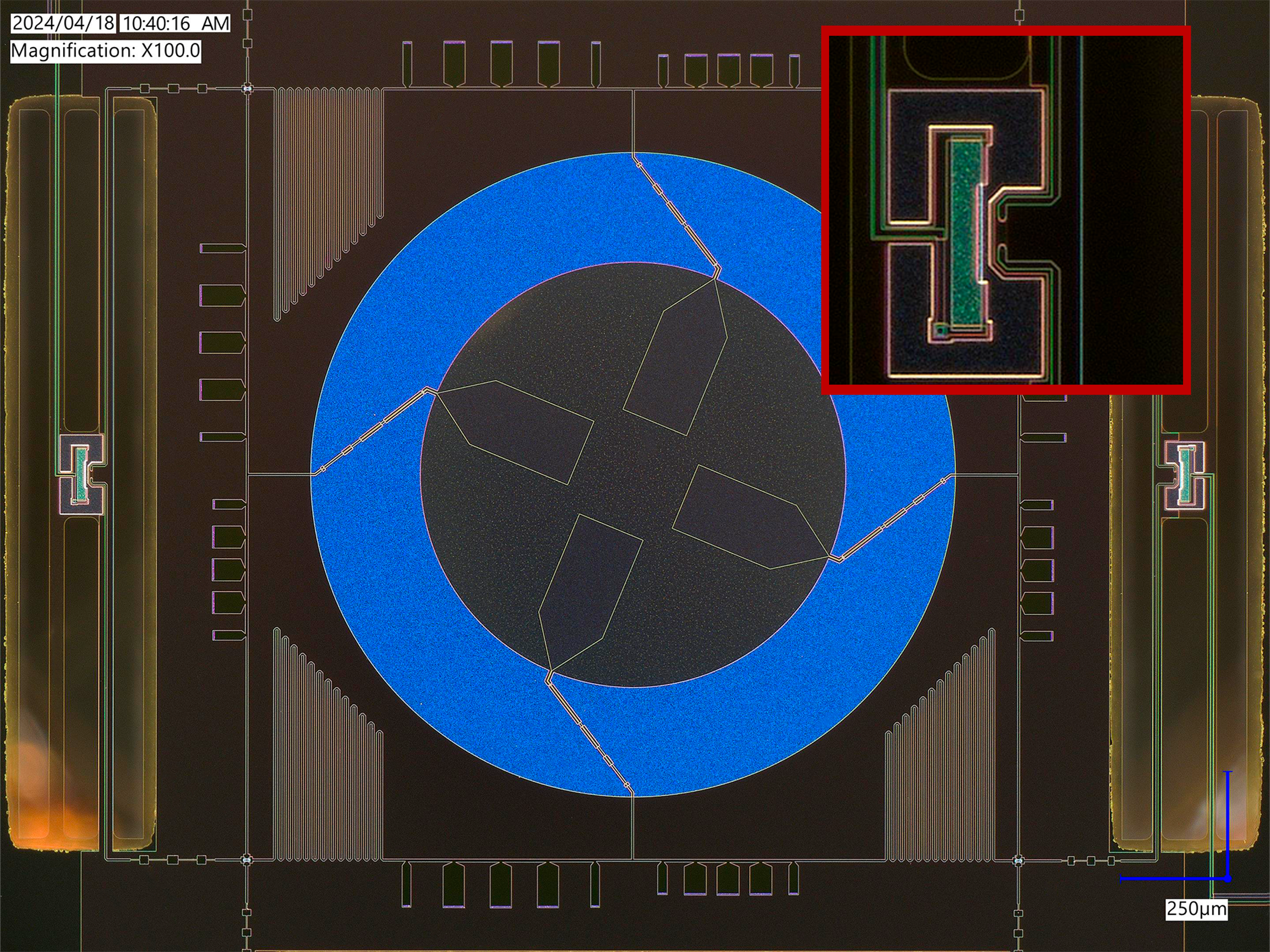}
    \caption{\textbf{An antenna-coupled TES}
    fabricated by NIST which will find application on the MHFT instrument of the LiteBIRD~spacecraft. The red inset shows a close up on the TES thermistor }
    \label{fig:teslitebird}
\end{figure}

The chip shown in Figure~\ref{fig:teslitebird} represents one of the polarimeters {(i.e., two detectors sensitive to the two polarizations of light)} of the Medium and High Frequencies Telescope (MHFT) that will be part of the payload of the LiteBIRD spacecraft. Each chip consists of two TESs and four membrane-suspended antennas for the detection of the two components  polarized separately~\cite{mcmahon2012multi}. The~radiation thus coupled to the antennas is fed through a filter-bank for multichroic sensing and is then absorbed in the form of dissipated heat on a membrane-suspended TES (shown in the inset). The~TESs, fabricated at the (American) National Institute of Standards and Technology (NIST), are made of Al/Mn films with a critical temperature of $200$ mK.

\unskip

\subsection{SQUID~Amplification} \label{sec:amp}
\subsubsection{DC-SQUID}
\begin{figure}[H]
  
    \includegraphics[width=0.7\linewidth]{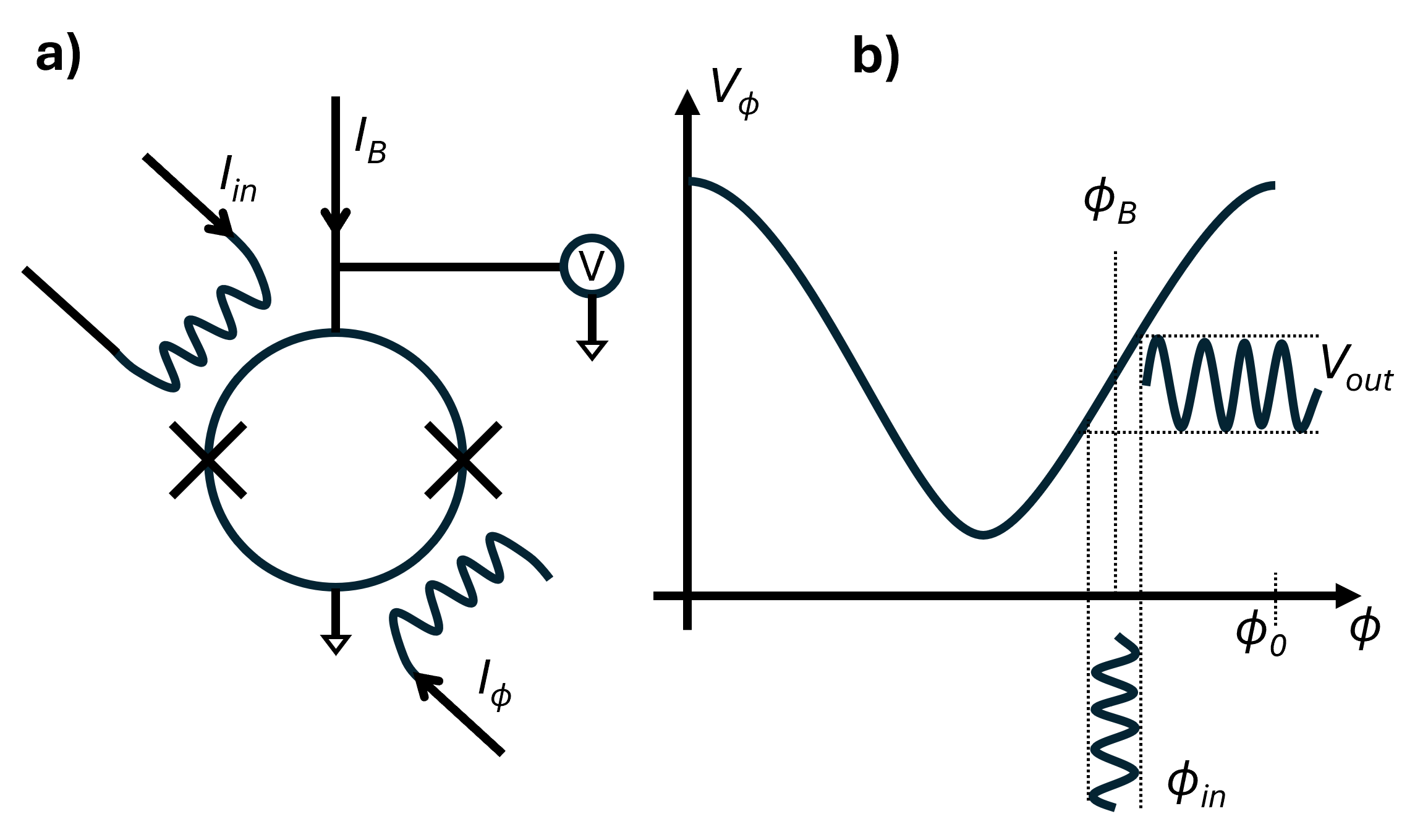}
    \caption{\textbf{\hl{Scheme of a SQUID amplifier.}} %MDPI: 1. We moved the Figure after where it is first mentioned in the text. Please confirm.  2. Please add the left bracket in the image, e.g., “a)” should be “(a)”
    (\textbf{a}) Electrical scheme of a SQUID amplifier. (\textbf{b}) $V_\phi$ characteristics of a SQUID amplifier and the response of the system related to the shape of the~characteristic.}
    \label{fig:SQUIDamp}
\end{figure}

{A SQUID is a superconducting ring, also known as a SQUID washer, interrupted by one (RF-SQUID) or two (DC-SQUID) weak links in  the  form of Superconductor--Insulator--Superconductor junction, Superconductor--Normal Metal--Superconductor junction, Dayem bridge, etc.~\cite{Likharev1979}. The~schematic of a DC-SQUID is shown in Figure~\ref{fig:SQUIDamp}a. Generally, when a SQUID is biased with a current $I_B$, a~variation in the magnetic flux coupled to the SQUID washer, it produces a voltage swing across the junctions,  which is periodic with  a fixed period  defined by the magnetic flux quantum $\phi_0$ ($\frac{h}{2e}= 2.065 \cdot 10^{-15}$ Wb). The~trans-characteristic curve of a DC-SQUID that correlates the magnetic flux ($\Phi$) with the voltage generated across the junctions is shown in Figure~\ref{fig:SQUIDamp}b. The~V--$\Phi$ curve is steepest when the flux applied is an even multiple of $\phi_0$: $\Phi=(2n - 1) \phi_0/4$. In~such a configuration, the~response of the DC-SQUID to a small magnetic flux $d\Phi<<\phi_0$ is linear and the transfer coefficient $V_\Phi= \partial V/\partial \Phi $ is maximum. In~order to use a DC-SQUID as an amplifier, usually, one fixes its working point around any one solution of $\Phi=(2n - 1) \phi_0/4$ by having a static flux coupled into the SQUID washer through an apposite flux bias coil and coupling the current signal that needs to be amplified through a separate closely coupled thin-film input coil. This is also recreated in SQUID Array amplifiers which combine a large number of SQUIDs connected in series and in parallel in order to optimize the trans-characteristics and the performance of the device.}

\subsubsection{SQUID Array Amplifiers}

The majority of TES devices leverage   the DC-SQUID as the most sensitive detector of magnetic flux currently available. To~date, SQUIDs have been used in innumerable low-frequency applications, including gravitational wave detection~\cite{pleikies2007squid}, susceptometry~\cite{ketchen1984miniature}, biomagnetism~\cite{wikswo1995squid}, non-destructive evaluation~\cite{jenks1997squids}  and~magnetic resonance imaging~\cite{clarke2007squid}. Recently, growing interest in low-noise radio frequency and microwave amplification for particle detection~\cite{muck2010radio}, infrared sensor readout and~superconducting quantum bit measurements~\cite{tanaka2002dc} has   formed. In~these applications, SQUIDs are one of the leading candidates due to their very low power dissipation and good noise~characteristics.

Connecting a number of identical DC-SQUIDs in series can considerably amplify the relatively small voltage signal produced by each device individually. In~case an array of identical SQUIDs is biased at an identical working point and the same magnetic flux is coupled into each SQUID, the~array acts like a single SQUID with an enhanced transimpedance which is ideally the product of the individual contributions. Series SQUID arrays provide outstanding slew rate performance because the linear flux range is preserved. Arrays of \mbox{100 SQUIDs} are commonplace, and~several mV output signals can be achieved. Although~the voltage noise density across the array increases linearly with the number of SQUIDs, a~100-SQUID array can drive a room-temperature preamplifier directly~\cite{irwin2001squid} and increase the system noise only slightly above the intrinsic noise of the SQUID array. The~flux noise density and input current noise density both scale inversely with the number of SQUIDs, conserving a constant coupled energy resolution. In~principle, a~pickup coil could be coupled to the input coil to form a magnetometer, but~series SQUID arrays are almost always used as transimpedance amplifiers in configurations like two-stage SQUIDs or readout devices for cryogenic particle detectors. Experimental results showed a very high bandwidth on the order of 100 MHz with 100-SQUID arrays~\cite{Tsang1975,huber2001dc}.

It is worth noticing that the performance of SQUID amplifiers can be boosted by having arrays with a similar number of SQUIDs in series as in parallel, in~particular, to~overcome the detrimental effect of a non-uniform bias configuration~\cite{GaliLabarias_2023} and to decrease the overall noise of the system~\cite{Mukhanov2014}.

In most applications or instruments, a~SQUID is used as a null-detector device linearized through  negative feedback~\cite{irwin2001squid}. The~negative feedback modes of operation include flux-lock modes. In~such operation modes, the~feedback signal is coupled to the SQUID: an external signal applied to the input coil generates a flux in the SQUID, which is countered by an opposing flux in the feedback coil coupled to the SQUID's inductance. If~the feedback and input coils are not screened adequately enough, the~feedback signal couples, together with the modulation signal, if~conventional readout electronics are used, to~the input circuit and interacts with the load. Also, in~experiments where the load inductance is usually superconducting and variable, the~feedback coupling changes with the load due to screening effects. Therefore, the~SQUID should be designed such that the coupling of the feedback and modulation signals to the input circuit is negligible. This can be achieved by designing the SQUID layout with separate secondary coils for the input and feedback flux~transformers.

For specifically critical applications,  multiple SQUID arrays can be cascaded~\cite{welty1993two}. Some applications include those cases in which the signal-to-noise ratio (SNR) needs to be pushed to its maximum, or~those in which the thermal budget at the cold state of the cryostat is limited or those in which the amplifier needs to be thermally decoupled from the detectors. While there is no unique cascading scheme~\cite{kiviranta2020two,welty1993two,drung2006low,cantor1997low}, the~general approach to a multi-stage SQUID amplification chain relies on a first stage which acts as a `front-end' amplifier, i.e.,~exhibits a limited amplification gain and a large bandwidth in order to not distort the signal. Unfortunately, such a SQUID amplifier also exhibits a rather limited saturation power which also limits the multiplexing factor of TESs read out in parallel through a multi stage SQUID chain.
The first stage acts as an emitter--follower transducing an impedance-sensitive signal such as the current produced by the TES into a much handier voltage signal which can be further amplified by the later stage SQUID amplifiers. Since the front-end amplifier produces a voltage signal, such voltage can be fed into a series of SQUID arrays anchored to a plate at a higher temperature. The~later stages, commonly referred to as the `booster'~\cite{kiviranta2020two}, can dissipate more heat because the cooling power increases largely at higher temperature and the amplification factor of a SQUID array is proportional to the power it dissipates~\cite{kiviranta2018low}.

\subsection{Multiplexing}\label{sec:mux}
In light of all the reasons discussed so far, the~operation of a TES occurs at extremely low temperatures well below 1 K. Even the most advanced commercial refrigerators only have a very limited cooling power, usually less than 100 mW~\cite{uhlig2012cryogen}, at~their milli-Kelvin stage. Such a small cooling power would severely hinder the up-scaling of a TES array if they were to be {read out} individually as the power dissipated by the SQUID arrays and the thermal-load of the signal lines would quickly saturate all the cooling power available. In~order to overcome this issue, several multiplexing strategies are available and are here discussed. More in general, the~multiplexing of analogue signals occurs in four defined and consequential~steps:
\begin{enumerate}
    \item{\textbf{Bandwidth limitation}}
    { is essential in order to prevent any degradation of the signal due to either out-of-band detector noise aliasing into the signal band or appearing as excess cross-talk}.
    \item{\textbf{Encoding or Modulation}}{ is the step whereby the signals from different pixels are encoded by multiplying them by orthonormal functions}.
    \item{\textbf{Summation}}{ of the encoded signals into one time-ordered data stream}.
    \item{\textbf{Decoding or De-modulation}}{ is achieved through knowing the encoding function and applying a decoding algorithm}.
\end{enumerate}

In the next few sections, we will present an overview of the different possible multiplexing schemes and we will demonstrate their working principle as well as show a simulated Time-Ordered Data (TOD) in the  case of a calorimetric detection. The~use of a TES as a bolometer is merely a simplification of such a case where all the signals are~constant.

\subsubsection{Time Division~Multiplexing}
Time Division Multiplexing (TDM) is a rather old technique dating back to the 1870s and  was developed in parallel to the rise of the telegraph~\cite{callahan1938time}. TDM allots a specific interval of time to each of the detectors for their readout. With~the use of a switch matrix, all the detectors are  read out in order in columns and rows. Ensuring a fast commuting time and a small time slot to each detector, frame rates of up to 20 kHz can be achieved~\cite{battistelli2008functional}. For~applications where the detector signals do not vary in time-scales of the order of 100 $\upmu$s, this results in an effectively `\textit{continuous}' monitoring of the~source.

\begin{figure}[H]
    
    \includegraphics[width=0.84\textwidth]{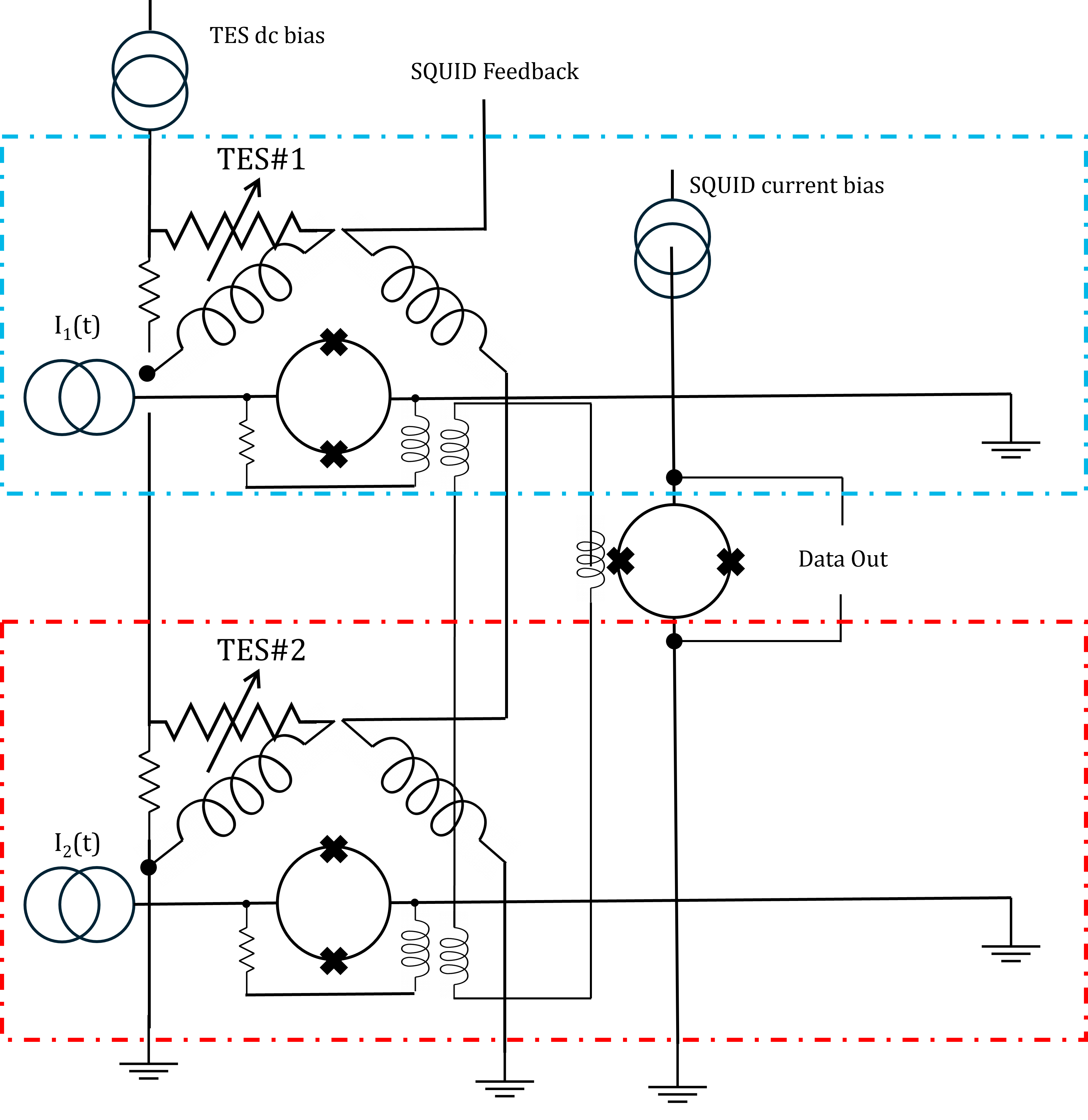}
    \caption{\textbf{Circuital schematic implementation of a two-TES TDM.} Each channel is colour coded: cyan for CH\#1 and red for CH\#2. The~switching occurs by turning on and off the different bias lines of the SQUID~amplifiers.}
    \label{fig:TDM-circuit}
\end{figure}

\begin{figure}[H]
    
    \includegraphics[width=0.9\textwidth]{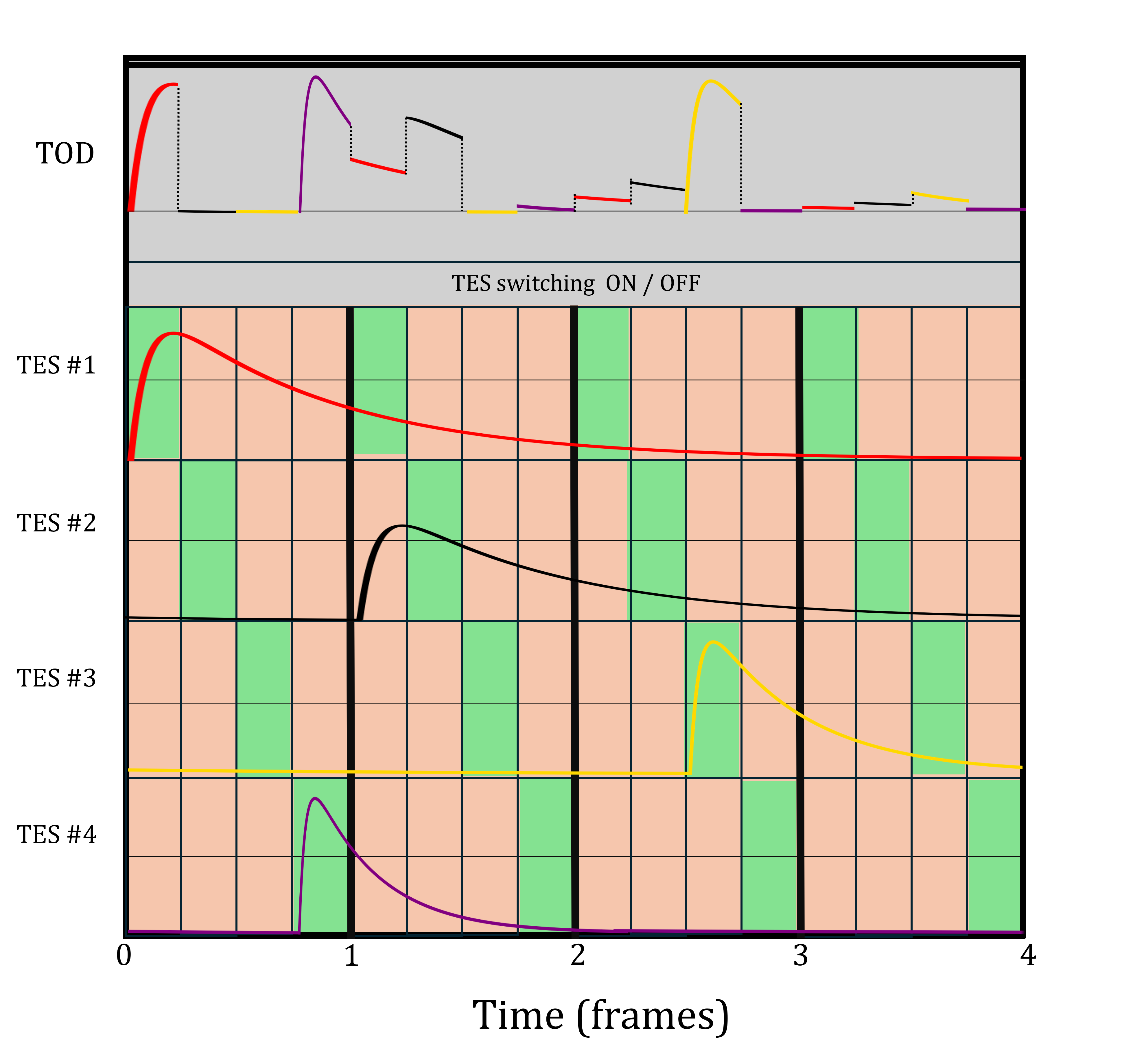}
    \caption{\textbf{Time Division Multiplexing scheme}. The~signal produced by the four different TESs is shown in red, black, yellow and violet. The~green (red) fields in the picture show, per each frame the time each channel is switched on (off). The~{TOD} read out by the electronics contains the data points of the TES signals at (convolved with) their respective own on time intervals. This picture represents an extreme case where the sampling interval is comparable with the characteristic time figures of the TES and shows how such a scheme may lead to data loss because of the difficulty in reconstructing the data curves (see TES\#2 in black). Ideally, the~sampling time and frequency is much faster than the rise and fall time of the TESs so that a proper sampling can be achieved without data loss. If, on~the other hand, the~sampling interval is significantly larger than the rise and fall time of the TESs, the~pulses are properly reconstructed, but~the dead time of the detectors is significantly~increased.}
    \label{fig:TDM}
\end{figure}

TDM has been widely implemented as a multiplexing scheme for Transition Edge Sensors in a large number of different schemes, all of which exploit the same working principle: a 2D array of TES ($M-rows\times N-columns$) with $M$ SQUIDs, one per each row, is read out by sequentially switching on the SQUIDs (1, 2, $\cdots$, M) during the on-phase of each of the $M$ SQUIDs. A~firmware-controlled DAC biases the individual $N$TESs on the M-th row one at the time for time intervals as short as 1 $\upmu$s. In~order to prevent aliasing, the~bandwidth of the TES is limited to below the Nyquist frequency defined by the switching interval (e.g., 500 {kHz} for the  1 $\upmu$s switching time in the example above). This limitation is achieved with a single-pole RL low-pass filter where  $R$ is the resistance of the TES and $L$ is the inductance of the input coil of the SQUID~\cite{wu2022multiplexing}. The~number of channels that can be read in a TDM scheme is limited by SQUID-noise aliasing and it scales as $\sqrt{N}$ \cite{niemack2010code,DORIESE2006808} and the energy resolution can be degraded by insufficient accuracy in the pulse arrival time. One possible implementation of a {TDM} scheme with only two TESs is shown in Figure~\ref{fig:TDM-circuit}. Figure~\ref{fig:TDM} shows the working principle of a TDM readout with four TESs read out in sequential time frames. During~each time frame, each TES is read out only for a fraction of the time frame. Each TES produces an individual pulse with different amplitudes and at different times. The~TOD read out at room temperature given by the concatenation of the signals on each channel during their own `on' period is shown at the top of the graph.  It is worth stating that this scheme only serves the purpose of explaining the working principle of TDM. In~all practical applications, the~switching frequency is much higher than the characteristic time-scales with which the TES signal varies. This results in a fine sampling of the signal without a large information loss on the signals thus~acquired. 

\unskip

\unskip

\subsubsection{Frequency Division~Multiplexing}
\begin{figure}[H]
    \centering
    \includegraphics[width=\textwidth]{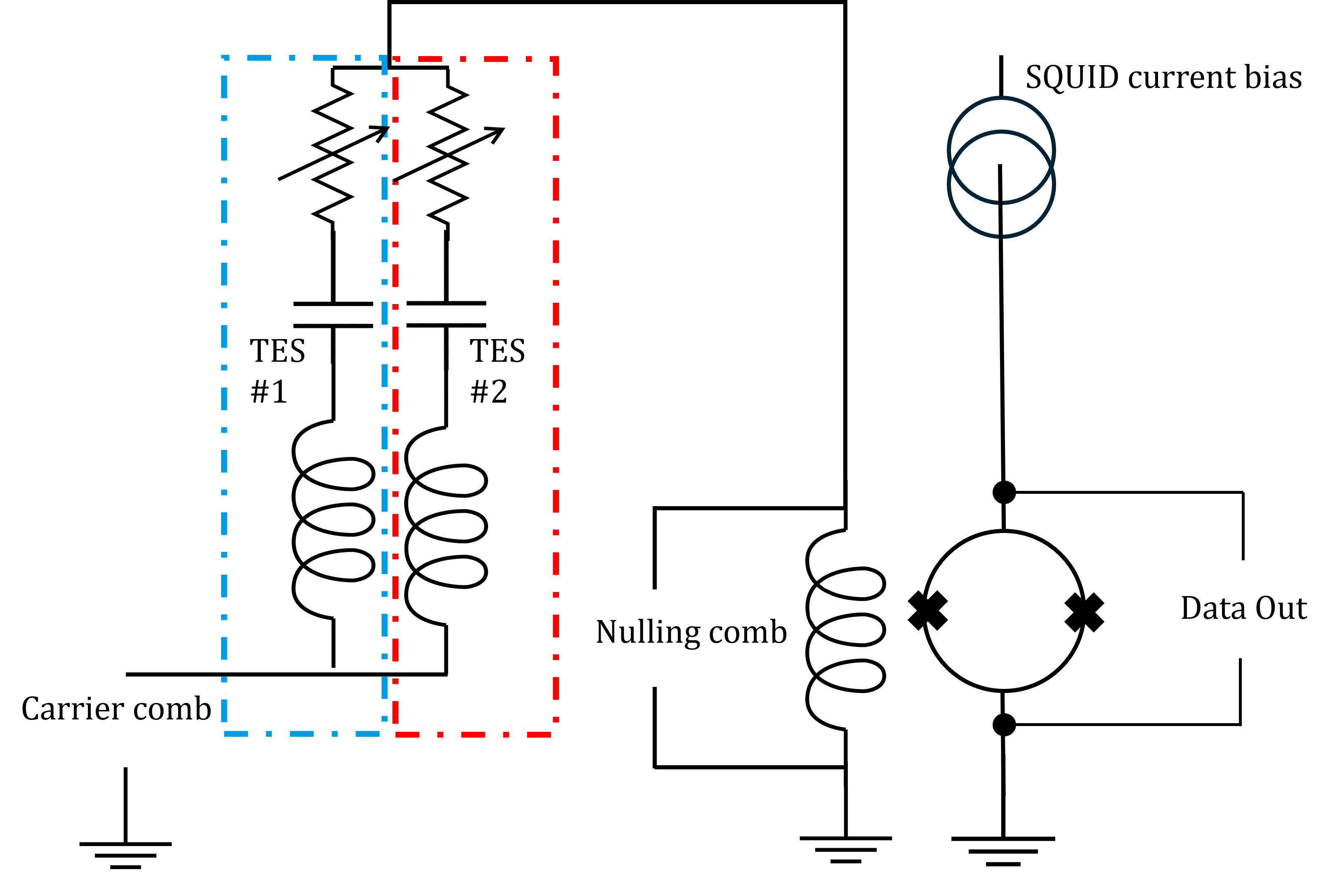}
    \caption{\textbf{Frequency Division Multiplexing}. In~a simple 2-TES scheme, each detector is connected to a band-pass LC resonator filter. The~comb  of carriers is generated and sent to the multiplexing circuit through only one wire. Each channel is color coded, cyan for CH\#1 and red for CH\#2. The~nulling occurs at the input coil of the SQUID~amplifier. }
    \label{fig:FDM-circuit}
\end{figure}

The Frequency Division Multiplexing (FDM) scheme exploits frequency ranges instead of time intervals, resulting in a continuous monitoring of the detectors, which solves the aliasing problem that is inherent to TDM. In~FDM, each TES is connected in series to an inductor and to a capacitor, resulting in a bank of band-pass filters, as~shown in Figure~\ref{fig:FDM-circuit}. The~circuit will be eventually composed by a fixed number of branches (multiplexing factor), each  of which includes the detector's variable resistance and the aforementioned LC filter. Each TES will be biased with a sine wave at a specific frequency, selected by the resonator geometry. This configuration is, alongside the TDM, the~most used for the TES~readout.

While ideally, the~arguments discussed in Section~\ref{sec:TES} still stands, it is worth stating that, following the discussion found in \citet{Dobbs_multiplexing}, the~responsivity of a TES biased with a sinusoidal voltage becomes
\begin{equation}\label{fdmsense}
    S = \sqrt{2}/V_{c}^{rms},
\end{equation}
\noindent
where $V_c^{rms}$ is the rms of the amplitude of the sine wave and the factor $\sqrt{2}$ arises from the average voltage carried by a sine wave over one~period.

\begin{figure}[H]
   
    \includegraphics[width=0.9\textwidth]{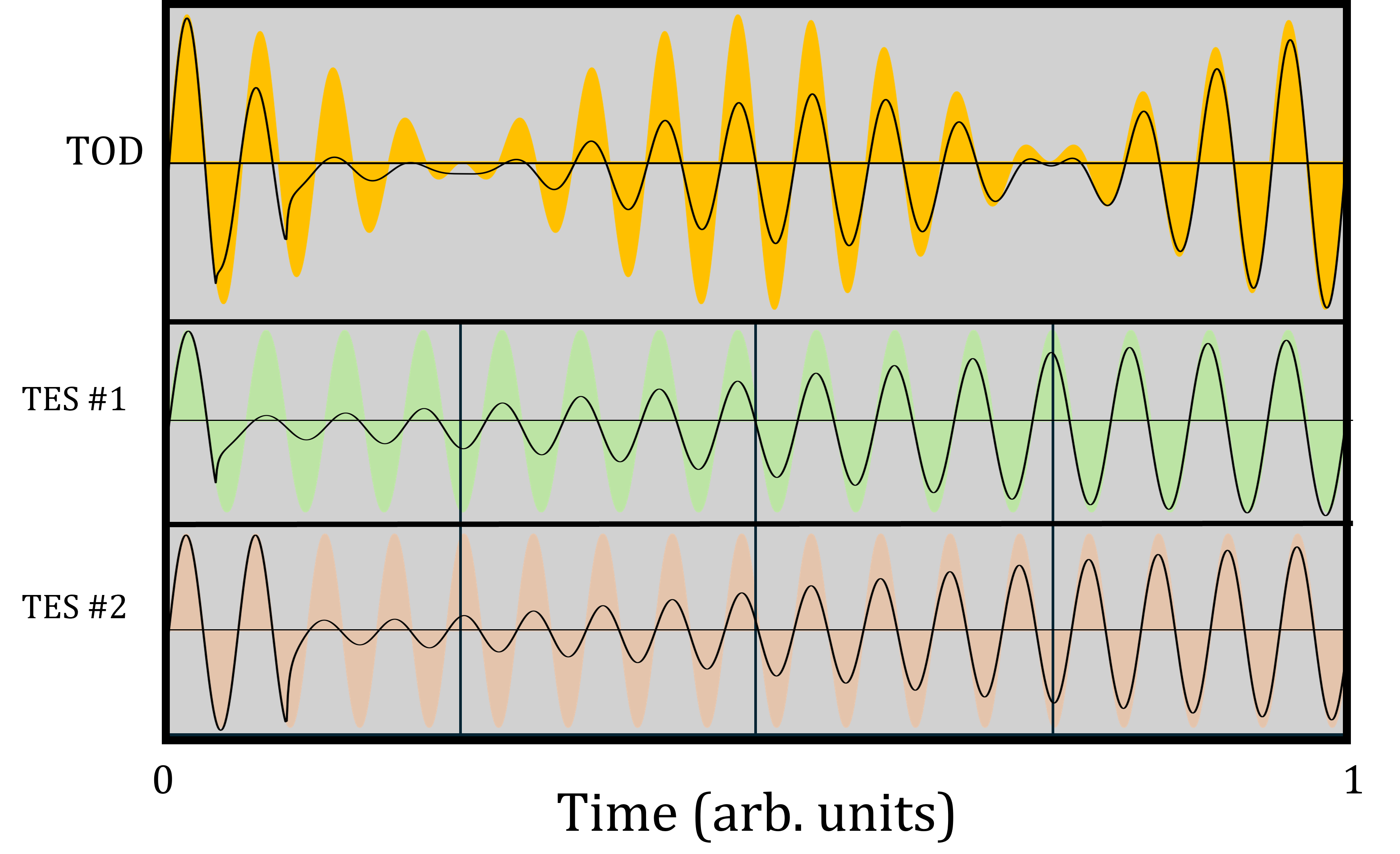}
    \caption{\textbf{Frequency Division Multiplexing scheme}. In~a simple 2-TES scheme, each detector is continuously monitored. By~de-convolving the TOD at the frequencies of each individual channel, it is possible to obtain information on the amplitude of the oscillations at each frequency. By~then plotting the amplitude of the pulse which is the envelope function that is convoluted with the sine wave at each specific frequency, it is possible to obtain a pulse-like shape typical of a TES detector. {The shaded area represents the response of the TESs in their idle state.}}
    \label{fig:FDM}
\end{figure}

The FDM is performed by generating a comb of sinusoidal signals at the selected resonance frequencies which are fed to the array on a single transmission line. It is up to the filter bank to distribute the bias signal to each TES, the variable resistance of which acts as an amplitude modulator to the sine waves. In~the bolometric regime, the~variation of the incoming radiation power leads to a variation in the resistance value of the TES and to an amplitude modulation of the bolometer bias current which can be measured. In~the calorimetric regime, the~impulsive signal is retrieved by de-convolving the TOD at the frequencies of each individual channel, as~shown in Figure~\ref{fig:FDM}.

The advantage FDM holds is linked to the possibility of using a single wire for transferring the bias current to the entire circuit, which can comprehend several branches and therefore  bolometers. The~only limit on the number of detectors that can be read {out} in this configuration is set by the required bandwidth and the capabilities of the readout electronics. Moreover, the~ signals coming from the TES channels at different frequencies are summed in a summing node and transferred to the rest of the readout chain on a single wire, in~particular to a low input impedance SQUID. This results in a decreased  thermal load and lower heat dissipation at the cryogenic~stages.

In general, the~physical signal produced by the TES is much smaller in amplitude than the bias sine wave, therefore an inverted carrier comb, called \textit{nulling}, is injected in the circuit at the input coil of the SQUID, as~shown in Figure~\ref{fig:FDM-circuit}. This signal cancels out the carrier current, leaving the physical signal modulation unaltered and prevents the saturation of the dynamic range of the SQUID~amplifier. 
\subsubsection{Code Division~Multiplexing}
\begin{figure}[H]  
    \includegraphics[width=0.9\textwidth]{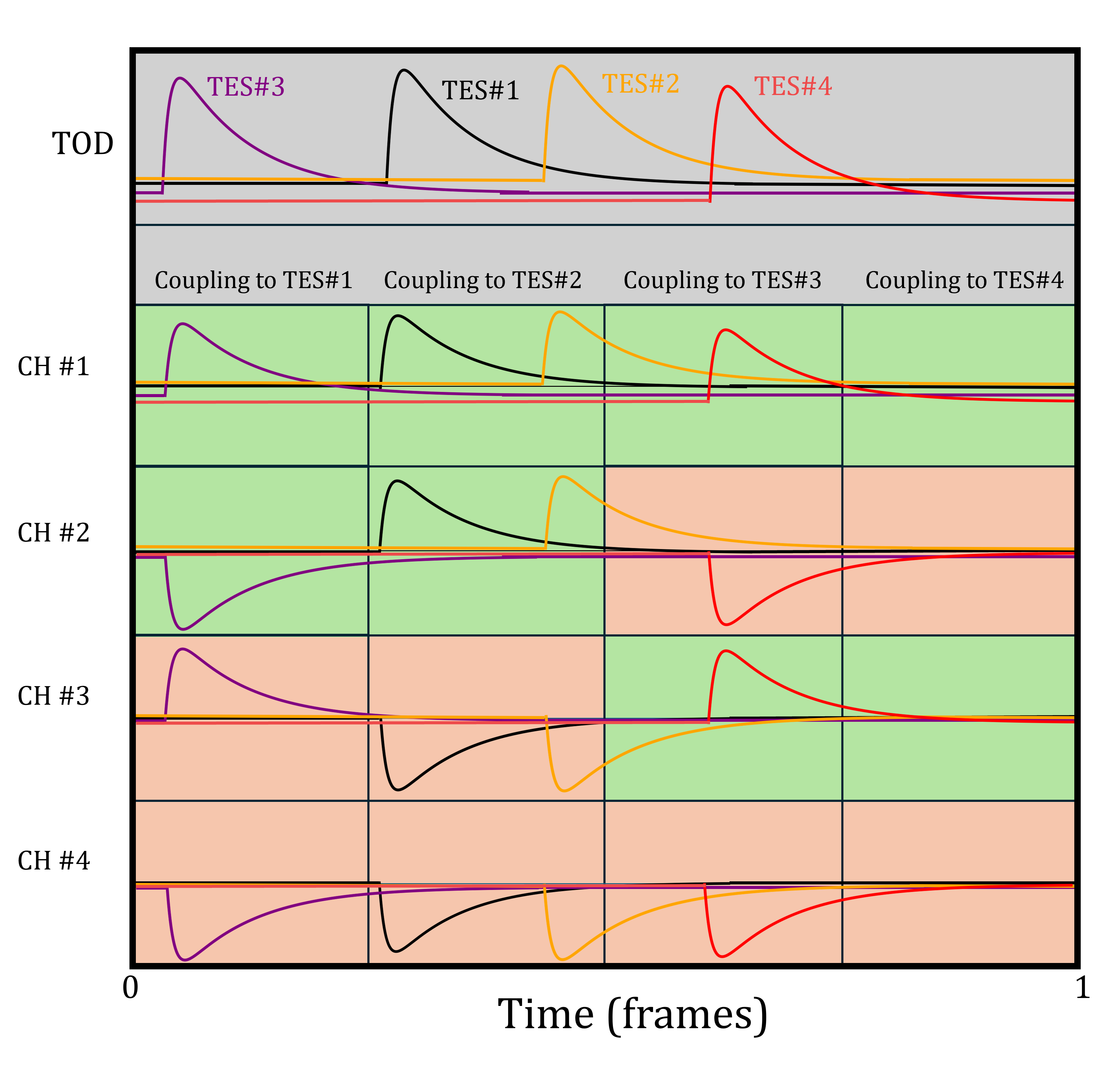}
    \caption{\textbf{Code Division Multiplexing scheme}. The~top of the figure shows the TOD of four different pulses as produced by four TESs (TES\#1, TES\#2, TES\#3, TES\#4). Such signals are equally distributed across the four channels (CH\#1, CH\#2, CH\#3, CH\#4) with a polarity that is unique to each TES-Channel pair. The~polarity is shown as different background color: green for positive and red for negative. The~de-multiplexing is achieved by looking at the polarity combination of each pulse on the different feedlines and comparing it with a look-up~table.}
    \label{fig:CDM}
\end{figure}

Code Division Multiplexing (CDM) is an architecture that combines the advantages of TDM and FDM. In~TDM, the modulating function that distinguishes the pulses produced by different detectors is a combination of $N$ square waves with low duty cycle whereas in FDM, the~modulation is achieved through $N$ sinusoidal waves at different frequencies. The~major breakthrough of CDM is that the detectors have the polarity of their coupling to the SQUID amplifier modulated by Walsh matrices~\cite{niemack2010code}. The~simplest case possible is that of a two-channel CDM: the sum and the difference of the signals from TES\#1 and TES\#2. Knowing the sum and difference of the two signals allows the unique identification of the two individual signals. The~advantage of CDM compared to TDM and FDM is that while the SQUID noise is degraded by a factor $\sqrt{N}$, $N$ samples of the $N$ pixels are read out at each frame, therefore resulting in a $\sqrt{N}$ improvement in signal-to-noise ratio~\cite{irwin2010code}. Unfortunately, the~very simplified case of using Walsh matrix modulation has yielded non-profitable results due to the complexity of achieving bandwidth limitation in such architecture. Two solutions have been proposed and developed: CDM with flux summation and CDM with current steering. Regardless of the architecture it is implemented in, the~working principle of {CDM} is explained in Figure~\ref{fig:CDM}. The~top of the figure shows the TOD of four different pulses as produced by four TESs. The~signals are equally distributed, with~1/4 amplitude across the four channels and the coupling to each SQUID that defines the channels is achieved with a polarity that is unique to each TES-Channel pair. The~de-multiplexing is achieved by comparing the polarity combination of each pulse on the different feedlines with a look-up~table. 
\vspace{-10pt}

\subsubsection*{CDM with Flux Summation}
The simplest way to achieve CDM is to sum the currents from $N$ TESs in $N$ different SQUIDs with different coupling polarities and then read them out sequentially with the use of a traditional TDM SQUID multiplexer. Going back to the two-TES example, this means that the sums of the currents arising from TES\#1 and TES\#2 are summed at the input of SQUID\#1 and subtracted at the input of SQUID\#2. In~general, through a set of inductors, the~fluxes thus produced {are} summed (with sign) into superconducting transformers that are then coupled to the individual SQUIDs. Compared to current-steering CDM, flux summation is extremely straightforward but it comes at the expense of the complexity of the routing scheme and the size of lithographic~elements.

\subsubsection*{CDM with Current Steering}
Current steering CDM exploits a superconducting Single Pole Double Throw (SPDT) switch to modulate the polarity with which each TES is coupled to the SQUID amplifier. First, the~bandwidth of signal from the TES is limited through an $L$/$R$ low-pass filter~\cite{irwin2012advanced}; the~signal thus limited is then modulated. The~two arms of the SPDT are routed to contain a superconducting-to-normal (S-N) which is a low-inductance SQUID operated as a flux-controlled variable resistor. Furthermore, the~second arm of the SPDT has an additional $\pi$ phase offset; thus, when no bias is applied to the SPDT, the~SN\#2 switch is {off} and SN\#1 is {on}. In contrast, when SN1 is on, SN2 is off. In~the first case, the~signal is coupled with positive polarity, whereas in the second, the~polarity is negative. When the polarity of each TES is switched, a~back-action voltage is generated which has two main effects: acting as a source of cross-talk and producing an additional resistive component. Over~a full frame, the~cross-talk induced by the switching is null because of the ortho-normality of the Walsh vectors~\cite{irwin2012advanced}. The~extra resistance $R_s =L_{sw}/\tau_{sw}$ (where $L_{sw}$ is the parasitic inductance of the SN switch and $\tau_{sw}$ is the interval between two switching times)  is in series with the bias voltage and needs to be small compared to the normal state resistance of the TES.   Because~the polarity modulation induced by the current-steering occurs at a frequency much larger than the bandwidth of the signal, no aliasing deteriorates the noise figure of the detector~\cite{irwin2012advanced}. In~principle, a hybrid scheme between current-steering and flux summation is possible and it involves coupling the signal from the TES into two superconducting transformers with opposite polarities and each branch is then switched on or off through SN switches. This last approach has the great advantage of reducing the number of address lines required to drive the commutation of the SN switches. Thanks to the periodic nature of the response of the SN switches, the~number of address lines required to read out $N_{rows}$ of TESs is $log_2 N_{rows}$ \cite{irwin2010code}; this means that 256 rows of TES can be read out with only 8 address lines reducing dramatically the power dissipated in the cryostat and the thermal~load.  
\begin{figure}[H]
    
    \includegraphics[width=\linewidth]{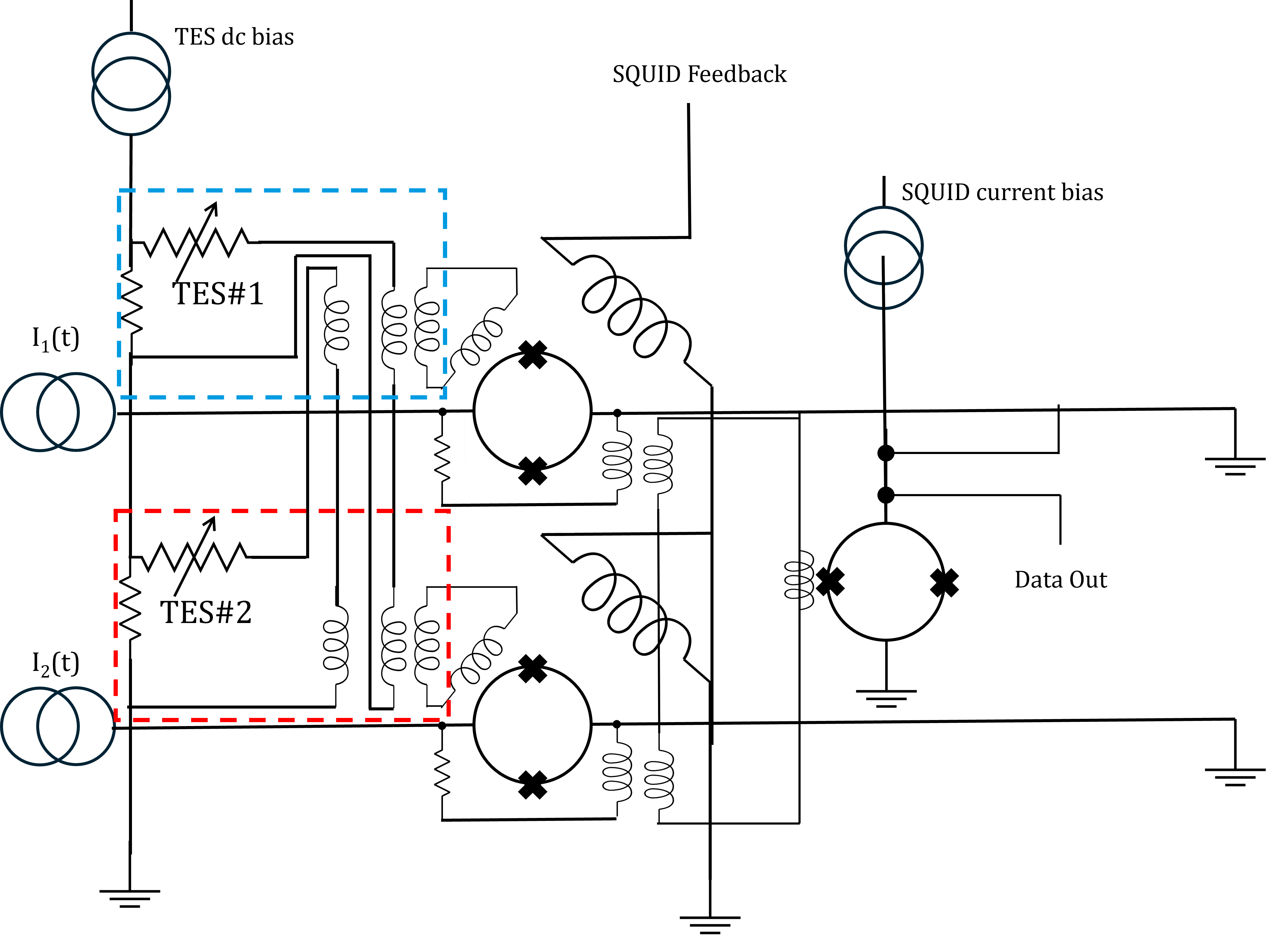}
    \caption{\textbf{Code Division Multiplexing}. %MDPI: Please cite the figure in the text and ensure that the first citation of each figure appears in numerical order.
    In~a simple 2-TES flux-summation scheme, the~two SQUIDs are fed with the signal produced by the two TESs with alternate polarities. The~two SQUIDs are alternatively read out on the `Data Out'~line.}
    \label{fig:enter-label}
\end{figure}
\unskip

\subsubsection{Microwave Multiplexing ($\upmu$-MUX)}
\begin{figure}[H]
  
    \includegraphics[width=0.9\textwidth]{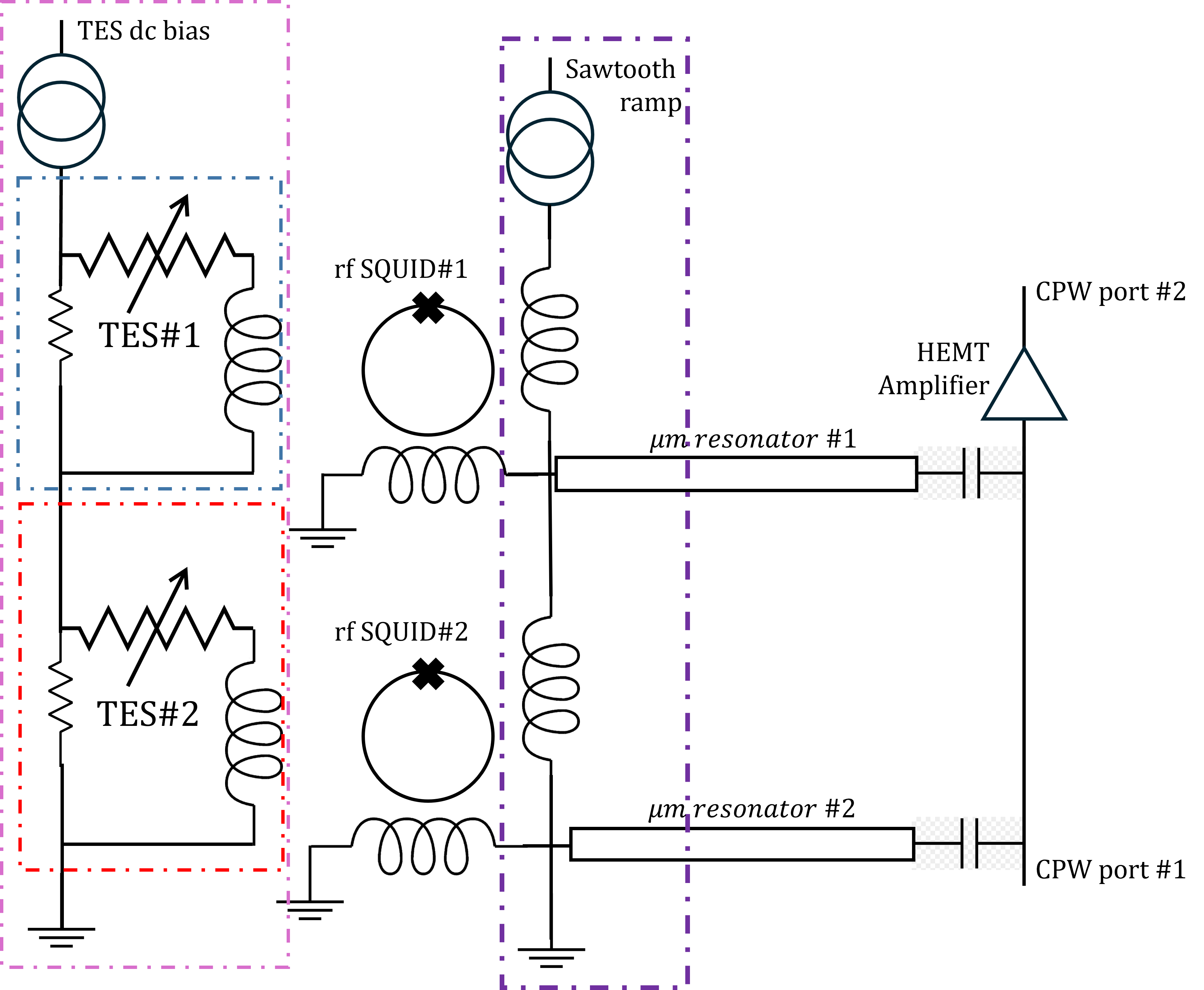}
    \caption{\textbf{Electronic schematic for the implementation of a 2-TES $\upmu$MUX.} The section in the violet box represents the two TESs and their bias circuit. They are individually coupled to a microwave resonator through an {RF-}SQUID  (blue box for \#1 and red box for \#2). The~section in the violet box represents the saw-tooth wave used as a signal frequency~up-converted.}
    \label{fig:mu-mux}
\end{figure}

The biggest challenge of both TDM TESs and FDM TESs is the maximum achievable multiplexing factor which, for~most practical implementations, is limited to less than 100. The~development of other cryogenic detectors which operate at frequencies up to 8~GHz contributed to the development of ever-improving microwave electronics. Microwave multiplexing ($\upmu$-MUX) exploits the same working principle of FDM of coupling each TES to a resonant circuit with {a} unique resonance frequency and reading out each element in parallel by modulating and demodulating the signals produced by the TES at different frequencies. Such coupling, in~the $\upmu$-MUX scheme, occurs through an  {RF-}SQUID, one per each TES, that couples the current flowing through the TES to the resonance frequency of an LC circuit through local variations in  magnetic flux~\cite{cukierman2020microwave}. All the resonating circuits are then coupled to the same coplanar waveguide (CPW) transmission line for their readout through the first amplification stage through a High Electron Mobility Transistor (HEMT) amplifier. Furthermore, there is often present a  saw-tooth wave coupled to the RF-SQUID in order to up-convert the TES signal to higher frequencies in the 20 kHz range where the 1/f noise becomes negligible.  A~typical scheme for the implementation of microwave multiplexing can be found in Figure~\ref{fig:mu-mux}.

In such a scheme, a~multiplexing factor of up to 2000 can be achieved in the 4--8~GHz octave, where the 2 MHz spacing between adjacent frequencies is an optimum both for the readout electronics, and~the fabrication of the multiplexer chip. Each resonant circuit has an estimated bandwidth of about 100 kHz and a depth
of about 10--20 dB, which result in a quality factor of the resonators in the order of 40,000--80,000.  In~such a configuration, \citet{dober2021microwave} demonstrated that the readout only contributes to less than 5\% of the overall noise of the detectors with an input-referred current noise of 45 pA/$\sqrt{\text{Hz}}$ and a $1/f$ knee at about 20~mHz.

\subsubsection{Comparison} 
The previous section on different multiplexing techniques is rather difficult to follow without previous hands-on experience on TES detectors and quite some familiarity with SQUID amplifiers and transducers. Table~\ref{tab:multiplexing-comparison} summarizes the main characteristics of the different multiplexing schemes. We hope such a table may help the reader choose the best option for the multiplexing of their TES~array.
\begin{table}[H]
\setlength{\tabcolsep}{6.4mm}

   \caption{\textbf{Comparison between different multiplexing schemes}. 
   All values are intended to be typical unless differently~stated.}
    \label{tab:multiplexing-comparison}
    \begin{tabular}{lcccc}
    \toprule
         & \textbf{TDM} & \textbf{FDM} & \textbf{CDM} & \boldmath{$\upmu$}\textbf{-MUX} \\
         \midrule
       Complexity  & $\circ$ & $\circ\circ$ & $\circ\circ\circ\circ$ & $\circ\circ\circ$\\ 
       Cost & $\circ$ & $\circ\circ$ & $\circ\circ\circ$ & $\circ\circ\circ\circ$\\
       Aliasing & Yes & No & No & No\\
       Dead Time & Yes & No  & Yes & No\\
       Noise Level (pA/$\sqrt{\mathrm{Hz}}$) & 10  & 10 & 19 & 45 \\
       MUX factor & <128 & 32 & $\sim$100& 2000\\
       \bottomrule
    \end{tabular}
   
\end{table}

\subsection{Noise~Sources}\label{sec:noise}
The main thermodynamical noise sources that affect TESs are the following:  Johnson noise arising from both TES and {load} resistance; and thermal noise due to fluctuations in power transfer between the reservoir (at a temperature $T_R$) and the TES through the thermal link.  One further noise contribution is introduced by both the  cold (SQUID) and warm (amplification chain) readout~electronics.

In addition to these intrinsic sources, TES bolometers exhibit  photon noise, which arises from the fluctuation in the photon occupation number of the impinging radiation. Photon noise  solely depends on the impinging radiation and {not} on the detector or its experimental setup and therefore cannot be further improved upon.
For this reason, the desired noise level of a bolometer is comparable to or smaller than the photon noise of the source. A~detector that operates in this regime is said to be photon noise~limited.

For many practical applications, it is convenient to define the figure of merit commonly referred to as Noise Equivalent Power (NEP). NEP is defined as the signal power that must be applied to the TES in order to produce a signal which is equivalent to the measured noise. It can be calculated as the square root of the power spectral density (PSD) of the noise
\begin{equation}
    NEP(\omega) = \sqrt{S_P (\omega)} \quad {\left(\mathrm{W/\sqrt{\mathrm{Hz}}}\right)}.
\end{equation}
In the next few pages, for~the sake of clarity we will indicate the power spectral density $S_P(\omega)$ as $NEP^2(\omega)$. In~general, it is fair to assume that the different noise sources are independent and uncorrelated; therefore, we can add in quadrature the different contributions and define a total noise equivalent power $NEP_{tot}$ as
\begin{equation}
    NEP_{tot}= \sqrt{\sum_{i=1}^n NEP^2_i}.
\end{equation}

Following the arguments presented throughout this review, we want to make a distinction between the case of a TES used as a bolometer and as a calorimeter. In~the former, it is the low-frequency component of the noise $\left(\omega_{noise}\leq 1/(2\pi\tau_{eff})\right)$ that leaks into the signal band, whereas in the latter, the~signal has a much wider band and therefore the noise components at much higher frequency feed into the signal.
This is because the response function of a TES to an instantaneous absorption of energy by the calorimeter is formally identical to the noise associated with thermal fluctuations; therefore,  their PSDs are identical in shape~\cite{mccammon}.
While the performance of a bolometer is defined by its $NEP_{tot}$, the~performance of a calorimeter is defined by its energy resolution, i.e.,~the smallest variation in energy it can measure. The~maximum energy resolution achievable by a calorimeter, under~the assumption that the noise that affects it is stationary, can be computed through an optimum filter and it can be computed  following \citet{mccammon} as
\begin{equation}
    \delta E_{FWHM} =2\sqrt{2ln2}\left(\int_{0}^{\infty}\frac{4}{NEP^2_{tot}(f)}\,df\right)^{-1/2} \quad \left(\text{J}\right).
\end{equation}

\subsubsection{Thermodynamic Fluctuation~Noise}
This component is associated with white shot noise due to the discrete energy carrier flow in the thermal link between the reservoir and the TES. An~estimate of this noise can be made by considering the average energy fluctuation associated with a thermal capacity $C$ in equilibrium with a thermal reservoir with a temperature $T_{R}$, $\langle\Delta E^2 \rangle =k_B T^2 C$. Since shot noise is white, we can calculate the energy fluctuation as the integral over all frequencies of noise spectral amplitude by the response function of the weak conductive link and the thermal mass assembly (a low-pass filter with cutoff frequency, $f_c=1/\tau$)  \cite{richards1994bolometers}. Thus, we obtain that the power fluctuations propagating through the thermal link exhibit a spectral density given by
\begin{equation}
    NEP_{tfn}^2\simeq 4 k_B T^2G \quad \left(\text{W}^2/\text{Hz}\right).
\end{equation}
This equation only stands if the reservoir and detector are in thermal equilibrium. Outside of equilibrium,  it is necessary to consider the thermal gradient across the link; therefore, the two limiting cases have been found considering the two extreme cases whereby the energy carriers have a mean free path which is  significantly larger than the length of the thermal link or much smaller.
In the former case, according to \citet{boyle1959performance}, the~NEP can be computed as
\begin{equation}
    NEP_{tfn}^2=4k_BT^2_RG_0\frac{t^{\beta_G+2}+2}{2} \left(\text{W}^2/\text{Hz}\right),
\end{equation}
\noindent
where $t= T/T_R$, $G_0$ is the thermal conductance at T = 0K, $G=|\partial P_{link}/\partial T | \propto T^{\beta_G}$ \cite{mccammon}.
In the latter, \citet{mather1982bolometer} derived
\begin{equation}
        NEP_{tfn}^2=4K_BT^2_RG_0\frac{\beta_G+1}{2\beta_G+3}\frac{t^{2\beta_G+3}+2}{t^{\beta_G+1}-1} \quad \left(\text{W}^2/\text{Hz}\right)
\end{equation}
with the aforementioned~nomenclature.

\subsubsection{Johnson~Noise}
Johnson noise~\cite{johnson1928thermal} is due to both voltage fluctuations across the bias resistor and across the TES. For~a general resistor,  these voltage fluctuations exhibit a power spectral density given by~\cite{perepelitsa2006johnson}
\begin{equation}\label{johnson}
    S_V=4k_B T R \quad \left(\text{W}^2/\text{Hz}\right).
\end{equation}
Since the TES is in NTEF, we must consider its effect when discussing the effects of both sources.
In the case of voltage fluctuations generated by the bias resistor, the~TES produces current signal as a response to the voltage change according to the electrical admittance of the TES~\cite{irwin2005transition}
\begin{equation}
    Y_{ext}=S(\omega)I_0\frac{\mathcal{L}-1}{\mathcal{L}}\left(1+i\omega\tau_I\right) \quad \left(\text{A/V}\right),
    \label{eq:admittance}
\end{equation}
\noindent
where $S(\omega)$ is the responsivity of the TES, $\mathcal{L}$ is the loop-gain, and~the time constant of a current-biased TES reads~\cite{irwin2005transition}
\begin{equation}
    \tau_I=\frac{\tau}{1-\mathcal{L}}. 
\end{equation}

The NEP can be calculated by multiplying the voltage Johnson noise of the bias resistance by Equation~(\ref{eq:admittance}) and dividing it by the responsivity of the TES according to \citet{irwin2005transition}
\begin{equation}
    NEP_{P_{R_{bias}}}^2(\omega)=4k_B T_{bias} I_0^2 R_{bias}\frac{\left(\mathcal{L}-1\right)^2}{\mathcal{L}^2}(1+\omega^2\tau_I^2) \quad \left(\text{W}^2/\text{Hz}\right),
    \label{eq:noise2}
\end{equation}
\noindent
where $T_{bias}$ is the electronic temperature of the bias impedance.
Whereas the voltage Johnson noise arises across the resistance of the TES, the~ Joule power generated is dissipated differently through the TES with respect to the previous scenario; hence,  \citet{irwin2005transition} computed the Johnson NEP as
\begin{equation}\label{tesjon}
    NEP_{P_{TES}}^2(\omega) =4k_B T I_0^2 R_{TES} \frac{\xi(I_0)}{\mathcal{L}_I^2}(1+\omega^2\tau^2) \quad \left(\text{W}^2/\text{Hz}\right).
\end{equation}
In addition, in~order to account for the non-linearity in the voltage drop across the superconductor which is due to  a $\Delta R$ induced by a change in the current flowing through it, we must include a factor ($\xi$)
\begin{equation}
    \xi(I)=1+2\beta_I,
\end{equation}
\noindent
where $\beta_I$ is the logarithmic derivative of the TES resistance with respect to the current flowing through it and  defined as $\beta_I = \frac{dR}{dI}\frac{I}{R}$.

\subsubsection{Readout~Noise}
The main contribution to the noise in the readout chain usually arises from the SQUID amplifier. For~most commercial applications, the~current equivalent noise at its input coil $i_{n_{SQUID}} \quad (\text{A}/\sqrt{\text{Hz}})$ is provided by the manufacturer. In~order to obtain its NEP, we can divide $i_{n_{SQUID}}$  by the detector responsivity
\begin{equation}
    NEP^2_{SQUID}=\frac{i_{n_{SQUID}}(\omega)^2}{S(\omega)^2} \quad \left(\text{W}^2/\text{Hz}\right).
\end{equation}
If well designed, the~noise produced by the  cold-amplification stage dominates over the noise introduced by any other element further down the readout chain. Typically, the~voltage noise $e_n$ of an operational amplifier operating at room temperature is of the order 1 nV/$\sqrt{\text{Hz}}$. Such an operational amplifier exhibits a current noise $i_n \sim$ pA$/\sqrt{\text{Hz}}$ coming out of its input terminals. If~the SQUID exhibits a dynamic resistance $R_{dyn}\sim 100 \;\Omega$ (which is typical for SQUID arrays with gain of a few hundred V/A), the~interplay between the noise current and the dynamic resistance of the SQUID becomes non-negligible and it induces a voltage drop across the SQUID amplifier. Under~these conditions, the NEP generated from the warm readout becomes
\begin{equation}
    NEP_{wr}^2(\omega)\simeq \frac{(e_n(\omega)^2 + i_n(\omega)^2 R_{dyn}^2)}{Z_{tran}^2 S(\omega)^2} \quad \left(\text{W}^2/\text{Hz}\right).
\end{equation}

\subsubsection{Photon~Noise}
As previously discussed, the~photon noise can be regarded as a noise floor which is only dependent on the detected radiation and not on the detector nor on its readout chain. It sets a fundamental limit to the sensitivity of any detector. It includes two terms~\cite{Lamarre:86}
\begin{equation}
    NEP_{opt}^2=2h\nu P_{opt} + \zeta \frac{ P_{opt}^2}{\delta\nu} \quad \left(\text{W}^2/\text{Hz}\right),
\end{equation}
\noindent
where $\nu$ is the band center of the detected radiation, $\delta\nu$ is the detection bandwidth, $P_{opt}$ is the total optical power and $\zeta$ is called the bunching factor~\cite{shirokoff2011south} and ranges between 0 and 1 and its nature will be discussed~later. 

The first term is simply due to the photon shot noise and derives from the simple counting of the number of photons with given energy per unit time, whereas the second term arises from the fact that photons are bosonic in nature and tend to distribute their energies according to Bose--Einstein's distribution. In~general, it is possible to state, with~good measure, that $0.3\leq\zeta\leq0.6$ \cite{shirokoff2011south}.
\subsubsection{Excess~Noise}
In the last few years, a number of research groups have reported noise figures with components that exceed the ones we have just discussed. In~most cases, such excess noise appears at high frequency, in~the proximity of the roll-off frequency of the TES~\cite{shirokoff2011south}. Such components appear to be correlated with the steepness of the superconducting-to-normal transition and the a-dimensional derivative $\alpha$ and it seems to improve if $\alpha$-mitigation strategies are put in place~\cite{ullom2004characterization,jethava2009dependence}. Furthermore, there is anecdotal mention of this effect being exacerbated by inhomogeneities in the superconducting film, but~the published data are  scarce~\cite{shirokoff2011south}. 

{Excess noise in X-ray calorimeters has been widely studied since \citet{galeazzi2010fundamental} and can be attributed to two different effects:}
\begin{itemize}
    \item \textbf{Flux Flow Noise} {is due to trapped magnetic field lines that create a vortex in the superconductor. A~vortex is free to move on the surface of the superconductor and while doing so, it generates a voltage drop. {The noise term arising from flux flow can be written as $\sqrt{\phi_0\frac{I R_{ff}}{1+\frac{IR}{V_0}}}$ where $\phi_0$ is the magnetic flux quantum, $I$ is the current flowing through the TES $R_{ff}$ and $V_0$; the latter represents the  voltage above which  the noise saturates ($V_0$), whereas the former is only a coefficient with units of Ohms.}}
    \item \textbf{Internal Thermal Fluctuation Noise}  {arises from the finite thermal conductance in the TES film. Different segments of the film may exhibit a different temperature simply because of geometrical and physical properties of the TES and the point where the TES is struck. These thermal fluctuations can lead to an extra noise component.  {Accounting for internal thermal fluctuations,  noise allows one  to scale the simple one-conductance approximation of a TES to a more realistic description of the detector.}}
\end{itemize}

{I}t is worth mentioning that in addition to this excess noise, there is a well known effect that arises in a TES when the heat capacity of the thermal link to the reservoir becomes comparable with the heat capacity of the TES itself. An~in-depth discussion can be found in \citet{gildemeister2001model}.

\subsubsection{Noise Comparison between Different Multiplexing~Schemes}
It is obvious that different multiplexing schemes, such as those presented in Section~\ref{sec:mux}, exhibit different noise levels and some extra contributions which are inherent to the multiplexing architecture may appear. In~this section, we will address the aforementioned noise sources and we will discuss how they interplay with different readout~schemes.

In general, wide-band noise from the SQUID is added to the signal after it is encoded and such noise is then filtered out during decoding~\cite{irwin2005transition}. The~noise that is added during the encoding strongly depends on the bandwidth of the encoded~signal.
\begin{itemize}
    \item \textbf{Time Division Multiplexing:}
    {In TDM, the encoding occurs through a modulation of the signal with a boxcar that represents the switching between different TES channels. Each pixel is read out for a `time-frame' which has a rate of $1/N\delta t_s$, where $N$ is the multiplexing factor and $\delta t_s$ is the time span for which the multiplexer dwells on one pixel~\cite{irwin2005transition}.
    
    The noise components with a frequency above the Nyquist frequency associated with the frame rate $f_{Low}=1/(2N\delta t_s)$ is embedded in the signal. Any noise above the boxcar Nyquist frequency is aliased and averages out $f_{High}=1/(2 \delta t_s)$. The~NEP of the SQUID is thus increased by a factor $N$ and therefore the gain of the SQUID needs to be tuned to be $N$ times larger than that of a non-multiplexed SQUID in order to retain the same {SNR} \cite{irwin2005transition}.}
    \item \textbf{Frequency Division Multiplexing:}
    {In an FDM scheme, two phenomena can occur: \mbox{(i) noise} sources can modulate the amplitude of the carrier tones and (ii) noise leakage in the band of the carrier tones. Category (i)  includes photon noise and thermal fluctuation noise. These sources  are accounted in an analogous fashion to  a single TES biased in DC. Category (ii)  includes Johnson noises and readout noises. The~current signal that these sources contributes must be  multiplied by a factor $\sqrt{2}$ as shown by \citet{dobbs2012frequency} in order to take into account the AC bias of the TES. Similarly, the~responsivity of the detector also includes a $\sqrt{2}$ factor (Equation (\ref{fdmsense})) which leaves the NEP  unchanged.
    Furthermore,  \citet{lueker2009thermal}  found that  the effects of NETF on Johnson noise sources when the TES is DC biased  (Equations (\ref{eq:noise2}) and (\ref{tesjon})) also apply when the TES AC was biased (besides from the aforementioned $\sqrt{2}$ factor).}
    \item \textbf{Code Division Multiplexing:} {In a CDM scheme, the~TESs are all effectively Frequency Domain Multiplexed, therefore the same consideration apply except their outputs are split and sent to the input coils of several SQUIDs with different polarities.}  
    \item \textbf{$\upmu$-MUX:} {Effectively, for~$\upmu$-MUX, all the previous considerations observed  about FDM TESs hold true. Except, the~SQUID noise component needs to be replaced by the far more detrimental noise figure of a HEMT~\cite{zmuidzinas2012superconducting,de2023limitations} and the NEP of an RF-SQUID~\cite{yoshihara2003rf} needs to be added in quadrature to the total NEP $NEP_{tot}$. } 
\end{itemize}

\subsection{Cross-Talk}\label{sec:cross-talk}
Cross-talk is a known issue of detector arrays where multiple pixels are read out simultaneously. A~large number of mitigation strategies can be put in place, but~overall inter-pixel cross-talk cannot be fully eradicated. In~{FDM}, and~likewise for CDM, {a} cross-talk occurs between the frequency channels. Different types of cross-talk arise from the LC filter configuration. Four major cross-talk effects can be measured  according to \citet{Dobbs_multiplexing} and \citet{xtalkmumux1}: the electro-magnetic coupling between the inductors, the~coupling between the oscillators, the~carrier current leakage and the common impedance.

The first term is related to the mutual inductance coupling $M_{ij}=k_{ij}L_iL_j$ between two inductors of self inductance $L_i$ and $L_j$. The~current flowing in the i-th leg $I_i(\omega_i)$ can induce a voltage drop $V_j=\omega_iM_{ij}I_i(\omega_i)$ at the ends of a coupled inductor~\cite{dobbs2012frequency}. This term can be eliminated or optimized by maximizing the physical distance between LC resonators which are neighbours in frequency~space.

\begin{figure}[H]
    
    \includegraphics[width=\linewidth]{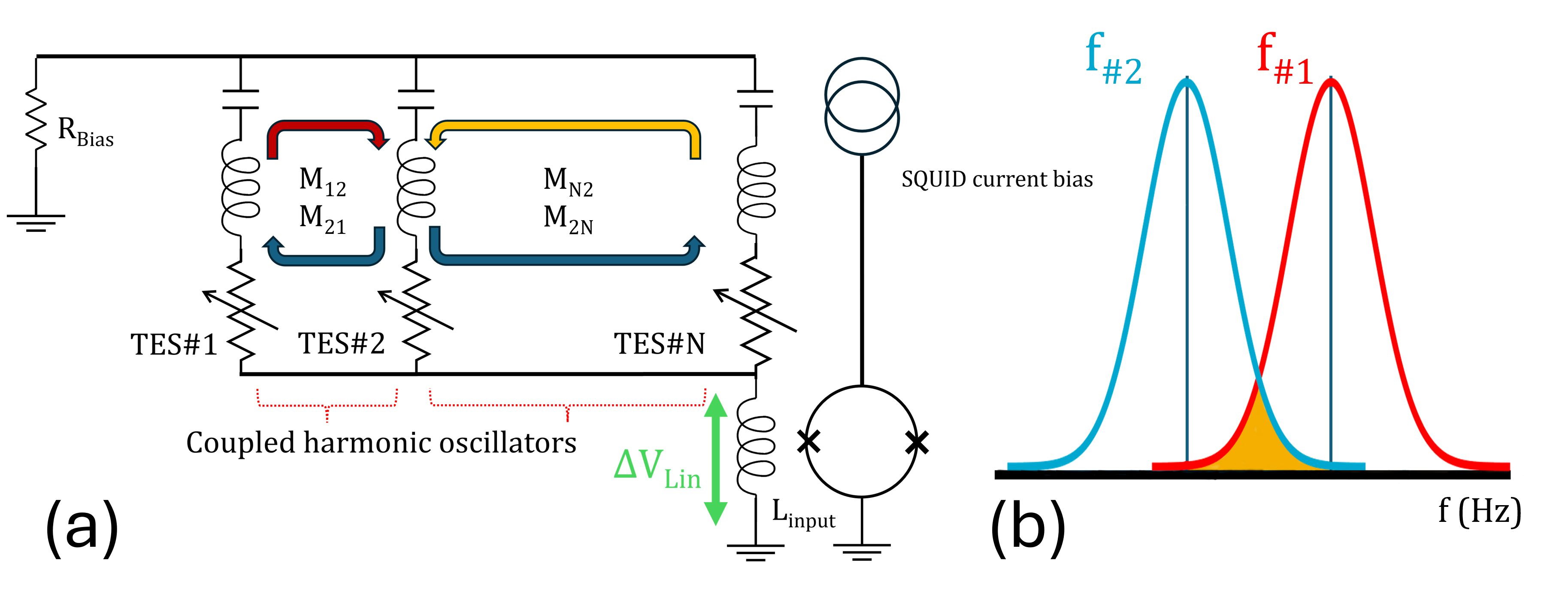}
    \caption{\textbf{Cross-talk scheme in FDM.}
     (\textbf{a}) Schematic of  $N$ frequency division multiplexed TESs with the different cross-talk components: (i) the mutual inductance between different channels, (ii) the coupled oscillators, (iii) the voltage drop due to the shared input impedance. (\textbf{b}) The cross-talk term due to to the overlap of the Lorentzian -- shown as Gaussians for simplicity. The~area shaded in orange represents the~cross-talk.}
    \label{fig:xtalk}
\end{figure}

One further cross-talk mechanism is due to the physics of tightly packed weakly coupled oscillators~\cite{noroozian2012crosstalk,hirayama2016interchannel} and has to do with the hybridization of the individual resonance frequencies which through this interaction repel each other. Further details are discussed by \citet{xtalkmumux1}, who evaluate this cross-talk component as
\begin{equation}
    \chi = \frac{16 f^4}{(f_2 - f_1)^2}\frac{M_{12}^2}{Z_0^2}\;,
\end{equation}
\noindent
where $f_1$ and $f_2$ are the two resonance frequencies of the two RLC channels, which are rather close in frequency; hence, $f \sim f_1 \sim f_2$, $M_{12}$ is the mutual inductance and $Z_0$ is their impedance. 
This cross-talk effect scales with the square of the
magnetic coupling $M_{12}$ and inversely with the square of the frequency separation. Therefore, it can be mitigated by both separating the coupled oscillators both in space and in frequency.
The carrier current leakage is an unavoidable form  of cross-talk. It occurs because two LC filters, in~frequency space, are described by two resonance peaks (described by two Lorentzian curves, see Figure~\ref{fig:xtalk}b), the~tails of which necessarily overlap. The~overlap of such ends represents the fraction of electrical power that the two resonators share. This cross-talk component describes the  modulation of the current flowing through one single RLC channel when the resistance in one of the frequency-neighbouring branches varies. This effect can only be mitigated by only increasing the frequency distance between the multiplexing frequencies within the limits given by the bandwidth. For~a generic j-th branch, the off-resonance current is defined as
\begin{equation}
    I_j(\omega_i)=\frac{V_{bias}(\omega_i)}{R_{TES,i}+i\omega_iL_j+1/(i\omega_iC_j)}\;.
    \label{eq:overlap}
\end{equation}
The cross-talk is due to the variation in this current $I_j(\omega_i)$ when the resistance $R_i$ varies: $\Delta I_j(\omega_i)/\Delta R_i$ \cite{dobbs2012frequency}. The~off-resonance component is infinite, in~the ideal scenario, but~for most practical applications it needs to be considered finite, and~therefore the current $I_j$ described by Equation~(\ref{eq:overlap}) is~finite. 
The common impedance term is linked to the presence of the wiring and of a preamplifier such as a SQUID. The~modulation of the current at the frequency $\omega_i$ causes the modulation of the current flowing through the input coil of the SQUID, which in turn causes the modulation of the current in all the neighbouring branches~\cite{dobbs2012frequency}.

In a rather similar fashion, the inter-pixel cross-talk in $\upmu$-MUX multiplexing schemes deals with the electromagnetic coupling between different harmonic oscillators. In~$\upmu$-MUX schemes,  some of the same coupling mechanisms already discussed for FDM apply: the widely discussed mutual electro-magnetic coupling between the inductors of different channels, the~innate effects of the coupled harmonic oscillators and~the overlap of the Lorentzian tails of the resonance shapes. In~addition to these, we must consider the non-linearity of the microwave components which act on the superposition of the channel tones and result in a weak mixing of the tones~\cite{xtalkmumux1}. In~particular, it is worth stating that while other cross-talk components fall off quite quickly with physical distance and frequency spacing, this last cross-talk component does not and {therefore it} sets an effective cross-talk floor. In~principle, it can be mitigated by an accurate selection of the microwave components, and~by driving the resonators at lower powers in a trade off between {SNR} and cross-talk. Potentially,  a tone-tracking readout~\cite{dober2021microwave} could also reduce the effects of such a cross-talk component~\cite{groh2024crosstalk}.

As a rule  of thumb, it is good practice to space the resonators in frequency by at least 10 times their line width in order to effectively reduce the effect of the overlap of the Lorentzian tails~\cite{xtalkmumux1}. The~final contribution to cross-talk that still needs to be addressed is the contribution arising from capacitive and inductive coupling between different transmission lines which carry the FDM signals. Unfortunately,  these strongly depend on a case-by-case basis and are severely influenced by the geometry of the problem. {Generally}, it is always good practice to keep the transmission lines as far apart as possible and include a ground plane in between each two different transmission~lines.

\section{Applications in Astronomy and~Astrophysics}\label{sec:Applications}
Despite that the original development of Transition Edge Sensor micro-calorimeters was not intended specifically for astronomical purposes, TESs find a plethora of applications in the field of astronomy: ranging from high-energy $\gamma$-ray astronomy to cosmological application at extremely low energies such as the investigations of the polarization modes of the cosmic microwave background (CMB).
Their large bandwidth coupled with the excellent energy resolution make TES promising detectors for most applications. Furthermore, TESs can be  optimized in order to be single-photon sensitive across a wide spectrum and can also be  operated as bolometers (coupled with lenses, feed-horns or antennas) at lower energies. In~this section,  we will discuss some of the most prominent applications of TESs in the field of~astronomy.

\subsection{X-Ray Astronomy}%MDPI: section order number added, please confirm, and please pay attention to the change of the order below.

\subsubsection{ESTREMO/WFXRT}
ESTREMO/WFXRT was a proposed mission for the ESA Cosmic Vision Programme 2015--2025~\cite{piro2006estremo, piro2007future}. This was a spacecraft mission for observation of very high-energy events such as gamma-ray bursts (GRBs) which often also exhibit an X-ray afterglow which can be detected directly from the ESTREMO/WFXRT detectors. The~importance of GRBs is that they can be studied to extract information on the first population of luminous sources ignited in the dark universe at high redshift $z$ > 7. Furthermore, GRB  measurement can help determine the cosmic history of metals in star-forming regions with metal absorption and edges while also allowing the probing of the Warm-Hot Interstellar Medium (WHIM) properties through high-resolution absorption studies. Observing WHIM can allow the identification of the missing baryonic matter. Dark Matter (DM) baryons are heated up to X-ray-emitting temperature $10^5$--$10^7$ K. Thus, X-ray observations play a fundamental role in the indirect measurement and characterization of~DM.

While the configuration of the payload and the instruments on board the ESTREMO/ WFXRT were never finalized, two main instruments were envisioned: the Wide Field Instrument (WFI) and the Narrow Field Instrument (NFI). The~first of which spans energies between 2 and 200~keV in order to catch X-ray Flashes (XRF) and galactic transients and supposedly would have developed CdZTe detectors. The~NFI, instead,  would have either deployed only TESs or a mix of Charge Coupled Devices and TESs. In~any case, the~desired energy resolution of such TESs would have been of 1 eV at 1~keV.

\subsubsection{Athena~X-IFU}
The Advanced Telescope for High-Energy Astrophysics (\textit{Athena}) is an ESA space-borne X-ray observatory in the energy range 0.2--12~keV~\cite{barrett2016athena}. The~main scientific goals of \textit{Athena} involve the study of highly energetic processes in space such as the physical and chemical properties of hot plasmas, such as those found in galaxy clusters and the study of black hole accretion disks with their jets, outflows and winds. It will also be able to perform measurements on exoplanets, stars, supernovae and interstellar~mediums.

\textit{Athena} will exhibit an angular resolution of 5'' for its two main instruments the Wide-Field Imager~\cite{meidinger2018development} and the X-ray Integral Field Unit (X-IFU) \cite{barrett2016athena}. \textit{Athena} X-IFU is a spatially resolved spectrometer with a field of view of 5' and an energy resolution of 2.5 eV at 10~keV~\cite{barret2023athena}.  The~focal plane consists of 3840 Mo/Au TESs~\cite{khosropanah2018development} with an Au/Bi absorber with an expected NEP below $10^{-17}$~W/$\sqrt{\mathrm{Hz}}$. Furthermore, located 1 mm underneath the main TES array, there is an Ir/Au TES-based cryogenic anti-coincidence detector (CryoAC) \cite{d2022athena} to reject the non X-ray-induced events to a rate below $5 \times 10^{-3}$ counts/s/cm$^{2}$/keV in the 2--10~keV range. The~X-IFU TESs are multiplexed in an FDM scheme with a multiplexing factor 40:1~\cite{durkin2019demonstration} allowing for the complete readout of the TES array with 96 channels~\cite{d2022athena}.  

\subsubsection{Lynx X-Ray~Observatory}
The Lynx X-ray Observatory (\textit{Lynx}) is a NASA-funded concept study for future missions with a proposed launch date of  2036~\cite{bandler2019lynx}. If~launched, \textit{Lynx} will have similar scientific goals to {\emph{Athena}}. It will, in~fact, investigate the birth and the evolution of supermassive black holes, observing in the X-ray high-redshift black holes (z $\sim$ 10) with a mass a few thousand times that of our sun. Furthermore, \textit{Lynx} will allow a large survey of black holes at redshifts $z$ = 2--6 to investigate whether all massive black holes emerge at $z \leq$ 6 or whether some low-mass black holes have developed at lower redshift~\cite{gaskin2019lynx}.

\textit{Lynx} will be capable of addressing several open questions in the field of star formation. In~conjunction with infra-red data from other telescopes, \textit{Lynx} with its census of star-forming regions in the X-rays will contribute to a better understanding of circumstellar grain growth, and~will help investigate the magnetic and non-gravitational effects in proto-planetary discs. Furthermore, X-ray imaging spectroscopy will revolutionize the field of Supernovae Remnants science,  providing three-dimensional spectroscopic images of metals synthesised in the explosions. The~full list of scientific goals of \textit{Lynx} can be found in its concept study paper~\cite{gaskin2018lynx}.

The Lynx X-ray Observatory will have a primary mirror with a diameter of 3 m and will be equipped with three main instruments: the High-Definition X-ray Imager (HIXI) \cite{falcone2019overview}, the~\textit{Lynx} X-ray Microcalorimeter (LXM) \cite{bandler2019lynx} and the X-ray Grating Spectrometer (XGS) \cite{mcentaffer2019reflection}. While HIXI is a wide-field high imager with high spatial resolution, the~spectroscopic measurements are performed with the LXM (with a resolving power R$\sim$2000 in the \mbox{0.2--7~keV range}) and the XGS (R$\sim$5000--7500 in the  0.2--2~keV range). In~particular, the~\textit{Lynx} X-ray Microcalorimeter will consist of a 100000 TES array with energy resolution of 2 eV up to 7 eV and 0.3 eV in the 0.2--0.75~keV range~\cite{smith2020toward}.  

\subsection{Infrared Astronomy}

\subsubsection{Origins Space~Telescope}
The NASA flagship mission Origins Space Telescope (\textit{Origins}) proposed for launch in 2035 is the largest spacecraft-borne mission so far with a mirror size of 5.9 m in diameter and a spectral sensitivity ranging from 2 $\upmu$m to 588 $\upmu$m~\cite{leisawitz2021origins}. The~current mission goals for \textit{Origins} \cite{battersby2018origins} include the study of galaxy formation, planet formation and the evolution of supermassive black holes. In~particular, the~scientific interest lies in the generation of metals, dust and organic molecules and how pulsars and supernovae affect the interstellar medium. A~second topic of interest is planet formation and the development of habitability: in particular for Earth and Earth-like exoplanets, how were water and other life's ingredients delivered to such planets? Finally, \textit{Origins} aims to perform transition spectroscopy of K- and M-Dwarf planets to investigate which, if~any, exhibit the conditions necessary for habitability. \textit{Origins} is expected to produce a broader and clearer picture on dwarf planet habitability than JWST~\cite{battersby2018origins}.

\textit{Origins} will be equipped with three main instruments: the Origins Survey Spectrometer (OSS) \cite{bradford2021origins} which will allow the observation of emission lines between 25 $\upmu$m and 588 $\upmu$m and probe galaxy evolution up to $z\sim8.5$. By~design, OSS requires detectors with an NEP $\sim 3\times10^{-20}$~W/$\sqrt{\mathrm{Hz}}$ and a saturation power of 0.2 fW. The~Far-infrared Imager and Polarimeter (FIP) \cite{meixner2020origins} will perform wide field polarimetric observations of astrophysical objects bridging the energy gap between ALMA {(Atacama Large Millimeter/submillimeter Array)} and JWST. Its detectors are less challenging with a required NEP $\sim 3\times10^{-19}$~W/$\sqrt{\mathrm{Hz}}$ and a saturation power of 20 fW. Finally, the~Mid-Infrared Spectrometer Camera Transit (MISC-T) \cite{roellig2020mid} will be responsible for the transit spectroscopy of exoplanets including K- and M-Dwarf planets. MISC-T will have a resolving power $R<300$ and for it TES in both the calorimetric and bolometric use are being investigated despite a strong suggestion towards their use as calorimeters~\cite{nagler2021transition}.

\subsubsection{SPICA-SAFARI}
SPICA (Space Infrared Telescope for Cosmology and Astrophysics) was a medium-class spacecraft mission proposed to ESA and cancelled in 2021. Like the Origins Space Telescope, SPICA intended to bridge the gap in energies between the observational capabilities of ALMA and JWST~\cite{Roelfsema_2018}. SPICA was intended to perform spectroscopy between 20 $\upmu$m and 230 $\upmu$m. Its scientific goals were aligned with the proposed science that will be performed with \textit{Origins}: galactic formation and evolution, black hole evolution and studies on metallicity in dust-covered galaxies and active galactic nuclei. SAFARI, the~Spica FAR-infrared instrument~\cite{goicoechea2009exoplanet}, was proposed as a spectrometer between 30 $\upmu$m and \mbox{210 $\upmu$m} with an NEP $\leq 3\times10^{-19}$ W$/\sqrt{\mathrm{Hz}}$ with a saturation power of 1 fW~\cite{mauskopf2010tes}. SAFARI was intended to sport 3600 FDM~\cite{audley2022optical} Ti/Au TESs with a critical temperature of 155 mK suspended on a 1 $\upmu$m SiN membrane~\cite{khosropanah2016ultra}. Despite substantial progress being made toward the development of ultra-low noise TESs~\cite{khosropanah2010low,khosropanah2016ultra}, ESA decided not to move forward with SPICA as its next medium-class~mission.

\subsection{Millimetre and Sub-Millimetre Astronomy}

\subsubsection{SCUBA-2}
SCUBA-2 (Submillimetre Common-User Bolometer Array 2) is a dual-band TES camera operating at 450 and 850 $\upmu$m with 10000 TES bolometers on the focal plane read out in a TDM scheme~\cite{walton2004design}. The~Transition Edge Sensors, developed by NIST, are Mo/Cu with a critical temperature of 130 mK for the TESs in the 850 $\upmu$m band and, in~order to account for higher optical powers in the sky, the~TESs in the 450 $\upmu$m band were designed with a critical temperature of 190 mK in mind. All the sub-assemblies of SCUBA-2 exhibited a dark NEP   in the range 1.6--3.2 $\times 10^{-16}$~W/ $\sqrt{\mathrm{Hz}}$ and a saturation power ranging between 87 and \mbox{541 pW~\cite{audley2004scuba,bintley2014scuba}}.

Ground-based sub-mm observations are meant to explore the cold universe by investigating dust and gases in the early stages of galaxy, star and planet formation. At~such wavelengths, the dust continuum is effectively optically thin and therefore it is possible to observe the critical processes that occur during 
galaxy/star/planet formation. The~commissioning of the instrument, deployed on the James Clerk Maxwell Telescope, involved the  characterization of selected galaxies with SCUBA-2 at both 450 and 850 $\upmu$m and   was reported by \citet{casey2013characterization}. A~review with the main scientific achievement of SCUBA-2 in its first decade (2011--2021) was published by~\cite{mairs2021decade}.
In 2023, SCUBA-2   performed a full extra-galactic survey at 850 $\upmu$m~\cite{garratt2023scuba} with an unprecedented angular resolution of \mbox{14 arcseconds},  which is a survey over three orders of magnitude more detailed compared to the previous survey at the same wavelengths (COBE) \cite{bennett1993scientific}. 

\subsubsection{GISMO}
The Goddard IRAM Superconducting Millimeter Observer (GISMO) was a guest instrument at the IRAM 30m telescope operating between 2012 and 2014. GISMO consisted of one single TES focal plane with  128 pixels (8 × 16) in a TDM configuration~\cite{staguhn2008instrument}. The~TESs that constituted the focal plane of GISMO were designed with an operating temperature of 400 mK that could be reached in the $^3$He/$^4$He evaporation cryostat and exhibited an NEP $ 4\times10^{-17}$~W/$\sqrt{\mathrm{Hz}}$. GISMO also exhibited a bandpass filter with 25~GHz bandwidth about its maximum at 150~GHz (2 mm). The~GISMO 2 mm survey~\cite{magnelli2019iram} provided valuable information on the bright end of the infrared luminosity function and the massive end of the dust mass function at $z \sim$ 4 for rare massive high-redshift highly star-forming galaxies. The~work GISMO carried out paved the way for the observations of ALMA 2 mm continuum survey~\cite{casey2021mapping}.

\subsection{Cosmic Microwave Background}

\subsubsection{CLOVER}
CLOVER was a project funded by the Science and Technology Facilities Council (STFC) of the United Kingdom and lead by the universities of Cambridge, Cardiff, Manchester, and~Oxford~\cite{giard2005clover,taylor2004clover,north2007clover}. Its main goal was the observation of primordial B-mode polarization of the CMB radiation with a sensitivity equivalent to a tensor-to-scalar ratio $r \geq 0.01$. CLOVER consisted of three separate instruments operating at 97, 150 and 220~GHz to provide discrimination between the CMB and foregrounds~\cite{piccirillo2008clover}. Each instrument would have deployed NIST TES polarimeters coupled to feed-horns which act as band-pass filters and focusing elements. Each polarimeter consists of a Mo/Cu Transition Edge Sensor with an NEP of {1.5$\times 10^{-17}$}, 2.5$\times 10^{-17}$, and~4.5 $\times 10^{-17}$~W/$\sqrt{\mathrm{Hz}}$, respectively, in each band~\cite{audley2008performance,audley2008microstrip}. Despite its ambitious goals and promising science goals, the~project was not funded to~completion.

\subsubsection{Atacama Cosmology~Telescope}
The Atacama Cosmology Telescope {(ACT)} was a telescope in the Atacama desert in Chile specifically designed for cosmological observations of the CMB operational between 2007 and 2022~\cite{fowler2004atacama}. During~its life span, it sported three main instruments, the Millimetre Bolometer Array Camera  (MBAC)  \cite{swetz2008instrument,marriage2006testing} and its two upgrades, ACTPol~\cite{niemack2010actpol} and Advanced ACTPol (AdvACT) \cite{simon2018advanced}. ACT has provided the scientific community with extremely important results, among~which are  the~first detection of gravitational lensing in a CMB map  by \citet{das2011detection} and the discovery of the El Gordo galaxy cluster (ACT-CL J0102-4915) \cite{marriage2011atacama}.
All instruments have  ACT mounted Transition Edge Sensors as the sensitive element. MBAC comprises three TES arrays in the frequency bands 145, 220 and~280~GHz. Each array contains 32 $\times$ 32 detectors read out in a TDM scheme. The~superconducting elements of the detectors are Mo/Au bilayers tuned so as to achieve a critical temperature of 0.5 K and a measured NEP below 8$~\times~10^{-17}$~W/$\sqrt{\mathrm{Hz}}$ \cite{zhao2008characterization}.

The first upgrade to MBAC, ACTPol, consists of three polarization-sensitive arrays, PA1 and PA2 which operate at 148~GHz, while PA3 operates both at 97 and 148~GHz~\cite{niemack2010actpol}. Each polarimeter  is coupled to a feed-horn, except~for PA3 which has 4 polarimeters per feed-horn, one in each band. Combined, there are 1279 feed horns and 3068 detectors in the ACTPol instrument. Each TES is constituted of a Mo/Cu bilayer with a critical temperature of 150 mK and an NEP below 2.0 $\times 10^{-17}$~W/$\sqrt{\mathrm{Hz}}$ and a saturation power of about 10 pW~\cite{grace2014characterization}.

The final upgrade to ACTPol, AdvACT,  produced a map of the sky in five bands between 28 and 230~GHz~\cite{louis2017atacama,choi2018characterization}. The~AdvACT TESs are fabricated from single-layer Al/Mn films~\cite{westbrook2024thermal} instead of bilayer Mo/Cu films as in ACTPol with a critical temperature of 160 mK and an NEP below  2.0$~\times~10^{-17}$ $~W/\sqrt{\mathrm{Hz}}$ \cite{li2016almn,crowley2018advanced,niemack2010actpol}.  
The final constraint set by the ACT on the tensor-to-scalar ratio was published by \citet{galloni2023updated} and it currently is $r\leq0.03$. 

\subsubsection{ABS}
The Atacama B-mode Search (ABS) was both an experiment and an instrument operating at 145~GHz~\cite{simon2014characterization}. As~per its name, it was deployed in the Atacama desert of Chile and its main purpose was the measurement of B-mode polarization of the CMB and  especially the determination of the tensor-to-scalar ratio of the primordial B-mode components. The~receiver of ABS consists of 240 feed-horn coupled polarimeters~\cite{appel2012detectors}. Each polarimeter consists of a planar orthomode transducer (OMT) that couples the orthogonal polarization components of light to two different TESs for detection and measurement~\cite{essinger2009atacama}. The~detectors are produced by NIST and are designed to operate at 300 mK in a $^3$He/$^4$He adsorption refrigerator.  The~upper limit ABS set for the tensor-to-scalar ratio was r $\leq2.3$ with a 95\% confidence level~\cite{kusaka2018results}. 

 \subsubsection{BICEP}
The Background Imaging of Cosmic Extragalactic Polarization (BICEP) along with the Keck Array are a series of experiments on the CMB polarization deployed in Antarctica~\cite{keating2003bicep}. During~the years, the experiments have evolved and were upgraded from  BICEP to BICEP2~\cite{ade2014bicep2}, the~Keck Array~\cite{ade2015bicep2}, BICEP3~\cite{karkare2014keck} and~eventually the BICEP Array~\cite{hui2018bicep}.
The first generation of BICEP (2006--2008) used NTD Ge thermistors as detectors~\cite{haller1994advanced}, but~the choice quickly moved to TESs starting from BICEP2 (2010--2012), which sported 500 Ti TESs coupled to 150~GHz radiation. The~TESs exhibited a critical temperature of about 450--500 mK and an NEP of 5--6 $~\times~10^{-17}$~W/$\sqrt{\mathrm{Hz}}$ \cite{brevik2010initial}. During~the lifespan of BICEP2, the~Keck Array which consists of five BICEP2-class receivers sharing the same mount was developed. The~Keck Array is also sensitive to radiation in different frequency bands (95, 150  and~220~GHz). The~data from the Keck Array alone set an upper limit to the tensor-to-scalar ratio  r < 0.09 at 95\% confidence {level}, which was further reduced to r < 0.07 when combining the Keck Array observations with external datasets~\cite{grayson2016bicep3,ade2015bicep2}.

BICEP 3 is the latest BICEP Telescope with 2560 TESs measuring radiation in the 95~GHz band. The~detector technology was not upgraded between BICEP2 and BICEP3, but~the number of detectors on the focal plane was increased over 5~fold. 

The Keck Array has been replaced by the BICEP Array which consists of four BICEP3-like receivers operating in the 30/40, 95, 150 and 220/270~GHz bands~\cite{hui2018bicep}.

\subsubsection{CLASS}
The Cosmology Large Angular Scale Surveyor (CLASS) is an array of microwave telescopes for CMB studies, including the detection of primordial B-mode polarization~\cite{essinger2014class}. CLASS is designed as follows: it exhibits one focal plane with 36 detector pairs operating at 38~GHz, two focal planes with 259 detector pairs each operating at 93~GHz, and~a dual-band telescope with 1000 detector pairs at each of 148 and 217~GHz~\cite{iuliano2018cosmology,chuss2016cosmology,eimer2012cosmology}. Each detector consists of a feed-horn coupled to an OMT coupled to two Mo/Au TESs with a critical temperature of 150 mK and an NEP of 2.1$~\times~10^{-17}$~W/$\sqrt{\mathrm{Hz}}$ \cite{dahal2018design,appel2014cosmology}. CLASS is currently in operation, carrying out a survey of 70\% of the sky from which it will not only infer the value of the tensor-to-scalar ratio, but~also carry out new physics while investigating both the Milky Way galaxy and the circular polarization of~CMB. 

\subsubsection{SPT}
The South Pole Telescope (SPT)  is a 10 m diameter off-axis Gregorian Telescope designed to observe the CMB from Antarctica in order to study its anisotropies. In~particular, the~SPT first observed B-mode polarized CMB radiation (albeit not being primordial). Three main instruments have been deployed at the SPT since its commissioning: the SPT-SZ camera with 960 TDM TESs sensitive in bands about  95, 150  and~220~GHz~\cite{shirokoff2009south}. The~main science goal of the SPT-SZ camera was a large-area survey of the Southern Sky. The~second camera mounted on the SPT was the SPTPol (SPT-Polarimeter) with 1560 Transition Edge Sensors in a polarimeter configuration with a feed-horn, an~OMT and two TESs, one per each polarized component of the light. The~780 polarimeters were divided in two frequency bands: 90 and 150~GHz~\cite{austermann2012sptpol}. 
The third generation camera, SPT-3G, \textls[-25]{was installed in 2018 with 16000 detectors on the focal plane, split evenly between 90, 150  and~220~GHz~\cite{benson2014spt,anderson2018spt,sobrin2022design}}.

\subsubsection{POLARBEAR}
The polarization of the background radiation (POLARBEAR) is an experiment set in the Atacama desert of Chile and {it} aims at measuring the B-mode polarization component of the CMB radiation~\cite{kermish2012polarbear}. The~POLARBEAR array consists of 637 polarization-sensitive pixels for a total of 1274 antenna-coupled TES bolometers. The~TESs are made of Ti/Al bilayers with a critical temperature of about 500 mK and exhibit an NEP of 1$~\times~10^{-16}$~W/$\sqrt{\text{Hz}}$, while the coupling at 148~GHz occurs through double-slot dipole antennas. By~the end of its lifetime, POLARBEAR was capable of performing a large angle survey of the Southern Sky with low enough noise that the lensing signal could be reconstructed with more precision from polarization than from CMB temperature. By~design, POLARBEAR was supposed to be capable of  detecting  B-mode polarization down to a tensor-to-scalar ratio of 0.025. An~updated POLARBEAR has been deployed in 2022 under the name    POLARBEAR 2b (PB-2b) and is currently a part of the Simons Observatory Array~\cite{suzuki2014polarbear}. PB-2b's focal plane consists of 7588 lenslet-antenna coupled polarization-sensitive Al/Mn TES bolometers which are simultaneously sensitive to 95~GHz and 150~GHz bands. The~7588 TESs are read out with 40 lines in an FDM scheme with a multiplexing factor of about \linebreak 40:1~\cite{barron2021integrated}. 
\subsubsection{QUBIC}
The Q and U Bolometric Interferometer for Cosmology (QUBIC) \cite{mennella2019qubic} employs an interferometric approach to detecting  the polarization of CMB radiation within two main bands: at 150 and 220~GHz~\cite{o2015qu}. The~deployment site for QUBIC is in Alto Chorrillos, Argentina~\cite{marnieros2020tes}. Bolometric interferometry allows for the production of maps of the CMB polarization with a resolution of 23 arcminutes. The~detector module consists of 256 NbSi TESs with a critical temperature of $\sim 410$ mK multiplexed 128:1 in a TDM scheme~\cite{piat2022qubic}. The~TESs exhibit an NEP below $2\times10^{-16}$ W$/\sqrt{\mathrm{Hz}}$  \cite{o2015qu}. 
\subsubsection{AliCPT}
The Ali CMB Polarization Telescope (AliCPT) is the first Chinese CMB telescope in the Tibetan plateau  to perform unprecedented CMB polarization measurements in the northern hemisphere. Its primary scientific goal is the detection of primordial gravitational waves with sensitivity $\sigma (r) \sim$ 0.02  and an \textit{a posteriori} maximum of $r$ in the range  \mbox{$0.012$--$0.029$} \cite{ghosh2022performance}.
AliCPT is a 72 cm aperture two-lens refracting telescope and will feature 32,376 TES polarimeters aimed at the detection of the 90 and 150~GHz bands~\cite{gao2017introduction}. The~AliCPT focal plane consists of 19 independent module polarization-sensitive feedhorn-coupled Al/Mn TESs ($T_C \sim 420$ mK) equally split between the 90 and the 150~GHz bands. The~TESs are read out in a $\upmu$-MUX scheme with a multiplexing factor of 2000:1. The~90~GHz band is expected to exhibit a  saturation power of 7 pW and an NEP of $3.8\times10^{-17}$ W $/\sqrt{\mathrm{Hz}}$, whereas the expected saturation power for the 150~GHz band is 12 pW with an NEP of {$5.3 \times10^{-17}$ W$/\sqrt{\mathrm{Hz}}$} \cite{salatino2020design}.
\subsubsection{Simons~Observatory}
The Simons Observatory, also located in the Atacama desert of Chile, is a ground-based CMB observatory funded by the Marilyn and Jim Simons foundation as well as the Heising--Simons foundation~\cite{lee2019simons}. The~Simons Observatory aims to observe the CMB polarization at six frequency bands centered around 30, 40, 90, 150, 230 and~280~GHz. It will consist of one Large Aperture Telescope (6 m primary mirror and 7.8$^{\circ}$ {field of view})  and three Small Aperture Telescopes (0.42 m aperture) \cite{ade2019simons}. While the former is sensitive to all bands mentioned before, two of the Small Aperture Telescopes observe in the  90 and 150~GHz bands while the remaining one observes at 220 and 270~GHz. The~TESs (Al/Mn) exhibit a critical temperature of $\sim$190 mK and an NEP $\sim 4.1\times10^{-17}$~W/$\sqrt{\text{Hz}}$ \cite{stevens2020characterization,galitzki2018simons}. The~primary scientific goal of the Simons Observatory is producing a polarization maps of the sky with a ten-fold improved sensitivity compared to Planck~\cite{ade2019simons}. As~secondary goals, the~Simons Observatory aims to measure the sum of neutrino masses, and investigate the galaxy evolution  and~the duration of the reionization epoch~\cite{bryan2018development}.
\subsubsection{CMB-S4}
The cosmic microwave background Stage 4 (CMB-S4) is a future ground-based CMB experiment with 19 telescopes between the Atacama desert in Chile and Antarctica aiming to investigate the existence of primordial B-mode polarized light to a tensor-to-scalar ratio as low as $10^{-3}$ \cite{abazajian2016cmb,abitbol2017cmb}. In~its preliminary form, CMB-S4 aims to  employ fourteen 0.55 m {Small Area Telescopes} (SATs) at 155~GHz and below and four 0.44 m SATs at 230/280~GHz, with~dichroic, horn-coupled Al/Mn~\cite{duff2024simons} TESs in each SAT, measuring two of the eight targeted frequency bands between 25 and 280~GHz. Furthermore, it aims to employ one 5 m class {Large Area Telescope} (LAT), for~'delensing' purposes, equipped with TESs distributed over seven bands from 20 to 280~GHz~\cite{abazajian2022snowmass}. 

\subsubsection{EBEX}
The E and B Experiment (EBEX) was a balloon-borne experiment for the detection of CMB anisotropies during a 27-day circumpolar flight in Antarctica~\cite{oxley2004ebex}. The~great advantage of a balloon-borne experiment over the aforementioned ground-based experiments is the fact that its flight altitude is $\sim$42 km, at~which the absorption of microwave radiation is almost negligible. This allows for higher SNR compared to ground-based experiments of the same size. EBEX  features 1432 TES bolometric detectors read out in an FDM scheme. EBEX observes in three frequency bands, namely 150~GHz, 250~GHz 
and 410~GHz, with~768, 384  and~280  TESs,  respectively~\cite{reichborn2010ebex}. Each TES is made of an Al/Ti bilayer with a critical temperature of 500 mK and as an array, the~NEP is below 1.7$~\times~10^{-17}$~W/$\sqrt{\mathrm{Hz}}$ \cite{oxley2004ebex}. It contributed to setting a constraint on the value of the tensor-to-scalar to $r \leq$ 0.034 ratio along with the measurements performed with {Planck} and the BICEP/Keck~Array.  

\subsubsection{SPIDER}
SPIDER is a balloon-borne CMB experiment~\cite{crill2008spider,gualtieri2018spider} deployed in a 17-day circumpolar flight from Antarctica in 2015~\cite{gambrel2018measurement}. The~payload consists of six monochromatic refracting telescopes mounted within the same cryostat for observations at 100, 150  and~280~GHz. Each telescope was equipped with a polarization-sensitive TES array for a total of \mbox{800 pixels} read out in a TDM scheme. The~TESs for such observations are made of pure Ti with a critical temperature of 500 mK and exhibit an overall NEP $\sim 4\times10^{-17}$~W/$\sqrt{\text{Hz}}$ \cite{hubmayr2016design}. With~just a 10\% sky coverage, SPIDER derived the upper limit on the tensor-to-scalar ratio as r $\leq 0.11$ with a confidence level of 95\% \cite{rahlin2014pre}.

\subsubsection{LSPE} \label{sec:lspe}
The Large-Scale Polarization Explorer (LSPE) is a proposed balloon-borne experiment for the measurement of CMB anisotropies at large angular scales~\cite{lamagna2020progress}. The~project, expected to fly in 2026, is funded by INFN and ASI and the circumpolar flight is in partnership with NASA. LSPE consists of two instruments: SWIPE (Short Wavelength Instrument for the Polarization Explorer)  and STRIP (Survey Tenerife Polarimeter). LSPE/SWIPE~\cite{columbro2019short} is a Stokes polarimeter with a cold  half-wave plate modulator and its two focal planes are equipped with  330 multi-mode Spiderweb TES bolometers~\cite{tartari2024characterization} at 140, 220 and 240~GHz. LSPE/STRIP~\cite{sandri2017lspe} is aimed at accurate measurements of the low-frequency polarized radiation  at 44  and 90~GHz, dominated by Galactic synchrotron emission and deploys a detection system based on OMTs and HEMT amplifiers. 

 The LSPE/SWIPE TES are made of a Ti/Au with a critical temperature of 500 mK \textls[-5]{and a saturation power of 10 pW. The~TESs are intended to exhibit an NEP below 1$~\times~10^{-17}$~W/$\sqrt{\text{Hz}}$} so that the experiment is limited by the photon noise of the~sources.

\subsubsection{LiteBIRD} \label{sec:litebird}
LiteBIRD, the~acronym for `The Lite (Light) satellite for the study of B-mode polarization and Inflation from cosmic background Radiation Detection' is a JAXA-led spacecraft mission which aims to measure the B-mode polarization of the CMB with sensitivity such that a tensor-to-scalar ratio as small as r$\sim0.001$ can be measured (LiteBIRD will be able to measure $r=0$ with $\delta r  \leq 0.001$) \cite{matsumura2014mission}. LiteBIRD will produce a fine CMB at large angular scales at 60, 78, 100, 140, 195 and 280~GHz, where the first two and last two bands are required to perform foreground subtraction and the 100 and 140~GHz bands are the CMB channels.

As of July 2024, the full engineering details of LiteBIRD are not defined yet, but~it will most likely be constituted of two main instruments:  the Low Frequency Telescope (LFT) \cite{sekimoto2020concept}, and~the Medium and High Frequency Telescope (MHFT) \cite{montier2020overview}. While the LFT is sensitive between 30 and 160~GHz in nine overlapping bands, the~MHFT spans between 80 and 500~GHz in nine more overlapping bands. LiteBIRD will be equipped with over 4000 multichroic TESs read out in an FDM scheme with a multiplexing factor $\sim 68$. The~TES bolometers, produced by NIST, are made of Al/Mn with a critical temperature of 200 mK, an~NEP of $\sim 1 \times10^{-17}$~W/$\sqrt{\text{Hz}}$ and a saturation power in the order of 1.9--4.8 pW~\cite{tominaga2022design}. 

\section{{Applications in Nuclear, Particle and Astroparticle~Physics}}\label{sec:particle}
The search of DM is moving towards joint ventures between astrophysical observations and laboratory experiments. Given the multitude of DM models currently available, different experiments are being proposed and carried out to extend the search of DM particles to masses below 10 GeV. These experiments rely on measuring very small fractions of energy deposited in the detector through DM-nucleus scattering~\cite{abdelhameed2019first,Alkhatib_2021,Lattaud_2022,dangelo2015darkside50resultsargonrun,Cao_2019}. Similarly, extremely sensitive detectors capable of measuring very small energies can enable new experiments that further progress our understanding of neutrino physics. These experiments include the search for neutrino-less double-beta ($0\nu2\beta$) decay~\cite{adams2022cuore,agrawal2024improvedlimitneutrinolessdouble} and other processes beyond the standard model, the~detection and measurement of Coherent Elastic Neutrino-Nucleus Scattering (CE$\nu$NS) \cite{https://doi.org/10.5281/zenodo.3903810}
 and probe {the absolute scale of the neutrino mass (e.g., goals of HOLMES and CUPID)} enabling the possibility to determine the neutrino mass~hierarchy.
 
Furthermore, neutrinos represent an unavoidable background for DM search~\cite{O_Hare_2021}  and sterile neutrinos are competitive DM candidates~\cite{Boyarsky_2019}.

TESs could be the cornerstone for experimental studies both for neutrino physics and DM search, thanks to their excellent sensitivity. In~this section, we will present current experiments exploiting TESs for neutrino studies and DM searches.
\subsection{Electron Capture Decay}.
\subsubsection{HOLMES}
HOLMES is an experiment which aims to determine the absolute value of electron neutrino mass with a sensitivity of $\sim$1 eV by measuring the de-excitation energy spectrum of $^{163}$Ho electron capture (EC) decay~\cite{alpert2015holmes}.
This approach requires a calorimetric measurement, which allows one to reconstruct all the EC de-excitation energy, and~a high energy resolution, especially for energies close to the kinematic endpoint of the decay ($Q_{EC}\sim 3\text{~keV}$).
In order to meet these requirements, HOLMES will use an array composed of a large number of TES~microcalorimeters.

Each single pixel of the HOLMES experiment consists of a ($125\times125\;\upmu\text{m}^2$) Mo-Cu bilayer TES sensor in thermal contact with a ($200\times200\times2\;\upmu\text{m}^3$) Au absorber which is directly implanted with $^{163}$Ho. Both the sensor and the absorber are suspended on a $\text{Si}_3\text{N}_4$ membrane. This device exhibits an energy resolution of 4.22 eV at 6~keV~\cite{borghesi2023updated}.

HOLMES TESs are read out by a microwave SQUID multiplexing chip, developed and fabricated by NIST, which allows for the parallel read out of 33~channels.

\subsubsection{NuMECS}
{The neutrino ($\nu$) Mass Electron-Capture Spectroscopy~\cite{croce2016development} collaboration aims at measuring the neutrino mass through calorimetric spectroscopy of EC events in decay isotopes such as $^{163}$Ho which are embedded in Mo/Au TES, similar to those fabricated for HOLMES~\cite{nucciotti2016use}.}

\subsection{Rare $\beta$ and Double-$\beta$ Decays}

\subsubsection{CUPID}
CUPID (CUORE Upgrade with Particle Identification) is the next generation experiment which aims to measure the lepton flavour-violating process  ($0\nu2\beta$) decay~\cite{group2019cupid, alfonso2023cupid}. It will be host in the same cryogenic infrastructure of its predecessor, and~state-of-art $0\nu2\beta$ decay experiment, CUORE~\cite{adams2022cuore} at the Gran Sasso National Laboratory (Italy). When deployed, it will utilise an array of approximately 1600 scintillating Li$_2$ $^{100}$MoO$_4$  crystals, each enriched with the $^{100}\text{Mo}$ isotope and having a mass of 280 g~\cite{alfonso2023cupid}. CUPID, compared to CUORE, will benefit from an improved background suppression achieved via a double read out of the thermal and optical signals produced by its~crystals.

{While the detectors of choice for CUPID are NTD Ge, TESs are considered as a promising technology for upgrades towards CUPID-1T. \citet{singh2023large} present a large-area TES photon detector which meets the requirement to be employed in the CUPID experiment for the light channel read out.
In this device, a TES is coupled to a large area (50.8 mm diameter) Si wafer which acts as a photon absorber. The~sensor is a bilayer (45 nm Ir and 20 nm Pt) superconducting film; it has an area of $330\;\upmu\text{m}\times 300\;\upmu\text{m}$ and exhibits a critical temperature of $\sim33\;\text{mK}$. Further R\&D efforts towards further upgrades of CUPID such as CUPID-1T involved the instrumentation of a Li$_2$$^{100}$MoO$_4$ crystal with an Al/Mn TES, such as described by \citet{bratrud2024demonstrationtesbasedcryogenic}. }
{\subsubsection{ACCESS~Project}
The ACCESS (Array of Cryogenic Calorimeters to Evaluate Spectral Shapes) intends to perform precision measurements of forbidden $\beta$-decays. A~testbed with a pilot array of calorimeters based on natural  and doped  crystal containing $\beta$-emitters ($^{113}$Cd and $^{115}$Id, and~$^{99}$Tc) will be performed. While the obvious choice of detectors for ACCESS is germanium NTDs such as those for CUORE and CUPID, the~possibility of employing TESs is currently being investigated~\cite{Pagnanini_2023}.}

\subsection{Coherent Elastic Neutrino-Nucleus Scattering}

\subsubsection{{NUCLEUS}}
NUCLEUS is a cryogenic experiment that will explore CE$\nu$NS by measuring nuclear recoil energies down to 10 eV~\cite{Strauss_2017}. It will be installed at an experimental site at the Chooz nuclear power plant (France) where it will benefit from a large flux of $O(\text{MeV})$ energy (anti)neutrinos~\cite{Nucleus_Chooz}.

NUCLEUS, as~a nuclear recoil detector, will employ an array of cryogenic calorimeters composed of nine CaWO$_4$ and nine Al$_2$O$_3$ crystals with a total mass of 10 g~\cite{Strauss_2017}. In~each calorimeter, a TES, similar to those developed for the CRESST experiment (see Section~\ref{sec:cresst}), is used as the sensor. This TES consists of a 200 nm tungsten film coupled to an Al phonon collector and is operated at a temperature of 15 mK. A~prototype 0.5 g Al$_2$O$_3$ calorimeter achieved an energy threshold below 20~eV.

In order to attain background reduction, the~NUCLEUS detector will be encapsulated in a cryogenic veto~\cite{GOUPY2024169383}. This will include an instrumented Si detector holder and a surrounding {kg-}scale detector. Tungsten TESs are employed in both veto detectors as phonon sensors.
{\subsubsection{MInER}
The Mitchell Institute Neutrino Experiment at Reactor (MInER) is a CE$\nu$NS experiment with a 1 MW reactor as the source of electron anti-neutrinos with an energy up to a few MeV~\cite{agnolet2017background}. The~detector consists of a sapphire Al$_2$O$_3$ absorber connected to SuperCDMS-like TES modules~\cite{adari2022excess}. Further details on the detectors can be found in Section~\ref{sec:SuperCDMS}.
\subsubsection{CRAB}
The CRAB (Calibrated nuclear Recoils for Accurate Bolometry) Collaboration suggests the use of nuclear recoil induced by radiative capture of thermal neutrons as an efficient way to calibrate detectors in the 100 eV range~\cite{angloher2023observation}. The~NUCLEUS and CRAB Collaborations reported on the observation of such a detected peak in a 0.75 g protoypal CaWO$_4$ detector at 112 eV irradiated by a neutron source arising from $^{252}$Cf. The~detector was built for the first phase of NUCLEUS and is based on TES developed for CRESST  (see Section~\ref{sec:cresst}) and described in~\cite{abdelhameed2019first}.
\subsubsection{RES-NOVA}
RES-NOVA is a proposed observatory which intends to detect cosmic neutrinos originating from core-collapsing supernovae through nuclear scattering (CE$\nu$NS) in  array detectors primarily made of Roman lead (Pb) exploiting the high CE$\nu$NS cross-section and the unmatched isotopic purity of archelogical Pb~\cite{pattavina2020neutrino}. The~detector is made up of 500 large-mass Pb single crystals paired with ultra-sensitive TESs to measure any possible temperature increases due to particle interaction~\cite{pattavina2021res}. RES-NOVA  aims at targeting energy thresholds of 1~keV or below in order to be sensitive to flavour-blind neutrinos  with an exposure of $620$ ton $\cdot$ y~\cite{pattavina2020neutrino}.}
\subsubsection{Ricochet}
The Ricochet experiment will performs an accurate measurement of CE$\nu$NS spectrum at the Institute Laue-Langevin nuclear reactor (France) \cite{augier2023ricochet}. For~this purpose, it will employ an array of cryogenic calorimeters based on two different technologies: NTD-Ge thermistors~\cite{ricochet2024first} and TESs~\cite{Richochet_Main}.
The TES-based detector will include the following: (1) an absorber that could be made of different material such as Si, Ge, Zn or Al; (2) a TES film deposited on a Si substrate separated from the latter; and (3) a readout and bias~circuit.

A prototype of the Ricochet TES-based detector, which employs an Al/Mn superconductive film and a 1 g Si absorber, is presented in~\cite{Ricochet_prototype_detector}. This device exhibits a baseline root-mean-square energy resolution lower than 40~eV.

\subsection{Dark Matter and Axions}

\subsubsection{CRESST}\label{sec:cresst}
CRESST (Cryogenic Rare Event Search using Superconducting Thermometers) is an experiment hosted at the Grans Sasso National Laboratory which searches for weakly interacting massive particles (WIMPs) of energy $<400$~keV~\cite{BRAVIN1999107} by taking advantage of their elastic scattering with CaWO$_4$ crystals~\cite{Angloher2016-da}. This interaction produces both phononic overheating of the crystal and scintillation light. The~latter strongly depends on the  type of the interacting particle. CRESST exploits TESs both as phonon and photon detectors to evaluate the type and energy of the incident particle. The~thermistor of these TESs is made of tungsten and operate at a temperature of about 15 mK. These detectors showed an energy resolution of 4.1 eV (6.7 eV) at an energy of 5.9~keV (6.5~keV) \cite{Rothe2018-sa}. 

\subsubsection{SuperCDMS} \label{sec:SuperCDMS}
The Cryogenic Dark Matter Search (SuperCDMS) collaboration intends to detect DM particles at the Sudbury Neutrino Observatory facility (SNOLAB) in Canada. The~detection principle relies on measuring the ionization and the phonons produced by the scattering of Weakly Interacting Massive Particles (WIMPs), a~DM candidate, with~the nucleus of  ultra-pure Ge and Si crystals. The~thermal energy thus deposited is read out with tungsten TESs.
 The~expected phonon energy resolution is better than 5 eV and recoil energy resolution is  better than 1 eV~\cite{kurinsky2016supercdms}.

\subsubsection{ALPS~II}
 Any Light Particle Search (ALPS) II is a light shining through a wall (LSW) experiment searching for axion-like particles (ALPs) which are possible candidates as weakly interactive sub-eV particles (WISPs) \cite{RBahre_2013}.
In an LSW experiment, a~laser beam is sent through a long magnet, allowing for the coherent
photon--axion conversion due to the Primakoff effect~\cite{Sikivie}. The~wall acts as a barrier for the light, while the almost zero cross-section for the axion/baryonic matter interaction enables the axion to pass through the  wall. The %please check text removed here: "A~second magnet placed after the wall causes. "
~photon--axion back conversion is produced by a second magnet after the wall. Since the conversion process is very rare, ALPS II exploits optical resonators to boost both the number of photons for ALP production and their reconversion probability to light. ALPS II aims at demonstrating the existence of particles of mass 1 $\upmu$eV$<m<1$ meV. To~this end, tungsten TESs operating at $\sim20$ mK have been developed. At~the moment, TESs with energy resolution of about 150 meV have been demonstrated at \mbox{1.165 eV~\cite{RubieraGimeno:2023J9}.}

\subsubsection{EDELWEISS}
Similarly to SuperCDMS, EDELWEISS (Expérience pour Detecter Les WIMPs En Site Souterrain) targets the detection of WIMPs through the ionization and the phononic signals produced in ultra-pure Ge crystals. While EDELWEISS relies on Ge crystals instrumented with NTD detectors~\cite{armengaud2017performance}, in~order to investigate light-WIMPs, the EDELWEISS collaboration is pursuing an R\&D activity to achieve 1 eV threshold Ge bolometers coupled to efficient background-rejection techniques. In~this scope, Nb$_x$/Si$_{1-x}$ TESs are currently being developed~\cite{marnieros2023high}. Further R\&D efforts are being put in place by the CRYOSEL collaboration which effectively demonstrated the operation of an NTD and a superconducting single-electron device, which operates in a similar way to the devices described in Section~\ref{sec:SNSPDs}. For~further details,  see~\cite{lattaud2024characterization}.

\subsubsection{STAX}
STAX is an LSW experiment proposing to search for ALPs at extremely low energies \mbox{$m<10\;\upmu$eV~\cite{CAPPARELLI201637}} by exploiting gyrotrons, which are very intense sub-THz photon \linebreak sources~\cite{Kat7780459}. In~particular, STAX proposed the use of gyrotrons operating at 30~GHz with a power of several tense  MW. The~main challenges of the project were as follows: (i) the realization of a high quality factor~GHz cavity operating under high inject power, (ii) the realization of a high quality factor~GHz cavity operating at cryogenic temperatures under strong magnetic fields (10 T), and~(iii) the development of TES calorimeters for extremely low energies. In~particular, STAX proposed the use of a molybdenum TES with an operating temperature of about 15 mK. Furthermore, the~miniaturization of the thermistor/absorber was proposed~\cite{Paolucci2020}.
{\subsubsection{COSINUS}
The COSINUS experiment (Cryogenic Observatory for Signals seen in Next-generation Underground Searches) is intended to perform  follow up measurements on the DAMA/ LIBRA~\cite{bernabei2020dama} results. DAMA/LIBRA detected an annual modulation of DM signal due to the Earth orbiting around the sun and thus modulating its relative velocity with respect to the DM particles~\cite{krishak2020robust}. The~novelty introduced by COSINUS consists in coupling the NaI targets of DAMA/LIBRA  with a dual channel (light and phonon) readout in a similar fashion to CRESST~\cite{angloher2016cosinus}. The~phonon channel consists of an Au pad on the NaI absorber which is wire bonded to a tungsten TES with a critical temperature of 28 mK and an energy resolution of 0.441~keV, while the light channel, which also employs the same tungsten TES, exhibits an energy resolution of 0.988~keV up to 200~keV~\cite{angloher2024deep}.
\subsubsection{BabyIAXO}
BabyIAXO is an intermediate stage of the \textls[-15]{International Axion Observatory (IAXO) and its primary goal is the testing of the different subsystems of the IAXO, BabyIAXO will also detect or reject solar axions with axion--photon coupling down to \mbox{$g_{a \gamma}\sim 1.5\times 10^{-11}$ GeV$^{-1}$ \cite{abeln2021conceptual}}}. As~of the present date, several detector technologies are being investigated including TESs as possible for implementation~\cite{abeln2021conceptual}. 
\subsubsection{TESSERACT}
TESSERACT represents the synergistic effort of searching DM through the exploitation of cryogenic targets and sub-eV TESs by two distinct  experiments HeRALD (Helium Roton Apparatus for Light Dark matter) and SPICE (Sub-eV Polar Interaction Cryogenic Experiment) complemented with EDELWEISS and Ricochet~\cite{BILLARD2024116465}.
HeRALD exploits the interaction of DM with a superfluid helium target, which produced scintillation light and evaporated atoms (an energy of 1 meV is necessary to evaporate an atom) \cite{herald}. Instead, SPICE uses sapphire or GaAs crystals that generate athermal phonons after interacting with DM~\cite{SPICE}.

TESSERACT exploits the same TESs with aluminum absorbers to detect light, atom evaporation and phonons. This approach ensures a different DM mass search window from meV-to-GeV allowing one to probe DM candidates over 12 orders of magnitude, thus testing different DM models~\cite{BILLARD2024116465}.}

\section{Biophysics and Medical Imaging with Transition Edge~Sensors}\label{sec:bio}
The technological advancement of Transition Edge Sensors as semi-commercial energy-resolving single-photon detectors has allowed several groups around the world to investigate the possibility of using {such} detectors for biomedical~imaging. 

The fundamental idea behind these applications is using a TES as the main detector of a Confocal Laser Scanning Microscope (CLSM). Differently from a conventional microscope, where the sample is evenly illuminated with a wide light source producing a focused image and an unfocused background,  a~confocal microscope uses point-illumination and a pin-hole aperture in order to effectively eliminate the out-of-focus light. By~raster-scanning the whole surface of the specimen under investigation, it is possible to achieve an effective focus-stack of the specimen as well as information on the thickness of the specimen. In~cases where a laser source is used to induce fluorescence in the specimen, a~confocal microscope has the extra advantage of only exciting a single point of the specimen. Since only the fluorescence light that is produced very close to the focal plane can be detected, this results in a higher optical resolution~image.

In this scenario, the~single-photon sensitivity of TESs comes as a great advantage when working with living cells, where light irradiation deteriorates the cells and causes the photo-bleaching of any fluorescent dyes used as markers~\cite{icha2017phototoxicity}. Moreover, the~incredible resolution of TESs allows for the identification of different biomarkers in the same specimen, hence the identification of different functions and features within the specimen. The~energy resolution in standard confocal microscopes is achieved through the use of a diffraction grating or optical filters which further increase the radiation dose to which the specimen is~exposed. 

\citet{niwa2017few} (2017) proposed the first confocal microscope with a TES detector which also exploits the optical fiber that is used to feed the light into the cryostat as the pinhole of the microscope, further reducing the complexity of the instrument. Such an instrument exhibited an astounding 98\% photon-detection efficiency at 850 nm~\cite{fukuda2011titanium} while being successfully able to capture photons in the 400--2800 nm range spanning the whole visible and near infra-red. 

Such an instrument was used to collect a sample image of  yellow, red and blue ink spots. Under~the same illumination, in~the few ($\sim 20$)  photons regime, a~raster scan was performed and the light was detected by both a photo-multiplier tube (PMT) and a TES. Associating a colored pixel in the image according to the response of the TES (blue for wavelengths shorter than 500 nm, green between 500 and 600 nm and red from 600 to 800 nm) \citet{niwa2017few} thus demonstrated the capability of the TES to distinguish the different ink pigments, impossible for the PMT which only produced a grey-scale image. They also demonstrated that a CMOS sensor needed one hundred-fold the illumination and a ten-fold exposure time to produce similar images. Furthermore, neither the CMOS sensors nor the PMTs exhibited any sensitivity to the near infra-red radiation to which the TES was~sensitive.

Building on these results, \citet{niwa2021few} demonstrated the use of a TES-based CLSM as a bio-imager for a fixed fluorescence-labelled cell, where three different dyes were simultaneously excited with the use of a 405 and a 488 nm laser. Using a single raster-scan, an~RGB color image and a near-infrared image were produced with the means of only a few tens of photons per pixel and an irradiation power of 100 nW. The~total exposure time was of 35 min at 20 ms per pixel and a raster step-size of 25 $\upmu$m. In~order to speed up the scan, two further improvements can be made: a faster X-Y micro-manipulator for the scan and the development of a multi-fiber--multi-TES array which contributes to the reduction of the scanning steps necessary. Further work in this direction by~\cite{Hattori2020complex}   consisted in the development of sub-$\upmu$s TESs. Such fast detectors can be combined into arrays with extremely short dead times, and hence may result in reduced scan~times. 

In the same research group, \citet{fukuda2019single} investigated the energy resolution of single pixel TESs and the possibility of employing them in a raster-scan microscope where the sample/specimen is wide-field illuminated with a low-intensity light source. Such an apparatus demonstrated the capability of discerning different materials such as copper, silk and resistance by their color. Furthermore, since each pixel contains the spectrum between 400 and 1800 nm, it is possible to infer the different materials by their optical properties.

As a final remark, the~work of \citet{hao2003inductive} is worth of a mention. In~said paper, they suggested the development of an Inductive Superconducting Transition Edge Detector (ISTED) which employs a thin patch of  a superconductor deposited within a SQUID loop and maintained below transition temperature. Any energy deposition produces an increase in the temperature of the film and thus a change in the loop inductance by   means of a change in the London penetration depth of the film. ISTEDs  are expected to be capable of detecting massive molecular or polymeric species. Such bolometers would allow for the detection of large (>200 Da) fragments of DNA or bio-markers such as immunoglobulin with an unprecedented energy and spectrographs of `difficult' molecular species~\cite{fukuda2019single}.

\section{TES for Quantum~Applications}\label{sec:quantum}
Quantum optics and quantum computing are on the rise as bleeding edge topics. In~the dynamic landscape of quantum applications, TESs can play a critical role thanks to their single photon sensitivity and high quantum efficiency at communication wavelengths. Their unmatched sensitivity and their rather limited dark count rate make TESs promising detectors for a veto that rejects calculations which are potentially disturbed by the~environment.
\subsection{Quantum Optics and Quantum~communication}\label{sec:quantumoptics}
Transition Edge Sensors with their near-unity quantum efficiency and their single-photon sensitivity are currently being investigated as possible photon-counting detectors for quantum information systems and quantum optics experiments. As~of currently, the~detection and counting of telecom wavelength (1550 nm) photons with near-unity efficiency can only be achieved with low-temperature superconducting detectors such as {Microwave Kinetic Inductance Detectors (MKIDs), Superconducting Nanowire Single Photon Detectors (SNSPDs)} and Transition Edge~Sensors. 

The first experiment to benefit from the photon-detection capabilities of TESs was a Hong-Ou-Mandel (HOM) interference~\cite{hominterference} experiment in which TESs were deployed to perform measurement of the output photon-number statistics. HOM interference is an effect that arises from the bosonic nature of photons and forbids two indistinguishable non-entangled photons that enter two different inputs of a beam splitter to exit at different outputs. \citet{di2003direct} successfully measured the photon-number statistics at the output of an HOM inteferometer when illuminated with a pair of orthogonally polarized photons arising from a  {Beta Barium Borate} (BBO) photonic crystal. As~the difference between the two photon paths approaches zero, the~probability of measuring one photon at either output of the interferometer drops to zero.  \citet{di2003direct} varied the optical path to create a delay between the two photons and demonstrated that, when then optical paths are identical, no single photon statistic can be measured at either end of the~interferometer.

The production and verification of non-classical light states is one of the most critical limitations to the further development of quantum optics, quantum metrology and quantum computing~\cite{knill2001scheme,gisin2007quantum,chunnilall2014metrology}. Photon-number-resolving tungsten TESs allow direct access to the photon statistics of the light field in quantum metrology of light sources~\cite{schmidt2018photon}.  Another application of TESs for quantum optics experiments is in the generation of Coherent State Superpositions (CSSs) also known as Schroedinger Cat States. Such states exist when each subsystem of the superposition contains a macroscopic number of photons~\cite{gerrits2016superconducting}. At~least two experiments aimed at generating CSSs through photon subtraction have been performed using tungsten TESs to detect and indicate the presence of the CSS. The~detection of such a CSS occurs through an optical homodyne readout scheme~\cite{gerrits2010generation, bartley2012multiphoton, sridhar2014direct}. The~TES used in such experiments exhibited an efficiency of $85\%$ at 860 nm and photon-counting capabilities of up to a few tens of photons~\cite{bartley2012multiphoton,laiho2010probing}.

One further example of CSS produced by photon subtraction was demonstrated by \citet{zhai2013photon}. They demonstrated the subtraction of up to eight photons from a thermal state. By~the use of a tungsten Transition Edge Sensor, it was possible to reconstruct the photon number statistics as a function of the subtracted photons, demonstrating that the photon number of the photon-subtracted states increased linearly with the number of photons subtracted from the thermal state. \citet{zhai2013photon} further demonstrated that a TES-based setup allows for the derivation of correlation functions of second and further order can be derived from the photon number statistics~\cite{gerrits2010generation,zhai2013photon}.

TESs have been used as photon detectors in experiments intended to demonstrate the violation of the CH Bell inequality~\cite{clauser1974experimental}, which is a modified version of the more famous Bell inequality. \citet{christensen2013detection} deployed TES optimized for a wavelength of 710 nm with a 95\% detection efficiency, achieving an overall 75\% system efficiency.  \mbox{\citet{giustina2013bell}} demonstrated the violation of the CH Bell inequality with a 78\% system  efficiency when using a TES optimized for the detection of 810 nm and a photon-detection efficiency of 95\%.

\citet{schmidt2018photon} developed a system for photon number-resolving experiment with the use of two fiber-coupled TESs optimized for the detection of photons at 1087 nm with a detection efficiency larger than 87\% and such a system has been deployed in experiments on single and twin-photon states~\cite{heindel2017bright} and the investigation of the emission statistics of bimodal lasers~\cite{schlottmann2018exploring} and polarization lasers~\cite{klaas2018photon}.

\citet{chunnilall2014metrology} suggest the use of TESs for Quantum Key Distribution (QKD). QKD  uses photons to generate a cyphered secret key between either end of a communication. QKD is a `secure' communication channel because if the photons that serve as the communication medium are intercepted, their state is changed and the users at either end of the communication will know that the communication channel was `hacked'. The~paper suggests using a TES-based radiometer for the detection of such QKD photons (1550 nm) with optical powers in the range 50 fW--20 nW with repeatability better than $\pm 0.3\%$ and an NEP of $5\times 10^{-15} $~W/$\sqrt{\text{Hz}}$.

\subsection{Quantum~Computing}\label{sec:quantumcomputing}
A concept device has been presented that combines a TES with a superconducting qubit on a shared silicon substrate. The~idea is that the phonons produced by any radiation can spread efficiently through the substrate and thus both the qubits and the TES will experience correlated disturbances induced by the radiation. This correlated disturbance has been directly measured with substrates on which multiple quibits~\cite{wilen2021correlated,mcewen2022resolving} and MKIDs~\cite{cardani2021reducing} are present. The~idea of pairing a qubit with a TES relies on the assumption that one single sensor is capable of detecting the condition of a radiation-induced error for all qubits on the same silicon substrate~\cite{orrell2021sensor} and exploits the signal from the TES as a veto to reject the calculations that could be potentially incorrect due to an environmental disturbance. \citet{orrell2021sensor} demonstrated that radiation-induced disturbances can be monitored with a TES and they appear as a current peak with an amplitude of $\sim 100$ nA, a~rising edge of 5 $\upmu$s and a 100 $\upmu$s decay time. In~reality, \citet{orrell2021sensor} also affirm that the choice of superconducting detector (in this case a TES) is 
potentially interchangable between TESs, MKIDs and~SNSPDs.

\section{TES as Detectors for X-Ray Spectroscopy and~Imaging}\label{sec:X-ray}
\unskip

\subsection{X-Ray Spectroscopy at Beam-Line~Facilities}
Transition Edge Sensors can be optimized to be excellent Soft X-ray detectors for many applications at beam-line facilities and in STEM equipment~\cite{maehata2016transition}. Near~unity quantum efficiency, an~energy resolution in the order of a few eV at 1--10~keV and their wide bandwidth makes them the obvious choice for many applications such as X-ray Emission and Absorption Spectroscopy (XES and XAS) and Resonant Inelastic X-ray Scattering (RIXS) \cite{lee2019soft}. TES detectors have been deployed at X-ray light sources since 2012 (U7A at Sinchrotron National Light Source - SNLS) \cite{ullom2014transition}. The~detector array consists of 240 Mo/Cu ($T_c=107$ mK) TESs coupled each to a Bi absorber and  read out in a 8$\times$30 TDM scheme with a switching interval of 160~ns.

\textls[-15]{The deployment of TESs as beam-line X-ray detectors has been shown by \citet{doriese2017practical}} who report  NIST spectrometers deployed at light sources such as NIST, SNLS, Paul Scherrer Institut (PSI), Jyväskylä Pelletron and Lund~Kemicentrum. 

At NIST, the demonstration scale spectrometer acquired the spectrum of ammonium ferrioxalate in water~\cite{miaja2015laser} as well as Fe$_{2}$O$_{3}$ and FeS$_{3}$, which are high-spin and low-spin compounds. {XES} allowed one to distinguish the different features of the different ferrous compounds through the intensity ratios of K$_{{\alpha}_{1}}$ and K$_{{\alpha}_{2}}$ lines as well as K$_{{\beta}'}$ features in the Fe$_{2}$O$_{3}$ specimen. 

The NIST system, like the one deployed in Lund, was proven to be capable of performing pump-probe XES to measure the spin state of different Fe vs time in photo-induced reactions~\cite{joe2015observation}.

At SNLS, the~spectrometers were deployed to measure the elemental compositions of different samples, in~particular the first tests included the spectrum of a 0.7\% C sample in SiO$_2$, but~most importantly it was successful at distinguishing ammonium nitrite (NH$_4$NO$_3$) from the explosive RDX (C$_3$H$_6$N$_6$O$_6$) based solely on the nitrogen content and in faster time than with more standard spectrometers~\cite{uhlig2015high}.

The application at PSI involved the spectroscopy of exotic atoms and is fully described in Section~\ref{sec:exotic}; at Jyväskylä Pelletron, the spectrometer was tested for Particle-Induced X-ray Emission (PIXE) spectroscopy. An~ion beam excites the samples in order to study its elemental composition through X-ray spectroscopy. TESs are excellent detectors for PIXE spectroscopy because of their high energy resolution that allows the resolution of almost all spectral-line overlaps of different atoms and because they exhibit better collection efficiency than {Silicon Drift Detectors (SDDs)} \cite{doriese2017practical}. \citet{palosaari2016broadband} demonstrated the capability of a TES-based PIXE spectrometer to distinguish eV-scale shifts in the Ti $K_\alpha$ line and $K_\beta$ structures in samples known to exhibit Ti atoms in different oxidation~states.
 
One further application of a 240 pixel TES array produced by NIST is in an instrument at the SPring-8 (Super Photon Ring – 8 GeV) synchrotron. Such a TES array was used to conduct X-ray Absorption Near-Edge Structure (XANES) analysis in fluorescence. TESs resulted in excellent detectors of fluorescence lines from diluted samples thanks to their energy resolution being better than 5 eV at 6~keV. As~a test example, the~NIST TES array could detect the presence of rare earths in a natural sample: europium (Eu), for~instance, could be detected as soon as its mass fraction exceeded 0.1\% of Mn~\cite{yamada2021broadband}. Finally, \mbox{\citet{yamada2021broadband}} demonstrated the capability of the instrument to distinguish between the spectrum of a blank target from the same target in a rare Fe aerosol.  
In 2019, the NIST EBIT TES Spectrometer (NETS) was commissioned, it improves on the capabilities of the spectrometer for the Electron Beam Ion Trap (EBIT). It consists of a 192 TES array optimized for sensitivity in the 0.5--8~keV range with energy resolution ranging between 3.7 and 5 eV. The~calibration of the instrument resulted in line accuracy to below 100 meV in narrow band~spectra.

Further development is in progress to develop a 1000 pixel spectrometer with 0.5 eV energy resolution for energies below 1~keV to be deployed at the Linac Coherent Light Source (LCLS-II) at SLAC National Laboratory~\cite{morgan2019use}. 

\subsection{X-Ray Spectroscopy for Exotic~Atoms}\label{sec:exotic}
Exotic atoms are atoms which are normal except for the replacement of one or more of the sub-atomic particles with different particles with the same electric charge. For~instance, Hadronic atoms are atoms which have one or more electrons replaced by hadrons such as $\pi$ (pionic atoms) or $K$ (kaonic atoms); hadrons such as $K$ can interact with the atomic nucleus not only through electromagnetic force, but~also with attractive short-range strong force. This is most relevant for hadrons in the most inner and tightly bound orbitals. When looking at the innermost orbitals of a hadronic atom, the~strong-force interaction produces a shift in their energy levels and a widening of their spectral lines due to the absorption of the hadron by the nucleus~\cite{okada2014high}. This interaction is not well established quantitatively due to a  lack of precise data and because the current data can be explained through both a phenomenological interaction potential ($\sim -180$ MeV) and a chirally motivated potential ($\sim-50$ MeV). The~shift and the broadening can be measured from the characteristic X-ray emission. Understanding such nuclear states would not only provide the community with better understanding of hadron properties, but~also enlarge the concept of matter and provide further ground for exotic-atoms based astrophysics and cosmological~models.
  
Traditionally, exotic atom X-ray spectroscopy was performed with SDDs but their energy resolution of about 200 eV at 6~keV was a major limitation to the experiments. The~HEATES collaboration~\cite{tatsuno2020mitigating,heates2016first} demonstrated that  Transition Edge Sensors can be multiplexed in a 240-pixel TDM array and optimized to exhibit an energy resolution of \mbox{$\sim$5.7 eV} at 6.9~keV, allowing the identification of spectral features with systematic uncertainties in the order of 0.1 eV as shown by~\cite{heates2016first}. With~such an  instrument, in~2014  the HEATES collaboration \citet{heates2016first} observed the $\pi-^{12}C$ $4f\xrightarrow{}3d$ transition and for the first time the $\pi-^{12}C$ $4d\xrightarrow{}3p$ which turned out to be compatible with the theoretical calculations through Seki--Masutani potential~\cite{seki1983unified}. Based off such promising results,  the 
$3d\xrightarrow{}2p$ transition was measured in kaonic-helium~\cite{hashimoto2016beamline} such as JPARC-E62, producing measurements with {sensitivities} ten times better than those achieved with SDDs~\cite{hashimoto2022measurements}. Further development is in progress to upgrade the test facility~\cite{aoki2021extension}.

\subsection{Nanoscale X-Ray Tomography of Integrated~Circuits}
All proposed instruments for X-ray tomography exploit the same working principle: the X-rays are generated in a metallic conversion layer (usually Pt or Cu) on which the electron beam of a Scanning Electron Microscope (SEM) is focused  in a  nanometre spot,  which results in an X-ray-generation volume with a diameter $\sim$130--160 nm. The~X-rays thus produced are attenuated by the sample and measured by an instrument which features a TES array. NIST is leading the development efforts of such instruments. The~first proposed instrument was NSENSE (Non-destructive Statistical Estimation of Nanoscale Structures and Electronics) \cite{pappas2019tes} which then morphed into TOMCAT (Tomographic Circuit Analysis Tool) \cite{szypryt2023tabletop} and finally into MINT (Microscope for Integrated circuit NanoTomography). Across the years, the detectors have changed from Mo/Cu TESs with a critical temperature of 130 mK~\cite{pappas2019tes, nakamura2024nanoscale} to Mo/Au TESs with the same critical temperature~\cite{szypryt2023tabletop} suspended on a SiN membrane and coupled to a bismuth X-ray absorber. The~number of detectors increased from 240 for the first prototype of NSENSE~\cite{pappas2019tes} and MINT~\cite{nakamura2024nanoscale} to about \mbox{1000 pixels} for the first assemblies of  TOMCAT~\cite{szypryt2023tabletop}. In~this picture, the~multiplexing technique   evolved from an array of TESs read out in a TDM scheme for NSENSE~\cite{pappas2019tes} and MINT~\cite{nakamura2024nanoscale} to a $\upmu$-MUX scheme for  the TOMCAT~\cite{szypryt2023tabletop}  prototype instrument. The~energy resolution of the TESs has been measured to be 17.9 eV at the Cu K$_\alpha$ line (8.04~keV) for MINT~\cite{nakamura2024nanoscale}, 14 eV for TOMCAT~\cite{szypryt2023tabletop} and 12 eV for NSENSE~\cite{pappas2019tes}. Regardless, this technology has been proven effective at resolving features as small as 160 nm in a planar Cu-SiO$_2$ IC~\cite{levine2023tabletop,nakamura2024nanoscale}. Further development is being carried out, especially towards the optimization of the TOMCAT instrument which is being expanded in order to accommodate 3000 pixels instead of the currently 1000 available, which will both increase the contrast of the tomography scans produced and reduce the photon-collection~times.

\citet{kikuchi} developed an 8-Pixel TES-spectrometer for 320~keV radiation, pushing the detection range of TES-based instruments into the $\gamma$-ray range. Each pixel consists of a 0.8 mm-thick Sn absorber on a SiO$_2$/Si$_x$N$_y$/SiO$_2$ membrane and a Ti/Au TES with a critical temperature $T_C$ $\sim 115$ mK. Such an instrument has been validated by carrying out $^{51}$Cr spectroscopy, yielding an impressive energy resolution of 159 eV at 320~keV. 
\subsection{Nuclear~Safety}
In the scope of global nuclear nonproliferation and safeguards, the~assay of plutonium-bearing materials can be carried out in a destructive and in a non-destructive fashion. The~state of the art involves High-Purity germanium (HPGe) detectors which still exhibit an uncertainty limit of about 1\% relative error for measured isotope ratios,  which is still about one order of magnitude larger than that for destructive assays.~\cite{Hoover2014Uncertainty}.         
In order to be able to measure the relative isotope abundance to 0.5\%, spectra with up to tens of millions of counts are required~\cite{bennett2012high}, and~therefore detectors capable of handling kHz count rates are necessary in order to perform such measurements in reasonable times. Transition Edge Sensors, therefore, fit perfectly the required technology.
Unfortunately, the~energy resolution of a TES calorimeter scales with the square root of its heat capacity~\cite{Hoover2014Uncertainty}; therefore, in~order to scale up the collection area while keeping the noise of the instrument to a minimum, it is necessary to employ a multi-pixel array: the individual heat capacity is small, whereas the total collection area~increases.

Since the early 2000s, the~ongoing development and optimization of Transition Edge Sensor technology has been trying to push the boundaries of $\gamma$-ray spectrometry, especially in the 20--200~keV range, which is that typical of applications such as the Non-Destructive Assay (NDA) of plutonium-enriched materials. The~seminal work in this direction was lead by \citet{zink2006array} and proved the feasibility of measuring a $\gamma$-ray spectrum at 103.5~keV with  $\Delta E$ as low as 42 eV~\cite{zink2006array} and later 27 eV~\cite{ullom2007multiplexed}. Such a calorimeter~\cite{zink2006array} was achieved through a 16-array   1x1 mm$^2$ Sn absorber each coupled to a Cu/Mo TES suspended on a SiN~membrane. 

In 2012, a NIST-LANL $\gamma$-spectrometer was first developed~\cite{bennett2012high,hoover2011large},  scaling up the aforementioned instrument~\cite{1596455,doriese2008toward,5204694}. It exhibited a 256-pixel array with a collecting area of 5 cm$^2$ and it featured an energy resolution of 53 eV at 97~keV. Each detector consisted of a tin (Sn) absorber (0.9$\times$0.9$\times$ 0.22 mm$^3$) coupled to a {Cu/Mo} TES with critical temperature in the 100--150 mK range. The~NIST-LANL instrument exhibited a $92$\% pixel yield and achieved optimal performance for {counting rates} below 2.5$\times$10$^3$ counts per second and, as~a proof of concept, was demonstrated to resolve the six most prominent $\gamma$-ray lines of a $^{153}$Gd radioisotope. {The capability of an array of TES micro-calorimeters to measure the isotopic abundance in different Pu measurement standards was shown in~\cite{hoover2014determination,winkler2015256}}. Using the peaks  at 129~keV and 203~keV, the isotopic ratios were calculated: $^{238}$ Pu/$^{239}$ Pu, $^{240}$ Pu/$^{239}$ Pu, $^{241}$ Pu/$^{239}$ Pu and $^{241}$Am/$^{239}$Pu. Such detectors were demonstrated to measure the correct value for such ratios, but~when compared to HPGe detectors, TES micro-calorimeters exhibit an uncertainty which is still a factor 2 higher, thus limiting the resolution of such isotopic ratios  to about $1\%$.

Unfortunately, a~relative uncertainty of about 1\% is not sufficient to achieve desired safeguard goals for large power plants. For~clarity, a~1\% Pu content in the fuel and a 1\% measurement uncertainty leads to an uncertainty in the mass of the Pu fuel larger than the International Atomic Energy Agency (IAEA) of 8 kg for all plants with a fuel throughput larger than 80,000 kg per~year. 

In 2019, \citet{becker2019advances} finally demonstrated isotope ratios with uncertainty levels below 0.25\%. This was possible using the latest NIST/LANL instrument developed in collaboration with the University of Colorado, SLEDGEHAMMER (Spectrometer to Leverage Extensive Development of Gamma-ray TESs for Huge Arrays using Microwave Multiplexed Enabled Readout), which consists of 125 TES micro-calorimeters with typical energy resolution of 75 eV at 100~keV. Such low uncertainties were achieved by acquiring high-count spectra (10$^8$ events in about 14 h) and feeding them through a pipeline. Such a pipeline at first operates to reject all data that exhibit event pile-up and also discards any events for which the baseline rms is larger than expected. Thus,  the~data were fed into an optimum filter pipeline~\cite{fomin2012optimal} before being calibrated against a $^{57}$Co-$^{166m}$Ho source which features nine calibration lines in the energy range of interest (50--200~keV). The~abundance ratio was evaluated at 103~keV with resolution on the area underneath the peak as low as 0.05\%. Further improvement would rely on the development of a section of the pipeline that corrects for drifts in the signal produced by the calorimeters. Currently, there is ongoing work towards the increase in the number of micro-calorimeters by a factor 10 in order to speed up the assay and also the development of real-time analysis software.

Only in 2021 were the 103~keV and 159~keV $^{242}$Pu lines   detected in a non-destructive measurement. The~instrument used was SOFIA (Spectromenter Optimized for Facility Integrated Applications) \cite{croce2021electrochemical,croce2022nuclear};  it was used to characterize a sample consisting of 113.6 g of PuO$_2$ (99.75g of Pu) in a steel `food-pack' and placed inside a SAVVY container. \citet{mercer2022quantification} demonstrated a very clear observation of the 103~keV peak from the $^{242}$Pu isotope with a precision on the position of the peak better than 0.5 eV. Furthermore, SOFIA is capable of resolving the $^{242}$Pu peak at 103.46~keV from {the ones} at 103.68~keV (arising from $^{241}$Pu) and the peak at 104.23~keV ($^{240}$Pu). Furthermore, in~the region around 159~keV, SOFIA is capable of distinguishing the 159.02~keV peak ($^{241}$Pu) from the ones at 159.96 kev  ($^{241}$Pu) and 160.31~keV ($^{241}$Pu $\alpha$-decay). SOFIA was capable of measuring a $^{242}$Pu/$^{240}$Pu ratio of (8.5 $\pm$ 2)\% after a 12.5-hour measurement with the counting statistics being the dominant factor in the uncertainty~level.

\section{Comparison with Other Superconducting~Detectors}\label{sec:comparison}
Transition Edge Sensors are the most mature technology when it comes to superconducting detectors {and} have been widely used for the better part of the last thirty years since \mbox{\citet{Kent_D_Irwin_1995}} suggested their use in the electro-thermal feedback. TESs (both as bolometers and calorimeters) have seen applications in a plethora of fields of research and civil applications, most of which are described in this review. Transition Edge Sensors have been developed to a state of the art of unprecedented performance. Modern TESs exhibit resolving powers up to $R$ = 4000~\cite{zink2006array} and NEP below 1 $\times 10^{-17} $~W/$\sqrt{\mathrm{Hz}}$. Large arrays of TES have been developed and deploy several thousand pixels with multiplexing factors of up to 2000:1~\cite{dober2021microwave}. 

In more recent years, other more-novel superconducting detectors have been invented and optimized. We intend to provide the reader with a brief overview of their characteristics and figures of merit. Our aim is to offer a helping hand in the troubled water of deciding what are the ideal superconducting detectors in many typical cases. Table~\ref{tab:comparison1} shows the main figures of merits of different superconducting~detectors.

\begin{table}[H]
   \setlength{\tabcolsep}{5.6mm}
    \caption{\textbf{Comparison between different kinds of superconducting detectors} %MDPI: 1. Is the bold for the first sentence necessary? please check it in all Table captions. 2. Please confirm if the explanation of the circles needs to be added in the table footer. 3. We moved the table after where it is first mentioned in the text. Please confirm.
     through some of their main figures of merit and other important~information.}
    \label{tab:comparison1}
    \begin{tabular}{cccc}
     \toprule
        & \textbf{TES} & \textbf{MKIDs} & \textbf{SNSPD}\\
    \midrule
      Fabrication complexity & $\circ \circ \circ$ & $\circ$ & $\circ \circ$  \\
      Max. R (X-ray)& $\leq$5000 & $\leq$590 & N/A \\
      Max. R (UVOIR)& $\sim$200 & 65 & $\leq$10 \\
      NEP (typical) (W$/\sqrt{\mathrm{Hz}})$ & $1\times10^{-19}$ & $3\times10^{-19}$ & $4.5\times10^{-19}$  \\
      Time resolution & $\sim$1 $\upmu$s & $\sim $1 $\upmu$s & $\sim $1 ps \\
      $\upmu$-MUX available? & Yes & Innate & Yes \\ 
      Spectral range & THz/$\gamma$-ray & THz/X-ray & infra-red/Vis \\
      Array size (typical) & 1 k & 20 k & 10 k\\
      Pixel Yield & 95\% & 75--80\% & >95\% \\
      \bottomrule
         
    \end{tabular}
   
\end{table}

\subsection{Microwave Kinetic Inductance~Detectors}
MKIDs are superconducting LC micro-resonators invented by \citet{day2003broadband} in 2003 and have since been employed as photo-detectors in instrumentation ranging from astronomical telescopes to particle physics, to~security applications~\cite{ulbricht2021applications}. Their working principle is rather simple: when a particle deposits its energy in the superconductor, it breaks down a number of Cooper pairs producing un-paired electrons. As~the superconductor is depleted of superconducting carriers, its kinetic inductance increases (see \citet{zmuidzinas2012superconducting}), which results in a reduced resonance frequency of the LC circuit. With~a time constant that is typical of the superconductor of choice, the~un-paired electrons recombine and the MKID is in its idle state ready for a new detection event. The~shift in frequency is proportional to the energy deposited in the superconductor; therefore,  it is possible to monitor the LC resonator in frequency and phase to identify particle-detection events and measure the energy deposited by the~particle.

Advantages include sensitivity from X-rays to THz and their very straightforward fabrication as well as their inherent capability of being read out in an FDM/$\upmu$-MUX scheme. So far, arrays with up to 20,000 pixels with a multiplexing factor of $2000:1$  have been demonstrated~\cite{MEC}. The~pixel yield still lies below 80\% and it can be due to inhomogeneities in the properties of the thin films~\cite{vissers2013characterization}  but it can be optimized through an accurate choice of the superconducting thin film~\cite{de2022high} cautiously simulated design~\cite{mcaleer2024automation} (as opposed to interpolated) and \textit{post-facto} leg trimming~\cite{mckenney2019tile} and DC-bias~\cite{de2020multiplexable}.

\subsection{Superconducting Nanowire Single Photon~Detectors}\label{sec:SNSPDs}
SNSPDs are thin and narrow superconducting nanowires patterned in a meander geometry to arrange them in a compact pixel. Quite like a TES, the~SNSPD is kept at a temperature below its critical temperature but it is biased with a current which is close to, but~smaller, than~its critical current. When a particle strikes on the superconductor, it depletes it locally of Cooper pairs and produces a resistive hotspot. Such finite resistance produces a voltage drop across the SNSPD which can be measured. SNSPDs exhibit incredible quantum efficiency and pico-second time resolution, but~exhibit no energy resolution and no photon-counting capabilities if not multiplexed (usually FDM) into arrays. As~of July 2024, kilo-pixel arrays of SNSPDs are commercially~available.

%\subsection{Metallic Magnetic Calorimeters}
%Metallic Magnetic Calorimeters (MMCs) consist of a metallic absorber and a paramagnetic sensor, which are in strong thermal contact with each other but are weakly coupled to a thermal bath (kept usually below 100 mK). When an energetic particle strikes on the metallic absorber, its temperature increases and this increase in temperature results in a variation in the magnetisation of the paramagnetic sensor. Slowly, the thermal contact to the bath brings the MMC back to its idle configuration ready for a new detection event. This change in magnetisation can be measured through a SQUID magnetometer in a similar fashion to what is done for TESs. FDM multiplexing is currently being optimized in order to scale MMC arrays to kilo and mega-pixel capabilities. State of the art MMC arrays exhibit a resolving power as high as 6000 for large energy bands~\cite{kempf2018physics,sikorsky2020measurement}.  

%%%%%%%%%%%%%%%%%%%%%%%%%%%%%%%%%%%%%%%%%%\section{Materials and Methods}

\vspace{6pt}

\authorcontributions{Conceptualization, {M.D.L}.;  writing---original draft preparation, {M.D.L.} except: {F.P.} (2.1--2.4; 5.0,5.4), %MDPI: Please confirm if these refer to section number, if yes, please add the right section ID for them.
{P.D.B.} (3.8), {E.D.G.} (3.7.2;3.9), {T.L.} (5.1--5-3), {C.P.} (3.6.2); writing---review and editing, {M.D.L.}; visualization, {M.D.L}. and {F.P.}; supervision, {M.D.L.}; project administration, {M.D.L.} All authors have read and agreed to the published version of the manuscript.}

\funding{This research was partially funded by Italian Space Agency under ASI Grants No. 2020-9-HH.0 and 2020-25-HH.0 ({M.D.L.)}, by~the Piano Nazionale di Ripresa e Resilienza, Ministero dell'Universit\'a e della Ricerca (PNRR MUR) Project under Grant PE0000023-NQSTI ({C.P.}), and~by the National Recovery and Resilience Plan (NRRP), Mission 4, Component 2, Investment 1.1, Call PRIN 2022 by the Italian Ministry of University and Research (MUR), funded by the European Union – NextGenerationEU – EQUATE Project, ``Defect engineered graphene for electro-thermal quantum technology''---Grant Assignment Decree No. 2022Z7RHRS ({F.P.}).}

\acknowledgments{The authors acknowledge Prof. F. Gatti for the fabrication of the TES shown in Figure~\ref{fig:teslspe} and the NIST, Boulder Quantum Sensors Division for the fabrication of the TES shown in Figure~\ref{fig:teslitebird}.}

\conflictsofinterest{The authors declare no conflicts of~interest.} 

%%%%%%%%%%%%%%%%%%%%%%%%%%%%%%%%%%%%%%%%%%
%% Optional

%% Only for journal Encyclopedia
%\entrylink{The Link to this entry published on the encyclopedia platform.}

\abbreviations{Abbreviations}{
~The following abbreviations are used in this manuscript:\\
\vspace{-6pt}
%\noindent 
\begin{longtable}{>{\raggedright}p{0.2\textwidth}>{\raggedright\arraybackslash}p{0.75\textwidth}}
ABS & Atacama B-mode~Search \\
ACT & Atacama Cosmology~Telescope\\
AliCPT & Ali CMB Polarization~Telescope\\
ALMA & Atacama Large Millimeter/submillimeter~Array\\
ALPS II & Any Light Particle Search~II\\
ASI & Agenzia Spaziale~Italiana \\
Athena & Advanced Telescope for High-ENergy~Astrophysics\\
BBO & Beta Barium~Borate\\
BCS & Bardeen Cooper~Schreifer\\
BICEP & Background Imaging of Cosmic Extragalactic~Polarization\\
CCD & Charge Coupled~Device\\
CDM & Code Division~Multiplexing\\
CE$\nu$NS & Coherent Elastic Neutrino Nucleus~Scattering\\
CLASS & Cosmology Large Angular Scale~Surveyor\\
CLSM & Confocal Laser Scanning~Microscope\\
CMB & Cosmic Microwave~Background\\
CMB-S4 & Cosmic Microwave Background - Stage~4\\
CMOS & Complimentary Metal-Oxide-Semiconductor\\
CSS & Coherent State~Superposition\\
CPW & Coplanar~Waveguide\\
CRESST & Cryogenic Rare Event Search using Superconducting~Thermometers\\
CUPID & CUORE Upgrade with Particle~IDentification\\
DAC & Digital-to-Analog~Converter\\
DM & Dark~Matter\\
EBEX & E and B~EXperiment\\
EBIT & Electron Beam Ion~Trap\\
EC & Electron~Capture\\
ESA & European Space~Agency\\
FIP & Far-Infrared~Polarimeter\\
FoV & Field of~View\\
FDM & Frequency Division~Multiplexing\\
GISMO & Goddard IRAM Superconducting Millimeter~Observer\\
HEMT & High Electron Mobility~Transistor\\
HeRALD & Helium Roton Apparatus for Light Dark~matter\\
HIXI &  High Definition X-ray~Imager\\
HOM & Hong-Ou-Mandel\\
HPGe & High Purity~germanium\\
IAEA & International Atomic Energy~Agency\\
IC & Integrated~Circuits\\
INFN & Istituto Nazionale Fisica~Nucleare\\
ISTED & Inductive Superconducting Transition Edge~Sensor\\
JPARC & Japan Proton Accelerator Research~Complex\\
JWST & James Webb Space~Telescope\\
LANL & Los Alamos National~Laboratory\\
LAT & Large Aperture~Telescope\\
LCLS & Linac Coherent Light~Source\\
LFT & Low Frequency~Telescope\\
LiteBIRD &  Lite (Light) satellite for the study of B-mode~polarization\\
& and Inflation from cosmic background Radiation~Detection\\
LSPE & Large-Scale Polarization~Explorer\\
LSW & Light Shining through a~Wall\\
LXM & Lynx X-ray~Microcalorimeter\\
$\upmu$-MUX & Microwave~Multiplexing\\
MBAC & Millimeter Bolometer Array~Camera\\
MHFT & Medium and High Frequency~Telescope\\
MINT & Microscope for Integrated circuit~NanoTomography\\
MISC-T & Mid-Infrared Spectrometer Camera~Transit\\
MKID & Microwave Kinetic Inductance~Detector\\
NDA & Non-Destructive~Assay\\
NEP & Noise Equivalent~Power\\
NETF & Negative electro-thermal~Feedback\\
NETS & NIST EBIT TES~Spectrometer\\
NIST & National Institute of Standards and~Technology\\
NSENSE & Non-destructive Statistical Estimation of Nanoscale Structures and~Electronics\\
NTD & Neutron Transmutation~Doped\\
OMT & Ortho-Mode~Transducer\\
OSS & Origins Survey~Spectrometer\\
PIXE & Particle-Induced X-ray~Emission\\
PMT & Photo-multiplier~Tube\\
POLARBEAR & POLARization of the Background~Radiation\\
PSD & Power Spectral~Density\\
PSI & Paul Scherrer~Institut\\
QKD & Quantum Key~Distribution\\
QUBIC & Q  and  U Bolometric Interferometer for~Cosmology \\
RIXS & Resonant Inelastic X-ray~Scattering\\
SAT & Small Aperture~Telescope\\
SCUBA-2 & Submillimetre Common-User Bolometer Array~2\\
SDD & Silicon Drift~Detector\\
SEM & Scanning Electron~Microscope\\
SLEDGEHAMMER & Spectrometer to Leverage Extensive Development of Gamma-ray\\
& TESs for Huge Arrays using Microwave Multiplexed Enabled~Readout\\
SNLS & Sinchrotron National Light~Source \\
SNSPD & Superconducting Nanowire Single Photon~Detector\\
SNR & Signal-to-Noise~Ratio\\
SPDT & Single Pole Double~Throw\\
SOFIA & Spectromenter Optimized for Facility Integrated~Applications \\
SPICA & Space Infrared Telescope for Cosmology and~Astrophysics\\
SPICE & Sub-eV Polar Interactions Cryogenic~Experiment\\
SPT & South Pole~Telescope\\
SPTPol & South Pole Telescope~Polarimeter\\
SQUID & Superconducting QUantum Interference~Device\\
STEM & Scanning Transmission Electron~Microscope\\
STFC & (UK) Science and Technology Facilities~Council\\
STRIP & Survey TeneRIfe~Polarimeter\\
SWIPE & Short Wavelength Instrument for the Polarization~Explorer\\
TDM & Time Division~Multiplexing\\
TES & Transition Edge~Sensor\\
TOD & Time-Ordered~Data\\
TOMCAT & Tomographic Circuit Analysis~Tool\\
WFI & Wide Field~Instrument\\
WHIM & Warm-Hot Interstellar~Medium\\
WIMP & Weakly Interacting Massive~Particle\\
X-IFU & X-ray Integral Field~Unit\\
XANES & X-ray Absorption Near-Edge~Structure\\
XAS & X-ray Absorption~Spectroscopy\\
XES & X-ray Emission~Spectroscopy\\
XGS & X-ray Grating Spectrometer
\end{longtable}
}

%%%%%%%%%%%%%%%%%%%%%%%%%%%%%%%%%%%%%%%%%%
%% Optional
%\appendixtitles{no} % Leave argument "no" if all appendix headings stay EMPTY (then no dot is printed after "Appendix A"). If the appendix sections contain a heading then change the argument to "yes".
%\appendixstart
%\appendix
%\section[\appendixname~\thesection]{}
%\subsection[\appendixname~\thesubsection]{}
%The appendix is an optional section that can contain details and data supplemental to the main text---for example, explanations of experimental details that would disrupt the flow of the main text but nonetheless remain crucial to understanding and reproducing the research shown; figures of replicates for experiments of which representative data are shown in the main text can be added here if brief, or as Supplementary Data. Mathematical proofs of results not central to the paper can be added as an appendix.

%\begin{table}[H] 
%\caption{This is a table caption.\label{tab5}}
%\begin{tabularx}{\textwidth}{CCC}
%\toprule
%\textbf{Title 1}	& \textbf{Title 2}	& \textbf{Title 3}\\
%\midrule
%Entry 1		& Data			& Data\\
%Entry 2		& Data			& Data\\
%\bottomrule
%\end{tabularx}
%\end{table}

%\section[\appendixname~\thesection]{}
%All appendix sections must be cited in the main text. In the appendices, Figures, Tables, etc. should be labeled, starting with ``A''---e.g., Figure A1, Figure A2, etc.

%%%%%%%%%%%%%%%%%%%%%%%%%%%%%%%%%%%%%%%%%%
%\begin{adjustwidth}{-\extralength}{0cm}

\begin{adjustwidth}{-\extralength}{0cm}
%\centering %% If there is a figure in wide page, please release command \centering
\reftitle{References}


%\bibliography{bibliography}

\begin{thebibliography}{999}

\bibitem[Enss(2005)]{enss2005cryogenic}
Enss, C.E.
\newblock {\em Cryogenic Particle Detection}; Springer-Verlag Berlin Heidelberg, Germany, 2005 %MDPI: Please add the publisher and location.
 Volume~99.

\bibitem[Enss and Mccammon(2008)]{enss2008physical}
Enss, C.; Mccammon, D.
\newblock Physical principles of low temperature detectors: Ultimate
  performance limits and current detector capabilities.
\newblock {\em J. Low Temp. Phys.} {\bf 2008}, {\em
  151},~5--24.

\bibitem[Gottardi and Nagayashi(2021)]{gottardi2021review}
Gottardi, L.; Nagayashi, K.
\newblock A review of X-ray microcalorimeters based on superconducting
  Transition Edge Sensors for astrophysics and particle physics.
\newblock {\em Appl. Sci.} {\bf 2021}, {\em 11},~3793.

\bibitem[Nucciotti(2016)]{nucciotti2016use}
Nucciotti, A.
\newblock The use of low temperature detectors for direct measurements of the
  mass of the electron neutrino.
\newblock {\em Adv. High Energy Phys.} {\bf 2016}, {\em
  2016},~9153024.

\bibitem[Pirro and Mauskopf(2017)]{pirro2017advances}
Pirro, S.; Mauskopf, P.
\newblock Advances in bolometer technology for fundamental physics.
\newblock {\em Annu. Rev. Nucl. Part. Sci.} {\bf 2017}, {\em
  67},~161--181.

\bibitem[Poda and Giuliani(2017)]{poda2017low}
Poda, D.; Giuliani, A.
\newblock Low background techniques in bolometers for double-beta decay search.
\newblock {\em Int. J. Mod. Phys. A} {\bf 2017}, {\em
  32},~1743012.

\bibitem[Poda(2021)]{poda2021scintillation}
Poda, D.
\newblock Scintillation in low-temperature particle detectors.
\newblock {\em Physics} {\bf 2021}, {\em 3},~473--535.

\bibitem[Koehler(2021)]{koehler2021low}
Koehler, K.E.
\newblock Low temperature microcalorimeters for decay energy spectroscopy.
\newblock {\em Appl. Sci.} {\bf 2021}, {\em 11},~4044.

\bibitem[Ullom and Bennett(2015)]{ullom2015review}
Ullom, J.N.; Bennett, D.A.
\newblock Review of superconducting transition-edge sensors for X-ray and
  gamma-ray spectroscopy.
\newblock {\em Supercond. Sci. Technol.} {\bf 2015}, {\em
  28},~084003.

\bibitem[Tinkham(2004)]{Tinkham2004}
Tinkham, M.
\newblock {\em {Introduction to Superconductivity}}; Dover Publications:
  Mineola, NY,  USA, 2004.

\bibitem[Meissner and Ochsenfeld(1933)]{Meissner1933}
Meissner, W.; Ochsenfeld, R.
\newblock Ein neuer Effekt bei Eintritt der Supraleitf{\"a}higkeit.
\newblock {\em Naturwissenschaften} {\bf 1933}, {\em 21},~787--788.
\newblock {\url{https://doi.org/10.1007/BF01504252}}.

\bibitem[London(1948)]{London1948}
London, F.
\newblock On the Problem of the Molecular Theory of Superconductivity.
\newblock {\em Phys. Rev.} {\bf 1948}, {\em 74},~562--573.
\newblock {\url{https://doi.org/10.1103/PhysRev.74.562}}.

\bibitem[Doll and N\"abauer(1961)]{Doll}
Doll, R.; N\"abauer, M.
\newblock Experimental Proof of Magnetic Flux Quantization in a Superconducting
  Ring.
\newblock {\em Phys. Rev. Lett.} {\bf 1961}, {\em 7},~51--52.
\newblock {\url{https://doi.org/10.1103/PhysRevLett.7.51}}.

\bibitem[Deaver and Fairbank(1961)]{Deaver}
Deaver, B.S.; Fairbank, W.M.
\newblock Experimental Evidence for Quantized Flux in Superconducting
  Cylinders.
\newblock {\em Phys. Rev. Lett.} {\bf 1961}, {\em 7},~43--46.
\newblock {\url{https://doi.org/10.1103/PhysRevLett.7.43}}.

\bibitem[Schrieffer(1999)]{schrieffer1999theory}
Schrieffer, J.
\newblock {\em Theory of Superconductivity}; Advanced Books Classics; Avalon
  Publishing: London, UK, %MDPI: newly  added the location of the publisher, please confirm.
  1999.

\bibitem[Keesom and Kok(1934)]{KEESOM1934175}
Keesom, W.; Kok, J.
\newblock Measurements of the specific heat of thallium at liquid helium
  temperatures.
\newblock {\em Physica} {\bf 1934}, {\em 1},~175--181.
\newblock
  {\url{https://doi.org/10.1016/S0031-8914(34)90022-7}}.

\bibitem[Giaever(1960)]{Giaever}
Giaever, I.
\newblock Electron Tunneling Between Two Superconductors.
\newblock {\em Phys. Rev. Lett.} {\bf 1960}, {\em 5},~464--466.
\newblock {\url{https://doi.org/10.1103/PhysRevLett.5.464}}.

\bibitem[Biondi and Garfunkel(1959)]{biondi}
Biondi, M.A.; Garfunkel, M.P.
\newblock Millimeter Wave Absorption in Superconducting Aluminum. I.
  Temperature Dependence of the Energy Gap.
\newblock {\em Phys. Rev.} {\bf 1959}, {\em 116},~853--861.
\newblock {\url{https://doi.org/10.1103/PhysRev.116.853}}.

\bibitem[Glover and Tinkham(1957)]{glover}
Glover, R.E.; Tinkham, M.
\newblock Conductivity of Superconducting Films for Photon Energies between 0.3
  and $40k{T}_{c}$.
\newblock {\em Phys. Rev.} {\bf 1957}, {\em 108},~243--256.
\newblock {\url{https://doi.org/10.1103/PhysRev.108.243}}.

\bibitem[Bardeen et~al.(1957)Bardeen, Cooper, and Schrieffer]{BCS}
Bardeen, J.; Cooper, L.N.; Schrieffer, J.R.
\newblock Microscopic Theory of Superconductivity.
\newblock {\em Phys. Rev.} {\bf 1957}, {\em 106},~162--164.
\newblock {\url{https://doi.org/10.1103/PhysRev.106.162}}.

\bibitem[Cooper(1956)]{Cooper1956}
Cooper, L.N.
\newblock Bound Electron Pairs in a Degenerate Fermi Gas.
\newblock {\em Phys. Rev.} {\bf 1956}, {\em 104},~1189--1190.
\newblock {\url{https://doi.org/10.1103/PhysRev.104.1189}}.

\bibitem[De~Gennes(1999)]{DeGennes}
De~Gennes, P.G.
\newblock {\em {Superconductivity of Metals and Alloys}}; Advanced Book
  Classics; Perseus: Cambridge, MA, USA, 1999.

\bibitem[Usadel(1970)]{Usadel1970}
Usadel, K.D.
\newblock Generalized Diffusion Equation for Superconducting Alloys.
\newblock {\em Phys. Rev. Lett.} {\bf 1970}, {\em 25},~507--509.
\newblock {\url{https://doi.org/10.1103/PhysRevLett.25.507}}.

\bibitem[Anderson(1959)]{anderson1959theory}
Anderson, P.W.
\newblock Theory of dirty superconductors.
\newblock {\em J. Phys. Chem. Solids} {\bf 1959}, {\em
  11},~26--30.

\bibitem[Fominov and Feigel’man(2001)]{fominov2001superconductive}
Fominov, Y.V.; Feigel’man, M.
\newblock Superconductive properties of thin dirty superconductor--normal-metal
  bilayers.
\newblock {\em Phys. Rev. B} {\bf 2001}, {\em 63},~094518.

\bibitem[Brammertz et~al.(2002)Brammertz, Golubov, Verhoeve, den Hartog,
  Peacock, and Rogalla]{Bram}
Brammertz, G.; Golubov, A.A.; Verhoeve, P.; den Hartog, R.; Peacock, A.;
  Rogalla, H.
\newblock {Critical temperature of superconducting bilayers: Theory and
  experiment}.
\newblock {\em Appl. Phys. Lett.} {\bf 2002}, {\em 80},~2955--2957.
  \newblock {\url{https://doi.org/10.1063/1.1470712}}.

\bibitem[Cooper(1961)]{Cooper1961}
Cooper, L.N.
\newblock Superconductivity in the Neighborhood of Metallic Contacts.
\newblock {\em Phys. Rev. Lett.} {\bf 1961}, {\em 6},~689--690.
\newblock {\url{https://doi.org/10.1103/PhysRevLett.6.689}}.

\bibitem[Martinis et~al.(2000)Martinis, Hilton, Irwin, and
  Wollman]{MARTINIS200023}
Martinis, J.M.; Hilton, G.; Irwin, K.; Wollman, D.
\newblock Calculation of TC in a normal-superconductor bilayer using the
  microscopic-based Usadel theory.
\newblock {\em Nucl. Instrum. Methods Phys. Res. Sect. A Accel. Spectrom. Detect. Assoc. Equip.} {\bf 2000},
  {\em 444},~23--27.
\newblock
  {\url{https://doi.org/10.1016/S0168-9002(99)01320-0}}.

\bibitem[Giazotto et~al.(2006)Giazotto, Heikkil\"a, Luukanen, Savin, and
  Pekola]{Giazotto2006}
Giazotto, F.; Heikkil\"a, T.T.; Luukanen, A.; Savin, A.M.; Pekola, J.P.
\newblock Opportunities for mesoscopics in thermometry and refrigeration:
  Physics and applications.
\newblock {\em Rev. Mod. Phys.} {\bf 2006}, {\em 78},~217--274.
\newblock {\url{https://doi.org/10.1103/RevModPhys.78.217}}.

\bibitem[Pekola et~al.(2004)Pekola, Heikkil\"a, Savin, Flyktman, Giazotto, and
  Hekking]{Pekola2004}
Pekola, J.P.; Heikkil\"a, T.T.; Savin, A.M.; Flyktman, J.T.; Giazotto, F.;
  Hekking, F.W.J.
\newblock Limitations in Cooling Electrons using Normal-Metal-Superconductor
  Tunnel Junctions.
\newblock {\em Phys. Rev. Lett.} {\bf 2004}, {\em 92},~056804.
\newblock {\url{https://doi.org/10.1103/PhysRevLett.92.056804}}.

\bibitem[Bardeen et~al.(1959)Bardeen, Rickayzen, and Tewordt]{Bardeen1959}
Bardeen, J.; Rickayzen, G.; Tewordt, L.
\newblock Theory of the Thermal Conductivity of Superconductors.
\newblock {\em Phys. Rev.} {\bf 1959}, {\em 113},~982--994.
\newblock {\url{https://doi.org/10.1103/PhysRev.113.982}}.

\bibitem[Pekola and Karimi(2021)]{Pekola2021}
Pekola, J.P.; Karimi, B.
\newblock Colloquium: Quantum heat transport in condensed matter systems.
\newblock {\em Rev. Mod. Phys.} {\bf 2021}, {\em 93},~041001.
\newblock {\url{https://doi.org/10.1103/RevModPhys.93.041001}}.

\bibitem[Bergeret et~al.(2018)Bergeret, Silaev, Virtanen, and
  Heikkil\"a]{Bergeret2018}
Bergeret, F.S.; Silaev, M.; Virtanen, P.; Heikkil\"a, T.T.
\newblock Colloquium: Nonequilibrium effects in superconductors with a
  spin-splitting field.
\newblock {\em Rev. Mod. Phys.} {\bf 2018}, {\em 90},~041001.
\newblock {\url{https://doi.org/10.1103/RevModPhys.90.041001}}.

\bibitem[Andreev(1964)]{Andreev1964}
Andreev, A.F.
\newblock The Thermal Conductivity of the Intermediate State in
  Superconductors.
\newblock {\em JETP} {\bf 1964}, {\em 19},~12284.

\bibitem[Paolucci et~al.(2020)Paolucci, Buccheri, Germanese, Ligato, Paoletti,
  Signorelli, Bitossi, Spagnolo, Falferi, Rajteri, Gatti, and
  Giazotto]{Paolucci2020}
Paolucci, F.; Buccheri, V.; Germanese, G.; Ligato, N.; Paoletti, R.;
  Signorelli, G.; Bitossi, M.; Spagnolo, P.; Falferi, P.; Rajteri, M.;  et~al.
\newblock {Development of highly sensitive nanoscale Transition Edge Sensors
  for gigahertz astronomy and dark matter search}.
\newblock {\em J. Appl. Phys.} {\bf 2020}, {\em 128},~194502.
\newblock {\url{https://doi.org/10.1063/5.0021996}}.

\bibitem[Andrews et~al.(1942)Andrews, Brucksch, Ziegler, and
  Blanchard]{Andrews_TES_1942}
Andrews, D.H.; Brucksch, W.F.J.; Ziegler, W.T.; Blanchard, E.R.
\newblock {Attenuated Superconductors I. For Measuring Infra‐Red Radiation}.
\newblock {\em Rev. Sci. Instrum.} {\bf 1942}, {\em
  13},~281--292.
\newblock {\url{https://doi.org/10.1063/1.1770037}}.

\bibitem[Irwin(1995)]{Kent_D_Irwin_1995}
Irwin, K.D.
\newblock {An application of electrothermal feedback for high resolution
  cryogenic particle detection}.
\newblock {\em Appl. Phys. Lett.} {\bf 1995}, {\em 66},~1998--2000.
\newblock {\url{https://doi.org/10.1063/1.113674}}.

\bibitem[Irwin and Hilton(2005)]{irwin2005transition}
Irwin, K.D.; Hilton, G.C.
\newblock Transition-edge sensors.
\newblock In {\em Cryogenic Particle Detection}; Springer:  Berlin/Heidelberg, Germany, %MDPI: newly added information, please confirm, same below.
 {2005}; pp.~63--150.

\bibitem[McMahon et~al.(2012)McMahon, Beall, Becker, Cho, Datta, Fox,
  Halverson, Hubmayr, Irwin, Nibarger, et~al.]{mcmahon2012multi}
McMahon, J.; Beall, J.; Becker, D.; Cho, H.; Datta, R.; Fox, A.; Halverson, N.;
  Hubmayr, J.; Irwin, K.; Nibarger, J.;  et~al.
\newblock Multi-chroic feed-horn coupled TES polarimeters.
\newblock {\em J. Low Temp. Phys.} {\bf 2012}, {\em
  167},~879--884.

\bibitem[Likharev(1979)]{Likharev1979}
Likharev, K.K.
\newblock Superconducting weak links.
\newblock {\em Rev. Mod. Phys.} {\bf 1979}, {\em 51},~101--159.
\newblock {\url{https://doi.org/10.1103/RevModPhys.51.101}}.

\bibitem[Pleikies et~al.(2007)Pleikies, Usenko, Kuit, Flokstra, De~Waard, and
  Frossati]{pleikies2007squid}
Pleikies, J.; Usenko, O.; Kuit, K.; Flokstra, J.; De~Waard, A.; Frossati, G.
\newblock SQUID developments for the Gravitational Wave antenna MiniGRAIL.
\newblock {\em IEEE Trans. Appl. Supercond.} {\bf 2007}, {\em
  17},~764--767.

\bibitem[Ketchen et~al.(1984)Ketchen, Kopley, and Ling]{ketchen1984miniature}
Ketchen, M.B.; Kopley, T.; Ling, H.
\newblock Miniature SQUID susceptometer.
\newblock {\em Appl. Phys. Lett.} {\bf 1984}, {\em 44},~1008--1010.

\bibitem[Wikswo(1995)]{wikswo1995squid}
Wikswo, J.P.
\newblock SQUID magnetometers for biomagnetism and nondestructive testing:
  important questions and initial answers.
\newblock {\em IEEE Trans. Appl. Supercond.} {\bf 1995}, {\em
  5},~74--120.

\bibitem[Jenks et~al.(1997)Jenks, Sadeghi, and Wikswo~Jr]{jenks1997squids}
Jenks, W.; Sadeghi, S.; Wikswo~Jr, J.P.
\newblock SQUIDs for nondestructive evaluation.
\newblock {\em J. Phys. D Appl. Phys.} {\bf 1997}, {\em
  30},~293.

\bibitem[Clarke et~al.(2007)Clarke, Hatridge, and
  M{\"o}{\ss}le]{clarke2007squid}
Clarke, J.; Hatridge, M.; M{\"o}{\ss}le, M.
\newblock SQUID-detected magnetic resonance imaging in microtesla fields.
\newblock {\em Annu. Rev. Biomed. Eng.} {\bf 2007}, {\em 9},~389--413.

\bibitem[M{\"u}ck and McDermott(2010)]{muck2010radio}
M{\"u}ck, M.; McDermott, R.
\newblock Radio-frequency amplifiers based on dc SQUIDs.
\newblock {\em Supercond. Sci. Technol.} {\bf 2010}, {\em
  23},~093001.

\bibitem[Tanaka et~al.(2002)Tanaka, Sekine, Saito, and
  Takayanagi]{tanaka2002dc}
Tanaka, H.; Sekine, Y.; Saito, S.; Takayanagi, H.
\newblock DC-SQUID readout for qubit.
\newblock {\em Phys. C Supercond.} {\bf 2002}, {\em 368},~300--304.

\bibitem[Irwin and Huber(2001)]{irwin2001squid}
Irwin, K.D.; Huber, M.
\newblock SQUID operational amplifier.
\newblock {\em IEEE Trans. Appl. Supercond.} {\bf 2001}, {\em
  11},~1265--1270.

\bibitem[Tsang and {Van Duzer}(1975)]{Tsang1975}
Tsang, W.T.; {Van Duzer}, T.
\newblock {Dc analysis of parallel arrays of two and three Josephson
  junctions}.
\newblock {\em J. Appl. Phys.} {\bf 1975}, {\em 46},~4573--4580.
\newblock {\url{https://doi.org/10.1063/1.321397}}.

\bibitem[Huber et~al.(2001)Huber, Neil, Benson, Burns, Corey, Flynn,
  Kitaygorodskaya, Massihzadeh, Martinis, and Hilton]{huber2001dc}
Huber, M.E.; Neil, P.A.; Benson, R.G.; Burns, D.A.; Corey, A.; Flynn, C.S.;
  Kitaygorodskaya, Y.; Massihzadeh, O.; Martinis, J.M.; Hilton, G.
\newblock DC SQUID series array amplifiers with 120 MHz bandwidth (corrected).
\newblock {\em IEEE Trans. Appl. Supercond.} {\bf 2001}, {\em
  11},~4048--4053.

\bibitem[Labarias et~al.(2023)Labarias, Müller, and
  Mitchell]{GaliLabarias_2023}
Labarias, M.A.G.; Müller, K.H.; Mitchell, E.E.
\newblock The effect of bias current configuration on the performance of SQUID
  arrays.
\newblock {\em Supercond. Sci. Technol.} {\bf 2023}, {\em
  36},~115016.
\newblock {\url{https://doi.org/10.1088/1361-6668/acfa7a}}.

\bibitem[Mukhanov et~al.(2014)Mukhanov, Prokopenko, and
  Romanofsky]{Mukhanov2014}
Mukhanov, O.; Prokopenko, G.; Romanofsky, R.
\newblock Quantum Sensitivity: Superconducting Quantum Interference
  Filter-Based Microwave Receivers.
\newblock {\em IEEE Microw. Mag.} {\bf 2014}, {\em 15},~57--65.
\newblock {\url{https://doi.org/10.1109/MMM.2014.2332421}}.

\bibitem[Welty and Martinis(1993)]{welty1993two}
Welty, R.P.; Martinis, J.M.
\newblock Two-stage integrated SQUID amplifier with series array output.
\newblock {\em IEEE Trans. Appl. Supercond.} {\bf 1993}, {\em
  3},~2605--2608.

\bibitem[Kiviranta(2020)]{kiviranta2020two}
Kiviranta, M.
\newblock Two-stage SQUID amplifier with bias current re-use.
\newblock {\em arXiv} {\bf 2020}, arXiv:2012.15362.

\bibitem[Drung et~al.(2006)Drung, Hinnrichs, and Barthelmess]{drung2006low}
Drung, D.; Hinnrichs, C.; Barthelmess, H.J.
\newblock Low-noise ultra-high-speed dc SQUID readout electronics.
\newblock {\em Supercond. Sci. Technol.} {\bf 2006}, {\em
  19},~S235.

\bibitem[Cantor et~al.(1997)Cantor, Lee, Matlashov, and
  Vinetskiy]{cantor1997low}
Cantor, R.; Lee, L.P.; Matlashov, A.; Vinetskiy, V.
\newblock A low-noise, two-stage DC SQUID amplifier with high bandwidth and
  dynamic range.
\newblock {\em IEEE Trans. Appl. Supercond.} {\bf 1997}, {\em
  7},~3033--3036.

\bibitem[Kiviranta(2018)]{kiviranta2018low}
Kiviranta, M.
\newblock Low-dissipating push-pull SQUID amplifier for TES detector readout.
\newblock {\em arXiv} {\bf 2018}, arXiv:1810.04706.

\bibitem[Uhlig(2012)]{uhlig2012cryogen}
Uhlig, K.
\newblock {Cryogen-free dilution refrigerator with separate 1K cooling
  circuit.}
\newblock {\em AIP Conf. Proc.} {\bf 2012}, {\em 1434},~1823--1829.
\newblock {\url{https://doi.org/10.1063/1.4707119}}.

\bibitem[Callahan et~al.(1938)Callahan, Mathes, and Kahn]{callahan1938time}
Callahan, J.; Mathes, R.; Kahn, A.
\newblock Time-Division Multiplex in Radiotelegraphic Practice.
\newblock {\em Proc. Inst. Radio Eng.} {\bf 1938},
  {\em 26},~55--75.

\bibitem[Battistelli et~al.(2008)Battistelli, Amiri, Burger, Halpern, Knotek,
  Ellis, Gao, Kelly, Macintosh, Irwin, et~al.]{battistelli2008functional}
Battistelli, E.S.; Amiri, M.; Burger, B.; Halpern, M.; Knotek, S.; Ellis, M.;
  Gao, X.; Kelly, D.; Macintosh, M.; Irwin, K.;  et~al.
\newblock Functional description of read out electronics for time-domain
  multiplexed bolometers for millimeter and sub-millimeter astronomy.
\newblock {\em J. Low Temp. Phys.} {\bf 2008}, {\em
  151},~908--914.

\bibitem[Wu et~al.(2022)Wu, Yu, He, Liu, and Chen]{wu2022multiplexing}
Wu, X.; Yu, Q.; He, Y.; Liu, J.; Chen, W.
\newblock Multiplexing technology based on SQUID for readout of superconducting
  transition-edge sensor arrays.
\newblock {\em Chin. Phys. B} {\bf 2022}, {\em 31},~108501.

\bibitem[Niemack et~al.(2010)Niemack, Beyer, Cho, Doriese, Hilton, Irwin,
  Reintsema, Schmidt, Ullom, and Vale]{niemack2010code}
Niemack, M.D.; Beyer, J.; Cho, H.; Doriese, W.; Hilton, G.; Irwin, K.;
  Reintsema, C.D.; Schmidt, D.R.; Ullom, J.N.; Vale, L.R.
\newblock Code-division SQUID multiplexing.
\newblock {\em Appl. Phys. Lett.} {\bf 2010}, {\em 96}, 163509.

\bibitem[Doriese et~al.(2006)Doriese, Beall, Duncan, Ferreira, Hilton, Irwin,
  Reintsema, Ullom, Vale, and Xu]{DORIESE2006808}
Doriese, W.; Beall, J.; Duncan, W.; Ferreira, L.; Hilton, G.; Irwin, K.;
  Reintsema, C.; Ullom, J.; Vale, L.; Xu, Y.
\newblock Progress toward kilopixel arrays: 3.8 eV microcalorimeter resolution
  in 8-channel SQUID multiplexer.
\newblock {\em Nucl. Instrum. Methods Phys. Res. Sect. A Accel. Spectrom. Detect. Assoc. Equip.} {\bf 2006},
  {\em 559},~808--810. {\url{https://doi.org/10.1016/j.nima.2005.12.146}}.

\bibitem[Dreyer et~al.(2007)Dreyer, Arnold, Lanting, Dobbs, Friedrich, Lee, and
  Spieler]{Dobbs_multiplexing}
Dreyer, J.G.; Arnold, K.; Lanting, T.M.; Dobbs, M.A.; Friedrich, S.; Lee, A.T.;
  Spieler, H.G.
\newblock Frequency-Domain Multiplexed Readout for Superconducting Gamma-Ray
  Detectors.
\newblock {\em IEEE Trans. Appl. Supercond.} {\bf 2007}, {\em
  17},~633--636.
\newblock {\url{https://doi.org/10.1109/TASC.2007.898249}}.

\bibitem[Irwin et~al.(2010)Irwin, Niemack, Beyer, Cho, Doriese, Hilton,
  Reintsema, Schmidt, Ullom, and Vale]{irwin2010code}
Irwin, K.; Niemack, M.; Beyer, J.; Cho, H.; Doriese, W.; Hilton, G.; Reintsema,
  C.; Schmidt, D.; Ullom, J.; Vale, L.
\newblock Code-division multiplexing of superconducting transition-edge sensor
  arrays.
\newblock {\em Supercond. Sci. Technol.} {\bf 2010}, {\em
  23},~034004.

\bibitem[Irwin et~al.(2012)Irwin, Cho, Doriese, Fowler, Hilton, Niemack,
  Reintsema, Schmidt, Ullom, and Vale]{irwin2012advanced}
Irwin, K.D.; Cho, H.M.; Doriese, W.B.; Fowler, J.W.; Hilton, G.C.; Niemack,
  M.D.; Reintsema, C.D.; Schmidt, D.R.; Ullom, J.N.; Vale, L.R.
\newblock Advanced code-division multiplexers for superconducting detector
  arrays.
\newblock {\em J. Low Temp. Phys.} {\bf 2012}, {\em
  167},~588--594.

\bibitem[Cukierman et~al.(2020)Cukierman, Ahmed, Henderson, Young, Yu, Barkats,
  Brown, Chaudhuri, Cornelison, D’Ewart, et~al.]{cukierman2020microwave}
Cukierman, A.; Ahmed, Z.; Henderson, S.; Young, E.; Yu, C.; Barkats, D.; Brown,
  D.; Chaudhuri, S.; Cornelison, J.; D’Ewart, J.M.;  et~al.
\newblock Microwave multiplexing on the Keck Array.
\newblock {\em J. Low Temp. Phys.} {\bf 2020}, {\em
  199},~858--866.

\bibitem[Dober et~al.(2021)Dober, Ahmed, Arnold, Becker, Bennett, Connors,
  Cukierman, D'Ewart, Duff, Dusatko, et~al.]{dober2021microwave}
Dober, B.; Ahmed, Z.; Arnold, K.; Becker, D.; Bennett, D.; Connors, J.;
  Cukierman, A.; D'Ewart, J.; Duff, S.; Dusatko, J.;  et~al.
\newblock A microwave SQUID multiplexer optimized for bolometric applications.
\newblock {\em Appl. Phys. Lett.} {\bf 2021}, {\em 118}, 062601.

\bibitem[McCammon(2005)]{mccammon}
McCammon, D.
\newblock Thermal equilibrium calorimeters---An introduction.
\newblock In {\em Cryogenic Particle Detection}; Springer: Berlin/Heidelberg, Germany, {2005}; pp. 1--34.

\bibitem[Richards(1994)]{richards1994bolometers}
Richards, P.L.
\newblock Bolometers for infrared and millimeter waves.
\newblock {\em J. Appl. Phys.} {\bf 1994}, {\em 76},~1--24.

\bibitem[Boyle and Rodgers~Jr(1959)]{boyle1959performance}
Boyle, W.; Rodgers~Jr, K.
\newblock Performance characteristics of a new low-temperature bolometer.
\newblock {\em J. Opt. Soc. Am.} {\bf 1959}, {\em
  49},~66--69.

\bibitem[Mather(1982)]{mather1982bolometer}
Mather, J.C.
\newblock Bolometer noise: Nonequilibrium theory.
\newblock {\em Appl. Opt.} {\bf 1982}, {\em 21},~1125--1129.

\bibitem[Johnson(1928)]{johnson1928thermal}
Johnson, J.B.
\newblock Thermal agitation of electricity in conductors.
\newblock {\em Phys. Rev.} {\bf 1928}, {\em 32},~97.

\bibitem[Perepelitsa(2006)]{perepelitsa2006johnson}
Perepelitsa, D.V.
\newblock \emph{Johnson Noise and Shot Noise};
\newblock MIT Department of Physics: Cambridge, MA, USA, {2006}.

\bibitem[Lamarre(1986)]{Lamarre:86}
Lamarre, J.M.
\newblock Photon noise in photometric instruments at far-infrared and
  submillimeter wavelengths.
\newblock {\em Appl. Opt.} {\bf 1986}, {\em 25},~870--876.
\newblock {\url{https://doi.org/10.1364/AO.25.000870}}.

\bibitem[Shirokoff(2011)]{shirokoff2011south}
Shirokoff, E.D.
\newblock The South Pole Telescope Bolometer Array and the Measurement of
  Secondary Cosmic Microwave Background Anisotropy at Small Angular Scales.
\newblock Ph.D. Thesis, UC Berkeley, Berkeley, CA, USA, 2011.

\bibitem[Ullom et~al.(2004)Ullom, Doriese, Hilton, Beall, Deiker, Duncan,
  Ferreira, Irwin, Reintsema, and Vale]{ullom2004characterization}
Ullom, J.N.; Doriese, W.B.; Hilton, G.C.; Beall, J.A.; Deiker, S.; Duncan, W.;
  Ferreira, L.; Irwin, K.D.; Reintsema, C.D.; Vale, L.R.
\newblock Characterization and reduction of unexplained noise in
  superconducting transition-edge sensors.
\newblock {\em Appl. Phys. Lett.} {\bf 2004}, {\em 84},~4206--4208.

\bibitem[Jethava et~al.(2009)Jethava, Ullom, Irwin, Doriese, Beall, Hilton,
  Vale, and Zink]{jethava2009dependence}
Jethava, N.; Ullom, J.N.; Irwin, K.D.; Doriese, W.; Beall, J.; Hilton, G.;
  Vale, L.; Zink, B.
\newblock Dependence of excess noise on the partial derivatives of resistance
  in superconducting Transition Edge Sensors.
\newblock \emph{AIP Conf. Proc.}  \textbf{2009}, \emph{1185}, 31--33.

\bibitem[Galeazzi(2010)]{galeazzi2010fundamental}
Galeazzi, M.
\newblock Fundamental noise processes in TES devices.
\newblock {\em IEEE Trans. Appl. Supercond.} {\bf 2010}, {\em
  21},~267--271.

\bibitem[Gildemeister et~al.(2001)Gildemeister, Lee, and
  Richards]{gildemeister2001model}
Gildemeister, J.M.; Lee, A.T.; Richards, P.L.
\newblock Model for excess noise in voltage-biased superconducting bolometers.
\newblock {\em Appl. Opt.} {\bf 2001}, {\em 40},~6229--6235.

\bibitem[Dobbs et~al.(2012)Dobbs, Lueker, Aird, Bender, Benson, Bleem,
  Carlstrom, Chang, Cho, Clarke, et~al.]{dobbs2012frequency}
Dobbs, M.; Lueker, M.; Aird, K.; Bender, A.; Benson, B.; Bleem, L.; Carlstrom,
  J.; Chang, C.; Cho, H.M.; Clarke, J.;  et~al.
\newblock Frequency multiplexed superconducting quantum interference device
  readout of large bolometer arrays for cosmic microwave background
  measurements.
\newblock {\em Rev. Sci. Instrum.} {\bf 2012}, {\em 83}, 073113.

\bibitem[Lueker et~al.(2009)Lueker, Benson, Chang, Cho, Dobbs, Holzapfel,
  Lanting, Lee, Mehl, Plagge, et~al.]{lueker2009thermal}
Lueker, M.; Benson, B.A.; Chang, C.L.; Cho, H.M.; Dobbs, M.; Holzapfel, W.L.;
  Lanting, T.; Lee, A.T.; Mehl, J.; Plagge, T.;  et~al.
\newblock Thermal design and characterization of transition-edge sensor (TES)
  bolometers for frequency-domain multiplexing.
\newblock {\em IEEE Trans. Appl. Supercond.} {\bf 2009}, {\em
  19},~496--500.

\bibitem[Zmuidzinas(2012)]{zmuidzinas2012superconducting}
Zmuidzinas, J.
\newblock Superconducting microresonators: Physics and applications.
\newblock {\em Annu. Rev. Condens. Matter Phys.} {\bf 2012}, {\em 3},~169--214.

\bibitem[De~Lucia et~al.(2023)De~Lucia, Ulbricht, Baldwin, Piercy, Creaner,
  Bracken, and Ray]{de2023limitations}
De~Lucia, M.; Ulbricht, G.; Baldwin, E.; Piercy, J.; Creaner, O.; Bracken, C.;
  Ray, T.
\newblock Limitations to the energy resolution of single-photon sensitive
  microwave kinetic inductance detectors.
\newblock {\em AIP Adv.} {\bf 2023}, {\em 13}, 125026.

\bibitem[Yoshihara et~al.(2003)Yoshihara, Kanno, and Shinada]{yoshihara2003rf}
Yoshihara, F.; Kanno, I.; Shinada, K.
\newblock Rf-SQUID microcalorimeter.
\newblock {\em Supercond. Sci. Technol.} {\bf 2003}, {\em
  16},~1257.

\bibitem[Mates et~al.(2019)Mates, Becker, Bennett, Dober, Gard, Hilton, Swetz,
  Vale, and Ullom]{xtalkmumux1}
Mates, J.; Becker, D.T.; Bennett, D.A.; Dober, B.J.; Gard, J.D.; Hilton, G.C.;
  Swetz, D.S.; Vale, L.R.; Ullom, J.N.
\newblock Crosstalk in microwave SQUID multiplexers.
\newblock {\em Appl. Phys. Lett.} {\bf 2019}, {\em 115}, 202601.

\bibitem[Noroozian et~al.(2012)Noroozian, Day, Eom, Leduc, and
  Zmuidzinas]{noroozian2012crosstalk}
Noroozian, O.; Day, P.K.; Eom, B.H.; Leduc, H.G.; Zmuidzinas, J.
\newblock Crosstalk reduction for superconducting microwave resonator arrays.
\newblock {\em IEEE Trans. Microw. Theory Tech.} {\bf
  2012}, {\em 60},~1235--1243.

\bibitem[Hirayama et~al.(2016)Hirayama, Irimatsugawa, Yamamori, Kohjiro, Sato,
  Nagasawa, Fukuda, Sasaki, Hidaka, Sato, et~al.]{hirayama2016interchannel}
Hirayama, F.; Irimatsugawa, T.; Yamamori, H.; Kohjiro, S.; Sato, A.; Nagasawa,
  S.; Fukuda, D.; Sasaki, H.; Hidaka, M.; Sato, Y.;  et~al.
\newblock Interchannel crosstalk and nonlinearity of microwave SQUID
  multiplexers.
\newblock {\em IEEE Trans. Appl. Supercond.} {\bf 2016}, {\em
  27},~2500205.

\bibitem[Groh et~al.(2024)Groh, Ahmed, Henderson, Hubmayr, Mates, Silva-Feaver,
  Ullom, and Yu]{groh2024crosstalk}
Groh, J.C.; Ahmed, Z.; Henderson, S.W.; Hubmayr, J.; Mates, J.A.; Silva-Feaver,
  M.; Ullom, J.; Yu, C.
\newblock Crosstalk effects in microwave SQUID multiplexed TES bolometer
  readout.
\newblock {\em J. Low Temp. Phys.} {\bf 2024}, \emph{216}, 225--236.

\bibitem[Piro et~al.(2006)Piro, Amati, Barbera, Borgani, Bazzano, Branchini,
  Brunetti, Campana, Caroli, Cocchi, et~al.]{piro2006estremo}
Piro, L.; Amati, L.; Barbera, M.; Borgani, S.; Bazzano, A.; Branchini, E.;
  Brunetti, G.; Campana, S.; Caroli, E.; Cocchi, M.;  et~al.
\newblock ESTREMO/WFXRT: Extreme physics in the transient and evolving cosmos.
\newblock In Proceedings of the Space Telescopes and Instrumentation II:
  Ultraviolet to Gamma Ray, Orlando, FL, USA, 24--31 May 2006; SPIE: Bellingham, WA, USA,  2006; Volume~6266, pp. 163--174.

\bibitem[Piro(2007)]{piro2007future}
Piro, L.
\newblock The future of GRB investigation from ground and space.
\newblock {\em Philos. Trans. R. Soc. A Math. Phys. Eng. Sci.} {\bf 2007}, {\em 365},~1399--1409.

\bibitem[Barret et~al.(2016)Barret, Trong, Den~Herder, Piro, Barcons, Huovelin,
  Kelley, Mas-Hesse, Mitsuda, Paltani, et~al.]{barrett2016athena}
Barret, D.; Trong, T.L.; Den~Herder, J.W.; Piro, L.; Barcons, X.; Huovelin, J.;
  Kelley, R.; Mas-Hesse, J.M.; Mitsuda, K.; Paltani, S.;  et~al.
\newblock The Athena X-ray integral field unit (X-IFU).
\newblock In Proceedings of the Space Telescopes and Instrumentation 2016:
  Ultraviolet to Gamma Ray, Edinburgh, UK, 18 July 2016; SPIE: Bellingham, WA, USA,  2016; Volume~9905, pp. 714--754.

\bibitem[Meidinger et~al.(2018)Meidinger, Nandra, and
  Plattner]{meidinger2018development}
Meidinger, N.; Nandra, K.; Plattner, M.
\newblock Development of the Wide Field Imager instrument for ATHENA.
\newblock In Proceedings of the Space Telescopes and Instrumentation 2018:
  Ultraviolet to Gamma Ray, Austin, TX, USA, 10--15 June 2018; SPIE: Bellingham, WA, USA,  2018; Volume~10699, pp. 312--323.

\bibitem[Barret et~al.(2023)Barret, Albouys, Herder, Piro, Cappi, Huovelin,
  Kelley, Mas-Hesse, Paltani, Rauw, et~al.]{barret2023athena}
Barret, D.; Albouys, V.; Herder, J.W.d.; Piro, L.; Cappi, M.; Huovelin, J.;
  Kelley, R.; Mas-Hesse, J.M.; Paltani, S.; Rauw, G.;  et~al.
\newblock The Athena X-ray Integral Field Unit: A consolidated design for the
  system requirement review of the preliminary definition phase.
\newblock {\em Exp. Astron.} {\bf 2023}, {\em 55},~373--426.

\bibitem[Khosropanah et~al.(2018)Khosropanah, Taralli, Gottardi, De~Vries,
  Nagayoshi, Ridder, Akamatsu, Bruijn, and Gao]{khosropanah2018development}
Khosropanah, P.; Taralli, E.; Gottardi, L.; De~Vries, C.; Nagayoshi, K.;
  Ridder, M.; Akamatsu, H.; Bruijn, M.; Gao, J.R.
\newblock Development of TiAu TES X-ray calorimeters for the X-IFU on ATHENA
  space observatory.
\newblock In Proceedings of the Space Telescopes and Instrumentation 2018:
  Ultraviolet to Gamma Ray, Austin, TX, USA, 10--15 June 2018; SPIE: Bellingham, WA, USA,  2018; Volume~10699, pp. 386--396.

\bibitem[D’Andrea et~al.(2022)D’Andrea, Ravensberg, Argan, Brienza, Lotti,
  Macculi, Minervini, Piro, Torrioli, Chiarello, et~al.]{d2022athena}
D’Andrea, M.; Ravensberg, K.; Argan, A.; Brienza, D.; Lotti, S.; Macculi, C.;
  Minervini, G.; Piro, L.; Torrioli, G.; Chiarello, F.;  et~al.
\newblock ATHENA X-IFU demonstration model: First joint operation of the main
  TES array and its cryogenic AntiCoincidence detector (CryoAC).
\newblock {\em J. Low Temp. Phys.} {\bf 2022}, {\em
  209},~433--440.

\bibitem[Durkin et~al.(2019)Durkin, Adams, Bandler, Chervenak, Chaudhuri,
  Dawson, Denison, Doriese, Duff, Finkbeiner, et~al.]{durkin2019demonstration}
Durkin, M.; Adams, J.S.; Bandler, S.R.; Chervenak, J.A.; Chaudhuri, S.; Dawson,
  C.S.; Denison, E.V.; Doriese, W.B.; Duff, S.M.; Finkbeiner, F.M.;  et~al.
\newblock Demonstration of Athena X-IFU compatible 40-row
  time-division-multiplexed readout.
\newblock {\em IEEE Trans. Appl. Supercond.} {\bf 2019}, {\em
  29},~2101005.

\bibitem[Bandler et~al.(2019)Bandler, Chervenak, Datesman, Devasia, DiPirro,
  Sakai, Smith, Stevenson, Yoon, Bennett, et~al.]{bandler2019lynx}
Bandler, S.R.; Chervenak, J.A.; Datesman, A.M.; Devasia, A.M.; DiPirro, M.;
  Sakai, K.; Smith, S.J.; Stevenson, T.R.; Yoon, W.; Bennett, D.;  et~al.
\newblock Lynx X-ray microcalorimeter.
\newblock {\em J. Astron. Telesc. Instrum. Syst.}
  {\bf 2019}, {\em 5},~021017.

\bibitem[Gaskin et~al.(2019)Gaskin, Swartz, Vikhlinin, {\"O}zel, Gelmis,
  Arenberg, Bandler, Bautz, Civitani, Dominguez, et~al.]{gaskin2019lynx}
Gaskin, J.A.; Swartz, D.A.; Vikhlinin, A.; {\"O}zel, F.; Gelmis, K.E.;
  Arenberg, J.W.; Bandler, S.R.; Bautz, M.W.; Civitani, M.M.; Dominguez, A.;
  et~al.
\newblock Lynx X-ray observatory: An overview.
\newblock {\em J. Astron. Telesc. Instrum. Syst.}
  {\bf 2019}, {\em 5},~021001.
\newpage
\bibitem[Gaskin et~al.(2018)Gaskin, Dominguez, Gelmis, Mulqueen, Swartz,
  McCarley, {\"O}zel, Vikhlinin, Schwartz, Tananbaum, et~al.]{gaskin2018lynx}
Gaskin, J.A.; Dominguez, A.; Gelmis, K.; Mulqueen, J.J.; Swartz, D.; McCarley,
  K.; {\"O}zel, F.; Vikhlinin, A.; Schwartz, D.; Tananbaum, H.;  et~al.
\newblock The Lynx X-ray Observatory: Concept study overview and status.
\newblock In Proceedings of the Space Telescopes and Instrumentation 2018:
  Ultraviolet to Gamma Ray, Austin, TX, USA, 10--15 June 2018; SPIE: Bellingham, WA, USA,  2018; Volume~10699, pp. 120--129.

\bibitem[Falcone et~al.(2019)Falcone, Kraft, Bautz, Gaskin, Mulqueen, Swartz, ,
  and Technology Definition~Team]{falcone2019overview}
Falcone, A.D.; Kraft, R.P.; Bautz, M.W.; Gaskin, J.A.; Mulqueen, J.A.; Swartz,
  D.A.; .; Technology Definition~Team, f.t.L.S.
\newblock Overview of the high-definition X-ray imager instrument on the Lynx
  X-ray surveyor.
\newblock {\em J. Astron. Telesc. Instrum. Syst.}
  {\bf 2019}, {\em 5},~021019.

\bibitem[McEntaffer(2019)]{mcentaffer2019reflection}
McEntaffer, R.L.
\newblock Reflection grating concept for the Lynx X-ray grating spectrograph.
\newblock {\em J. Astron. Telesc. Instrum. Syst.}
  {\bf 2019}, {\em 5},~021002.

\bibitem[Smith et~al.(2020)Smith, Adams, Bandler, Beaumont, Chervenak,
  Datesman, Finkbeiner, Hummatov, Kelly, Kilbourne, et~al.]{smith2020toward}
Smith, S.J.; Adams, J.; Bandler, S.; Beaumont, S.; Chervenak, J.; Datesman, A.;
  Finkbeiner, F.; Hummatov, R.; Kelly, R.; Kilbourne, C.;  et~al.
\newblock Toward 100,000-pixel microcalorimeter arrays using multi-absorber
  transition-edge sensors.
\newblock {\em J. Low Temp. Phys.} {\bf 2020}, {\em
  199},~330--338.

\bibitem[Leisawitz et~al.(2021)Leisawitz, Amatucci, Allen, Arenberg, Armus,
  Battersby, Bauer, Beaman, Bell, Beltran, et~al.]{leisawitz2021origins}
Leisawitz, D.; Amatucci, E.; Allen, L.; Arenberg, J.; Armus, L.; Battersby, C.;
  Bauer, J.; Beaman, B.G.; Bell, R.; Beltran, P.;  et~al.
\newblock Origins Space Telescope: Baseline mission concept.
\newblock {\em J. Astron. Telesc. Instrum. Syst.}
  {\bf 2021}, {\em 7},~011002.

\bibitem[Battersby et~al.(2018)Battersby, Armus, Bergin, Kataria, Meixner,
  Pope, Stevenson, Cooray, Leisawitz, Scott, et~al.]{battersby2018origins}
Battersby, C.; Armus, L.; Bergin, E.; Kataria, T.; Meixner, M.; Pope, A.;
  Stevenson, K.B.; Cooray, A.; Leisawitz, D.; Scott, D.;  et~al.
\newblock The Origins space telescope.
\newblock {\em Nat. Astron.} {\bf 2018}, {\em 2},~596--599.

\bibitem[Bradford et~al.(2021)Bradford, Cameron, Moore, Hailey-Dunsheath,
  Amatucci, Bradley, Corsetti, Leisawitz, DiPirro, Tuttle,
  et~al.]{bradford2021origins}
Bradford, C.M.; Cameron, B.; Moore, B.; Hailey-Dunsheath, S.; Amatucci, E.;
  Bradley, D.; Corsetti, J.; Leisawitz, D.; DiPirro, M.; Tuttle, J.;  et~al.
\newblock Origins Survey Spectrometer: Revealing the hearts of distant galaxies
  and forming planetary systems with far-IR spectroscopy.
\newblock {\em J. Astron. Telesc. Instrum. Syst.}
  {\bf 2021}, {\em 7},~011017.

\bibitem[Meixner et~al.(2020)Meixner, Staguhn, Vieira, Amatucci, DiPirro,
  Bradley, Cooray, Battersby, Sandstrom, STDT, et~al.]{meixner2020origins}
Meixner, M.; Staguhn, J.; Vieira, J.; Amatucci, E.; DiPirro, M.; Bradley, D.;
  Cooray, A.; Battersby, C.; Sandstrom, K.; STDT, O.S.T.;  et~al.
\newblock Origins Space Telescope (Origins): Far-infrared Imager and Polarimeter (FIP).
\newblock In \emph{American Astronomical Society Meeting
  Abstracts\# 235};  Bulletin of the American Astronomical Society: Washington, DC, USA, 2020; Volume~235, p. 171.05.

\bibitem[Roellig et~al.(2020)Roellig, McMurtry, Greene, Matsuo, Sakon, and
  Staguhn]{roellig2020mid}
Roellig, T.L.; McMurtry, C.; Greene, T.; Matsuo, T.; Sakon, I.; Staguhn, J.
\newblock Mid-infrared detector development for the Origins Space Telescope.
\newblock {\em J. Astron. Telesc. Instrum. Syst.}
  {\bf 2020}, {\em 6},~041503.

\bibitem[Nagler et~al.(2021)Nagler, Sadleir, and Wollack]{nagler2021transition}
Nagler, P.C.; Sadleir, J.E.; Wollack, E.J.
\newblock Transition-edge sensor detectors for the Origins space telescope.
\newblock {\em J. Astron. Telesc. Instrum. Syst.}
  {\bf 2021}, {\em 7},~011005.

\bibitem[Roelfsema et~al.(2018)Roelfsema, Shibai, Armus, Arrazola, Audard,
  Audley, Bradford, Charles, Dieleman, Doi, Duband, Eggens, Evers, Funaki, Gao,
  Giard, di~Giorgio, Fernández, Griffin, Helmich, Hijmering, Huisman,
  Ishihara, Isobe, Jackson, Jacobs, Jellema, Kamp, Kaneda, Kawada, Kemper,
  Kerschbaum, Khosropanah, Kohno, Kooijman, Krause, van~der Kuur, Kwon,
  Laauwen, de~Lange, Larsson, van Loon, Madden, Matsuhara, Najarro, Nakagawa,
  Naylor, Ogawa, Onaka, Oyabu, Poglitsch, Reveret, Rodriguez, Spinoglio, Sakon,
  Sato, Shinozaki, Shipman, Sugita, Suzuki, van~der Tak, Redondo, Wada, Wang,
  Wafelbakker, van Weers, Withington, Vandenbussche, Yamada, and
  Yamamura]{Roelfsema_2018}
Roelfsema, P.R.; Shibai, H.; Armus, L.; Arrazola, D.; Audard, M.; Audley, M.D.;
  Bradford, C.; Charles, I.; Dieleman, P.; Doi, Y.;  et~al.
\newblock SPICA---A Large Cryogenic Infrared Space Telescope: Unveiling the
  Obscured Universe.
\newblock {\em Publ. Astron. Soc. Aust.} {\bf
  2018}, {\em 35}.
\newblock {\url{https://doi.org/10.1017/pasa.2018.15}}.

\bibitem[Goicoechea et~al.(2009)Goicoechea, Isaak, and
  Swinyard]{goicoechea2009exoplanet}
Goicoechea, J.R.; Isaak, K.; Swinyard, B.
\newblock Exoplanet research with SAFARI: A far-IR imaging spectrometer for
  SPICA. {\em ArXiV} \textbf{ 2009},  arXiv:astro-ph.EP/0901.3240.

\bibitem[Mauskopf et~al.(2010)Mauskopf, Ade, Beyer, Bruijn, Gao, Glowacka,
  Goldie, Griffin, Griffin, Hoevers, Khosrapanah, Kooijma, Korte, Morozov,
  Murphy, O’Sullivan, Ridder, Trappe, Weers, and Withington]{mauskopf2010tes}
Mauskopf, P.; Ade, P.; Beyer, J.; Bruijn, M.; Gao, J.; Glowacka, D.; Goldie,
  D.; Griffin, D.; Griffin, M.; Hoevers, H.;  et~al.
\newblock A TES focal plane for SPICA-SAFARI.
\newblock {In Proceedings of the 21st International Symposium on Space Terahertz Technology 2010,   ISSTT 2010, Oxford and Didcot, UK, 23--25 March 2010}; pp. 23--25.

\bibitem[Audley et~al.(2022)Audley, Huijser, de~Lange, and
  Orlando]{audley2022optical}
Audley, M.D.; Huijser, G.F.; de~Lange, G.; Orlando, A.
\newblock Optical measurements of ultra-sensitive far-infrared TES bolometers
  with FDM readout.
\newblock In Proceedings of the Millimeter, Submillimeter, and Far-Infrared
  Detectors and Instrumentation for Astronomy XI, Montreal, QC, Canada, 31 August 2022; SPIE: Bellingham, WA, USA,  2022; Volume~12190, pp. 715--719.

\bibitem[Khosropanah et~al.(2016)Khosropanah, Suzuki, Ridder, Hijmering,
  Akamatsu, Gottardi, Van Der~Kuur, Gao, and Jackson]{khosropanah2016ultra}
Khosropanah, P.; Suzuki, T.; Ridder, M.; Hijmering, R.; Akamatsu, H.; Gottardi,
  L.; Van Der~Kuur, J.; Gao, J.; Jackson, B.
\newblock Ultra-low noise TES bolometer arrays for SAFARI instrument on SPICA.
\newblock In Proceedings of the Millimeter, Submillimeter, and Far-Infrared
  Detectors and Instrumentation for Astronomy VIII, Edinburgh, UK, 26 June--1 July 2016; SPIE: Bellingham, WA, USA,  2016; Volume~9914, pp.
  49--53.

\bibitem[Khosropanah et~al.(2010)Khosropanah, Dirks, Parra-Border{\'\i}as,
  Ridder, Hijmering, Van~der Kuur, Gottardi, Bruijn, Popescu, Gao,
  et~al.]{khosropanah2010low}
Khosropanah, P.; Dirks, B.; Parra-Border{\'\i}as, M.; Ridder, M.; Hijmering,
  R.; Van~der Kuur, J.; Gottardi, L.; Bruijn, M.; Popescu, M.; Gao, J.;  et~al.
\newblock Low-noise Transition Edge Sensor (TES) for SAFARI instrument on
  SPICA.
\newblock In Proceedings of the Millimeter, Submillimeter, and Far-Infrared
  Detectors and Instrumentation for Astronomy V, San Diego, CA, USA, 27 June--2 July 2010; SPIE: Bellingham, WA, USA,  2010; Volume~7741, pp.
  163--171.

\bibitem[Walton et~al.(2004)Walton, Parkes, Terry, Dunare, Stevenson, Gundlach,
  Hilton, Irwin, Ullom, Holland, et~al.]{walton2004design}
Walton, A.; Parkes, W.; Terry, J.; Dunare, C.; Stevenson, J.; Gundlach, A.;
  Hilton, G.; Irwin, K.; Ullom, J.; Holland, W.;  et~al.
\newblock Design and fabrication of the detector technology for SCUBA-2.
\newblock {\em IEEE Proc.-Sci. Meas. Technol.} {\bf
  2004}, {\em 151},~110--120.

\bibitem[Audley et~al.(2004)Audley, Holland, Duncan, Atkinson, Cliffe, Ellis,
  Gao, Gostick, Hodson, Kelly, et~al.]{audley2004scuba}
Audley, M.; Holland, W.; Duncan, W.; Atkinson, D.; Cliffe, M.; Ellis, M.; Gao,
  X.; Gostick, D.; Hodson, T.; Kelly, D.;  et~al.
\newblock SCUBA-2: A large-format TES array for submillimetre astronomy.
\newblock {\em Nucl. Instrum. Methods Phys. Res. Sect. A Accel. Spectrom. Detect. Assoc. Equip.} {\bf 2004},
  {\em 520},~479--482.

\bibitem[Bintley et~al.(2014)Bintley, Holland, MacIntosh, Friberg, Bell, Berke,
  Berry, Berthold, Cookson, Coulson, et~al.]{bintley2014scuba}
Bintley, D.; Holland, W.S.; MacIntosh, M.J.; Friberg, P.; Bell, G.S.; Berke,
  D.A.; Berry, D.S.; Berthold, R.M.; Cookson, J.L.; Coulson, I.M.;  et~al.
\newblock SCUBA-2: An update on the performance of the 10,000 pixel bolometer
  camera after two years of science operation at the JCMT.
\newblock In Proccedings of the Millimeter, Submillimeter, and Far-Infrared
  Detectors and Instrumentation for Astronomy VII, Montréal, QC, Canada, 22--27 June 2014; SPIE: Bellingham, WA, USA,  2014; Volume~9153, pp.
  42--56.

\bibitem[Casey et~al.(2013)Casey, Chen, Cowie, Barger, Capak, Ilbert, Koss,
  Lee, Le~Floc'h, Sanders, et~al.]{casey2013characterization}
Casey, C.M.; Chen, C.C.; Cowie, L.L.; Barger, A.J.; Capak, P.; Ilbert, O.;
  Koss, M.; Lee, N.; Le~Floc'h, E.; Sanders, D.B.;  et~al.
\newblock Characterization of Scuba-2 450 $\mu$m and 850 $\mu$m selected
  galaxies in the COSMOS field.
\newblock {\em Mon. Not. R. Astron. Soc.} {\bf 2013},
  {\em 436},~1919--1954.

\bibitem[Mairs et~al.(2021)Mairs, Dempsey, Bell, Parsons, Currie, Friberg,
  Jiang, Tetarenko, Bintley, Cookson, et~al.]{mairs2021decade}
Mairs, S.; Dempsey, J.T.; Bell, G.S.; Parsons, H.; Currie, M.J.; Friberg, P.;
  Jiang, X.J.; Tetarenko, A.J.; Bintley, D.; Cookson, J.;  et~al.
\newblock A Decade of SCUBA-2: A Comprehensive Guide to Calibrating 450 $\mu$m
  and 850 $\mu$m Continuum Data at the JCMT.
\newblock {\em  Astron. J.} {\bf 2021}, {\em 162},~191.

\bibitem[Garratt et~al.(2023)Garratt, Geach, Tamura, Coppin, Franco, Ao, Chen,
  Cheng, Clements, Dai, et~al.]{garratt2023scuba}
Garratt, T.; Geach, J.; Tamura, Y.; Coppin, K.; Franco, M.; Ao, Y.; Chen, C.C.;
  Cheng, C.; Clements, D.; Dai, Y.;  et~al.
\newblock The SCUBA-2 Large eXtragalactic Survey: 850$\mu$m map, catalogue and
  the bright-end number counts of the XMM-LSS field.
\newblock {\em Mon. Not. R. Astron. Soc.} {\bf 2023},
  {\em 520},~3669--3687.

\bibitem[Bennett et~al.(1993)Bennett, Boggess, Cheng, Hauser, Kelsall, Mather,
  Moseley~Jr, Murdock, Shafer, Silverberg, et~al.]{bennett1993scientific}
Bennett, C.L.; Boggess, N.; Cheng, E.; Hauser, M.; Kelsall, T.; Mather, J.;
  Moseley~Jr, S.; Murdock, T.L.; Shafer, R.A.; Silverberg, R.F.;  et~al.
\newblock Scientific results from the Cosmic Background Explorer (COBE)
  (microwave/infrared).
\newblock {\em Proc. Natl. Acad. Sci. USA} {\bf 1993},
  {\em 90},~4766--4773.

\bibitem[Staguhn et~al.(2008)Staguhn, Benford, Allen, Maher, Sharp, Ames,
  Arendt, Chuss, Dwek, Fixsen, et~al.]{staguhn2008instrument}
Staguhn, J.G.; Benford, D.J.; Allen, C.A.; Maher, S.F.; Sharp, E.H.; Ames,
  T.J.; Arendt, R.G.; Chuss, D.T.; Dwek, E.; Fixsen, D.J.;  et~al.
\newblock Instrument performance of GISMO: A 2 millimeter TES bolometer camera
  used at the IRAM 30 m Telescope.
\newblock In Proccedings of the Millimeter and Submillimeter Detectors and
  Instrumentation for Astronomy IV, Marseille, France, 26--28 June 2008; SPIE: Bellingham, WA, USA,  2008; Volume~7020, pp. 52--60.

\bibitem[Magnelli et~al.(2019)Magnelli, Karim, Staguhn, Kov{\'a}cs,
  Jim{\'e}nez-Andrade, Casey, Zavala, Schinnerer, Sargent, Aravena,
  et~al.]{magnelli2019iram}
Magnelli, B.; Karim, A.; Staguhn, J.; Kov{\'a}cs, A.; Jim{\'e}nez-Andrade, E.;
  Casey, C.; Zavala, J.; Schinnerer, E.; Sargent, M.; Aravena, M.;  et~al.
\newblock The IRAM/GISMO 2 mm Survey in the COSMOS Field.
\newblock {\em  Astrophys. J.} {\bf 2019}, {\em 877},~45.

\bibitem[Casey et~al.(2021)Casey, Zavala, Manning, Aravena, B{\'e}thermin,
  Caputi, Champagne, Clements, Drew, Finkelstein, et~al.]{casey2021mapping}
Casey, C.M.; Zavala, J.A.; Manning, S.M.; Aravena, M.; B{\'e}thermin, M.;
  Caputi, K.I.; Champagne, J.B.; Clements, D.L.; Drew, P.; Finkelstein, S.L.;
  et~al.
\newblock Mapping Obscuration to Reionization with ALMA (MORA): 2 mm
  Efficiently Selects the Highest-redshift Obscured Galaxies.
\newblock {\em  Astrophys. J.} {\bf 2021}, {\em 923},~215.

\bibitem[Giard et~al.(2005)Giard, Casoli, Paletou, Maffei, Ade, Calderon,
  Challinor, De~Bernardis, Dunlop, Gear, et~al.]{giard2005clover}
Giard, M.; Casoli, F.; Paletou, F.; Maffei, B.; Ade, P.; Calderon, C.;
  Challinor, A.; De~Bernardis, P.; Dunlop, L.; Gear, W.;  et~al.
\newblock CLOVER: The CMB polarization observer.
\newblock {\em Eur. Astron. Soc. Publ. Ser.} {\bf 2005},
  {\em 14},~251--256.

\bibitem[Taylor et~al.(2004)Taylor, Challinor, Goldie, Grainge, Jones, Lasenby,
  Withington, Yassin, Gear, Piccirillo, et~al.]{taylor2004clover}
Taylor, A.C.; Challinor, A.; Goldie, D.; Grainge, K.; Jones, M.; Lasenby, A.;
  Withington, S.; Yassin, G.; Gear, W.K.; Piccirillo, L.;  et~al.
\newblock CLOVER---A new instrument for measuring the B-mode polarization of
  the CMB.
\newblock {\em ArXiV} {\bf 2004}, arXiv:astro-ph/0407148.

\bibitem[North et~al.(2007)North, Ade, Audley, Baines, Battye, Brown, Cabella,
  Calisse, Challinor, Duncan, et~al.]{north2007clover}
North, C.; Ade, P.A.; Audley, M.; Baines, C.; Battye, R.; Brown, M.; Cabella,
  P.; Calisse, P.G.; Challinor, A.; Duncan, W.;  et~al.
\newblock Clover-measuring the CMB B-mode polarization.
\newblock In Proccedings of the 18th International Symposium on Space Terahertz
  Technology 2007, ISSTT 2007,  Pasadena, CA, USA, 21--23 March 2007; pp. 238--243.

\bibitem[Piccirillo et~al.(2008)Piccirillo, Ade, Audley, Baines, Battye, Brown,
  Calisse, Challinor, Duncan, Ferreira, et~al.]{piccirillo2008clover}
Piccirillo, L.; Ade, P.; Audley, M.; Baines, C.; Battye, R.; Brown, M.;
  Calisse, P.; Challinor, A.; Duncan, W.; Ferreira, P.;  et~al.
\newblock The CLOVER experiment.
\newblock In Proccedings of the Millimeter and Submillimeter Detectors and
  Instrumentation for Astronomy IV, Marseille, France, 26--28 June 2008; SPIE: Bellingham, WA, USA,  2008; Volume~7020, pp. 385--394.

\bibitem[Audley et~al.(2008{\natexlab{a}})Audley, Glowacka, Goldie, Tsaneva,
  Withington, Grimes, North, Yassin, Piccirillo, Ade,
  et~al.]{audley2008performance}
Audley, M.D.; Glowacka, D.; Goldie, D.J.; Tsaneva, V.; Withington, S.; Grimes,
  P.K.; North, C.E.; Yassin, G.; Piccirillo, L.; Ade, P.;  et~al.
\newblock Performance of microstrip-coupled TES bolometers with finline
  transitions.
\newblock In Proccedings of the Millimeter and Submillimeter Detectors and
  Instrumentation for Astronomy IV, Marseille, France, 26--28 June 2008; SPIE: Bellingham, WA, USA,  2008; Volume~7020, pp. 205--216.

\bibitem[Audley et~al.(2008{\natexlab{b}})Audley, Glowacka, Goldie, Tsaneva,
  Withington, Grimes, North, Yassin, Piccirillo, Pisano,
  et~al.]{audley2008microstrip}
Audley, M.D.; Glowacka, D.; Goldie, D.J.; Tsaneva, V.N.; Withington, S.;
  Grimes, P.K.; North, C.; Yassin, G.; Piccirillo, L.; Pisano, G.;  et~al.
\newblock Microstrip-coupled TES bolometers for CLOVER.
\newblock {\em Gravitat. Waves} {\bf 2008}, {\em 2},~3.

\bibitem[Fowler(2004)]{fowler2004atacama}
Fowler, J.W.
\newblock The atacama cosmology telescope project.
\newblock In Proccedings of the Millimeter and Submillimeter Detectors for
  Astronomy II, Glasgow, UK, 21--25 June 2004; SPIE: Bellingham, WA, USA,  2004; Volume~5498, pp. 1--10.

\bibitem[Swetz et~al.(2008)Swetz, Ade, Allen, Amiri, Appel, Battistelli,
  Burger, Chervenak, Dahlen, Das, et~al.]{swetz2008instrument}
Swetz, D.; Ade, P.A.; Allen, C.; Amiri, M.; Appel, J.; Battistelli, E.S.;
  Burger, B.; Chervenak, J.; Dahlen, A.; Das, S.;  et~al.
\newblock Instrument design and characterization of the millimeter Bolometer
  Array Camera on the Atacama Cosmology Telescope.
\newblock In Proccedings of the Millimeter and Submillimeter Detectors and
  Instrumentation for Astronomy IV, Marseille, France, 26--28 June 2008; SPIE: Bellingham, WA, USA,  2008; Volume~7020, pp. 84--95.

\bibitem[Marriage et~al.(2006)Marriage, Chervenak, and
  Doriese]{marriage2006testing}
Marriage, T.; Chervenak, J.; Doriese, W.
\newblock Testing and assembly of the detectors for the Millimeter Bolometer
  Array Camera on ACT.
\newblock {\em Nucl. Instrum. Methods Phys. Res. Sect. A Accel. Spectrom. Detect. Assoc. Equip.} {\bf 2006},
  {\em 559},~551--553.

\bibitem[Niemack et~al.(2010)Niemack, Ade, Aguirre, Barrientos, Beall, Bond,
  Britton, Cho, Das, Devlin, et~al.]{niemack2010actpol}
Niemack, M.D.; Ade, P.A.; Aguirre, J.; Barrientos, F.; Beall, J.; Bond, J.;
  Britton, J.; Cho, H.; Das, S.; Devlin, M.;  et~al.
\newblock ACTPol: A polarization-sensitive receiver for the Atacama Cosmology
  Telescope.
\newblock In Proccedings of the Millimeter, Submillimeter, and Far-Infrared
  Detectors and Instrumentation for Astronomy V, San Diego, CA, USA, 27 June--2 July 2010; SPIE: Bellingham, WA, USA,  2010; Volume~7741, pp.
  537--557.

\bibitem[Simon et~al.(2018)Simon, Beall, Cothard, Duff, Gallardo, Ho, Hubmayr,
  Koopman, McMahon, Nati, et~al.]{simon2018advanced}
Simon, S.; Beall, J.; Cothard, N.; Duff, S.; Gallardo, P.; Ho, S.; Hubmayr, J.;
  Koopman, B.; McMahon, J.; Nati, F.;  et~al.
\newblock The advanced ACTPol 27/39~GHz array.
\newblock {\em J. Low Temp. Phys.} {\bf 2018}, {\em
  193},~1041--1047.

\bibitem[Das et~al.(2011)Das, Sherwin, Aguirre, Appel, Bond, Carvalho, Devlin,
  Dunkley, Dunner, Essinger-Hileman, et~al.]{das2011detection}
Das, S.; Sherwin, B.D.; Aguirre, P.; Appel, J.W.; Bond, J.R.; Carvalho, C.S.;
  Devlin, M.J.; Dunkley, J.; Dunner, R.; Essinger-Hileman, T.;  et~al.
\newblock Detection of the power spectrum of cosmic microwave background
  lensing by the atacama cosmology telescope.
\newblock {\em arXiv} {\bf 2011}, arXiv:1103.2124.

\bibitem[Marriage et~al.(2011)Marriage, Acquaviva, Ade, Aguirre, Amiri, Appel,
  Barrientos, Battistelli, Bond, Brown, et~al.]{marriage2011atacama}
Marriage, T.A.; Acquaviva, V.; Ade, P.A.; Aguirre, P.; Amiri, M.; Appel, J.W.;
  Barrientos, L.F.; Battistelli, E.S.; Bond, J.R.; Brown, B.;  et~al.
\newblock The Atacama Cosmology Telescope: Sunyaev--Zel'Dovich-Selected Galaxy
  Clusters at 148~GHz in the 2008 Survey.
\newblock {\em  Astrophys. J.} {\bf 2011}, {\em 737},~61.

\bibitem[Zhao et~al.(2008)Zhao, Allen, Amiri, Appel, Battistelli, Burger,
  Chervenak, Dahlen, Denny, Devlin, et~al.]{zhao2008characterization}
Zhao, Y.; Allen, C.; Amiri, M.; Appel, J.; Battistelli, E.S.; Burger, B.;
  Chervenak, J.; Dahlen, A.; Denny, S.; Devlin, M.;  et~al.
\newblock Characterization of Transition Edge Sensors for the Millimeter
  Bolometer Array Camera on the Atacama Cosmology Telescope.
\newblock In Proccedings of the Millimeter and Submillimeter Detectors and
  Instrumentation for Astronomy IV, Marseille, France, 26--28 June 2008; SPIE: Bellingham, WA, USA,  2008; Volume~7020, pp. 228--238.

\bibitem[Grace et~al.(2014)Grace, Beall, Cho, Devlin, Fox, Hilton, Hubmayr,
  Irwin, Klein, Li, et~al.]{grace2014characterization}
Grace, E.; Beall, J.; Cho, H.; Devlin, M.; Fox, A.; Hilton, G.; Hubmayr, J.;
  Irwin, K.; Klein, J.; Li, D.;  et~al.
\newblock Characterization and performance of a kilo-TES sub-array for ACTPol.
\newblock {\em J. Low Temp. Phys.} {\bf 2014}, {\em
  176},~705--711.

\bibitem[Louis et~al.(2017)Louis, Grace, Hasselfield, Lungu, Maurin, Addison,
  Ade, Aiola, Allison, Amiri, et~al.]{louis2017atacama}
Louis, T.; Grace, E.; Hasselfield, M.; Lungu, M.; Maurin, L.; Addison, G.E.;
  Ade, P.A.; Aiola, S.; Allison, R.; Amiri, M.;  et~al.
\newblock The Atacama Cosmology Telescope: Two-season ACTPol spectra and
  parameters.
\newblock {\em J. Cosmol. Astropart. Phys.} {\bf 2017}, {\em
  2017},~031.

\bibitem[Choi et~al.(2018)Choi, Austermann, Beall, Crowley, Datta, Duff,
  Gallardo, Ho, Hubmayr, Koopman, et~al.]{choi2018characterization}
Choi, S.K.; Austermann, J.; Beall, J.A.; Crowley, K.T.; Datta, R.; Duff, S.M.;
  Gallardo, P.A.; Ho, S.; Hubmayr, J.; Koopman, B.J.;  et~al.
\newblock Characterization of the mid-frequency arrays for advanced ACTPol.
\newblock {\em J. Low Temp. Phys.} {\bf 2018}, {\em
  193},~267--275.

\bibitem[Westbrook et~al.(2024)Westbrook, Prasad, Raum, Lee, Suzuki, Hubmayr,
  Duff, Link, and Lucas]{westbrook2024thermal}
Westbrook, B.; Prasad, B.; Raum, C.R.; Lee, A.T.; Suzuki, A.; Hubmayr, J.;
  Duff, S.M.; Link, M.J.; Lucas, T.J.
\newblock Thermal Annealing of AlMn Transition Edge Sensors for Optimization in
  Cosmic Microwave Background Experiments.
\newblock {\em J. Low Temp. Phys.} {\bf 2024}, \emph{216}, 264--272.

\bibitem[Li et~al.(2016)Li, Austermann, Beall, Becker, Duff, Gallardo,
  Henderson, Hilton, Ho, Hubmayr, et~al.]{li2016almn}
Li, D.; Austermann, J.E.; Beall, J.A.; Becker, D.T.; Duff, S.M.; Gallardo,
  P.A.; Henderson, S.W.; Hilton, G.C.; Ho, S.P.; Hubmayr, J.;  et~al.
\newblock AlMn Transition Edge Sensors for advanced ACTPol.
\newblock {\em J. Low Temp. Phys.} {\bf 2016}, {\em
  184},~66--73.

\bibitem[Crowley et~al.(2018)Crowley, Austermann, Choi, Duff, Gallardo, Ho,
  Hubmayr, Koopman, Nati, Niemack, et~al.]{crowley2018advanced}
Crowley, K.T.; Austermann, J.E.; Choi, S.K.; Duff, S.M.; Gallardo, P.A.; Ho,
  S.P.P.; Hubmayr, J.; Koopman, B.J.; Nati, F.; Niemack, M.D.;  et~al. %MDPI: Refs. 145 and 146 are duplicated. Please remove duplicated references and rearrange all the references to appear in numerical order. Please ensure that there are no duplicated references
\newblock Advanced ACTPoL TES device parameters and noise performance in
  fielded arrays.
\newblock {\em J. Low Temp. Phys.} {\bf 2018}, {\em
  193},~328--336.


\bibitem[Galloni et~al.(2023)Galloni, Bartolo, Matarrese, Migliaccio,
  Ricciardone, and Vittorio]{galloni2023updated}
Galloni, G.; Bartolo, N.; Matarrese, S.; Migliaccio, M.; Ricciardone, A.;
  Vittorio, N.
\newblock Updated constraints on amplitude and tilt of the tensor primordial
  spectrum.
\newblock {\em J. Cosmol. Astropart. Phys.} {\bf 2023}, {\em
  2023},~062.

\bibitem[Simon et~al.(2014)Simon, Raghunathan, Appel, Becker, Campusano, Cho,
  Essinger-Hileman, Ho, Irwin, Jarosik, et~al.]{simon2014characterization}
Simon, S.; Raghunathan, S.; Appel, J.; Becker, D.; Campusano, L.; Cho, H.;
  Essinger-Hileman, T.; Ho, S.; Irwin, K.; Jarosik, N.;  et~al.
\newblock Characterization of the Atacama B-mode Search.
\newblock In Proccedings of the Millimeter, Submillimeter, and Far-Infrared
  Detectors and Instrumentation for Astronomy VII, Montréal, QC, Canada, 22--27 June 2014; SPIE: Bellingham, WA, USA,  2014; Volume~9153, \mbox{pp.
  283--297.}

\bibitem[Appel(2012)]{appel2012detectors}
Appel, J.W.
\newblock Detectors for the Atacama B-Mode Search Experiment.
\newblock Ph.D. Thesis, Princeton University,  Princeton, NJ, USA, 2012.

\bibitem[Essinger‐Hileman et~al.(2009)Essinger‐Hileman, Appel, Beal, Cho,
  Fowler, Halpern, Hasselfield, Irwin, Marriage, Niemack, Page, Parker, Pufu,
  Staggs, Stryzak, Visnjic, Yoon, and Zhao]{essinger2009atacama}
Essinger‐Hileman, T.; Appel, J.W.; Beal, J.A.; Cho, H.M.; Fowler, J.;
  Halpern, M.; Hasselfield, M.; Irwin, K.D.; Marriage, T.A.; Niemack, M.D.;
  et~al.
\newblock {The Atacama B‐Mode Search: CMB Polarimetry with
  Transition‐Edge‐Sensor Bolometers}.
\newblock {\em AIP Conf. Proc.} {\bf 2009}, {\em 1185},~494--497.
\newblock {\url{https://doi.org/10.1063/1.3292387}}.

\bibitem[Kusaka et~al.(2018)Kusaka, Appel, Essinger-Hileman, Beall, Campusano,
  Cho, Choi, Crowley, Fowler, Gallardo, et~al.]{kusaka2018results}
Kusaka, A.; Appel, J.; Essinger-Hileman, T.; Beall, J.A.; Campusano, L.E.; Cho,
  H.M.; Choi, S.K.; Crowley, K.; Fowler, J.W.; Gallardo, P.;  et~al.
\newblock Results from the Atacama B-mode Search (ABS) experiment.
\newblock {\em J. Cosmol. Astropart. Phys.} {\bf 2018}, {\em
  2018},~005.

\bibitem[Keating et~al.(2003)Keating, Ade, Bock, Hivon, Holzapfel, Lange,
  Nguyen, and Yoon]{keating2003bicep}
Keating, B.G.; Ade, P.A.; Bock, J.J.; Hivon, E.; Holzapfel, W.L.; Lange, A.E.;
  Nguyen, H.; Yoon, K.W.
\newblock BICEP: A large angular-scale CMB polarimeter.
\newblock In \emph{Proccedings of the Polarimetry in Astronomy}; SPIE: Bellingham, WA, USA,  2003; Volume
  4843, pp. 284--295.

\bibitem[Ade et~al.(2014)Ade, Aikin, Amiri, Barkats, Benton, Bischoff, Bock,
  Brevik, Buder, Bullock, et~al.]{ade2014bicep2}
Ade, P.A.; Aikin, R.; Amiri, M.; Barkats, D.; Benton, S.; Bischoff, C.A.; Bock,
  J.; Brevik, J.; Buder, I.; Bullock, E.;  et~al.
\newblock BICEP2. II. Experiment and three-year Data Set.
\newblock {\em  Astrophys. J.} {\bf 2014}, {\em 792},~62.

\bibitem[Ade et~al.(2015)Ade, Aikin, Barkats, Benton, Bischoff, Bock, Bradford,
  Brevik, Buder, Bullock, et~al.]{ade2015bicep2}
Ade, P.A.; Aikin, R.; Barkats, D.; Benton, S.; Bischoff, C.A.; Bock, J.;
  Bradford, K.; Brevik, J.; Buder, I.; Bullock, E.;  et~al.
\newblock BICEP2/Keck array. IV. Optical characterization and performance of
  the BICEP2 and Keck array experiments.
\newblock {\em  Astrophys. J.} {\bf 2015}, {\em 806},~206.

\bibitem[Karkare et~al.(2014)Karkare, Ade, Ahmed, Aikin, Alexander, Amiri,
  Barkats, Benton, Bischoff, Bock, et~al.]{karkare2014keck}
Karkare, K.; Ade, P.A.; Ahmed, Z.; Aikin, R.; Alexander, K.; Amiri, M.;
  Barkats, D.; Benton, S.; Bischoff, C.; Bock, J.;  et~al.
\newblock Keck array and BICEP3: Spectral characterization of 5000+ detectors.
\newblock In Proccedings of the Millimeter, Submillimeter, and Far-Infrared
  Detectors and Instrumentation for Astronomy VII, Montréal, QC, Canada, 22--27 June 2014; SPIE: Bellingham, WA, USA,  2014; Volume~9153, pp.
  1027--1037.

\bibitem[Hui et~al.(2018)Hui, Ade, Ahmed, Aikin, Alexander, Barkats, Benton,
  Bischoff, Bock, Bowens-Rubin, et~al.]{hui2018bicep}
Hui, H.; Ade, P.; Ahmed, Z.; Aikin, R.; Alexander, K.D.; Barkats, D.; Benton,
  S.J.; Bischoff, C.A.; Bock, J.J.; Bowens-Rubin, R.;  et~al.
\newblock BICEP Array: A multi-frequency degree-scale CMB polarimeter.
\newblock In Proccedings of the Millimeter, Submillimeter, and Far-Infrared
  Detectors and Instrumentation for Astronomy IX, Austin, TX, USA, 12--15 June 2018; SPIE: Bellingham, WA, USA,  2018; Volume~10708, pp.
  75--89.

\bibitem[Haller(1994)]{haller1994advanced}
Haller, E.
\newblock Advanced far-infrared detectors.
\newblock {\em Infrared Phys. Technol.} {\bf 1994}, {\em 35},~127--146.

\bibitem[Brevik et~al.(2010)Brevik, Aikin, Amiri, Benton, Bock, Bonetti,
  Burger, Dowell, Duband, Filippini, et~al.]{brevik2010initial}
Brevik, J.; Aikin, R.; Amiri, M.; Benton, S.; Bock, J.; Bonetti, J.; Burger,
  B.; Dowell, C.; Duband, L.; Filippini, J.;  et~al.
\newblock Initial performance of the BICEP2 antenna-coupled superconducting
  bolometers at the South Pole.
\newblock In Proccedings of the Millimeter, Submillimeter, and Far-Infrared
  Detectors and Instrumentation for Astronomy V, San Diego, CA, USA, 27 June--2 July 2010; SPIE: Bellingham, WA, USA,  2010; Volume~7741, pp.
  426--435.

\bibitem[Grayson et~al.(2016)Grayson, Ade, Ahmed, Alexander, Amiri, Barkats,
  Benton, Bischoff, Bock, Boenish, et~al.]{grayson2016bicep3}
Grayson, J.A.; Ade, P.; Ahmed, Z.; Alexander, K.D.; Amiri, M.; Barkats, D.;
  Benton, S.; Bischoff, C.A.; Bock, J.; Boenish, H.;  et~al.
\newblock BICEP3 performance overview and planned Keck Array upgrade.
\newblock In Proccedings of the Millimeter, Submillimeter, and Far-Infrared
  Detectors and Instrumentation for Astronomy VIII, Edinburgh, UK, 26 June--1 July 2016; SPIE: Bellingham, WA, USA,  2016; Volume~9914, pp.
  157--173.

\bibitem[Essinger-Hileman et~al.(2014)Essinger-Hileman, Ali, Amiri, Appel,
  Araujo, Bennett, Boone, Chan, Cho, Chuss, et~al.]{essinger2014class}
Essinger-Hileman, T.; Ali, A.; Amiri, M.; Appel, J.W.; Araujo, D.; Bennett,
  C.L.; Boone, F.; Chan, M.; Cho, H.M.; Chuss, D.T.;  et~al.
\newblock CLASS: The cosmology large angular scale surveyor.
\newblock In Proccedings of the Millimeter, Submillimeter, and Far-Infrared
  Detectors and Instrumentation for Astronomy VII, Montréal, QC, Canada, 22--27 June 2014; SPIE: Bellingham, WA, USA,  2014; Volume~9153, pp.
  491--513.

\bibitem[Iuliano et~al.(2018)Iuliano, Eimer, Parker, Rhoades, Ali, Appel,
  Bennett, Brewer, Bustos, Chuss, et~al.]{iuliano2018cosmology}
Iuliano, J.; Eimer, J.; Parker, L.; Rhoades, G.; Ali, A.; Appel, J.W.; Bennett,
  C.; Brewer, M.; Bustos, R.; Chuss, D.;  et~al.
\newblock The cosmology large angular scale surveyor receiver design.
\newblock In Proccedings of the Millimeter, Submillimeter, and Far-Infrared
  Detectors and Instrumentation for Astronomy IX, Austin, TX, USA, 12--15 June 2018; SPIE: Bellingham, WA, USA,  2018; Volume~10708, \mbox{pp.
  259--277.}

\bibitem[Chuss et~al.(2016)Chuss, Ali, Amiri, Appel, Bennett, Colazo, Denis,
  D{\"u}nner, Essinger-Hileman, Eimer, et~al.]{chuss2016cosmology}
Chuss, D.T.; Ali, A.; Amiri, M.; Appel, J.; Bennett, C.; Colazo, F.; Denis, K.;
  D{\"u}nner, R.; Essinger-Hileman, T.; Eimer, J.;  et~al.
\newblock Cosmology large angular scale surveyor (CLASS) focal plane
  development.
\newblock {\em J. Low Temp. Phys.} {\bf 2016}, {\em
  184},~759--764.

\bibitem[Eimer et~al.(2012)Eimer, Bennett, Chuss, Marriage, Wollack, and
  Zeng]{eimer2012cosmology}
Eimer, J.R.; Bennett, C.L.; Chuss, D.T.; Marriage, T.; Wollack, E.J.; Zeng, L.
\newblock The cosmology large angular scale surveyor (CLASS): 40~GHz optical
  design.
\newblock In Proccedings of the Millimeter, Submillimeter, and Far-Infrared
  Detectors and Instrumentation for Astronomy VI, Amsterdam, The Netherlands, 1--6 July 2012; SPIE: Bellingham, WA, USA,  2012; Volume~8452, pp.
  619--633.

\bibitem[Dahal et~al.(2018)Dahal, Ali, Appel, Essinger-Hileman, Bennett,
  Brewer, Bustos, Chan, Chuss, Cleary, et~al.]{dahal2018design}
Dahal, S.; Ali, A.; Appel, J.W.; Essinger-Hileman, T.; Bennett, C.; Brewer, M.;
  Bustos, R.; Chan, M.; Chuss, D.T.; Cleary, J.;  et~al.
\newblock Design and characterization of the cosmology large angular scale
  surveyor (CLASS) 93~GHz focal plane.
\newblock In Proccedings of the Millimeter, Submillimeter, and Far-Infrared
  Detectors and Instrumentation for Astronomy IX, Austin, TX, USA, 12--15 June 2018; SPIE: Bellingham, WA, USA,  2018; Volume~10708, pp.
  230--245.

\bibitem[Appel et~al.(2014)Appel, Ali, Amiri, Araujo, Bennet, Boone, Chan, Cho,
  Chuss, Colazo, et~al.]{appel2014cosmology}
Appel, J.W.; Ali, A.; Amiri, M.; Araujo, D.; Bennet, C.L.; Boone, F.; Chan, M.;
  Cho, H.M.; Chuss, D.T.; Colazo, F.;  et~al.
\newblock The cosmology large angular scale surveyor (CLASS): 38-GHz detector
  array of bolometric polarimeters.
\newblock In Proccedings of the Millimeter, Submillimeter, and Far-Infrared
  Detectors and Instrumentation for Astronomy VII, Montréal, QC, Canada, 22--27 June 2014; SPIE: Bellingham, WA, USA,  2014; Volume~9153, pp.
  514--528.

\bibitem[Shirokoff et~al.(2009)Shirokoff, Benson, Bleem, Chang, Cho, Crites,
  Dobbs, Holzapfel, Lanting, Lee, et~al.]{shirokoff2009south}
Shirokoff, E.; Benson, B.A.; Bleem, L.E.; Chang, C.L.; Cho, H.M.; Crites, A.T.;
  Dobbs, M.A.; Holzapfel, W.L.; Lanting, T.; Lee, A.T.;  et~al.
\newblock The south pole telescope SZ-receiver detectors.
\newblock {\em IEEE Trans. Appl. Supercond.} {\bf 2009}, {\em
  19},~517--519.

\bibitem[Austermann et~al.(2012)Austermann, Aird, Beall, Becker, Bender,
  Benson, Bleem, Britton, Carlstrom, Chang, et~al.]{austermann2012sptpol}
Austermann, J.E.; Aird, K.; Beall, J.; Becker, D.; Bender, A.; Benson, B.;
  Bleem, L.; Britton, J.; Carlstrom, J.; Chang, C.;  et~al.
\newblock SPTpol: An instrument for CMB polarization measurements with the
  South Pole Telescope.
\newblock In Proccedings of the Millimeter, Submillimeter, and Far-Infrared
  Detectors and Instrumentation for Astronomy VI, Amsterdam, The Netherlands, 1--6 July 2012; SPIE: Bellingham, WA, USA,  2012; Volume~8452, pp.
  393--410.

\bibitem[Benson et~al.(2014)Benson, Ade, Ahmed, Allen, Arnold, Austermann,
  Bender, Bleem, Carlstrom, Chang, et~al.]{benson2014spt}
Benson, B.A.; Ade, P.; Ahmed, Z.; Allen, S.; Arnold, K.; Austermann, J.;
  Bender, A.; Bleem, L.; Carlstrom, J.; Chang, C.;  et~al.
\newblock SPT-3G: A next-generation cosmic microwave background polarization
  experiment on the South Pole telescope.
\newblock In Proccedings of the Millimeter, Submillimeter, and Far-Infrared
  Detectors and Instrumentation for Astronomy VII, Montréal, QC, Canada, 22--27 June 2014; SPIE: Bellingham, WA, USA,  2014; Volume~9153, pp.
  552--572.

\bibitem[Anderson et~al.(2018)Anderson, Ade, Ahmed, Austermann, Avva, Barry,
  Thakur, Bender, Benson, Bleem, et~al.]{anderson2018spt}
Anderson, A.; Ade, P.; Ahmed, Z.; Austermann, J.; Avva, J.; Barry, P.; Thakur,
  R.B.; Bender, A.; Benson, B.; Bleem, L.;  et~al.
\newblock Spt-3g: A multichroic receiver for the south pole telescope.
\newblock {\em J. Low Temp. Phys.} {\bf 2018}, {\em
  193},~1057--1065.

\bibitem[Sobrin et~al.(2022)Sobrin, Anderson, Bender, Benson, Dutcher, Foster,
  Goeckner-Wald, Montgomery, Nadolski, Rahlin, et~al.]{sobrin2022design}
Sobrin, J.; Anderson, A.; Bender, A.; Benson, B.; Dutcher, D.; Foster, A.;
  Goeckner-Wald, N.; Montgomery, J.; Nadolski, A.; Rahlin, A.;  et~al.
\newblock The design and integrated performance of SPT-3G.
\newblock {\em  Astrophys. J. Suppl. Ser.} {\bf 2022}, {\em
  258},~42.

\bibitem[Kermish et~al.(2012)Kermish, Ade, Anthony, Arnold, Barron, Boettger,
  Borrill, Chapman, Chinone, Dobbs, et~al.]{kermish2012polarbear}
Kermish, Z.D.; Ade, P.; Anthony, A.; Arnold, K.; Barron, D.; Boettger, D.;
  Borrill, J.; Chapman, S.; Chinone, Y.; Dobbs, M.A.;  et~al.
\newblock The POLARBEAR experiment.
\newblock In Proccedings of the Millimeter, Submillimeter, and Far-Infrared
  Detectors and Instrumentation for Astronomy VI, Amsterdam, The Netherlands, 1--6 July 2012; SPIE: Bellingham, WA, USA,  2012; Volume~8452, pp.
  366--380.

\bibitem[Suzuki et~al.(2014)Suzuki, Ade, Akiba, Aleman, Arnold, Atlas, Barron,
  Borrill, Chapman, Chinone, et~al.]{suzuki2014polarbear}
Suzuki, A.; Ade, P.; Akiba, Y.; Aleman, C.; Arnold, K.; Atlas, M.; Barron, D.;
  Borrill, J.; Chapman, S.; Chinone, Y.;  et~al.
\newblock The POLARBEAR-2 experiment.
\newblock {\em J. Low Temp. Phys.} {\bf 2014}, {\em
  176},~719--725.

\bibitem[Barron et~al.(2021)Barron, Mitchell, Groh, Arnold, Elleflot, Howe,
  Ito, Lee, Lowry, Anderson, et~al.]{barron2021integrated}
Barron, D.; Mitchell, K.; Groh, J.; Arnold, K.; Elleflot, T.; Howe, L.; Ito,
  J.; Lee, A.T.; Lowry, L.N.; Anderson, A.;  et~al.
\newblock Integrated electrical properties of the frequency multiplexed
  cryogenic readout system for POLARBEAR/Simons Array.
\newblock {\em IEEE Trans. Appl. Supercond.} {\bf 2021}, {\em
  31},~2101805.

\bibitem[Mennella et~al.(2019)Mennella, Ade, Amico, Auguste, Aumont, Banfi,
  Barbar{\`a}n, Battaglia, Battistelli, Ba{\`u}, et~al.]{mennella2019qubic}
Mennella, A.; Ade, P.; Amico, G.; Auguste, D.; Aumont, J.; Banfi, S.;
  Barbar{\`a}n, G.; Battaglia, P.; Battistelli, E.; Ba{\`u}, A.;  et~al.
\newblock QUBIC: Exploring the Primordial Universe with the Q\&U Bolometric
  Interferometer.
\newblock {\em Universe} {\bf 2019}, {\em 5},~42.

\bibitem[O'Sulivan et~al.(2015)O'Sulivan, Scully, Gayer, Gradziel, Murphy,
  De~Petris, Buzi, Gervasi, Zannoni, Hamilton, et~al.]{o2015qu}
O'Sulivan, C.; Scully, S.; Gayer, D.; Gradziel, M.; Murphy, J.; De~Petris, M.;
  Buzi, D.; Gervasi, M.; Zannoni, M.; Hamilton, J.;  et~al.
\newblock The QU Bolometric Interferometer for Cosmology (QUBIC).
\newblock In Proccedings of the 36th ESA Antenna Workshop on Antennas and RF
  Systems for Space Science, Noordwijk, The Netherlands, 6--9 October 2015; ESA Publications Division c/o ESTEC.

\bibitem[Marnieros et~al.(2020)Marnieros, Ade, Alberro, Almela, Amico, Arnaldi,
  Auguste, Aumont, Azzoni, Banfi, et~al.]{marnieros2020tes}
Marnieros, S.; Ade, P.; Alberro, J.; Almela, A.; Amico, G.; Arnaldi, L.;
  Auguste, D.; Aumont, J.; Azzoni, S.; Banfi, S.;  et~al.
\newblock TES bolometer arrays for the QUBIC B-mode CMB experiment.
\newblock {\em J. Low Temp. Phys.} {\bf 2020}, {\em
  199},~955--961.

\bibitem[Piat et~al.(2022)Piat, Stankowiak, Battistelli, De~Bernardis,
  d'Alessandro, De~Petris, Grandsire, Hamilton, Hoang, Marnieros,
  et~al.]{piat2022qubic}
Piat, M.; Stankowiak, G.; Battistelli, E.; De~Bernardis, P.; d'Alessandro, G.;
  De~Petris, M.; Grandsire, L.; Hamilton, J.C.; Hoang, T.; Marnieros, S.;
  et~al.
\newblock QUBIC IV: Performance of TES bolometers and readout electronics.
\newblock {\em J. Cosmol. Astropart. Phys.} {\bf 2022}, {\em
  2022},~037.

\bibitem[Ghosh et~al.(2022)Ghosh, Liu, Zhang, Li, Zhang, Wang, Dou, Chen,
  Delabrouille, Remazeilles, et~al.]{ghosh2022performance}
Ghosh, S.; Liu, Y.; Zhang, L.; Li, S.; Zhang, J.; Wang, J.; Dou, J.; Chen, J.;
  Delabrouille, J.; Remazeilles, M.;  et~al.
\newblock Performance forecasts for the primordial Gravitational Wave detection
  pipelines for AliCPT-1.
\newblock {\em J. Cosmol. Astropart. Phys.} {\bf 2022}, {\em
  2022},~063.

\bibitem[Gao et~al.(2017)Gao, Liu, Li, Liu, Li, Li, Li, Gao, Lu, and
  Zhang]{gao2017introduction}
Gao, H.; Liu, C.; Li, Z.; Liu, Y.; Li, Y.; Li, S.; Li, H.; Gao, G.; Lu, F.;
  Zhang, X.
\newblock Introduction to the detection technology of Ali CMB polarization
  telescope.
\newblock {\em Radiat. Detect. Technol. Methods} {\bf 2017}, {\em
  1},~12.

\bibitem[Salatino et~al.(2020)Salatino, Austermann, Thompson, Ade, Bai, Beall,
  Becker, Cai, Chang, Chen, et~al.]{salatino2020design}
Salatino, M.; Austermann, J.; Thompson, K.L.; Ade, P.A.; Bai, X.; Beall, J.A.;
  Becker, D.T.; Cai, Y.; Chang, Z.; Chen, D.;  et~al.
\newblock The design of the Ali CMB Polarization Telescope receiver.
\newblock In Proccedings of the Millimeter, Submillimeter, and Far-Infrared
  Detectors and Instrumentation for Astronomy X, Virtual, 14--18 December 2020; SPIE: Bellingham, WA, USA,  2020; Volume~11453, pp.
  341--360.

\bibitem[Lee et~al.(2019)Lee, Abitbol, Adachi, Ade, Aguirre, Ahmed, Aiola, Ali,
  Alonso, Alvarez, et~al.]{lee2019simons}
Lee, A.; Abitbol, M.H.; Adachi, S.; Ade, P.; Aguirre, J.; Ahmed, Z.; Aiola, S.;
  Ali, A.; Alonso, D.; Alvarez, M.A.;  et~al.
\newblock The Simons Observatory.
\newblock {\em Bull. Am. Astron. Soc} {\bf 2019}, {\em 51},~147.

\bibitem[Ade et~al.(2019)Ade, Aguirre, Ahmed, Aiola, Ali, Alonso, Alvarez,
  Arnold, Ashton, Austermann, et~al.]{ade2019simons}
Ade, P.; Aguirre, J.; Ahmed, Z.; Aiola, S.; Ali, A.; Alonso, D.; Alvarez, M.A.;
  Arnold, K.; Ashton, P.; Austermann, J.;  et~al.
\newblock The Simons Observatory: Science goals and forecasts.
\newblock {\em J. Cosmol. Astropart. Phys.} {\bf 2019}, {\em
  2019},~056.

\bibitem[Stevens et~al.(2020)Stevens, Cothard, Vavagiakis, Ali, Arnold,
  Austermann, Choi, Dober, Duell, Duff, et~al.]{stevens2020characterization}
Stevens, J.R.; Cothard, N.F.; Vavagiakis, E.M.; Ali, A.; Arnold, K.;
  Austermann, J.E.; Choi, S.K.; Dober, B.J.; Duell, C.; Duff, S.M.;  et~al.
\newblock Characterization of Transition Edge Sensors for the Simons
  observatory.
\newblock {\em J. Low Temp. Phys.} {\bf 2020}, {\em
  199},~672--680.
\newpage
\bibitem[Galitzki et~al.(2018)Galitzki, Ali, Arnold, Ashton, Austermann,
  Baccigalupi, Baildon, Barron, Beall, Beckman, et~al.]{galitzki2018simons}
Galitzki, N.; Ali, A.; Arnold, K.S.; Ashton, P.C.; Austermann, J.E.;
  Baccigalupi, C.; Baildon, T.; Barron, D.; Beall, J.A.; Beckman, S.;  et~al.
\newblock The Simons observatory: Instrument overview.
\newblock In Proccedings of the Millimeter, Submillimeter, and Far-Infrared
  Detectors and Instrumentation for Astronomy IX, Austin, TX, USA, 12--15 June 2018; SPIE: Bellingham, WA, USA,  2018; Volume~10708, pp.
  40--52.

\bibitem[Bryan et~al.(2018)Bryan, Simon, Gerbino, Teply, Ali, Chinone, Crowley,
  Fabbian, Gallardo, Goeckner-Wald, et~al.]{bryan2018development}
Bryan, S.A.; Simon, S.M.; Gerbino, M.; Teply, G.; Ali, A.; Chinone, Y.;
  Crowley, K.; Fabbian, G.; Gallardo, P.A.; Goeckner-Wald, N.;  et~al.
\newblock Development of calibration strategies for the Simons Observatory.
\newblock In Proccedings of the Millimeter, Submillimeter, and Far-Infrared
  Detectors and Instrumentation for Astronomy IX, Austin, TX, USA, 12--15 June 2018; SPIE: Bellingham, WA, USA,  2018; Volume~10708, pp.
  988--1000.

\bibitem[Abazajian et~al.(2016)Abazajian, Adshead, Ahmed, Allen, Alonso,
  Arnold, Baccigalupi, Bartlett, Battaglia, Benson, et~al.]{abazajian2016cmb}
Abazajian, K.N.; Adshead, P.; Ahmed, Z.; Allen, S.W.; Alonso, D.; Arnold, K.S.;
  Baccigalupi, C.; Bartlett, J.G.; Battaglia, N.; Benson, B.A.;  et~al.
\newblock CMB-S4 science book.
\newblock {\em arXiv} {\bf 2016}, arXiv:1610.02743.

\bibitem[Abitbol et~al.(2017)Abitbol, Ahmed, Barron, Thakur, Bender, Benson,
  Bischoff, Bryan, Carlstrom, Chang, et~al.]{abitbol2017cmb}
Abitbol, M.H.; Ahmed, Z.; Barron, D.; Thakur, R.B.; Bender, A.N.; Benson, B.A.;
  Bischoff, C.A.; Bryan, S.A.; Carlstrom, J.E.; Chang, C.L.;  et~al.
\newblock CMB-S4 technology book.
\newblock {\em arXiv} {\bf 2017}, arXiv:1706.02464.

\bibitem[Duff et~al.(2024)Duff, Austermann, Beall, Daniel, Hubmayr, Jaehnig,
  Johnson, Jones, Link, Lucas, et~al.]{duff2024simons}
Duff, S.M.; Austermann, J.; Beall, J.A.; Daniel, D.P.; Hubmayr, J.; Jaehnig,
  G.C.; Johnson, B.R.; Jones, D.; Link, M.J.; Lucas, T.J.;  et~al.
\newblock The Simons observatory: Production-level fabrication of the mid-and
  ultra-high-frequency wafers.
\newblock {\em J. Low Temp. Phys.} {\bf 2024}, \emph{216}, 135--143.

\bibitem[Abazajian et~al.(2022)Abazajian, Abdulghafour, Addison, Adshead,
  Ahmed, Ajello, Akerib, Allen, Alonso, Alvarez, et~al.]{abazajian2022snowmass}
Abazajian, K.; Abdulghafour, A.; Addison, G.E.; Adshead, P.; Ahmed, Z.; Ajello,
  M.; Akerib, D.; Allen, S.W.; Alonso, D.; Alvarez, M.;  et~al.
\newblock Snowmass 2021 CMB-S4 white paper.
\newblock {\em arXiv} {\bf 2022}, arXiv:2203.08024.

\bibitem[Oxley et~al.(2004)Oxley, Ade, Baccigalupi, deBernardis, Cho, Devlin,
  Hanany, Johnson, Jones, Lee, et~al.]{oxley2004ebex}
Oxley, P.; Ade, P.A.; Baccigalupi, C.; deBernardis, P.; Cho, H.M.; Devlin,
  M.J.; Hanany, S.; Johnson, B.; Jones, T.; Lee, A.T.;  et~al.
\newblock The EBEX experiment.
\newblock In Proccedings of the Infrared Spaceborne Remote Sensing XII, Denver, CO, USA, 2--6 August 2004; SPIE: Bellingham, WA, USA,
  2004; Volume~5543, pp. 320--331.

\bibitem[Reichborn-Kjennerud et~al.(2010)Reichborn-Kjennerud, Aboobaker, Ade,
  Aubin, Baccigalupi, Bao, Borrill, Cantalupo, Chapman, Didier,
  et~al.]{reichborn2010ebex}
Reichborn-Kjennerud, B.; Aboobaker, A.M.; Ade, P.; Aubin, F.; Baccigalupi, C.;
  Bao, C.; Borrill, J.; Cantalupo, C.; Chapman, D.; Didier, J.;  et~al.
\newblock EBEX: A balloon-borne CMB polarization experiment.
\newblock In Proccedings of the Millimeter, Submillimeter, and Far-Infrared
  Detectors and Instrumentation for Astronomy V, San Diego, CA, USA, 27 June--2 July 2010; SPIE: Bellingham, WA, USA,  2010; Volume~7741, pp.
  381--392.

\bibitem[Crill et~al.(2008)Crill, Ade, Battistelli, Benton, Bihary, Bock, Bond,
  Brevik, Bryan, Contaldi, et~al.]{crill2008spider}
Crill, B.; Ade, P.A.; Battistelli, E.S.; Benton, S.; Bihary, R.; Bock, J.;
  Bond, J.; Brevik, J.; Bryan, S.; Contaldi, C.;  et~al.
\newblock SPIDER: A balloon-borne large-scale CMB polarimeter.
\newblock In {Proccedings of the Space Telescopes and Instrumentation 2008:
  Optical, Infrared, and Millimeter, Marseille, France, 23--28 June 2008}; SPIE: Bellingham, WA, USA,  2008; Volume~7010, pp. 800--811.

\bibitem[Gualtieri et~al.(2018)Gualtieri, Filippini, Ade, Amiri, Benton,
  Bergman, Bihary, Bock, Bond, Bryan, et~al.]{gualtieri2018spider}
Gualtieri, R.; Filippini, J.; Ade, P.; Amiri, M.; Benton, S.; Bergman, A.;
  Bihary, R.; Bock, J.; Bond, J.; Bryan, S.;  et~al.
\newblock SPIDER: CMB Polarimetry from the Edge of Space.
\newblock {\em J. Low Temp. Phys.} {\bf 2018}, {\em
  193},~1112--1121.

\bibitem[Gambrel(2018)]{gambrel2018measurement}
Gambrel, A.E.
\newblock Measurement of the Polarization of the Cosmic Microwave Background
  with the SPIDER Instrument.
\newblock Ph.D. Thesis, Princeton University, Princeton, NJ, USA, 2018.

\bibitem[Hubmayr et~al.(2016)Hubmayr, Austermann, Beall, Becker, Benton,
  Bergman, Bond, Bryan, Duff, Duivenvoorden, et~al.]{hubmayr2016design}
Hubmayr, J.; Austermann, J.E.; Beall, J.A.; Becker, D.T.; Benton, S.J.;
  Bergman, A.S.; Bond, J.R.; Bryan, S.; Duff, S.M.; Duivenvoorden, A.J.;
  et~al.
\newblock Design of 280~GHz feedhorn-coupled TES arrays for the balloon-borne
  polarimeter SPIDER.
\newblock In Proccedings of the Millimeter, Submillimeter, and Far-Infrared
  Detectors and Instrumentation for Astronomy VIII, Edinburgh, UK, 26 June--1 July 2016; SPIE: Bellingham, WA, USA,  2016; Volume~9914, pp.
  185--198.

\bibitem[Rahlin et~al.(2014)Rahlin, Ade, Amiri, Benton, Bock, Bond, Bryan,
  Chiang, Contaldi, Crill, et~al.]{rahlin2014pre}
Rahlin, A.; Ade, P.; Amiri, M.; Benton, S.; Bock, J.; Bond, J.; Bryan, S.;
  Chiang, H.C.; Contaldi, C.; Crill, B.;  et~al.
\newblock Pre-flight integration and characterization of the SPIDER
  balloon-borne telescope.
\newblock In Proccedings of the Millimeter, Submillimeter, and Far-Infrared
  Detectors and Instrumentation for Astronomy VII, Montréal, QC, Canada, 22--27 June 2014; SPIE: Bellingham, WA, USA,  2014; Volume~9153, pp.
  336--360.

\bibitem[Lamagna et~al.(2020)Lamagna, Addamo, Ade, Baccigalupi, Baldini,
  Battaglia, Battistelli, Ba{\`u}, Bersanelli, Biasotti,
  et~al.]{lamagna2020progress}
Lamagna, L.; Addamo, G.; Ade, P.; Baccigalupi, C.; Baldini, A.; Battaglia, P.;
  Battistelli, E.; Ba{\`u}, A.; Bersanelli, M.; Biasotti, M.;  et~al.
\newblock Progress report on the large-scale polarization explorer.
\newblock {\em J. Low Temp. Phys.} {\bf 2020}, {\em
  200},~374--383.

\bibitem[Columbro et~al.(2019)Columbro, Battistelli, Coppolecchia,
  D'Alessandro, de~Bernardis, Lamagna, Masi, Pagano, Paiella, Piacentini,
  et~al.]{columbro2019short}
Columbro, F.; Battistelli, E.; Coppolecchia, A.; D'Alessandro, G.;
  de~Bernardis, P.; Lamagna, L.; Masi, S.; Pagano, L.; Paiella, A.; Piacentini,
  F.;  et~al.
\newblock The short wavelength instrument for the polarization explorer
  balloon-borne experiment: Polarization modulation issues.
\newblock {\em Astron. Nachrichten} {\bf 2019}, {\em 340},~83--88.

\bibitem[Tartari et~al.(2024)Tartari, Baldini, Cei, Celasco, Dal~Bo, Di~Giorgi,
  Ferrari~Barusso, Galli, Gatti, Grosso, et~al.]{tartari2024characterization}
Tartari, A.; Baldini, A.; Cei, F.; Celasco, E.; Dal~Bo, P.; Di~Giorgi, E.;
  Ferrari~Barusso, L.; Galli, L.; Gatti, F.; Grosso, D.;  et~al.
\newblock A Characterization Procedure for Large Area Spiderweb TES.
\newblock {\em J. Low Temp. Phys.} {\bf 2024}, {\em 216},~112--118.

\bibitem[Sandri(2017)]{sandri2017lspe}
Sandri, M.
\newblock LSPE: The STRIP Instrument.
\newblock In  Proccedings of the {Workshop sull'Astronomia Millimetrica in
  Italia, Bologna, Italy, 7--10 November 2017}; p.~35.

\bibitem[Matsumura et~al.(2014)Matsumura, Akiba, Borrill, Chinone, Dobbs, Fuke,
  Ghribi, Hasegawa, Hattori, Hattori, et~al.]{matsumura2014mission}
Matsumura, T.; Akiba, Y.; Borrill, J.; Chinone, Y.; Dobbs, M.; Fuke, H.;
  Ghribi, A.; Hasegawa, M.; Hattori, K.; Hattori, M.;  et~al.
\newblock Mission design of LiteBIRD.
\newblock {\em J. Low Temp. Phys.} {\bf 2014}, {\em
  176},~733--740.

\bibitem[Sekimoto et~al.(2020)Sekimoto, Ade, Adler, Allys, Arnold, Auguste,
  Aumont, Aurlien, Austermann, Baccigalupi, et~al.]{sekimoto2020concept}
Sekimoto, Y.; Ade, P.A.; Adler, A.; Allys, E.; Arnold, K.; Auguste, D.; Aumont,
  J.; Aurlien, R.; Austermann, J.; Baccigalupi, C.;  et~al.
\newblock Concept design of low frequency telescope for CMB B-mode polarization
  satellite LiteBIRD.
\newblock In Proccedings of the Millimeter, Submillimeter, and Far-Infrared
  Detectors and Instrumentation for Astronomy X, Virtual, 14--18 December 2020; SPIE: Bellingham, WA, USA,  2020; Volume~11453, pp.
  189--209.

\bibitem[Montier et~al.(2020)Montier, Mot, de~Bernardis, Maffei, Pisano,
  Columbro, Gudmundsson, Henrot-Versill{\'e}, Lamagna, Montgomery,
  et~al.]{montier2020overview}
Montier, L.; Mot, B.; de~Bernardis, P.; Maffei, B.; Pisano, G.; Columbro, F.;
  Gudmundsson, J.E.; Henrot-Versill{\'e}, S.; Lamagna, L.; Montgomery, J.;
  et~al.
\newblock Overview of the medium and high frequency telescopes of the LiteBIRD
  space mission.
\newblock In {Proccedings of the Space Telescopes and Instrumentation 2020:
  Optical, Infrared, and Millimeter Wave,  Virtual, 14--18 December 2020}; SPIE: Bellingham, WA, USA,  2020; Volume~11443, pp.
  451--471.

\bibitem[Tominaga et~al.(2022)Tominaga, Tsujimoto, Smecher, Ishino, and
  Group]{tominaga2022design}
Tominaga, M.; Tsujimoto, M.; Smecher, G.; Ishino, H.; Group, L.J.S.
\newblock Design of the on-board data compression for the bolometer data of
  LiteBIRD.
\newblock {\em J. Low Temp. Phys.} {\bf 2022}, {\em
  209},~686--692.

\bibitem[Abdelhameed et~al.(2019)Abdelhameed, Angloher, Bauer, Bento, Bertoldo,
  Bucci, Canonica, D’Addabbo, Defay, Di~Lorenzo,
  et~al.]{abdelhameed2019first}
Abdelhameed, A.H.; Angloher, G.; Bauer, P.; Bento, A.; Bertoldo, E.; Bucci, C.;
  Canonica, L.; D’Addabbo, A.; Defay, X.; Di~Lorenzo, S.;  et~al.
\newblock First results from the CRESST-III low-mass dark matter program.
\newblock {\em Phys. Rev. D} {\bf 2019}, {\em 100},~102002.

\bibitem[Alkhatib et~al.(2021)Alkhatib, Amaral, Aralis, Aramaki, Arnquist,
  Ataee~Langroudy, Azadbakht, Banik, Barker, Bathurst, Bauer, Bezerra,
  Bhattacharyya, Binder, Bowles, Brink, Bunker, Cabrera, Calkins, Cameron,
  Cartaro, Cerdeño, Chang, Chaudhuri, Chen, Chott, Cooley, Coombes, Corbett,
  Cushman, De~Brienne, di~Vacri, Diamond, Fascione, Figueroa-Feliciano, Fink,
  Fouts, Fritts, Gerbier, Germond, Ghaith, Golwala, Harris, Herbert, Hines,
  Hollister, Hong, Hoppe, Hsu, Huber, Iyer, Jardin, Jastram, Kashyap, Kelsey,
  Kubik, Kurinsky, Lawrence, Li, Loer, Lopez~Asamar, Lukens, MacDonell,
  MacFarlane, Mahapatra, Mandic, Mast, Mayer, Meyer~zu Theenhausen, Michaud,
  Michielin, Mirabolfathi, Mohanty, Morales~Mendoza, Nagorny, Nelson, Neog,
  Novati, Orrell, Oser, Page, Pakarha, Partridge, Podviianiuk, Ponce, Poudel,
  Pyle, Rau, Reid, Ren, Reynolds, Roberts, Robinson, Saab, Sadoulet, Sander,
  Sattari, Schnee, Scorza, Serfass, Sincavage, Stanford, Street, Toback,
  Underwood, Verma, Villano, von Krosigk, Watkins, Wills, Wilson, Wilson,
  Winchell, Wright, Yellin, Young, Yu, Zhang, Zhang, Zhao, Zheng, Camilleri,
  Kolomensky, and Zuber]{Alkhatib_2021}
Alkhatib, I.; Amaral, D.; Aralis, T.; Aramaki, T.; Arnquist, I.;
  Ataee~Langroudy, I.; Azadbakht, E.; Banik, S.; Barker, D.; Bathurst, C.;
  et~al.
\newblock Light Dark Matter Search with a High-Resolution Athermal Phonon
  Detector Operated above Ground.
\newblock {\em Phys. Rev. Lett.} {\bf 2021}, {\em 127}, 061801.
\newblock {\url{https://doi.org/10.1103/physrevlett.127.061801}}.

\bibitem[Lattaud et~al.(2022)Lattaud, Armengaud, Arnaud, Augier, Benoit,
  Bergé, Billard, Broniatowski, Camus, Cazes, Chapellier, Charlieux, De.Jesus,
  Dumoulin, Eitel, Fillipini, Filosofov, Gascon, Giuliani, Gros, Jin, Juillard,
  Kleifges, Marnieros, Misiak, Navick, Nones, Olivieri, Oriol, Pari, Paul,
  Poda, Rozov, Salagnac, Sanglard, Siebenborn, Vagneron, Weber, Yakushev, and
  Zolotarova]{Lattaud_2022}
Lattaud, H.; Armengaud, E.; Arnaud, Q.; Augier, C.; Benoit, A.; Bergé, L.;
  Billard, J.; Broniatowski, A.; Camus, P.; Cazes, A.;  et~al.
\newblock Sub-MeV Dark Matter Searches with EDELWEISS: Results and prospects.
\newblock In Proccedings of the  European Physical Society
  Conference on High Energy Physics---PoS(EPS-HEP2021), Sissa Medialab,
  2022, EPS-HEP2021, Virtual, 26--30 July 2021; p. 153.
\newblock {\url{https://doi.org/10.22323/1.398.0153}}.

\bibitem[D'Angelo(2015)]{dangelo2015darkside50resultsargonrun}
D'Angelo, D.
\newblock DarkSide50 results from first argon run. \emph{ArXiV} \textbf{2015}, arXiv:hep-ex/1501.03541.

\bibitem[Cao et~al.(2019)Cao, He, Shang, Zhang, and Zhu]{Cao_2019}
Cao, J.; He, Y.; Shang, L.; Zhang, Y.; Zhu, P.
\newblock Current status of a natural NMSSM in light of LHC 13 TeV data and
  XENON-1T results.
\newblock {\em Phys. Rev. D} {\bf 2019}, {\em 99}, 075020.
\newblock {\url{https://doi.org/10.1103/physrevd.99.075020}}.

\bibitem[Adams et~al.(2022)Adams, Alduino, Alessandria, Alfonso, Andreotti,
  Avignone~III, Azzolini, Balata, Bandac, Banks, et~al.]{adams2022cuore}
Adams, D.; Alduino, C.; Alessandria, F.; Alfonso, K.; Andreotti, E.;
  Avignone~III, F.; Azzolini, O.; Balata, M.; Bandac, I.; Banks, T.;  et~al.
\newblock CUORE opens the door to tonne-scale cryogenics experiments.
\newblock {\em Prog. Part. Nucl. Phys.} {\bf 2022}, {\em
  122},~103902.

\bibitem[Agrawal et~al.(2024)Agrawal, Alenkov, Aryal, Beyer, Bhandari, Boiko,
  Boonin, Buzanov, Byeon, Chanthima, Cheoun, Choe, Choi, Choudhury, Chung,
  Danevich, Djamal, Drung, Enss, Fleischmann, Gangapshev, Gastaldo, Gavrilyuk,
  Gezhaev, Gileva, Grigorieva, Gurentsov, Ha, Ha, Ha, Hwang, Jeon, Jeon, Jo,
  Kaewkhao, Kang, Kang, Kazalov, Kempf, Khan, Khan, Kim, Kim, Kim, Kim, Kim,
  Kim, Kim, Kim, Kim, Kim, Kim, Kim, Kim, Kim, Kirdsiri, Ko, Kobychev,
  Kornoukhov, Kuzminov, Kwon, Lee, Lee, Lee, Lee, Lee, Lee, Lee, Lee, Lee, Lee,
  Lee, Lee, Leonard, Lim, Mailyan, Makarov, Nyanda, Oh, Olsen, Panasenko, Park,
  Park, Park, Park, Polischuk, Prihtiadi, Ra, Ratkevich, Rooh, Sari, Seo, Seo,
  Sharma, Shin, Shlegel, Siyeon, So, Sokur, Son, Song, Srisittipokakun,
  Tretyak, Wirawan, Woo, Yeon, Yoon, and
  Yue]{agrawal2024improvedlimitneutrinolessdouble}
Agrawal, A.; Alenkov, V.V.; Aryal, P.; Beyer, J.; Bhandari, B.; Boiko, R.S.;
  Boonin, K.; Buzanov, O.; Byeon, C.R.; Chanthima, N.;  et~al.
\newblock Improved limit on neutrinoless double beta decay of $^{100}$Mo from
  AMoRE-I. \emph{ArXiV}  \textbf{2024}, arXiv:nucl-ex/2407.05618.

\bibitem[Akimov et~al.(2020)Akimov, Albert, An, Awe, Barbeau, Becker, Belov,
  Blackston, Blokland, Bolozdynya, Cabrera-Palmer, Chen, Chernyak, Conley,
  Cooper, Daughhetee, del Valle~Coello, Detwiler, Durand, Efremenko, Elliott,
  Fabris, Febbraro, Fox, Galindo-Uribarri, Green, Hansen, Heath, Hedges,
  Hughes, Johnson, Kaemingk, Kaufman, Khromov, Konovalov, Kozlova, Kumpan, Li,
  Librande, Link, Liu, Mann, Markoff, McGoldrick, Moreno, Mueller, Newby,
  Parno, Penttila, Pershey, Radford, Rapp, Ray, Raybern, Razuvaeva, Reyna,
  Rich, Rudik, Runge, Salvat, Scholberg, Shakirov, Simakov, Sinev, Snow,
  Sosnovtsev, Suh, Tayloe, Tellez-Giron-Flores, Thornton, Tolstukhin,
  Vanderwerp, Varner, Virtue, Visser, Wiseman, Wongjirad, Yang, Yen, Yoo, Yu.,
  and Zettlemoyer]{https://doi.org/10.5281/zenodo.3903810}
Akimov, D.; Albert, J.; An, P.; Awe, C.; Barbeau, P.; Becker, B.; Belov, V.;
  Blackston, M.; Blokland, L.; Bolozdynya, A.;  et~al.
\newblock COHERENT Collaboration data release from the first detection of
  coherent elastic neutrino-nucleus scattering on argon. \emph{ArXiV}  \textbf{2020},  arXiv:2006.12659.

\bibitem[O’Hare(2021)]{O_Hare_2021}
O’Hare, C.A.J.
\newblock New Definition of the Neutrino Floor for Direct Dark Matter Searches.
\newblock {\em Phys. Rev. Lett.} {\bf 2021}, {\em 127}, 251802.
\newblock {\url{https://doi.org/10.1103/physrevlett.127.251802}}.

\bibitem[Boyarsky et~al.(2019)Boyarsky, Drewes, Lasserre, Mertens, and
  Ruchayskiy]{Boyarsky_2019}
Boyarsky, A.; Drewes, M.; Lasserre, T.; Mertens, S.; Ruchayskiy, O.
\newblock Sterile neutrino Dark Matter.
\newblock {\em Prog. Part. Nucl. Phys.} {\bf 2019}, {\em
  104},~1–45.
\newblock {\url{https://doi.org/10.1016/j.ppnp.2018.07.004}}.

\bibitem[Alpert et~al.(2015)Alpert, Balata, Bennett, Biasotti, Boragno,
  Brofferio, Ceriale, Corsini, Day, De~Gerone, et~al.]{alpert2015holmes}
Alpert, B.; Balata, M.; Bennett, D.; Biasotti, M.; Boragno, C.; Brofferio, C.;
  Ceriale, V.; Corsini, D.; Day, P.K.; De~Gerone, M.;  et~al.
\newblock HOLMES: The electron capture decay of $^{163}$Ho to measure the
  electron neutrino mass with sub-eV sensitivity.
\newblock {\em  Eur. Phys. J. C} {\bf 2015}, {\em 75},~112.

\bibitem[Borghesi et~al.(2023)Borghesi, Alpert, Balata, Becker, Bennet,
  Celasco, Cerboni, De~Gerone, Dressler, Faverzani,
  et~al.]{borghesi2023updated}
Borghesi, M.; Alpert, B.; Balata, M.; Becker, D.; Bennet, D.; Celasco, E.;
  Cerboni, N.; De~Gerone, M.; Dressler, R.; Faverzani, M.;  et~al.
\newblock An updated overview of the HOLMES status.
\newblock {\em Nucl. Instrum. Methods Phys. Res. Sect. A Accel. Spectrom. Detect. Assoc. Equip.} {\bf 2023},
  {\em 1051},~168205.

\bibitem[Croce et~al.(2016)Croce, Rabin, Mocko, Kunde, Birnbaum, Bond, Engle,
  Hoover, Nortier, Pollington, et~al.]{croce2016development}
Croce, M.P.; Rabin, M.W.; Mocko, V.; Kunde, G.J.; Birnbaum, E.R.; Bond, E.;
  Engle, J.W.; Hoover, A.S.; Nortier, F.M.; Pollington, A.D.;  et~al.
\newblock Development of holmium-163 electron-capture spectroscopy with
  transition-edge sensors.
\newblock {\em J. Low Temp. Phys.} {\bf 2016}, {\em
  184},~958--968.

\bibitem[collaboration(2019)]{group2019cupid}
collaboration, C.
\newblock CUPID pre-CDR.
\newblock {\em arXiv} {\bf 2019}, arXiv:1907.09376.

\bibitem[Alfonso et~al.(2023)Alfonso, Armatol, Augier, Avignone~III, Azzolini,
  Balata, Barabash, Bari, Barresi, Baudin, et~al.]{alfonso2023cupid}
Alfonso, K.; Armatol, A.; Augier, C.; Avignone~III, F.; Azzolini, O.; Balata,
  M.; Barabash, A.; Bari, G.; Barresi, A.; Baudin, D.;  et~al.
\newblock CUPID: The next-generation neutrinoless double beta decay experiment.
\newblock {\em J. Low Temp. Phys.} {\bf 2023}, {\em
  211},~375--383.

\bibitem[Singh et~al.(2023)Singh, Beretta, Hansen, Vetter, Benato, Marini,
  Capelli, Fujikawa, Schmidt, Chang, et~al.]{singh2023large}
Singh, V.; Beretta, M.; Hansen, E.; Vetter, K.; Benato, G.; Marini, L.;
  Capelli, C.; Fujikawa, B.; Schmidt, B.; Chang, C.;  et~al.
\newblock Large-area photon calorimeter with Ir-Pt bilayer transition-edge
  sensor for the CUPID experiment.
\newblock {\em Phys. Rev. Appl.} {\bf 2023}, {\em 20},~064017.

\bibitem[Bratrud et~al.(2024)Bratrud, Chang, Chen, Cudmore, Figueroa-Feliciano,
  Hong, Kennard, Lewis, Lisovenko, Mateo, Novati, Novosad, Oliveri, Ren,
  Scarpaci, Schmidt, Wang, Winslow, Yefremenko, Zhang, Baxter, Hollister,
  James, Lukens, and Temples]{bratrud2024demonstrationtesbasedcryogenic}
Bratrud, G.; Chang, C.L.; Chen, R.; Cudmore, E.; Figueroa-Feliciano, E.; Hong,
  Z.; Kennard, K.T.; Lewis, S.; Lisovenko, M.; Mateo, L.O.;  et~al.
\newblock First demonstration of a TES based cryogenic Li$_2$MoO$_4$ detector
  for neutrinoless double beta decay search. \emph{ArXiV} \textbf{2024}, arXiv:hep-ex/2406.02025.

\bibitem[Pagnanini et~al.(2023)Pagnanini, Benato, Carniti, Celi, Chiesa,
  Corbett, Dafinei, Di~Domizio, Di~Stefano, Ghislandi, Gotti, Helis, Knobel,
  Kostensalo, Kotila, Nagorny, Pessina, Pirro, Pozzi, Puiu, Quitadamo, Sisti,
  Suhonen, and Kuznetsov]{Pagnanini_2023}
Pagnanini, L.; Benato, G.; Carniti, P.; Celi, E.; Chiesa, D.; Corbett, J.;
  Dafinei, I.; Di~Domizio, S.; Di~Stefano, P.; Ghislandi, S.;  et~al.
\newblock Array of cryogenic calorimeters to evaluate the spectral shape of
  forbidden $\beta$-decays: The ACCESS project.
\newblock {\em  Eur. Phys. J. Plus} {\bf 2023}, {\em 138}, 445.
\newblock {\url{https://doi.org/10.1140/epjp/s13360-023-03946-x}}.

\bibitem[Strauss et~al.(2017)Strauss, Rothe, Angloher, Bento, Gütlein, Hauff,
  Kluck, Mancuso, Oberauer, Petricca, Pröbst, Schieck, Schönert, Seidel, and
  Stodolsky]{Strauss_2017}
Strauss, R.; Rothe, J.; Angloher, G.; Bento, A.; Gütlein, A.; Hauff, D.;
  Kluck, H.; Mancuso, M.; Oberauer, L.; Petricca, F.;  et~al.
\newblock The Nucleus experiment: A gram-scale fiducial-volume cryogenic
  detector for the first detection of coherent neutrino–nucleus scattering.
\newblock {\em  Eur. Phys. J. C} {\bf 2017}, {\em 77}, 506.
\newblock {\url{https://doi.org/10.1140/epjc/s10052-017-5068-2}}.

\bibitem[Collaboration et~al.(2019)Collaboration, Angloher, Ardellier-Desages,
  Bento, Canonica, Erhart, Ferreiro, Friedl, Ghete, Hauff,
  et~al.]{Nucleus_Chooz}
Collaboration, N.; Angloher, G.; Ardellier-Desages, F.; Bento, A.; Canonica,
  L.; Erhart, A.; Ferreiro, N.; Friedl, M.; Ghete, V.; Hauff, D.;  et~al.
\newblock Exploring CE$\nu$NS with NUCLEUS at the Chooz nuclear power plant.
\newblock {\em Eur. Phys. J. C} {\bf 2019}, {\em 79},~1018.

\bibitem[Goupy et~al.(2024)Goupy, Marnieros, Mauri, Nones, and
  Vivier]{GOUPY2024169383}
Goupy, C.; Marnieros, S.; Mauri, B.; Nones, C.; Vivier, M.
\newblock Prototyping a High Purity Germanium cryogenic veto system for a
  bolometric detection experiment.
\newblock {\em Nucl. Instrum. Methods Phys. Res. Sect. A Accel. Spectrom. Detect. Assoc. Equip.} {\bf 2024},
  {\em 1064},~169383.
\newblock {\url{https://doi.org/10.1016/j.nima.2024.169383}}.

\bibitem[Agnolet et~al.(2017)Agnolet, Baker, Barker, Beck, Carroll, Cesar,
  Cushman, Dent, De~Rijck, Dutta, et~al.]{agnolet2017background}
Agnolet, G.; Baker, W.; Barker, D.; Beck, R.; Carroll, T.; Cesar, J.; Cushman,
  P.; Dent, J.; De~Rijck, S.; Dutta, B.;  et~al.
\newblock Background studies for the MINER coherent neutrino scattering reactor
  experiment.
\newblock {\em Nucl. Instrum. Methods Phys. Res. Sect. A Accel. Spectrom. Detect. Assoc. Equip.} {\bf 2017},
  {\em 853},~53--60.

\bibitem[Adari et~al.(2022)Adari, Aguilar-Arevalo, Amidei, Angloher, Armengaud,
  Augier, Balogh, Banik, Baxter, Beaufort, et~al.]{adari2022excess}
Adari, P.; Aguilar-Arevalo, A.A.; Amidei, D.; Angloher, G.; Armengaud, E.;
  Augier, C.; Balogh, L.; Banik, S.; Baxter, D.; Beaufort, C.;  et~al.
\newblock EXCESS workshop: Descriptions of rising low-energy spectra.
\newblock {\em SciPost Phys. Proc.} {\bf 2022}, \emph{9}, 001. \url{https://doi.org/10.21468/SciPostPhysProc.9.001}.

\bibitem[Angloher et~al.(2023)Angloher, Banik, Benato, Bento, Bertolini,
  Breier, Bucci, Burkhart, Canonica, D’Addabbo,
  et~al.]{angloher2023observation}
Angloher, G.; Banik, S.; Benato, G.; Bento, A.; Bertolini, A.; Breier, R.;
  Bucci, C.; Burkhart, J.; Canonica, L.; D’Addabbo, A.;  et~al.
\newblock Observation of a low energy nuclear recoil peak in the neutron
  calibration data of the CRESST-III experiment.
\newblock {\em Phys. Rev. D} {\bf 2023}, {\em 108},~022005.

\bibitem[Pattavina et~al.(2020)Pattavina, Ferreiro~Iachellini, and
  Tamborra]{pattavina2020neutrino}
Pattavina, L.; Ferreiro~Iachellini, N.; Tamborra, I.
\newblock Neutrino observatory based on archaeological lead.
\newblock {\em Phys. Rev. D} {\bf 2020}, {\em 102},~063001.

\bibitem[Pattavina et~al.(2021)Pattavina, Iachellini, Pagnanini, Canonica,
  Celi, Clemenza, Ferroni, Fiorini, Garai, Gironi, et~al.]{pattavina2021res}
Pattavina, L.; Iachellini, N.F.; Pagnanini, L.; Canonica, L.; Celi, E.;
  Clemenza, M.; Ferroni, F.; Fiorini, E.; Garai, A.; Gironi, L.;  et~al.
\newblock RES-NOVA sensitivity to core-collapse and failed core-collapse
  supernova neutrinos.
\newblock {\em J. Cosmol. Astropart. Phys.} {\bf 2021}, {\em
  2021},~064.

\bibitem[Augier et~al.(2023)Augier, Beaulieu, Belov, Berge, Billard, Bres,
  Bret, Broniatowski, Calvo, Cazes, et~al.]{augier2023ricochet}
Augier, C.; Beaulieu, G.; Belov, V.; Berge, L.; Billard, J.; Bres, G.; Bret,
  J.L.; Broniatowski, A.; Calvo, M.; Cazes, A.;  et~al.
\newblock Ricochet progress and status.
\newblock {\em J. Low Temp. Phys.} {\bf 2023}, {\em
  212},~127--137.

\bibitem[ric(2024)]{ricochet2024first}
First demonstration of 30 eVee ionization energy resolution with Ricochet
  germanium cryogenic bolometers.
\newblock {\em  Eur. Phys. J. C} {\bf 2024}, {\em 84},~186.

\bibitem[Chen et~al.(2023)Chen, Pinckney, Figueroa-Feliciano, Hong, and
  Schmidt]{Richochet_Main}
Chen, R.; Pinckney, H.D.; Figueroa-Feliciano, E.; Hong, Z.; Schmidt, B.
\newblock Transition Edge Sensor Chip Design of a Modular CE$\nu$NS Detector
  for the Ricochet Experiment.
\newblock {\em J. Low Temp. Phys.} {\bf 2023}, {\em
  211},~237--247.

\bibitem[Augier et~al.(2023)Augier, Baulieu, Belov, Berg{\'e}, Billard, Bres,
  Broniatowski, Calvo, Cazes, Chaize, et~al.]{Ricochet_prototype_detector}
Augier, C.; Baulieu, G.; Belov, V.; Berg{\'e}, L.; Billard, J.; Bres, G.;
  Broniatowski, A.; Calvo, M.; Cazes, A.; Chaize, D.;  et~al.
\newblock Results from a prototype TES detector for the Ricochet experiment.
\newblock {\em Nucl. Instrum. Methods Phys. Res. Sect. A Accel. Spectrom. Detect. Assoc. Equip.} {\bf 2023},
  {\em 1057},~168765.

\bibitem[Bravin et~al.(1999)Bravin, Bruckmayer, Bucci, Cooper, Giordano, {von
  Feilitzsch}, Höhne, Jochum, Jörgens, Keeling, Kraus, Loidl, Lush,
  Macallister, Marchese, Meier, Meunier, Nagel, Nüssle, Pröbst, Ramachers,
  Sarsa, Schnagl, Seidel, Sergeyev, Sisti, Stodolsky, Uchaikin, and
  Zerle]{BRAVIN1999107}
Bravin, M.; Bruckmayer, M.; Bucci, C.; Cooper, S.; Giordano, S.; {von
  Feilitzsch}, F.; Höhne, J.; Jochum, J.; Jörgens, V.; Keeling, R.;  et~al.
\newblock The CRESST dark matter search.
\newblock {\em Astropart. Phys.} {\bf 1999}, {\em 12},~107--114.
\newblock
  {\url{https://doi.org/10.1016/S0927-6505(99)00073-0}}.

\bibitem[Angloher et~al.(2016)Angloher, Bento, Bucci, Canonica, Defay, Erb, von
  Feilitzsch, Iachellini, Gorla, G{\"u}tlein, Hauff, Jochum, Kiefer, Kluck,
  Kraus, Lanfranchi, Loebell, M{\"u}nster, Pagliarone, Petricca, Potzel,
  Pr{\"o}bst, Reindl, Sch{\"a}ffner, Schieck, Sch{\"o}nert, Seidel, Stodolsky,
  Strandhagen, Strauss, Tanzke, Trinh~Thi, T{\"u}rko{\u g}lu, Uffinger, Ulrich,
  Usherov, Wawoczny, Willers, W{\"u}strich, and Z{\"o}ller]{Angloher2016-da}
Angloher, G.; Bento, A.; Bucci, C.; Canonica, L.; Defay, X.; Erb, A.; von
  Feilitzsch, F.; Iachellini, N.F.; Gorla, P.; G{\"u}tlein, A.;  et~al.
\newblock Results on light dark matter particles with a low-threshold
  {CRESST-II} detector.
\newblock {\em Eur. Phys. J. C Part. Fields} {\bf 2016}, {\em 76},~25.

\bibitem[Rothe et~al.(2018)Rothe, Angloher, Bauer, Bento, Bucci, Canonica,
  D'Addabbo, Defay, Erb, Feilitzsch, Ferreiro~Iachellini, Gorla, G{\"u}tlein,
  Hauff, Jochum, Kiefer, Kluck, Kraus, Lanfranchi, Langenk{\"a}mper, Loebell,
  Mancuso, Mondragon, M{\"u}nster, Pagliarone, Petricca, Potzel, Pr{\"o}bst,
  Puig, Reindl, Sch{\"a}ffner, Schieck, Schipperges, Sch{\"o}nert, Seidel,
  Stahlberg, Stodolsky, Strandhagen, Strauss, Tanzke, Trinh~Thi, T{\"u}rko{\u
  g}lu, Ulrich, Usherov, Wawoczny, Willers, and W{\"u}strich]{Rothe2018-sa}
Rothe, J.; Angloher, G.; Bauer, P.; Bento, A.; Bucci, C.; Canonica, L.;
  D'Addabbo, A.; Defay, X.; Erb, A.; Feilitzsch, F.v.;  et~al.
\newblock {TES-based} light detectors for the {CRESST} direct dark matter
  search.
\newblock {\em J. Low Temp. Phys.} {\bf 2018}, {\em 193},~1160--1166.

\bibitem[Kurinsky et~al.(2016)Kurinsky, Brink, Partridge, Cabrera, and
  Pyle]{kurinsky2016supercdms}
Kurinsky, N.; Brink, P.; Partridge, R.; Cabrera, B.; Pyle, M.
\newblock SuperCDMS SNOLAB low-mass detectors: Ultra-sensitive phonon
  calorimeters for a sub-GeV dark matter search.
\newblock {\em arXiv} {\bf 2016}, arXiv:1611.04083.

\bibitem[Bähre et~al.(2013)Bähre, Döbrich, Dreyling-Eschweiler, Ghazaryan,
  Hodajerdi, Horns, Januschek, Knabbe, Lindner, Notz, Ringwald, von Seggern,
  Stromhagen, Trines, and Willke]{RBahre_2013}
Bähre, R.; Döbrich, B.; Dreyling-Eschweiler, J.; Ghazaryan, S.; Hodajerdi,
  R.; Horns, D.; Januschek, F.; Knabbe, E.A.; Lindner, A.; Notz, D.;  et~al.
\newblock Any light particle search II---Technical Design Report.
\newblock {\em J. Instrum.} {\bf 2013}, {\em 8},~T09001.
\newblock {\url{https://doi.org/10.1088/1748-0221/8/09/T09001}}.

\bibitem[Sikivie(1983)]{Sikivie}
Sikivie, P.
\newblock Experimental Tests of the "Invisible" Axion.
\newblock {\em Phys. Rev. Lett.} {\bf 1983}, {\em 51},~1415--1417.
\newblock {\url{https://doi.org/10.1103/PhysRevLett.51.1415}}.

\bibitem[Rubiera~Gimeno et~al.(2023)Rubiera~Gimeno, Januschek, Isleif, Lindner,
  Meyer, Othman, Schwemmbauer, and Shah]{RubieraGimeno:2023J9}
Rubiera~Gimeno, J.A.; Januschek, F.; Isleif, K.S.; Lindner, A.; Meyer, M.;
  Othman, G.; Schwemmbauer, C.; Shah, R.
\newblock {A TES system for ALPS II - Status and Prospects}.
\newblock {\em Proc. Sci.} {\bf 2023}, {\em 449},~567.
\newblock {\url{https://doi.org/10.22323/1.449.0567}}.

\bibitem[Armengaud et~al.(2017)Armengaud, Arnaud, Augier, Beno{\^\i}t,
  Berg{\'e}, Bergmann, Billard, De~Boissi{\`e}re, Bres, Broniatowski,
  et~al.]{armengaud2017performance}
Armengaud, E.; Arnaud, Q.; Augier, C.; Beno{\^\i}t, A.; Berg{\'e}, L.;
  Bergmann, T.; Billard, J.; De~Boissi{\`e}re, T.; Bres, G.; Broniatowski, A.;
  et~al.
\newblock Performance of the EDELWEISS-III experiment for direct dark matter
  searches.
\newblock {\em J. Instrum.} {\bf 2017}, {\em 12},~P08010.

\bibitem[Marnieros et~al.(2023)Marnieros, Armengaud, Arnaud, Augier,
  Beno{\^\i}t, Berg{\'e}, Billard, Broniatowski, Camus, Cazes,
  et~al.]{marnieros2023high}
Marnieros, S.; Armengaud, E.; Arnaud, Q.; Augier, C.; Beno{\^\i}t, A.;
  Berg{\'e}, L.; Billard, J.; Broniatowski, A.; Camus, P.; Cazes, A.;  et~al.
\newblock High impedance TES bolometers for EDELWEISS.
\newblock {\em J. Low Temp. Phys.} {\bf 2023}, {\em
  211},~214--219.

\bibitem[Lattaud et~al.(2024)Lattaud, Guy, Billard, Colas, J{\'e}sus, Gascon,
  Juillard, Marnieros, and Oriol]{lattaud2024characterization}
Lattaud, H.; Guy, E.; Billard, J.; Colas, J.; J{\'e}sus, M.D.; Gascon, J.;
  Juillard, A.; Marnieros, S.; Oriol, C.
\newblock Characterization of the Phonon Sensor of the CRYOSEL Detector with IR
  Photons.
\newblock {\em J. Low Temp. Phys.} {\bf 2024}, {\em
  215},~268--275.

\bibitem[Capparelli et~al.(2016)Capparelli, Cavoto, Ferretti, Giazotto, Polosa,
  and Spagnolo]{CAPPARELLI201637}
Capparelli, L.; Cavoto, G.; Ferretti, J.; Giazotto, F.; Polosa, A.; Spagnolo,
  P.
\newblock Axion-like particle searches with sub-THz photons.
\newblock {\em Phys. Dark Universe} {\bf 2016}, {\em 12},~37--44.
\newblock {\url{https://doi.org/10.1016/j.dark.2016.01.003}}.

\bibitem[He et~al.(2016)He, Zhang, Ren, and Sun]{Kat7780459}
He, K.; Zhang, X.; Ren, S.; Sun, J.
\newblock Deep Residual Learning for Image Recognition.
\newblock In Proccedings of the 2016 IEEE Conference on Computer Vision and
  Pattern Recognition (CVPR),  Las Vegas, NV, USA, 27--30 June 2016; pp. 770--778.
\newblock {\url{https://doi.org/10.1109/CVPR.2016.90}}.

\bibitem[Bernabei et~al.(2020)Bernabei, Belli, Bussolotti, Cappella,
  Caracciolo, Cerulli, Dai, d’Angelo, Di~Marco, Ferrari,
  et~al.]{bernabei2020dama}
Bernabei, R.; Belli, P.; Bussolotti, A.; Cappella, F.; Caracciolo, V.; Cerulli,
  R.; Dai, C.; d’Angelo, A.; Di~Marco, A.; Ferrari, N.;  et~al.
\newblock The DAMA project: Achievements, implications and perspectives.
\newblock {\em Prog. Part. Nucl. Phys.} {\bf 2020}, {\em
  114},~103810.

\bibitem[Krishak et~al.(2020)Krishak, Dantuluri, and Desai]{krishak2020robust}
Krishak, A.; Dantuluri, A.; Desai, S.
\newblock Robust model comparison tests of DAMA/LIBRA annual modulation.
\newblock {\em J. Cosmol. Astropart. Phys.} {\bf 2020}, {\em
  2020},~007.

\bibitem[Angloher et~al.(2016)Angloher, Carniti, Cassina, Gironi, Gotti,
  G{\"u}tlein, Hauff, Maino, Nagorny, Pagnanini, et~al.]{angloher2016cosinus}
Angloher, G.; Carniti, P.; Cassina, L.; Gironi, L.; Gotti, C.; G{\"u}tlein, A.;
  Hauff, D.; Maino, M.; Nagorny, S.; Pagnanini, L.;  et~al.
\newblock The COSINUS project: Perspectives of a NaI scintillating calorimeter
  for dark matter search.
\newblock {\em  Eur. Phys. J. C} {\bf 2016}, {\em 76},~441.

\bibitem[Angloher et~al.(2024)Angloher, Bharadwaj, Dafinei, Di~Marco, Einfalt,
  Ferroni, Fichtinger, Filipponi, Frank, Friedl, et~al.]{angloher2024deep}
Angloher, G.; Bharadwaj, M.; Dafinei, I.; Di~Marco, N.; Einfalt, L.; Ferroni,
  F.; Fichtinger, S.; Filipponi, A.; Frank, T.; Friedl, M.;  et~al.
\newblock Deep-underground dark matter search with a COSINUS detector
  prototype.
\newblock {\em Phys. Rev. D} {\bf 2024}, {\em 110},~043010.

\bibitem[Abeln et~al.(2021)Abeln, Altenm{\"u}ller, Arguedas~Cuendis, Armengaud,
  Atti{\'e}, Aune, Basso, Berg{\'e}, Biasuzzi, Borges De~Sousa,
  et~al.]{abeln2021conceptual}
Abeln, A.; Altenm{\"u}ller, K.; Arguedas~Cuendis, S.; Armengaud, E.; Atti{\'e},
  D.; Aune, S.; Basso, S.; Berg{\'e}, L.; Biasuzzi, B.; Borges De~Sousa, P.;
  et~al.
\newblock Conceptual design of BabyIAXO, the intermediate stage towards the
  International Axion Observatory.
\newblock {\em J. High Energy Phys.} {\bf 2021}, {\em 2021},~137.

\bibitem[Billard et~al.(2024)Billard, Gascon, Marnieros, and
  Scorza]{BILLARD2024116465}
Billard, J.; Gascon, J.; Marnieros, S.; Scorza, S.
\newblock Transition Edge Sensors with Sub-eV Resolution and Cryogenic Targets
  (TESSERACT) at the underground laboratory of Modane (LSM).
\newblock {\em Nucl. Phys. B} {\bf 2024}, {\em 1003},~116465.
 {\url{https://doi.org/10.1016/j.nuclphysb.2024.116465}}.

\bibitem[Hertel et~al.(2019)Hertel, Biekert, Lin, Velan, and McKinsey]{herald}
Hertel, S.A.; Biekert, A.; Lin, J.; Velan, V.; McKinsey, D.N.
\newblock Direct detection of sub-GeV dark matter using a superfluid
  $^{4}\mathrm{He}$ target.
\newblock {\em Phys. Rev. D} {\bf 2019}, {\em 100},~092007.
\newblock {\url{https://doi.org/10.1103/PhysRevD.100.092007}}.

\bibitem[Anthony-Petersen et~al.(2022)Anthony-Petersen, Biekert, Bunker, Chang,
  Chang, Chaplinsky, Fascione, Fink, Garcia-Sciveres, Germond, Guo, Hertel,
  Hong, Kurinsky, Li, Lin, Lisovenko, Mahapatra, Mayer, McKinsey, Mehrotra,
  Mirabolfathi, Neblosky, Page, Patel, Penning, Pinckney, Platt, Pyle, Reed,
  Romani, Queiroz, Sadoulet, Serfass, Smith, Sorensen, Suerfu, Suzuki,
  Underwood, Velan, Wang, Wang, Watkins, Williams, Yefremenko, and
  Zhang]{SPICE}
Anthony-Petersen, R.; Biekert, A.; Bunker, R.; Chang, C.L.; Chang, Y.Y.;
  Chaplinsky, L.; Fascione, E.; Fink, C.W.; Garcia-Sciveres, M.; Germond, R.;
  et~al.
\newblock A stress induced source of phonon bursts and quasiparticle poisoning.
\newblock {\em arXiv} {\bf 2022}, ArXiV:2208.02790.

\bibitem[Icha et~al.(2017)Icha, Weber, Waters, and
  Norden]{icha2017phototoxicity}
Icha, J.; Weber, M.; Waters, J.C.; Norden, C.
\newblock Phototoxicity in live fluorescence microscopy, and how to avoid it.
\newblock {\em BioEssays} {\bf 2017}, {\em 39},~1700003.

\bibitem[Niwa et~al.(2017)Niwa, Numata, Hattori, and Fukuda]{niwa2017few}
Niwa, K.; Numata, T.; Hattori, K.; Fukuda, D.
\newblock Few-photon color imaging using energy-dispersive superconducting
  transition-edge sensor spectrometry.
\newblock {\em Sci. Rep.} {\bf 2017}, {\em 7},~45660.

\bibitem[Fukuda et~al.(2011)Fukuda, Fujii, Numata, Amemiya, Yoshizawa,
  Tsuchida, Fujino, Ishii, Itatani, Inoue, et~al.]{fukuda2011titanium}
Fukuda, D.; Fujii, G.; Numata, T.; Amemiya, K.; Yoshizawa, A.; Tsuchida, H.;
  Fujino, H.; Ishii, H.; Itatani, T.; Inoue, S.;  et~al.
\newblock Titanium-based transition-edge photon number resolving detector with
  98\% detection efficiency with index-matched small-gap fiber coupling.
\newblock {\em Opt. Express} {\bf 2011}, {\em 19},~870--875.

\bibitem[Niwa et~al.(2021)Niwa, Hattori, and Fukuda]{niwa2021few}
Niwa, K.; Hattori, K.; Fukuda, D.
\newblock Few-photon spectral confocal microscopy for cell imaging using
  superconducting Transition Edge Sensor.
\newblock {\em Front. Bioeng. Biotechnol.} {\bf 2021}, {\em
  9},~789709.

\bibitem[Hattori et~al.(2020)Hattori, Kobayashi, Takasu, and
  Fukuda]{Hattori2020complex}
Hattori, K.; Kobayashi, R.; Takasu, S.; Fukuda, D.
\newblock {Complex impedance of a transition-edge sensor with sub-$\mu$s time
  constant}.
\newblock {\em AIP Adv.} {\bf 2020}, {\em 10},~035004.
\newblock {\url{https://doi.org/10.1063/1.5127100}}.

\bibitem[Fukuda(2019)]{fukuda2019single}
Fukuda, D.
\newblock Single-photon measurement techniques with a superconducting
  Transition Edge Sensor.
\newblock {\em IEICE Trans. Electron.} {\bf 2019}, {\em
  102},~230--234.

\bibitem[Hao et~al.(2003)Hao, Gallop, Gardiner, Josephs-Franks, Macfarlane,
  Lam, and Foley]{hao2003inductive}
Hao, L.; Gallop, J.; Gardiner, C.; Josephs-Franks, P.; Macfarlane, J.; Lam, S.;
  Foley, C.
\newblock Inductive superconducting transition-edge detector for single-photon
  and macro-molecule detection.
\newblock {\em Supercond. Sci. Technol.} {\bf 2003}, {\em
  16},~1479.

\bibitem[Hong et~al.(1987)Hong, Ou, and Mandel]{hominterference}
Hong, C.K.; Ou, Z.Y.; Mandel, L.
\newblock Measurement of subpicosecond time intervals between two photons by
  interference.
\newblock {\em Phys. Rev. Lett.} {\bf 1987}, {\em 59},~2044--2046.
\newblock {\url{https://doi.org/10.1103/PhysRevLett.59.2044}}.

\bibitem[Di~Giuseppe et~al.(2003)Di~Giuseppe, Atat{\"u}re, Shaw, Sergienko,
  Saleh, Teich, Miller, Nam, and Martinis]{di2003direct}
Di~Giuseppe, G.; Atat{\"u}re, M.; Shaw, M.D.; Sergienko, A.V.; Saleh, B.E.;
  Teich, M.C.; Miller, A.J.; Nam, S.W.; Martinis, J.
\newblock Direct observation of photon pairs at a single output port of a
  beam-splitter interferometer.
\newblock {\em Phys. Rev. A} {\bf 2003}, {\em 68},~063817.

\bibitem[Knill et~al.(2001)Knill, Laflamme, and Milburn]{knill2001scheme}
Knill, E.; Laflamme, R.; Milburn, G.J.
\newblock A scheme for efficient quantum computation with linear optics.
\newblock {\em Nature} {\bf 2001}, {\em 409},~46--52.

\bibitem[Gisin and Thew(2007)]{gisin2007quantum}
Gisin, N.; Thew, R.
\newblock Quantum communication.
\newblock {\em Nat. Photonics} {\bf 2007}, {\em 1},~165--171.

\bibitem[Chunnilall et~al.(2014)Chunnilall, Degiovanni, K{\"u}ck, M{\"u}ller,
  and Sinclair]{chunnilall2014metrology}
Chunnilall, C.J.; Degiovanni, I.P.; K{\"u}ck, S.; M{\"u}ller, I.; Sinclair,
  A.G.
\newblock Metrology of single-photon sources and detectors: A review.
\newblock {\em Opt. Eng.} {\bf 2014}, {\em 53},~081910.

\bibitem[Schmidt et~al.(2018)Schmidt, Von~Helversen, L{\'o}pez, Gericke,
  Schlottmann, Heindel, K{\"u}ck, Reitzenstein, and Beyer]{schmidt2018photon}
Schmidt, M.; Von~Helversen, M.; L{\'o}pez, M.; Gericke, F.; Schlottmann, E.;
  Heindel, T.; K{\"u}ck, S.; Reitzenstein, S.; Beyer, J.
\newblock Photon-number-resolving transition-edge sensors for the metrology of
  quantum light sources.
\newblock {\em J. Low Temp. Phys.} {\bf 2018}, {\em
  193},~1243--1250.

\bibitem[Gerrits et~al.(2016)Gerrits, Lita, Calkins, and
  Nam]{gerrits2016superconducting}
Gerrits, T.; Lita, A.; Calkins, B.; Nam, S.W.
\newblock Superconducting Transition Edge Sensors for quantum optics.
\newblock In {\em Superconducting Devices in Quantum Optics}; Springer: Berlin/Heidelberg, Germany, {2016}; \mbox{pp.
  31--60.}

\bibitem[Gerrits et~al.(2010)Gerrits, Glancy, Clement, Calkins, Lita, Miller,
  Migdall, Nam, Mirin, and Knill]{gerrits2010generation}
Gerrits, T.; Glancy, S.; Clement, T.S.; Calkins, B.; Lita, A.E.; Miller, A.J.;
  Migdall, A.L.; Nam, S.W.; Mirin, R.P.; Knill, E.
\newblock Generation of optical coherent-state superpositions by
  number-resolved photon subtraction from the squeezed vacuum.
\newblock {\em Phys. Rev. A---At. Mol. Opt. Phys.} {\bf
  2010}, {\em 82},~031802.

\bibitem[Bartley et~al.(2012)Bartley, Donati, Spring, Jin, Barbieri, Datta,
  Smith, and Walmsley]{bartley2012multiphoton}
Bartley, T.J.; Donati, G.; Spring, J.B.; Jin, X.M.; Barbieri, M.; Datta, A.;
  Smith, B.J.; Walmsley, I.A.
\newblock Multiphoton state engineering by heralded interference between single
  photons and coherent states.
\newblock {\em Phys. Rev. A---At. Mol. Opt. Phys.} {\bf
  2012}, {\em 86},~043820.

\bibitem[Sridhar et~al.(2014)Sridhar, Shahrokhshahi, Miller, Calkins, Gerrits,
  Lita, Nam, and Pfister]{sridhar2014direct}
Sridhar, N.; Shahrokhshahi, R.; Miller, A.J.; Calkins, B.; Gerrits, T.; Lita,
  A.; Nam, S.W.; Pfister, O.
\newblock Direct measurement of the Wigner function by photon-number-resolving
  detection.
\newblock {\em J. Opt. Soc. Am. B} {\bf 2014}, {\em 31},~B34--B40.

\bibitem[Laiho et~al.(2010)Laiho, Cassemiro, Gross, and
  Silberhorn]{laiho2010probing}
Laiho, K.; Cassemiro, K.N.; Gross, D.; Silberhorn, C.
\newblock Probing the negative Wigner function of a pulsed single photon point
  by point.
\newblock {\em Phys. Rev. Lett.} {\bf 2010}, {\em 105},~253603.

\bibitem[Zhai et~al.(2013)Zhai, Becerra, Glebov, Wen, Lita, Calkins, Gerrits,
  Fan, Nam, and Migdall]{zhai2013photon}
Zhai, Y.; Becerra, F.E.; Glebov, B.L.; Wen, J.; Lita, A.E.; Calkins, B.;
  Gerrits, T.; Fan, J.; Nam, S.W.; Migdall, A.
\newblock Photon-number-resolved detection of photon-subtracted thermal light.
\newblock {\em Opt. Lett.} {\bf 2013}, {\em 38},~2171--2173.

\bibitem[Clauser and Horne(1974)]{clauser1974experimental}
Clauser, J.F.; Horne, M.A.
\newblock Experimental consequences of objective local theories.
\newblock {\em Phys. Rev. D} {\bf 1974}, {\em 10},~526.

\bibitem[Christensen et~al.(2013)Christensen, McCusker, Altepeter, Calkins,
  Gerrits, Lita, Miller, Shalm, Zhang, Nam, et~al.]{christensen2013detection}
Christensen, B.G.; McCusker, K.T.; Altepeter, J.B.; Calkins, B.; Gerrits, T.;
  Lita, A.E.; Miller, A.; Shalm, L.K.; Zhang, Y.; Nam, S.W.;  et~al.
\newblock Detection-loophole-free test of quantum nonlocality, and
  applications.
\newblock {\em Phys. Rev. Lett.} {\bf 2013}, {\em 111},~130406.

\bibitem[Giustina et~al.(2013)Giustina, Mech, Ramelow, Wittmann, Kofler, Beyer,
  Lita, Calkins, Gerrits, Nam, et~al.]{giustina2013bell}
Giustina, M.; Mech, A.; Ramelow, S.; Wittmann, B.; Kofler, J.; Beyer, J.; Lita,
  A.; Calkins, B.; Gerrits, T.; Nam, S.W.;  et~al.
\newblock Bell violation using entangled photons without the fair-sampling
  assumption.
\newblock {\em Nature} {\bf 2013}, {\em 497},~227--230.

\bibitem[Heindel et~al.(2017)Heindel, Thoma, von Helversen, Schmidt, Schlehahn,
  Gschrey, Schnauber, Schulze, Strittmatter, Beyer, et~al.]{heindel2017bright}
Heindel, T.; Thoma, A.; von Helversen, M.; Schmidt, M.; Schlehahn, A.; Gschrey,
  M.; Schnauber, P.; Schulze, J.H.; Strittmatter, A.; Beyer, J.;  et~al.
\newblock A bright triggered twin-photon source in the solid state.
\newblock {\em Nat. Commun.} {\bf 2017}, {\em 8},~14870.

\bibitem[Schlottmann et~al.(2018)Schlottmann, von Helversen, Leymann, Lettau,
  Kr{\"u}ger, Schmidt, Schneider, Kamp, H{\"o}fling, Beyer,
  et~al.]{schlottmann2018exploring}
Schlottmann, E.; von Helversen, M.; Leymann, H.A.; Lettau, T.; Kr{\"u}ger, F.;
  Schmidt, M.; Schneider, C.; Kamp, M.; H{\"o}fling, S.; Beyer, J.;  et~al.
\newblock Exploring the photon-number distribution of bimodal microlasers with
  a Transition Edge Sensor.
\newblock {\em Phys. Rev. Appl.} {\bf 2018}, {\em 9},~064030.

\bibitem[Klaas et~al.(2018)Klaas, Schlottmann, Flayac, Laussy, Gericke,
  Schmidt, Helversen, Beyer, Brodbeck, Suchomel, et~al.]{klaas2018photon}
Klaas, M.; Schlottmann, E.; Flayac, H.; Laussy, F.; Gericke, F.; Schmidt, M.;
  Helversen, M.v.; Beyer, J.; Brodbeck, S.; Suchomel, H.;  et~al.
\newblock Photon-number-resolved measurement of an exciton-polariton
  condensate.
\newblock {\em Phys. Rev. Lett.} {\bf 2018}, {\em 121},~047401.

\bibitem[Wilen et~al.(2021)Wilen, Abdullah, Kurinsky, Stanford, Cardani,
  d’Imperio, Tomei, Faoro, Ioffe, Liu, et~al.]{wilen2021correlated}
Wilen, C.D.; Abdullah, S.; Kurinsky, N.; Stanford, C.; Cardani, L.;
  d’Imperio, G.; Tomei, C.; Faoro, L.; Ioffe, L.; Liu, C.;  et~al.
\newblock Correlated charge noise and relaxation errors in superconducting
  qubits.
\newblock {\em Nature} {\bf 2021}, {\em 594},~369--373.

\bibitem[McEwen et~al.(2022)McEwen, Faoro, Arya, Dunsworth, Huang, Kim,
  Burkett, Fowler, Arute, Bardin, et~al.]{mcewen2022resolving}
McEwen, M.; Faoro, L.; Arya, K.; Dunsworth, A.; Huang, T.; Kim, S.; Burkett,
  B.; Fowler, A.; Arute, F.; Bardin, J.C.;  et~al.
\newblock Resolving catastrophic error bursts from cosmic rays in large arrays
  of superconducting qubits.
\newblock {\em Nat. Phys.} {\bf 2022}, {\em 18},~107--111.

\bibitem[Cardani et~al.(2021)Cardani, Valenti, Casali, Catelani, Charpentier,
  Clemenza, Colantoni, Cruciani, D’Imperio, Gironi,
  et~al.]{cardani2021reducing}
Cardani, L.; Valenti, F.; Casali, N.; Catelani, G.; Charpentier, T.; Clemenza,
  M.; Colantoni, I.; Cruciani, A.; D’Imperio, G.; Gironi, L.;  et~al.
\newblock Reducing the impact of radioactivity on quantum circuits in a
  deep-underground facility.
\newblock {\em Nat. Commun.} {\bf 2021}, {\em 12},~2733.

\bibitem[Orrell and Loer(2021)]{orrell2021sensor}
Orrell, J.L.; Loer, B.
\newblock Sensor-assisted fault mitigation in quantum computation.
\newblock {\em Phys. Rev. Appl.} {\bf 2021}, {\em 16},~024025.

\bibitem[Maehata et~al.(2016)Maehata, Hara, Mitsuda, Hidaka, Tanaka, and
  Yamanaka]{maehata2016transition}
Maehata, K.; Hara, T.; Mitsuda, K.; Hidaka, M.; Tanaka, K.; Yamanaka, Y.
\newblock A Transition Edge Sensor microcalorimeter system for the energy
  dispersive spectroscopy performed on a scanning-transmission electron
  microscope.
\newblock {\em J. Low Temp. Phys.} {\bf 2016}, {\em
  184},~5--10.

\bibitem[Lee et~al.(2019)Lee, Titus, Alonso~Mori, Baker, Bennett, Cho, Doriese,
  Fowler, Gaffney, Gallo, et~al.]{lee2019soft}
Lee, S.J.; Titus, C.J.; Alonso~Mori, R.; Baker, M.L.; Bennett, D.A.; Cho, H.M.;
  Doriese, W.B.; Fowler, J.W.; Gaffney, K.J.; Gallo, A.;  et~al.
\newblock Soft X-ray spectroscopy with transition-edge sensors at Stanford
  Synchrotron Radiation Lightsource beamline 10-1.
\newblock {\em Rev. Sci. Instrum.} {\bf 2019}, {\em 90}, 113101.

\bibitem[Ullom et~al.(2014)Ullom, Doriese, Fischer, Fowler, Hilton, Jaye,
  Reintsema, Swetz, and Schmidt]{ullom2014transition}
Ullom, J.; Doriese, W.; Fischer, D.; Fowler, J.; Hilton, G.; Jaye, C.;
  Reintsema, C.; Swetz, D.; Schmidt, D.
\newblock Transition-edge sensor microcalorimeters for X-ray beamline science.
\newblock {\em Synchrotron Radiat. News} {\bf 2014}, {\em 27},~24--27.

\bibitem[Doriese et~al.(2017)Doriese, Abbamonte, Alpert, Bennett, Denison,
  Fang, Fischer, Fitzgerald, Fowler, Gard, et~al.]{doriese2017practical}
Doriese, W.B.; Abbamonte, P.; Alpert, B.K.; Bennett, D.; Denison, E.; Fang, Y.;
  Fischer, D.; Fitzgerald, C.; Fowler, J.; Gard, J.;  et~al.
\newblock A practical superconducting-microcalorimeter X-ray spectrometer for
  beamline and laboratory science.
\newblock {\em Rev. Sci. Instrum.} {\bf 2017}, {\em 88}, 053108.

\bibitem[Miaja-Avila et~al.(2015)Miaja-Avila, O'Neil, Uhlig, Cromer, Dowell,
  Jimenez, Hoover, Silverman, and Ullom]{miaja2015laser}
Miaja-Avila, L.; O'Neil, G.C.; Uhlig, J.; Cromer, C.L.; Dowell, M.L.; Jimenez,
  R.; Hoover, A.; Silverman, K.L.; Ullom, J.N.
\newblock Laser plasma X-ray source for ultrafast time-resolved X-ray
  absorption spectroscopy.
\newblock {\em Struct. Dyn.} {\bf 2015}, {\em 2}, 024301.

\bibitem[Joe et~al.(2015)Joe, O’Neil, Miaja-Avila, Fowler, Jimenez,
  Silverman, Swetz, and Ullom]{joe2015observation}
Joe, Y.I.; O’Neil, G.C.; Miaja-Avila, L.; Fowler, J.W.; Jimenez, R.;
  Silverman, K.; Swetz, D.; Ullom, J.
\newblock Observation of iron spin-states using tabletop X-ray emission
  spectroscopy and microcalorimeter sensors.
\newblock {\em J. Phys. B At. Mol. Opt. Phys.}
  {\bf 2015}, {\em 49},~024003.

\bibitem[Uhlig et~al.(2015)Uhlig, Doriese, Fowler, Swetz, Jaye, Fischer,
  Reintsema, Bennett, Vale, Mandal, et~al.]{uhlig2015high}
Uhlig, J.; Doriese, W.; Fowler, J.; Swetz, D.; Jaye, C.; Fischer, D.;
  Reintsema, C.; Bennett, D.; Vale, L.; Mandal, U.;  et~al.
\newblock High-resolution X-ray emission spectroscopy with transition-edge
  sensors: Present performance and future potential.
\newblock {\em J. Synchrotron Radiat.} {\bf 2015}, {\em
  22},~766--775.

\bibitem[Palosaari et~al.(2016)Palosaari, K{\"a}yhk{\"o}, Kinnunen, Laitinen,
  Julin, Malm, Sajavaara, Doriese, Fowler, Reintsema,
  et~al.]{palosaari2016broadband}
Palosaari, M.; K{\"a}yhk{\"o}, M.; Kinnunen, K.; Laitinen, M.; Julin, J.; Malm,
  J.; Sajavaara, T.; Doriese, W.; Fowler, J.; Reintsema, C.;  et~al.
\newblock Broadband ultrahigh-resolution spectroscopy of particle-induced x
  rays: Extending the limits of nondestructive analysis.
\newblock {\em Phys. Rev. Appl.} {\bf 2016}, {\em 6},~024002.

\bibitem[Yamada et~al.(2021)Yamada, Ichinohe, Tatsuno, Hayakawa, Suda, Ohashi,
  Ishisaki, Uruga, Sekizawa, Nitta, et~al.]{yamada2021broadband}
Yamada, S.; Ichinohe, Y.; Tatsuno, H.; Hayakawa, R.; Suda, H.; Ohashi, T.;
  Ishisaki, Y.; Uruga, T.; Sekizawa, O.; Nitta, K.;  et~al.
\newblock Broadband high-energy resolution hard X-ray spectroscopy using
  Transition Edge Sensors at SPring-8.
\newblock {\em Rev. Sci. Instrum.} {\bf 2021}, {\em 92}, 013103.

\bibitem[Morgan et~al.(2019)Morgan, Becker, Bennett, Doriese, Gard, Irwin, Lee,
  Li, Mates, Pappas, et~al.]{morgan2019use}
Morgan, K.M.; Becker, D.T.; Bennett, D.A.; Doriese, W.B.; Gard, J.D.; Irwin,
  K.D.; Lee, S.J.; Li, D.; Mates, J.A.; Pappas, C.G.;  et~al.
\newblock Use of transition models to design high performance TESs for the
  LCLS-II soft X-ray spectrometer.
\newblock {\em IEEE Trans. Appl. Supercond.} {\bf 2019}, {\em
  29},~2100605.

\bibitem[Okada et~al.(2014)Okada, Bennett, Doriese, Fowler, Irwin, Ishimoto,
  Sato, Schmidt, Swetz, Tatsuno, et~al.]{okada2014high}
Okada, S.; Bennett, D.; Doriese, W.; Fowler, J.; Irwin, K.; Ishimoto, S.; Sato,
  M.; Schmidt, D.; Swetz, D.; Tatsuno, H.;  et~al.
\newblock High-resolution kaonic-atom X-ray spectroscopy with
  transition-edge-sensor microcalorimeters.
\newblock {\em J. Low Temp. Phys.} {\bf 2014}, {\em
  176},~1015--1021.

\bibitem[Tatsuno et~al.(2020)Tatsuno, Bennett, Doriese, Durkin, Fowler, Gard,
  Hashimoto, Hayakawa, Hayashi, Hilton, et~al.]{tatsuno2020mitigating}
Tatsuno, H.; Bennett, D.; Doriese, W.; Durkin, M.; Fowler, J.; Gard, J.;
  Hashimoto, T.; Hayakawa, R.; Hayashi, T.; Hilton, G.;  et~al.
\newblock Mitigating the effects of charged particle strikes on TES arrays for
  exotic atom X-ray experiments.
\newblock {\em J. Low Temp. Phys.} {\bf 2020}, {\em
  200},~247--254.

\bibitem[collaboration et~al.(2016)collaboration, Okada, Bennett, Curceanu,
  Doriese, Fowler, Gard, Gustafsson, Hashimoto, Hayano,
  et~al.]{heates2016first}
collaboration, H.; Okada, S.; Bennett, D.; Curceanu, C.; Doriese, W.; Fowler,
  J.; Gard, J.; Gustafsson, F.; Hashimoto, T.; Hayano, R.;  et~al.
\newblock First application of superconducting transition-edge sensor
  microcalorimeters to hadronic atom X-ray spectroscopy.
\newblock {\em Prog. Theor. Exp. Phys.} {\bf 2016},
  {\em 2016},~091D01.

\bibitem[Seki and Masutani(1983)]{seki1983unified}
Seki, R.; Masutani, K.
\newblock Unified analysis of pionic atoms and low-energy pion-nucleus
  scattering: Phenomenological analysis.
\newblock {\em Phys. Rev. C} {\bf 1983}, {\em 27},~2799.

\bibitem[Hashimoto et~al.(2016)Hashimoto, Bazzi, Bennett, Berucci, Bosnar,
  Curceanu, Doriese, Fowler, Fujioka, Guaraldo, et~al.]{hashimoto2016beamline}
Hashimoto, T.; Bazzi, M.; Bennett, D.; Berucci, C.; Bosnar, D.; Curceanu, C.;
  Doriese, W.; Fowler, J.; Fujioka, H.; Guaraldo, C.;  et~al.
\newblock Beamline test of a transition-edge-sensor spectrometer in preparation
  for kaonic-atom measurements.
\newblock {\em IEEE Trans. Appl. Supercond.} {\bf 2016}, {\em
  27},~2100905.

\bibitem[Hashimoto et~al.(2022)Hashimoto, Aikawa, Akaishi, Asano, Bazzi,
  Bennett, Berger, Bosnar, Butt, Curceanu, et~al.]{hashimoto2022measurements}
Hashimoto, T.; Aikawa, S.; Akaishi, T.; Asano, H.; Bazzi, M.; Bennett, D.;
  Berger, M.; Bosnar, D.; Butt, A.; Curceanu, C.;  et~al.
\newblock Measurements of strong-interaction effects in kaonic-helium isotopes
  at sub-eV precision with X-ray microcalorimeters.
\newblock {\em Phys. Rev. Lett.} {\bf 2022}, {\em 128},~112503.

\bibitem[Aoki et~al.(2021)Aoki, Fujioka, Gogami, Hidaka, Hiyama, Honda, Hosaka,
  Ichikawa, Ieiri, Isaka, et~al.]{aoki2021extension}
Aoki, K.; Fujioka, H.; Gogami, T.; Hidaka, Y.; Hiyama, E.; Honda, R.; Hosaka,
  A.; Ichikawa, Y.; Ieiri, M.; Isaka, M.;  et~al.
\newblock Extension of the J-PARC hadron experimental facility: Third white
  paper.
\newblock {\em arXiv} {\bf 2021}, arXiv:2110.04462.

\bibitem[Pappas et~al.(2019)Pappas, Durkin, Fowler, Morgan, Ullom, Doriese,
  Hilton, O’Neil, Schmidt, Szypryt, et~al.]{pappas2019tes}
Pappas, C.G.; Durkin, M.; Fowler, J.W.; Morgan, K.M.; Ullom, J.N.; Doriese,
  W.B.; Hilton, G.C.; O’Neil, G.C.; Schmidt, D.R.; Szypryt, P.;  et~al.
\newblock A TES X-ray spectrometer for NSENSE.
\newblock {\em Proc. GOMACTech.} {\bf 2019}, 300--305. %MDPI: Please add the volume.


\bibitem[Szypryt et~al.(2023)Szypryt, Nakamura, Becker, Bennett, Dagel,
  Doriese, Fowler, Gard, Harris, Hilton, et~al.]{szypryt2023tabletop}
Szypryt, P.; Nakamura, N.; Becker, D.T.; Bennett, D.A.; Dagel, A.L.; Doriese,
  W.B.; Fowler, J.W.; Gard, J.D.; Harris, J.Z.; Hilton, G.C.;  et~al.
\newblock A tabletop X-ray tomography instrument for nanometer-scale imaging:
  demonstration of the 1000-element transition-edge sensor subarray.
\newblock {\em IEEE Trans. Appl. Supercond.} {\bf 2023}, {\em
  33},~2100705.

\bibitem[Nakamura et~al.(2024)Nakamura, Szypryt, Dagel, Alpert, Bennett,
  Doriese, Durkin, Fowler, Fox, Gard, et~al.]{nakamura2024nanoscale}
Nakamura, N.; Szypryt, P.; Dagel, A.L.; Alpert, B.K.; Bennett, D.A.; Doriese,
  W.B.; Durkin, M.; Fowler, J.W.; Fox, D.T.; Gard, J.D.;  et~al.
\newblock Nanoscale Three-Dimensional Imaging of Integrated Circuits Using a
  Scanning Electron Microscope and Transition-Edge Sensor Spectrometer.
\newblock {\em Sensors} {\bf 2024}, {\em 24},~2890.

\bibitem[Levine et~al.(2023)Levine, Alpert, Dagel, Fowler, Jimenez, Nakamura,
  Swetz, Szypryt, Thompson, and Ullom]{levine2023tabletop}
Levine, Z.H.; Alpert, B.K.; Dagel, A.L.; Fowler, J.W.; Jimenez, E.S.; Nakamura,
  N.; Swetz, D.S.; Szypryt, P.; Thompson, K.R.; Ullom, J.N.
\newblock A tabletop X-ray tomography instrument for nanometer-scale imaging:
  reconstructions.
\newblock {\em Microsyst. Nanoeng.} {\bf 2023}, {\em 9},~47.

\bibitem[Kikuchi et~al.(2023)Kikuchi, Fujii, Hayakawa, Smith, Hirayama, Sato,
  Kohjiro, Ukibe, Ohno, Sato, and Yamamori]{kikuchi}
Kikuchi, T.; Fujii, G.; Hayakawa, R.; Smith, R.; Hirayama, F.; Sato, Y.;
  Kohjiro, S.; Ukibe, M.; Ohno, M.; Sato, A.;  et~al.
\newblock A 320-keV Spectrometer Based on 8-Pixel Transition Edge Sensor With
  Trilayer Membrane and Novel Numerical Analysis.
\newblock {\em IEEE Trans. Appl. Supercond.} {\bf 2023}, {\em
  33},~2101706.
\newblock {\url{https://doi.org/10.1109/TASC.2023.3269002}}.

\bibitem[Hoover et~al.(2014)Hoover, Winkler, Rabin, Bennett, Doriese, Fowler,
  Hayes-Wehle, Horansky, Reintsema, Schmidt, Vale, Ullom, and
  Schaffer]{Hoover2014Uncertainty}
Hoover, A.S.; Winkler, R.; Rabin, M.W.; Bennett, D.A.; Doriese, W.B.; Fowler,
  J.W.; Hayes-Wehle, J.; Horansky, R.D.; Reintsema, C.D.; Schmidt, D.R.;
  et~al.
\newblock Uncertainty of Plutonium Isotopic Measurements with Microcalorimeter
  and High-Purity Germanium Detectors.
\newblock {\em IEEE Trans. Nucl. Sci.} {\bf 2014}, {\em
  61},~2365--2372.
\newblock {\url{https://doi.org/10.1109/TNS.2014.2332275}}.

\bibitem[Bennett et~al.(2012)Bennett, Horansky, Schmidt, Hoover, Winkler,
  Alpert, Beall, Doriese, Fowler, Fitzgerald, et~al.]{bennett2012high}
Bennett, D.A.; Horansky, R.D.; Schmidt, D.R.; Hoover, A.; Winkler, R.; Alpert,
  B.K.; Beall, J.A.; Doriese, W.B.; Fowler, J.W.; Fitzgerald, C.;  et~al.
\newblock A high resolution gamma-ray spectrometer based on superconducting
  microcalorimeters.
\newblock {\em Rev. Sci. Instrum.} {\bf 2012}, {\em 83}, 093113.

\bibitem[Zink et~al.(2006)Zink, Ullom, Beall, Irwin, Doriese, Duncan, Ferreira,
  Hilton, Horansky, Reintsema, et~al.]{zink2006array}
Zink, B.L.; Ullom, J.; Beall, J.A.; Irwin, K.D.; Doriese, W.B.; Duncan, W.;
  Ferreira, L.; Hilton, G.C.; Horansky, R.; Reintsema, C.D.;  et~al.
\newblock Array-compatible transition-edge sensor microcalorimeter $\gamma$-ray
  detector with 42 eV energy resolution at 103 keV.
\newblock {\em Appl. Phys. Lett.} {\bf 2006}, {\em 89}, 124101.

\bibitem[Ullom et~al.(2007)Ullom, Doriese, Beall, Duncan, Ferreira, Hilton,
  Horansky, Irwin, Jach, Mates, et~al.]{ullom2007multiplexed}
Ullom, J.; Doriese, W.B.; Beall, J.A.; Duncan, W.; Ferreira, L.; Hilton, G.C.;
  Horansky, R.; Irwin, K.D.; Jach, T.; Mates, B.;  et~al.
\newblock Multiplexed microcalorimeter arrays for precision measurements from
  microwave to gamma-ray wavelengths.
\newblock {\em Nucl. Instrum. Methods Phys. Res. Sect. A Accel. Spectrom. Detect. Assoc. Equip.} {\bf 2007},
  {\em 579},~161--164.

\bibitem[Hoover et~al.(2011)Hoover, Hoteling, Rabin, Ullom, Bennett, Karpius,
  Vo, Doriese, Hilton, Horansky, et~al.]{hoover2011large}
Hoover, A.; Hoteling, N.; Rabin, M.; Ullom, J.; Bennett, D.; Karpius, P.; Vo,
  D.; Doriese, W.; Hilton, G.; Horansky, R.;  et~al.
\newblock Large microcalorimeter arrays for high-resolution X-and
  gamma-rayspectroscopy.
\newblock {\em Nucl. Instrum. Methods Phys. Res. Sect. A Accel. Spectrom. Detect. Assoc. Equip.} {\bf 2011},
  {\em 652},~302--305.

\bibitem[Ullom et~al.(2005)Ullom, Zink, Beall, Doriese, Duncan, Ferreira,
  Hilton, Irwin, Reintsema, Vale, Rabin, Hoover, Rudy, Smith, Tournear, and
  Vo]{1596455}
Ullom, J.; Zink, B.; Beall, J.; Doriese, W.; Duncan, W.; Ferreira, L.; Hilton,
  G.; Irwin, K.; Reintsema, C.; Vale, L.;  et~al.
\newblock Development of large arrays of microcalorimeters for precision
  gamma-ray spectroscopy.
\newblock In Proccedings of the IEEE Nuclear Science Symposium Conference Record,   Fajardo, PR, USA, 23--29 October 2005; Volume~2, pp. 1154--1158.
\newblock {\url{https://doi.org/10.1109/NSSMIC.2005.1596455}}.

\bibitem[Doriese et~al.(2008)Doriese, Ullom, Beall, Duncan, Ferreira, Hilton,
  Horansky, Irwin, Mates, Reintsema, et~al.]{doriese2008toward}
Doriese, W.B.; Ullom, J.; Beall, J.A.; Duncan, W.; Ferreira, L.; Hilton, G.C.;
  Horansky, R.; Irwin, K.D.; Mates, J.; Reintsema, C.D.;  et~al.
\newblock Toward a 256-pixel array of gamma-ray microcalorimeters for
  nuclear-materials analysis.
\newblock {\em J. Low Temp. Phys.} {\bf 2008}, {\em
  151},~754--759.

\bibitem[Bacrania et~al.(2009)Bacrania, Hoover, Karpius, Rabin, Rudy, Vo,
  Beall, Bennett, Doriese, Hilton, Horansky, Irwin, Jethava, Sassi, Ullom, and
  Vale]{5204694}
Bacrania, M.K.; Hoover, A.S.; Karpius, P.J.; Rabin, M.W.; Rudy, C.R.; Vo, D.T.;
  Beall, J.A.; Bennett, D.A.; Doriese, W.B.; Hilton, G.C.;  et~al.
\newblock Large-Area Microcalorimeter Detectors for Ultra-High-Resolution X-Ray
  and Gamma-Ray Spectroscopy.
\newblock {\em IEEE Trans. Nucl. Sci.} {\bf 2009}, {\em
  56},~2299--2302.
\newblock {\url{https://doi.org/10.1109/TNS.2009.2022754}}.

\bibitem[Hoover et~al.(2013)Hoover, Winkler, Rabin, Vo, Ullom, Bennett,
  Doriese, Fowler, Horansky, Schmidt, Vale, and
  Schaffer]{hoover2014determination}
Hoover, A.S.; Winkler, R.; Rabin, M.W.; Vo, D.T.; Ullom, J.N.; Bennett, D.A.;
  Doriese, W.B.; Fowler, J.W.; Horansky, R.D.; Schmidt, D.R.;  et~al.
\newblock Determination of Plutonium Isotopic Content by Microcalorimeter
  Gamma-Ray Spectroscopy.
\newblock {\em IEEE Trans. Nucl. Sci.} {\bf 2013}, {\em
  60},~681--688.
\newblock {\url{https://doi.org/10.1109/TNS.2013.2249091}}.

\bibitem[Winkler et~al.(2015)Winkler, Hoover, Rabin, Bennett, Doriese, Fowler,
  Hays-Wehle, Horansky, Reintsema, Schmidt, et~al.]{winkler2015256}
Winkler, R.; Hoover, A.; Rabin, M.; Bennett, D.; Doriese, W.; Fowler, J.;
  Hays-Wehle, J.; Horansky, R.; Reintsema, C.; Schmidt, D.;  et~al.
\newblock 256-pixel microcalorimeter array for high-resolution $\gamma$-ray
  spectroscopy of mixed-actinide materials.
\newblock {\em Nucl. Instrum. Methods Phys. Res. Sect. A Accel. Spectrom. Detect. Assoc. Equip.} {\bf 2015},
  {\em 770},~203--210.

\bibitem[Becker et~al.(2019)Becker, Alpert, Bennett, Croce, Fowler, Gard,
  Hoover, Joe, Koehler, Mates, et~al.]{becker2019advances}
Becker, D.T.; Alpert, B.K.; Bennett, D.A.; Croce, M.P.; Fowler, J.W.; Gard,
  J.D.; Hoover, A.S.; Joe, Y.I.; Koehler, K.E.; Mates, J.A.;  et~al.
\newblock Advances in analysis of microcalorimeter gamma-ray spectra.
\newblock {\em IEEE Trans. Nucl. Sci.} {\bf 2019}, {\em
  66},~2355--2363.

\bibitem[Fomin(2012)]{fomin2012optimal}
Fomin, V.N.
\newblock {\em Optimal Filtering: Volume~I: Filtering of Stochastic Processes};
  Springer Science \& Business Media:  Berlin/Heidelberg, Germany, %newly added information, please confirm
  2012; Volune 457.

\bibitem[Croce et~al.(2021)Croce, Henzlova, Menlove, Becker, and
  Ullom]{croce2021electrochemical}
Croce, M.; Henzlova, D.; Menlove, H.; Becker, D.; Ullom, J.
\newblock Electrochemical Safeguards Measurement Technology Development at
  LANL.
\newblock {\em J. Nucl. Mater. Manag.} {\bf 2021}, {\em
  49},~116--135.

\bibitem[Croce et~al.(2022)Croce, Becker, Bennett, Cantor, Carpenter, Feissle,
  Friedrich, Gard, Imrek, Kim, et~al.]{croce2022nuclear}
Croce, M.; Becker, D.; Bennett, D.; Cantor, R.; Carpenter, M.; Feissle, E.;
  Friedrich, S.; Gard, J.; Imrek, J.; Kim, G.;  et~al.
\newblock \emph{Nuclear Facility Experience with the SOFIA Ultra-High-Resolution
  Microcalorimeter Gamma Spectrometer}; 
\newblock Technical Report; Lawrence Livermore National Lab. (LLNL): Livermore,
  CA, USA,  2022.

\bibitem[Mercer et~al.(2022)Mercer, Winkler, Koehler, Becker, Bennett,
  Carpenter, Croce, de~Castro, Feissle, Fowler,
  et~al.]{mercer2022quantification}
Mercer, D.J.; Winkler, R.; Koehler, K.E.; Becker, D.T.; Bennett, D.A.;
  Carpenter, M.H.; Croce, M.P.; de~Castro, K.I.; Feissle, E.A.; Fowler, J.W.;
  et~al.
\newblock Quantification of 242Pu with a Microcalorimeter Gamma Spectrometer.
\newblock {\em arXiv} {\bf 2022}, arXiv:2202.02933.

\bibitem[Day et~al.(2003)Day, LeDuc, Mazin, Vayonakis, and
  Zmuidzinas]{day2003broadband}
Day, P.K.; LeDuc, H.G.; Mazin, B.A.; Vayonakis, A.; Zmuidzinas, J.
\newblock A broadband superconducting detector suitable for use in large
  arrays.
\newblock {\em Nature} {\bf 2003}, {\em 425},~817--821.

\bibitem[Ulbricht et~al.(2021)Ulbricht, De~Lucia, and
  Baldwin]{ulbricht2021applications}
Ulbricht, G.; De~Lucia, M.; Baldwin, E.
\newblock Applications for microwave kinetic induction detectors in advanced
  instrumentation.
\newblock {\em Appl. Sci.} {\bf 2021}, {\em 11},~2671.

\bibitem[Walter et~al.(2018)Walter, Mazin, Bockstiegel, Fruitwala, Szypryt,
  Lipartito, Meeker, Zobrist, Collura, Coiffard, Strader, Guyon, Lozi, and
  Jovanovic]{MEC}
Walter, A.; Mazin, B.B.; Bockstiegel, C.; Fruitwala, N.; Szypryt, P.;
  Lipartito, I.; Meeker, S.; Zobrist, N.; Collura, G.; Coiffard, G.;  et~al.
\newblock {MEC: The MKID exoplanet camera for high contrast astronomy at Subaru
  (Conference Presentation)}.
\newblock In Proccedings of the Ground-Based and Airborne Instrumentation for
  Astronomy VII, Austin, TX, USA, 10--15 June 2018; Evans, C.J.; Simard, L.; Takami, H., Eds.; International
  Society for Optics and Photonics SPIE: Bellingham, WA, USA,  2018; Volume~10702.
\newblock {\url{https://doi.org/10.1117/12.2311586}}.

\bibitem[Vissers et~al.(2013)Vissers, Gao, Kline, Sandberg, Weides, Wisbey, and
  Pappas]{vissers2013characterization}
Vissers, M.R.; Gao, J.; Kline, J.S.; Sandberg, M.; Weides, M.P.; Wisbey, D.S.;
  Pappas, D.P.
\newblock Characterization and in-situ monitoring of sub-stoichiometric
  adjustable superconducting critical temperature titanium nitride growth.
\newblock {\em Thin Solid Film.} {\bf 2013}, {\em 548},~485--488.

\bibitem[De~Lucia et~al.(2022)De~Lucia, Baldwin, Ulbricht, Piercy, Creaner,
  Bracken, and Ray]{de2022high}
De~Lucia, M.; Baldwin, E.; Ulbricht, G.; Piercy, J.; Creaner, O.; Bracken, C.;
  Ray, T.
\newblock High-uniformity TiN/Ti/TiN multilayers for the development of
  Microwave Kinetic Inductance Detectors.
\newblock In Proccedings of the X-Ray, Optical, and Infrared Detectors for
  Astronomy X, Montréal, QC, Canada, 17--23 July 2022; SPIE: Bellingham, WA, USA,  2022; Volume~12191, pp. 21--31.

\bibitem[McAleer et~al.(2024)McAleer, Creaner, Bracken, Ulbricht, De~Lucia,
  Piercy, and Ray]{mcaleer2024automation}
McAleer, C.; Creaner, O.; Bracken, C.; Ulbricht, G.; De~Lucia, M.; Piercy, J.;
  Ray, T.
\newblock Automation of MKID Simulations for Array Building with AEM (Automated
  Electromagnetic MKID Simulations).
\newblock {\em J. Low Temp. Phys.} {\bf 2024},  \emph{216}, 57--66.

\bibitem[McKenney et~al.(2019)McKenney, Austermann, Beall, Dober, Duff, Gao,
  Hilton, Hubmayr, Li, Ullom, et~al.]{mckenney2019tile}
McKenney, C.M.; Austermann, J.E.; Beall, J.A.; Dober, B.J.; Duff, S.M.; Gao,
  J.; Hilton, G.C.; Hubmayr, J.; Li, D.; Ullom, J.N.;  et~al.
\newblock Tile-and-trim micro-resonator array fabrication optimized for high
  multiplexing factors.
\newblock {\em Rev. Sci. Instrum.} {\bf 2019}, {\em 90}, 023908.

\bibitem[De~Lucia et~al.(2020)De~Lucia, Baldwin, Ulbricht, Bracken, Stamenov,
  and Ray]{de2020multiplexable}
De~Lucia, M.; Baldwin, E.; Ulbricht, G.; Bracken, C.; Stamenov, P.; Ray, T.
\newblock Multiplexable frequency retuning of MKID arrays using their non-linear kinetic inductance.
\newblock In \emph{X-Ray, Optical, and Infrared Detectors for
  Astronomy IX;} SPIE: Bellingham, WA, USA,  2020; Volume~11454, pp. 580--590.

\end{thebibliography}
%\tableofcontents

\PublishersNote{}
\end{adjustwidth}

\end{document}